\documentclass[%
reprint,
 amsmath,amssymb,
 aps,
prx,
]{revtex4-1}
\usepackage[colorlinks=true,    
    urlcolor=blue,linkcolor=blue, citecolor=blue]{hyperref}
\usepackage{graphicx}
\usepackage{bm}
\begin{document} 
%\preprint{APS/123-QED}

\title{Scaling theory for Mott-Hubbard transitions} 

\author
{Anirban Mukherjee}
\email{am14rs016@iiserkol.ac.in}
\author{Siddhartha Lal}
\email{slal@iiserkol.ac.in}
\affiliation{Department of Physical Sciences, Indian Institute of Science Education and Research-Kolkata, W.B. 741246, India}
\begin{abstract}
We present a $T=0K$ renormalization group (RG) phase diagram for the electronic Hubbard model in two dimensions on the square lattice at, and away from, half filling. The RG procedure treats quantum fluctuations in the single particle occupation number nonperturbatively via the unitarily decoupling of one electronic state at every RG step. The resulting phase diagram thus possess the quantum fluctuation energy scale ($\omega$) as one of its axes. A relation is derived between $\omega$ and the effective temperature scale upto which gapless, as well as emergent gapped, phases can be obtained. We find that the transition in the half-filled Hubbard model involves, for any on-site repulsion, passage from a marginal Fermi liquid to a topologically-ordered gapped Mott liquid through a pseudogapped phase bookended by Fermi surface topology-changing Lifshitz transitions. Using effective Hamiltonians and wavefunctions for the low-energy many-body eigenstates for the doped Mott liquid obtained from the stable fixed point of the RG flow, we demonstrate the collapse of the pseudogap for charge excitations (Mottness) at a quantum critical point possessing a nodal non-Fermi liquid with superconducting fluctuations, and spin-pseudogapping near the antinodes. d-wave Superconducting order is shown to arise from this quantum critical state of matter. Benchmarking of the ground state energy per particle and the double-occupancy fraction against existing numerical results also yields excellent agreement. We present detailed insight into the $T=0$ origin of several experimentally observed findings in the cuprates, including Homes law and Planckian dissipation. We also establish that the heirarchy of temperature scales for the pseudogap ($T_{PG}$), onset temperature for pairing ($T_{ons}$), formation of the Mott liquid ($T_{ML}$) and superconductivity ($T_{C}$) obtained from our analysis is quantitatively consistent with that observed experimentally for some members of the cuprates. Our results offer insight on the ubiquitous origin of superconductivity in doped Mott insulating states, and pave the way towards a systematic search for higher superconducting transition temperatures in such systems.
\end{abstract}
\maketitle 
\tableofcontents
\section{Introduction}
The nature of, and the transition into, the Mott insulating state defines a central problem in strongly correlated quantum matter. An analytically exact solution for the electronic Mott-Hubbard metal-insulator transition (MIT) exists only in one spatial dimension~\cite{lieb1968absence}, while the status of the problem remains open in general. While the Mott insulator is often associated with a ($T=0$) first order transition leading to a Ne\'{e}l antiferromagnetic (AFM) ground state~\cite{imada1998metal}, the search continues for quantum liquid-like ground states corresponding to an insulating state that breaks no lattice or spin-space symmetries and is reached via a continuous transition. Indeed, there exist some theoretical proposals~\cite{anderson1987RVB, grosEdgerMuthukumar2007RVB, paramekanti2004RVB, PWA2004} as well as some experimental evidence for insulating spin-liquid ground states in layered organic conductors~\cite{kurosaki2005mott} and Herbertsmithite  ~\cite{helton2010dynamic}. Recently, the metal-organic compound $Cu(DCOO)_{2}.4D_{2}O$, an unfrustrated quasi two-dimensional antiferromagnet, was found to contain features of a resonating valence bond (RVB) like spin-liquid ground state~\cite{piazza2015TopologicalOrder}.
\par\noindent
Theoretical studies have not, however, identified unambiguously the order parameter for such correlation-driven metal-insulator transitions. The difficulties appear to be associated with an interplay of several complications: the fermion-sign problem limits some non-perturbative numerical investigations at low-temperatures \cite{iazzi2016topological}, while many other numerical methods are either limited to small sizes or certain ranges in the coupling $U/t$ (the ratio of the Hubbard repulsion to the nearest-neighbour hopping amplitude). It is, therefore, remarkable that a benchmarking exercise conducted on the 2D Hubbard model identified ranges in the values for the ground state energy per particle and the double occupancy fraction at specific values of the filling and $U/t$~\cite{leblanc2015solutions}. At the same time, a lack of an identifiable small parameter makes most analytic approaches beyond various mean-field schemes intractable when studying the problem at strong coupling. 
\par\noindent
Amidst these difficulties, several important questions related to the nature of the $T=0$ phase diagram of the Mott-Hubbard transition, as well as the nature of the ground state, continue to be debated. For instance, we may ask: is there a critical value of the ratio $U/t$ for the $T=0$ Mott transition in the half- filled unfrustrated (i.e., with nearest neighbour hopping only) Hubbard model on a square lattice that corresponds to a paramagnetic state (i.e., with no magnetic order)? Studies using dynamical mean field theory (DMFT)~\cite{kotliar1992, kotliar1996,werner2007} and quantum Monte Carlo~\cite{joo2001} approaches indicate a first order transition at $T>0$ ending at a critical $(U/t)_{c}$ at finite $T$. The status of the $T=0$ metal-insulator transition remains to be understood. Further, the paramagnetic calculations can be interpreted as solutions for the case of vanishing inter-site correlations, and can presumably be trusted within only the (single-site) DMFT framework for the case of infinite dimensions. 
\par\noindent 
The question, therefore, of whether the ground state of the two-dimensional Mott insulator at $T=0$ possesses magnetic ordering or not needs further consideration. A reduction in the value of $(U/t)_{c}$ has also been observed in dynamical cluster approximation (DCA) studies~\cite{jarell2001, gull2013, merino2014} as well as in cluster DMFT (CDMFT) studies~\cite{park2008}. Recent studies involving the dynamical vertex approximation (D$\Gamma$A)~\cite{toschi2007, held2008, schafer2015}, auxillary field quantum Monte Carlo (AFQMC)~\cite{bss1981, schafer2015}, density matrix embedding theory (DMET)~\cite{zheng2016} and ladder dual-fermion approach (LDFA)~\cite{loon2018} have instead supported the existence of a gapped antiferromagnetic Neel state for all $U>0$. Variational Monte Carlo studies using Gutzwiller projected wavefunctions have shown that a symmetry-preserved resonating valence bond (RVB) state is energetically close to the symmetry-broken Neel antiferromagnetic state ~\cite{grosEdgerMuthukumar2007RVB}. Upon including backflow correlations in the Gutzwiller projected wavefunctions, a nonmagnetic ground state~\cite{tocchio2008} was found to exist over a large range of $U/t$. This variety of results obtained from various numerical methods demands an analytical approach that yields unambiguous insight into the nature of Mott insulating ground state of the 2D Hubbard model at $T=0$, as well as the effective low-energy Hamiltonian that governs the low-lying excitations above this ground state. At the same time, a better view of the quantum metal-insulator transition involves understanding the nature of parent metallic state of the Mott insulator: is it a Fermi liquid with coherent Landau quasiparticle excitations, or some form of non-Fermi liquid involving collective excitations? 
\par
Similar issues exist for the case of doped Mott insulators. 
The correlation-induced Mott insulator has localized charge degrees of freedom at filling commensurate with the lattice (typically half-filling for the 2D square lattice). Upon doping with holes, the physics behind charge localization competes strongly with hopping-induced electronic delocalization. This competition has been studied extensively within the realms of the 2D Hubbard model with a finite chemical potential. A large number of numerical methods have been brought to bear on this problem. These include, for instance, quantum Monte Carlo simulations at finite-temperature (QMC)~\cite{cosentini1998, becca2000, bemmel1994, zhang1997, chang2008, chang2010, yokoyama1987, eichenberger2007, yamaji1998, giamarchi1991}, density matrix renormalization group (DMRG)~\cite{white2000, scalapino2001, white2003}, dynamical cluster approximation~
\cite{hettler1998,hettler2000,
khatami2010quantum,vidhyadhiraja2009quantum,
mikelsons2009thermodynamics}, cluster DMFT (CDMFT)\cite{lichtenstein2000,  kotliar2001,civelli2008nodal,ferrero2009pseudogap,sakai2010doped,
imada2010unconventional,
gull2010momentum} and the variational cluster approximation VCA \cite{potthoff2003, dahnken2003}. 
Together with several others, these techniques suggest a rich and detailed temperature-doping phase diagram that includes numerous phases including the antiferromagnetic Mott insulator, d-wave superconductivity, the pseudogap (or nodal-antinodal dichotomy) arising from electronic differentiation, non-Fermi liquid, stripes etc.~\cite{chang2008,chang2010,hettler1998,hettler2000,
khatami2010quantum,vidhyadhiraja2009quantum,
mikelsons2009thermodynamics,lichtenstein2000,  kotliar2001,civelli2008nodal,ferrero2009pseudogap,sakai2010doped,
imada2010unconventional,
gull2010momentum,schmitt2010, huang2018, kaczmarczyk2016, senechal2005, aichhorn2006, halboth2000, schulz1990, white1989, chubukov1995, capone2006, imada2013,lin2010,wang2009,gull2012,maier2005}. These findings are, by and large, in keeping with the experimental phenomenology of the doped cuprate Mott insulators (see~\cite{keimer2015quantum} for a recent review). A significant drawback remains, however, in the fact that detailed resolution of the low-energy neighbourhood of the Fermi surface is not available from these theoretical methods. Unfortunately, this leaves several critical questions unanswered on the nature and origins of the $T=0$ phenomenology of the doped Mott-Hubbard insulator. 
\par\noindent
Noteworthy among efforts towards resolving this problem involves the application of the functional renormalization group (FRG) technique to the Mott transitions in the 2D Hubbard model~(for reviews, see Refs.\cite{metzner2012,tagliavini2019} and references therein). Results from FRG studies provide evidence for nodal-antinodal dichotomy, as well as spectral weight transfer between Hubbard bands in the half-filled Hubbard model~\cite{fu2006}. For the case of doping-induced Mott transitions, FRG studies show the co-existence and interplay of antiferromagnetism with $d-$wave superconductivity~ \cite{wang2009, vilardi2019} over a wide doping range, in agreement with results obtained from CDMFT~\cite{lichenstein2000}. In this range of doping, signatures of stripes~\cite{yamase2016} and nematicity~\cite{husemann2012} have also been reported, as well as their interplay with $d-$ wave superconductivity. These findings are in consensus with results from diagrammatic mean field theory~\cite{zeyher2018} and CDMFT ~\cite{tuomas2018}. Signatures of the strange metal, the pseudogap, and a quantum critical region have also been reported within the FRG scheme~\cite{yamase2016, husemann2012, tuomas2018, zeyher2018, lange2017, liu2017, giering2012}. While the method is non-perturbative in principle, numerical implementations of the FRG have typically needed truncations at finite orders in the loop expansion. Thus, despite much success, the FRG is limited thus far to studying weak-to-intermediate values of $U/t$. 
\par\noindent
In this work, we present a novel Hamiltonian RG formalism in momentum space based on unitary transformations, and then employ it to develop a scaling theory for the 2D Hubbard model on a square lattice. We note that this model has earlier been studied using the continuous unitary transformation (CUT) RG formalism~ \cite{glazek1993,glazek1994,wegner1994,grote2002}, rendered perturbative via a truncation at 1 loop order. In contrast, we are able to conduct a nonperturbative study of the same model via our RG formalism. The groundwork for the RG is laid out in Sec. \ref{RGSetupSection}. We initially derive the exact analytical form for the unitary operator that completely decouples one electronic state from the every other electronic degree of freedom. This is carried out by the removal of the appropriate off-diagonal blocks of the Hamiltonian represented in the occupation number basis of the state to be decoupled. The renormalized Hamiltonian is then shown to become block diagonal. For the case of the Hubbard model, the unitary rotations are applied iteratively on electronic states farthest away from the Fermi surface of the tight-binding part of problem, and gradually leading towards its Fermi surface. This leads to an RG evolution in terms of an effective Hamiltonian, from which we have derived the RG flow equations for the 1-particle self energies, 2-particle vertices and 3-particle vertices. In comparison to the loop truncation approximations of the FRG
scheme~\cite{tagliavini2019}, we find non-perturbative RG equations of all 2-, 4-, and 6-point vertices, i.e., that have contributions from all loops resummed into closed-form expressions. Importantly, we find that the vertex RG flow happens across a family of quantum fluctuation scales ($\omega$'s) that arise out of the noncommutavity between the kinetic energy term and the Hubbard onsite repulsion term. Indeed, this non-commutativity leads to number-density fluctuations of electronic states in momentum-space. As a direct outcome of the non-perturbative nature of the RG equations, we obtain stable fixed points of the flows at any given fluctuation scale $\omega$. In this way, we are able to perform the RG analysis of the Hubbard model all the way from weak to strong coupling (in terms of the ratio of the Hubbard repulsion strength to the hopping amplitude, $U/t$). The effective Hamiltonian, and associated low-energy eigenstates, obtained at a stable fixed point then provides further avenues for analyses. This method is inspired by the strong-disorder RG approaches adopted by Dasgupta et al.~\cite{ma1979sk}, Fisher~\cite{fisher1992random}, Rademaker-Ortuno~\cite{rademaker2016explicit} and You et al.~\cite{you2016entanglement}.
We have also recently used this RG technique to obtain a zero temperature phase diagram for the Kagome spin-1/2 XXZ antiferromagnet at non-zero magnetic field~\cite{pal2019}.
\par\noindent
In Section \ref{NFLSection}, we present the marginal Fermi liquid as the parent metal of the Mott insulating phase of the 2D Hubbard model at half-filling. We follow this in Section \ref{PGSection} by detailing the journey through the pseudogap phase at half-filling, and the nature of the Mott-Hubbard MIT. In Section \ref{mottliquid}, we present some features of topological order for the insulating Mott liquid phase obtained from the RG, as well as benchmark some of its properties with the numerical results obtained from Refs.\cite{leblanc2015solutions,ehlers2017hybridDMRG,dagotto1992}. We also demonstrate how a RG relevant symmetry breaking perturbation leads to the Neel antiferromagnetic Mott state. We then extend the formalism to treating the case of the hole-doped 2D Hubbard model in Section \ref{MottCollapseSection}, unveiling the presence of a quantum critical point (QCP) reached at a critical doping. We also present results in this section for the numerical benchmarking of the doped Mott liquid against results obtained from Refs.\cite{leblanc2015solutions,ehlers2017hybridDMRG,dagotto1992}, as well as provided analytical results that explain a large body of experimental phenomenology observed in the hole doped cuprates. The latter includes, for instance, a detailed analysis of the origin of Homes law~\cite{homes2004} and Planckian dissipation~\cite{legros2019}. We then present results in Section \ref{SymmetryBreakingSection} for the presence of symmetry-broken states of matter (including superconductivity) in the RG analysis, and compare once more the results obtained with the well-known phenomenology of the cuprates. Finally, we conclude our presentation in Section \ref{ConclusionsSection} by a detailed discussion of the relevance of our work to the cuprates, and by presenting future perspectives. Further details of the derivations of various RG relations are presented in several appendices. We begin, however, in Sec.\ref{prelims} by first offering a guidemap to the work by summarising the RG method and the main results. 
\section{Preliminaries and Summary of main results}\label{prelims}
\subsection{Preliminaries}
\begin{figure}
\includegraphics[width=10.2cm,height=13cm]{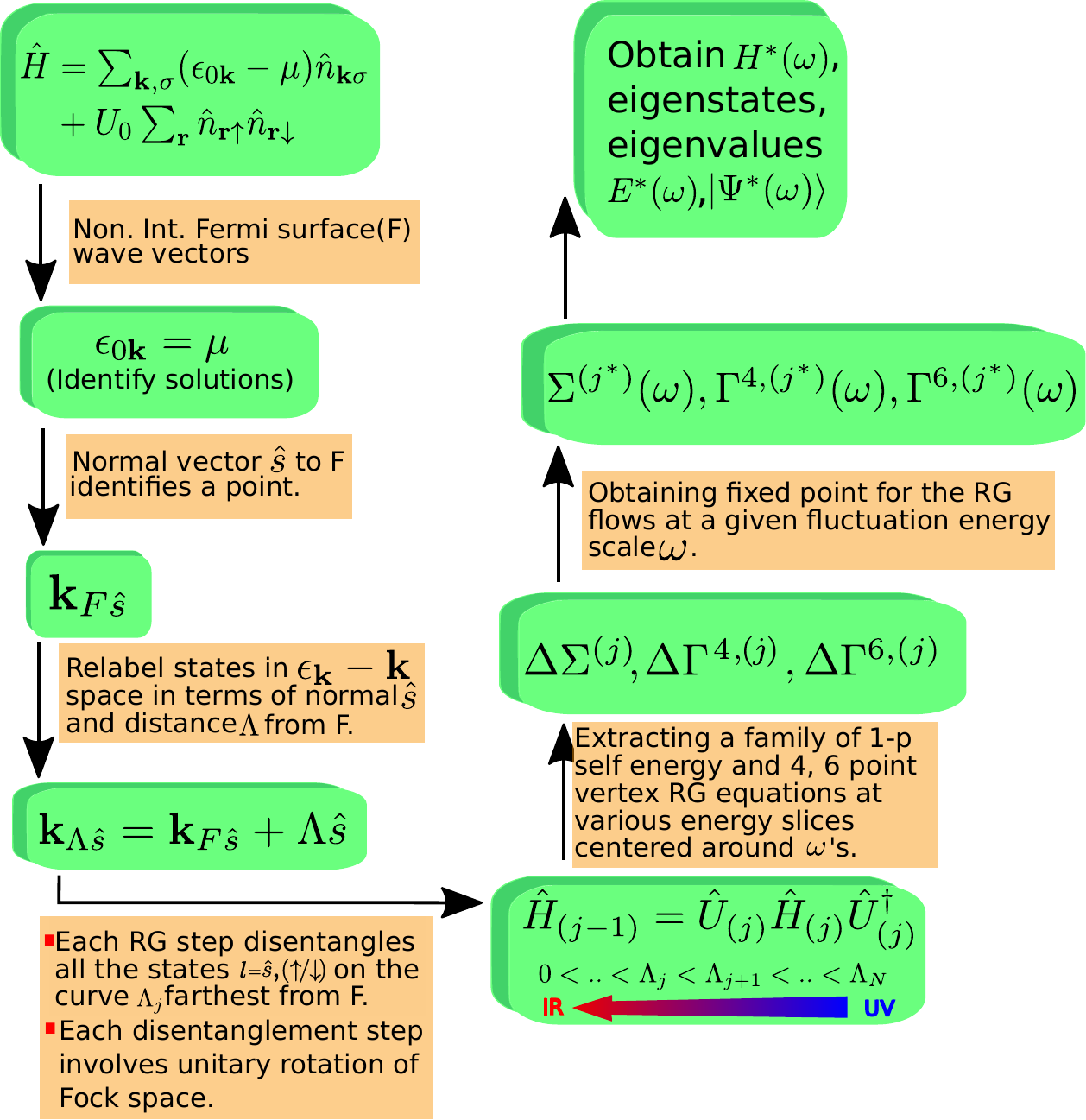}
\caption{Flowchart for the renormalization group methodology.}\label{RG-method}
\end{figure}

In this section, we provide a holistic overview of the method developed in this work and summarise the major results obtained therefrom for the Hubbard model. Readers interested in the technical details may skip this section. To begin with, we briefly outline the RG methodology shown in Fig.~\ref{RG-method} (described in detail in Sec.~\ref{RGSetupSection} A~-~F).
Given that the problem at hand (the 2D Hubbard model on a square lattice) possesses discrete translational invariance, the initial steps involve setting up the labelling scheme for the states in energy-momentum space with reference to the non-interacting Fermi surface (F) wave-vectors $\mathbf{k}_{F}$ (i.e., solution to the equation $\epsilon_{0\mathbf{k}}=\mu$). We mark the states in terms of unit vectors normal to F ($\hat{s}=\frac{\nabla\epsilon_{\mathbf{k}}}{|\nabla\epsilon_{\mathbf{k}}|}|_{\epsilon_{\mathbf{k}}=E_{F}}$) and lying at a distance $\Lambda$ from F: $\mathbf{k}_{\Lambda\hat{s}}=\mathbf{k}_{F\hat{s}}+\Lambda\hat{s}$. States are thus ordered in terms of distances $\Lambda_{N}>..>\Lambda_{j}>\Lambda_{j-1}>..>0$, with the largest lying near the Brillouin zone edge and the smallest proximate to the Fermi surface (i.e., also a monotonic variation in the electronic dispersion $\epsilon_{\mathbf{k}}$). The iterative RG procedure is then carried out such that, at step $j$, all the states (labelled by $\{ l=\hat{s},\sigma=\uparrow,\downarrow \}$ on the curve $\Lambda_{j}$ are completely disentangled via unitary rotations. The resulting  Hamiltonian 
$H_{(j-1)}=U_{(j)}H_{(j)}U_{(j)}^{\dagger}$ possesses off-diagonal terms (and therefore entanglement in the members of the eigenbasis) only for states at distances $\Lambda<\Lambda_{j}$. 
\par\noindent
The disentanglement of an entire curve at distance $\Lambda_{j}$ is represented via an unitary rotation 
$U_{(j)}=\prod_{l\in(\hat{s},\sigma=\uparrow/\downarrow)}U_{(j,l)}$, where $U_{(j,l)}$ disentangles one state $\mathbf{k}_{\Lambda_{j}\hat{s}}\sigma$ on the curve $\Lambda_{j}$.
The unitary RG evolution of the Hamiltonian 
then leads to a family of RG flow equations projected along multiple quantum fluctuation energyscales ($\omega$'s), enabling a study of the RG evolution of different parts of the many-body spectrum for the Hamiltonian $\hat{H}$. Importantly, quantum fluctuations of the occupation-number diagonal many-body configurations arise directly from the non-commutativity between various parts of the Hamiltonian. The formalism obtains a hierarchy of non-pertubative RG flow equations for various $n$-particle (or $2n$-point) interaction vertices. In this work, we have restricted ourselves in accounting for the 
contribution from the 1-particle self-energy ($\Sigma$), 4-point ($\Gamma^{4}$) and 6-point ($\Gamma^{6}$) vertices. From the stable fixed point of these vertex RG flow equations at a given energy scale $\omega$, we obtain 
an effective Hamiltonian $H^{*}(\omega)$. Whenever possible, we analytically diagonalize the spectrum of the effective Hamiltonian to obtain many-body eigenstates and eigenvalues.
We have also shown in Sec.\ref{meaning_of_omega} that the quantum fluctuation scale $\omega$ is equivalent to an effective thermal scale upto which a given effective Hamiltonian is valid.
\subsection{Summary of main results}
We now summarise the key results obtained, as well as provide a guidemap for the work. This begins with Sec.~\ref{mixing}, where we perform an unbiased RG study of all the 4-point scattering vertex diagrams arising out of the Hubbard repulsion $U$~: backward ($K_{c,l}^{(j)},K_{s,l}^{(j)}$) and forward ($V_{c,l}^{(j)},V_{s,l}^{(j)}$) scattering (eqns.\eqref{Long RG equations with doping}) and tangential ($L^{(j)}$) scattering (eq.\eqref{tang_scatt_RG_flows}). Importantly, we also obtain the RG flows of the emergent 6-point scattering vertices (eq.\eqref{six_point_RG_flows}) that arise out of non-commutativity between various 4-point vertices. The closed-form analytic expression for the RG equations have the following noteworthy features:
\begin{itemize}
\item noncommutivity between different four-fermion scattering terms causes quantum superposition between opposite spin-aligned $2$-electron and electron-hole pairs (eqns.\eqref{ee_eh_mixed_configuration} and \eqref{hh_he_mixed_configuration}); this superposition is manifested through the parameter $p$ in the Greens function eq.\eqref{determining p} that appears in the set of RG equations eq.\eqref{Long RG equations with doping},
\item sensitivity to the geometry of the Fermi surface (here square), reflected via an explicit dependence of the RG equations on $\hat{s}$ and the electronic differentiation in the dispersion ranging from the corners of the FS at $(\pm\pi,0)~\&~(0,\pm\pi)$ to the mid-points of the straight stretches at $(\pm\pi/2,\pm\pi/2)$ in eq.\eqref{dispersion_differentiation},
\item an explicit dependence on the energy scale for quantum fluctuations ($\omega_{(j)}$). The non-perturbative nature of the RG equations is manifest in the structure of their denominators (eqs.\eqref{Long RG equations with doping}, \eqref{tang_scatt_RG_flows}, \eqref{six_point_RG_flows}), with IR fixed points obtained for a projected energy subspace centered around a energy scale $\omega^{*}$~(see Sec.\ref{decoupling single-particle states}),
\item explicit dependence on the effective chemical potential $\Delta\mu_{eff}$ in the $2$-electron/electron-hole superposed Green function (eq.\eqref{Green_func_long}), and 
\item an unbiased treatment of various pair-scattering processes dependent on the non-zero offset ($\delta$) from the pairing momentum $(\pi,\pi)$. The scattering processes for pairing momentum $(\pi,\pi)$ lead to a log-singularity near the non-interacting problem's Fermi surface, seen via a 2nd order T-matrix perturbation theory calculation (eq.\eqref{T-matrix_argument}). The scattering processes with a finite offset  ($\delta$) in the pair momentum from $(\pi,\pi)$ contain subleading log-divergences observed at higher energies. 
The unbiased treatment reveals that different kinds of pairs dominate in the fixed point theory as a function of $\omega$ and $\Delta\mu_{eff}$.
\end{itemize} 
\par\noindent
\textbf{Normal State corresponding to the $1/2$-filled Mott insulator}~ In Sec.\ref{NFLSection}, we study the properties of the parent metal associated with the Mott transitions at half- filling ($\mu^{0}_{eff}=\mu-\frac{U_{0}}{2}=0$) as a function of bare $U_{0}$ and quantum fluctuation energyscale $\frac{W}{2}-\omega$ (where $W$ is the bandwidth of the electronic kinetic energy). At large $\frac{W}{2}-\omega$ (in the range eq.\eqref{NFL_cond}), we find a parent metallic state with a connected Fermi surface and
\begin{itemize}
\item characterized by a topological invariant $N^{\Downarrow}_{\hat{s}}(\omega)=1$ everywhere on the Fermi surface (eq.\eqref{Top_term}),
\item a 2-electron 1-hole effective Hamiltonian is obtained (eq.\eqref{fixed point Hamiltonian 2nd step}) within a window centered around the Fermi surface at the RG fixed point,
\item with the scaling forms for the 1-particle self-energy and quasiparticle residue (eq.\eqref{1-p self energy}) are found to resemble that  proposed on phenomenonlogical grounds for the marginal Fermi liquid (MFL)~\cite{varma-PhysRevLett.63.1996}. The quasiparticle inverse lifetime, and hence the electronic resistivity, is found to be linearly proportional to an equivalent temperature scale $T$ (eq.\eqref{Thermal scale}). This arises from a linear dependence on the quantum fluctuation scale $\frac{W}{2}-\omega$, and is numerically verified in Fig. \ref{passage_through_half_fill_ph_diag},
\item as shown in Fig. \ref{qp-residue}, with the fluctuation energy scale $\omega\to 0$, the Landau quasiparticle residue is found decay (eq.\eqref{1-p self energy}), while the 2-electron 1-hole residue (eq.\eqref{2-p 1-h self energy}) is found to rise to unity. This displays the replacement of Landau quasiparticles by newer composite degrees of freedom with net charge $e$ and net spin $1/2$.
\end{itemize}
\par\noindent
Our calculations enable the computation of the Luttinger sum in terms of 1-electron and 2-electron 1-hole green function (eq.\eqref{modified-Luttinger-sum}), providing a full resolution of the spectral function for the normal state in both momentum and frequency spaces. The spectral function reveals 
\begin{itemize}
\item a MFL (with effective Hamiltonian eq.\eqref{fixed point Hamiltonian 2nd step}) lying within a window near to the Fermi surface whose width ($\Lambda^{**}$) is obtained from the RG, 
\item a correlated Fermi liquid described by the effective Hamiltonian eq.\eqref{decoupled_states_Hamiltonian_2nd_level} corresponding to admixture of the marginal Fermi liquid 2-electron 1-hole composites + Landau quasiparticle excitations, found lying further away from the Fermi surface, i.e., in the momentum space window eq.\eqref{momentum space window_FL_MFL_admix}, and
\item Landau Fermi liquid (with effective Hamiltonian eq.\eqref{FL_Ham}) lying towards the Brillouin zone edge in the momentum space window eq.\eqref{fp_condition_MFL}.
\end{itemize}
\textbf{Main result:} Our analysis establishes the MFL as the parent normal state of the $1/2$-filled Mott insulating state of the 2D Hubbard model, as well as provides comprehensive insight into the microscopic origins of the MFL phenomenology.\\
\par\noindent
\textbf{Pseudogap state at $1/2$-filling}~ In Sec.\ref{PGSection}, we study the RG fixed point theories describing the pseudogap (PG) phase beyond the energy scale $\omega>0$. This phase is characterized by an effective Hamiltonian (eq.\eqref{Hamiltonian_pseudogap}) consisting of two parts: 
\begin{itemize}
\item gapped antinodal (AN) regions of the FS that describe
the condensation of bound states formed from pseudospins constituted by the backscattering vertices $K^{*}_{c/s,\hat{s}}$ of opposite spin-paired electron-electron/ electron-hole, and
\item gapless nodal (N) stretches of the FS with the MFL fixed point Hamiltonian (eq.\eqref{fixed point Hamiltonian 2nd step}).
\end{itemize}
We quantify the PG phase via a distinct topological invariant $N^{PG}_{\hat{s}}=1$ (eq.\eqref{Glob-PG}) which becomes trivial in both the MFL and Mott insulating phases. The PG phase is observe to involve the continuous gapping of the Fermi surface, starting from the AN with lowering quantum fluctuation scale $\frac{W}{2}-\omega$, and ending with the gapping of the N. The gapping process is linked to pole-to-zero conversion of the 2-electron 1-hole Greens function in 
eq.\eqref{2-electron 1-hole green function}, and is associated with an upturn of the resistivity computed numerically (see Fig. \ref{half_filled_phase_diag}).\\
\textbf{Main result:} 
By tracking the dynamical transfer of spectral weight under RG, we demonstrate that the Mott transition is continuous in nature and involves passage through a PG phase, bookended by two interacting Lifshitz transitions of the marginal Fermi liquid at the AN and N points of the FS respectively. Additional evidence is presented in Video S1.\\ 
\par\noindent
\textbf{Mott liquid insulating state $1/2$-filling}~
The Mott insulating state in Sec.~\ref{mottliquid} appears below the quantum fluctuation energy scale $\frac{W}{2}-\omega_{ins}$. This is arrived via a gapping of the nodal points in the PG phase, causing the Fermi surface to disappear entirely. The resulting symmetry-preserved quantum liquid state 
\begin{itemize}
\item is characterized by a global topological invariant $N^{\Uparrow}_{\hat{s}}(\omega)=1$ (eq.\eqref{Top_term}),
\item is described by the fixed point Hamiltonian eq.\eqref{fixed_point_Ham_mott} for a condensate of bound states formed from pseudospin degrees of freedom describing backscattering between opposite spin-paired electron-electron/ electron-hole pairs,
\item has a ground state manifold (eq.\eqref{ground_state}) found to be twofold degenerate in the thermodynamic limit, protected by a many body gap ($\Delta E$) 
(eq.\eqref{exc-1}, eq.\eqref{exc-2}), and possesses topological properties: non-trivial anticommutation relation between nonlocal gauge transformations, and low-lying charge $1/2$ topological excitations (eq.\eqref{nontriv-anticommute}) that interpolate between the two ground states,
\item possesses signatures of a subdominant phase-fluctutating Cooper pair order parameter (i.e., lacks off-diagonal long-ranged order, see Sec.\ref{cooperflucMOtt}), and
\item develops into a familiar Neel antiferromagnetic spin-ordered insulating phase when the RG in recomputed in the presence of a staggering spin rotational symmetry breaking field at weak coupling $U/t<<1$.
\end{itemize}
We also find further insight into the Mott liquid phase through analytic expressions for the ground state of the insulating state, as well as a large family of low energy eigenstates and their energy eigenvalues 
in eq.\eqref{class_1_wvfn}, eq.\eqref{energy-1}, eq.\eqref{class_2_wvfn} and eq.\eqref{energy-2}. In order to test the quantitative accuracy of the effective Hamiltonian and ground state wavefunction, we benchmark the ground state energy per site computed for the Mott liquid ($E_{g}$ at $U_{0}/t=8$, upon finite-size scaling to large system sizes) with that obtained from various numerical methods~\cite{leblanc2015solutions,ehlers2017hybridDMRG,dagotto1992}. The values for $E_{g}^{*}=-0.526$ (Fig.\ref{groundstateenergy}) and doublon fraction $D=0.051$ (Fig.\ref{windowwidth}) obtained from the finite-size scaling analysis is in excellent agreement with the ranges obtained from Refs.\cite{leblanc2015solutions,ehlers2017hybridDMRG,dagotto1992}:
$-0.51<E_{g}^{*}<-0.53$ and $0.0535< D< 0.0545$. Further benchmarking results for $2\leq U/t\leq 12$ are presented in Appendix \ref{further benchmarking}; equally close agreement is obtained throughout this range of $U/t$.\\
\textbf{Main result:} We demonstrate the existence of a symmetry preserved Mott liquid insulating state at $1/2$-filling with signatures of topological order. We also show that this quantum liquid develops into the Neel antiferromagnet Mott insulator upon symmetry breaking. This appears to provide, within the context of a Hubbard model, an explicit and detailed substantiation of Anderson's conjecture for the cuprate Mott insulator~\cite{anderson-science-1987}. The ground state energy per site obtained for the Mott liquid is benchmarked against existing numerical results for $2\leq U/t\leq 12$, displaying excellent agreement and imparting confidence in the effective Hamiltonian and ground state wavefunction obtained from the RG analysis. Codes used in the benchmarking have been made available electronically~\cite{anirbanhubbard2019code}.\\
\par\noindent
\textbf{Hole-doping the Mott liquid: Mottness collapse}~
In Sec.\ref{MottCollapseSection}, we study the effects of hole-doping (i.e., a non-zero negative change in the chemical potential, $\Delta\mu_{eff}=\mu^{0}_{eff}-\frac{\Delta U_{0}}{2}$) on the Mott insulating liquid at strong-coupling ($U_{0}=8t$). As shown in Fig.\ref{Phase_diagram_with_doping-1}, we find three major features upon increasing hole-doping: first, a doped Mott insulating region at low-doping ($f_{h}<0.25$), a quantum critical point (QCP) at a hole-doping fraction of $0.25$ and finally, a region corresponding to a correlated Fermi liquid lying at yet higher hole-doping ($f_{h}>0.25$). The hole-doped Mott liquid system possesses an effective Hamiltonian given by eq.\eqref{dopedMottLiquidHam}, and we obtain the eigenstates (eq.\eqref{manyBodyState}) and eigenvalues (eq.\eqref{minimizeE}) upon analytically diagonalizing it. From these we obtain 
\begin{itemize}
\item the ground state energy $E_{g}(\Delta\mu_{eff})$ and the hole doping fraction $f_{h}(\Delta\mu_{eff})$. Both the ground state energy $E_{g}(f_{h})=-0.776t$ (Fig.\ref{GSdopedML}) and the double occupancy $D=0.045$ (Fig.\ref{doublonwithdoping}) at hole doping fraction $f_{h}=0.125$ is found to be in agreement with that obtained from other numerical methods in Refs.\cite{leblanc2015solutions,ehlers2017hybridDMRG,dagotto1992}: $-0.74<E_{g}<-0.77$ and $0.04<D<0.045$. Further benchmarking results for $2\leq U/t\leq 12$ and $f_{h}=0.125$ are presented in Appendix \ref{further benchmarking}; equally close agreement is obtained throughout this range of $U/t$.
\item a closed-form expression for the variation of $E_{g}$ with $\Delta\mu_{eff}$, as well as $f_{h}$ (see Fig.\ref{GSwithDoping}). This reveals a clear signature of a QCP through a non-analytic behaviour of $E_{g}$ at a value of $\Delta\mu_{eff}^{*}=-4$, $f_{h}^{*}=0.25$,
\item the variation of $f_{h}$ and number compressibility ($\kappa$) with $\Delta\mu_{eff}$ (see Fig.\ref{DopingwithChemPot}). Again, this displays clear signatures of the QCP through abrupt changes in $f_{h}$ and $\kappa$ (corresponding to large number fluctuations at the QCP),
\item the nodal many-body gap ($\Delta E$) as a function of $\Delta\mu_{eff}$ (Fig.\ref{GapwithChemPot}). The non-Landau nature of the QCP can be seen from the sudden collapse of the gap $\Delta E$ to zero, and is observed to arise from the RG irrelevance of the charge backscattering vertex along the nodes (Mottness collapse).
\end{itemize}
\textbf{Main result:} The ground state energy ($E_{g}$) for the doped Mott liquid, obtained from the effective Hamiltonian and ground state wavefunction, again benchmarks very well with existing numerical results for the range of $2\leq U/t\leq 12$. Unambiguous signatures of a QCP associated with the collapse of Mottness~\cite{phillips2011mottness,zaanen2011mottness} are observed in $E_{g}$, $f_{h}$ and $\kappa$. The abrupt collapse of the many-body gap along the nodal directions shows that the QCP cannot be described within the Landau paradigm of phase transitions. Additional evidence is presented in Video S4. Codes used in the benchmarking have been made available electronically~\cite{anirbanhubbard2019code}.\\
\par\noindent
\textbf{Theory for the Mottness collapse QCP}~ 
The RG analyses (eq.\eqref{spin_cooper_RG}) in the vicinity of the QCP in Sec.\ref{QCPtheory} incorporates two major competing pairing momentum channels: (a) pairs of electronic states with net momentum ($\pi,\pi$) which dominate deep in the underdoped regime ($\Delta\mu_{eff}\to 0$), and (b) pairs of electronic states with net momnetum $0$ which dominate near optimal doping ($\Delta\mu_{eff}>-4$). This change in the dominant scattering mechanism with changing $\Delta\mu_{eff}$ describes the growth in superconducting fluctuations, as well as the presence of a nodal marginal Fermi liquid at the QCP opening up into an arc above it. Indeed, we find that the proliferation of dominant attractive spin pseudospin scattering processes near the QCP 
are equivalent to repulsive Cooper pair scattering-induced gapping of the FS (eq.\eqref{transmutation-1}). Within the conical V-shaped region starting from the QCP (Fig.\ref{Phase_diagram_with_doping-1}), the effective Hamiltonian obtained at the RG fixed point (eq.\eqref{H_V}) contains 
\begin{itemize}
\item a 2-electron 1-hole dispersion term of a marginal Fermi liquid metal for the nodal stretches (eq.\eqref{nodal_MFL}),
\item Cooper pair backscattering terms that lead to the gapping of the FS along the AN stretches,
\item a shifted effective chemical potential about the QCP for the Cooper pair degrees,
\item a modified electronic dispersion.
\end{itemize}
The effective Hamiltonian accounts for several features of the low energy spectrum. First, it displays the gaplessness of the nodal directions supporting a MFL; this is as an outcome of the Cooper instability being RG marginal (eq.\eqref{RG_equations_AN_N}) and the Mott instability RG irrelevant. Second, we find that the highest superfluid weight for preformed Cooper pairs within the spin pseudogapped AN regions appears at the largest energyscale right above QCP ($\omega^{sc}_{onset}  = 0$), linking optimal doping with the QCP (eq.\eqref{optcriterion}). This onset energyscale for pairing reduces upon both increasing or reducing $\Delta\mu_{eff}$ (see Fig. \ref{Phase_diagram_with_doping-1}).
\par\noindent
These observations suggest a relation between $T_{c}$ for superconductivity and the superfluid weight $\rho_{s}$, along the lines of the empirically observed Homes law~\cite{homes2004}. Our analysis provides microscopic insight into this relation. For this, we first show that the onset thermal scale for superconducting fluctuations at the AN is related to the superfluid weight carried by the Cooper pair degrees of freedom (eq.\eqref{onset_superconductivity}).
Then, we show a linear relationship between the onset scale for superconducting fluctuations without ODLRO and critical temperature below which Cooper pairs with ODLRO condense (see Appendix \ref{symmbreakAppendix}). Together, these two results help derive a $T=0$ orign for Homes law.\\
\textbf{Main result:} The theory for the QCP and its conical-shaped neighbourhood in Phase diagram Fig.\ref{Phase_diagram_with_doping-1} reveal Cooper pairing along the AN stretches of the FS at high quantum fluctuation energyscales~\cite{emery1995,anderson-advphys-1997}, together with gapless MFL regions along the nodal stretches. This reveals a universal relation between the superfluid stiffness and the onset scale for superconducting fluctuations, and is observed to be the $T=0$ origin of Homes law. Additional evidence is presented in Videos S2 and S3.\\  
\par\noindent
\textbf{The correlated Fermi liquid and emergent symmetry broken orders}~ 
In Sec.\ref{mixed_optical}, we show that there exists a crossover between the MFL at high $\omega$ energyscales and the correlated Fermi liquid lying beyond the QCP. We find that the crossover is characterized by a mixed form of optical conductivity 
arising from an imaginary part of the single-particle self-energy/ inverse lifetime (eq.\eqref{mixed_lifetime}) having a contribution from a Landau Fermi liquid as well as an additional logarithmic contribution characaterising a crossover from the MFL. This result is in consensus with 
experimental observations on overdoped Mott insulators~\cite{van2003}. 
\par\noindent
Finally, in Sec.\ref{SymmetryBreakingSection}, we demonstrate the $T=0$ existence of several symmetry broken phases of matter that emerge upon hole-doping the Mott insulator (Fig.\ref{phasediagwithsymmbreak}). One of our major findings is the detailed derivation of an effective theory for d-wave superconductivity arrived from within our RG approach (eq.\eqref{U(1) symmetry broken theory}). We demonstrate that the d-wave nature of the superconducting order parameter is tied to the gapless nodal-point $k$-space structure at the QCP. We also find  SDW, CDW and spin-nematic symmetry broken phases appear in the phase diagram (Fig.\ref{phasediagwithsymmbreak}) in regimes of hole-doping that are in broad agreement with that found in the cuprates phase diagram~(see,e.g., Ref.\cite{keimer2015quantum} for a review). We also present computations of various spectral signatures and transport properties at underdoping, optimal doping and overdoping in Figs.\ref{CFLAndSDW} - \ref{arc_to_point_spinon_holon_Arc}. Again, the results presented in these figures is in broad consensus with several
experimental observations on doped cuprate Mott insulators (see references given in Sec.\ref{SymmetryBreakingSection}). Additionally, using values of $t$ and $U$ obtained from first-principles calculations, we estimate some typical temperature scales from our formalism for the cuprate Mott insulating materials HBCO and LCO, finding reasonable upper bounds for, e.g., the superconducting $T_{C}$.\\
\textbf{Main result:} A detailed $T=0$ analysis of the correlated Fermi liquid and various symmetry broken phases of the doped Mott liquid offer considerable insight into their origins (e.g., the d-wave SC phase is observed to be tied to the $k$-space structure of the QCP). Qualitative comparisons of our findings with known experimental observations on the cuprates are found to be favourable, prompting the extension of our analyses to finite temperatures. Additional evidence is presented in Videos S5 and S6. 
\par\noindent
\textbf{Overview:} Before passing to the next section, we offer an overview of our work. It is important to note that, due to their non-perturbative nature, only some of the results discussed above have been observed in various numerical investigations of the 2D Hubbard model discussed in the introduction. Even in those cases, an overarching understanding of their origins and significance often continues to be debated. We stress, therefore, that a major achievement of our work is that it provides a \emph{comprehensive and unified analytic framework} for the exploration and analysis of such non-perturbative phenomena.
 \section{Renormalization Group Scheme}\label{RGSetupSection}
We analyze the Hubbard model on the two-dimensional square lattice with nearest neighbour hopping (strength $t$) and on-site repulsion (strength $U_{0}$)
\begin{eqnarray}
\hat{H}&=&\sum_{\mathbf{k}, \sigma}(\epsilon_{0\mathbf{k}}-\mu_{eff})c^{\dagger}_{\mathbf{k}\sigma}c_{\mathbf{k}\sigma}+U_{0}\sum_{\mathbf{r}}\hat{\tau}_{\mathbf{r}\uparrow}\hat{\tau}_{\mathbf{r}\downarrow}\label{Hubbard Hamiltonian}~,
\end{eqnarray}
where $c^{\dagger}_{\mathbf{k}\sigma}/c_{\mathbf{k}\sigma}$ is the electron creation/annihilation operator with wave-vector $\mathbf{k}$ and spin $\sigma$, $\hat{\tau}_{\mathbf{r}\sigma}=\hat{n}_{\mathbf{r}\sigma}-\frac{1}{2}$, $\hat{n}_{\mathbf{r}\sigma}=c^{\dagger}_{\mathbf{r}\sigma}c_{\mathbf{r}\sigma}$ is the number operator at lattice site $\mathbf{r}=j_{1}\hat{x}+j_{2}\hat{y}$, and $\epsilon_{0\mathbf{k}}$ is the bare dispersion. The effective chemical potential, $\mu_{eff} =\mu -\frac{U_{0}}{2}$, accounts for the energy imbalance between doublons (doubly occupied sites) and holons (empty sites). The hopping term is clearly diagonal in momentum-space, with a dispersion for the square lattice given by $\epsilon_{0\mathbf{k}}=-2t(\cos k_{x}+\cos k_{y})$~. 
On the other hand, the Hubbard repulsion term is diagonal in real-space, i.e., it contains off-diagonal elements in the momentum basis, causing fluctuations ($\Delta(\epsilon_{\mathbf{k}\sigma}\hat{n}_{\mathbf{k}\sigma})$) of the dispersion term $\epsilon_{\mathbf{k}\sigma}\hat{n}_{\mathbf{k}\sigma}$. This can be seen from the non-commutativity 
\begin{eqnarray}
\epsilon_{\mathbf{k}\sigma}[\hat{n}_{\mathbf{k}\sigma}, \hat{H}]&=&\epsilon_{\mathbf{k}\sigma}U_{0}[\hat{n}_{\mathbf{k}\sigma}, \sum_{\mathbf{r}}\hat{\tau}_{\mathbf{r}\uparrow}\hat{\tau}_{\mathbf{r}\downarrow}]\label{non-commutativity}~.
\end{eqnarray} 
Below, we will study the effects of such quantum fluctuations via a Hamiltonian renormalization group (RG) method. Further, we will study the Mott metal-insulator transition (MIT) at $1/2$-filling, i.e., by setting the doublon-holon chemical potential $\mu^{0}_{eff}=\mu-\frac{U_{0}}{2}=0$~\cite{krishnamurthy-PhysRevLett.64.950, kemeny1967}, as well as phase transitions induced by hole doping, $\Delta\mu_{eff}=\mu^{0}_{eff}-\frac{\Delta U_{0}}{2}$. 
In subsection~\ref{derive-U}, we first derive the form of the exact unitary disentanglement operator that causes the one-step decoupling of a single electronic state. Then, in subsection~\ref{derive-rot-H}, we compute the form of the rotated Hamiltonian resulting from this transformation. We follow this in subsection~\ref{decoupling single-particle states} by adapting a successive set of such unitary operations on the Hamiltonian into a RG scheme.  
In this scheme, the states with the highest bare electronic kinetic energy $\epsilon_{\mathbf{k}\sigma}$ 
are the first to be exactly decoupled. This is followed by exactly decoupling the next highest $\epsilon_{\mathbf{k}\sigma}$, thus gradually scaling towards the Fermi surface. In subsection~\ref{meaning_of_omega}, we give a detailed description of the relation between the quantum fluctuation energyscale $\omega$ that appears in our RG formalism, and an equivalent thermal scale. We then present a discussion of instabilities of the Fermi surface in subsection~\ref{FSinstab}, and follow it with a detailed derivation of various RG relations for the 2D Hubbard model in subsection~\ref{RG_flow_long_tan}.
\subsection{Derivation of unitary operator for one-step decoupling of an electronic state}\label{derive-U}
The RG procedure is carried out by decoupling one single-particle state $|\mathbf{k}\sigma\rangle$ at every RG step via a unitary operation $U_{\mathbf{k}\sigma}$
\begin{eqnarray}
\hat{n}_{\mathbf{k}\sigma}U_{\mathbf{k}\sigma}\hat{H}U^{\dagger}_{\mathbf{k}\sigma}(1-\hat{n}_{\mathbf{k}\sigma})=0\to [\hat{n}_{\mathbf{k}\sigma}, U_{\mathbf{k}\sigma}\hat{H}U^{\dagger}_{\mathbf{k}\sigma}]=0~,~~\label{decoupling_condition}
\end{eqnarray}
thereby trivializing the non-commutativity relation (eq.\eqref{non-commutativity}) for the decoupled state. In this subsection, we will derive the form of the unitary operator $U_{\mathbf{k}\sigma}$ that satisfies the decoupling condition. The equation eq.\eqref{decoupling_condition} can equivalently be written as
\begin{eqnarray}
P_{\mathbf{k}\sigma}H(1-P_{\mathbf{k}\sigma})=0~,~\label{equivalent-decoupling-cond}
\end{eqnarray}
where $P_{\mathbf{k}\sigma}=U^{\dagger}_{\mathbf{k}\sigma}\hat{n}_{\mathbf{k}\sigma}U_{\mathbf{k}\sigma}$, $1-P_{\mathbf{k}\sigma}$ are the rotated many-body projection operators on orthogonal subspaces. We define a new Hamiltonian $H'$ such that
\begin{equation}
P_{\mathbf{k}\sigma}HP_{\mathbf{k}\sigma}=P_{\mathbf{k}\sigma}H'P_{\mathbf{k}\sigma}~~,~~[H',\hat{n}_{\mathbf{k}\sigma} ]=0~.
\end{equation} 
Then, eq.\eqref{equivalent-decoupling-cond} amounts to solving
\begin{eqnarray}
\hat{H}|\Psi\rangle = H'|\Psi\rangle~,
\end{eqnarray}
where $|\Psi\rangle$ satisfies the condition: $P_{\mathbf{k}\sigma}|\Psi\rangle =|\Psi\rangle$.
In order to show clearly that certain terms in the rotated Hamiltonian vanish and lead to eq.\eqref{decoupling_condition}, we decompose the Hamiltonian $H$ into diagonal and off-diagonal pieces: $\hat{H} = H^{D}+H^{X}_{\mathbf{k}\sigma}+H^{X}_{\bar{\mathbf{k}\sigma}}$. The diagonal piece ($H^{D}$) constitutes the 1-particle dispersion and 2-particle density-density (Hartree-Fock) terms. The second term, $H^{X}_{\mathbf{k}\sigma}$, represents the off-diagonal coupling between state $|\mathbf{k}\sigma\rangle$ and other momentum states states $|\mathbf{k}'\sigma'\rangle$. Finally, the third piece ($H^{X}_{\bar{\mathbf{k}\sigma}}$) represents the off-diagonal coupling among all momentum states other than $|\mathbf{k}\sigma\rangle$. Solving eq.\eqref{decoupling_condition} is then equivalent to finding a state $|\Psi\rangle$ such that
\begin{equation}
(H^{D}+H^{X}_{\mathbf{k}\sigma}+H^{X}_{\bar{\mathbf{k}\sigma}})|\Psi\rangle = H^{'}|\Psi\rangle~,\label{decoupling_equation_wave_operator}
\end{equation}
where $H^{'}=H^{'D}+H^{'X}_{\bar{\mathbf{k}\sigma}}$. Here, $H^{'D}$ is the renormalized operator that satisfies the condition eq.\eqref{decoupling_condition}. Further, $H^{'D}$ and $H^{'X}_{\bar{\mathbf{k}\sigma}}$ have similar definitions as given above for $H^{D}$ and $H^{X}$ respectively. To proceed with solving this equation, we write $|\Psi\rangle$ in a Schmidt decomposed form
\begin{eqnarray}
|\Psi\rangle = a_{1}|\Psi_{1},1_{\mathbf{k}\sigma}\rangle + a_{0}|\Psi_{0},0_{\mathbf{k}\sigma}\rangle~.\label{superposition}
\end{eqnarray}
In the above expression, the occupation number states $|1_{\mathbf{k}\sigma}\rangle$, $|0_{\mathbf{k}\sigma}\rangle$ comprise a 2-dimensional Hilbert space and the orthogonal states $|\Psi_{1}\rangle$ and $|\Psi_{0}\rangle$ ($\langle\Psi_{0}|\Psi_{1}\rangle=0$) belong to the remnant $2^{N-1}$ dimensional Hilbert space of $1,..,N-1$ single-electron degrees of freedom.  Then, the decoupling equation (eq\eqref{decoupling_condition}) connects the elements in $|\Psi\rangle$ as follows
\begin{eqnarray}
a_{1}|\Psi_{1},1_{\mathbf{k}\sigma}\rangle &=& a_{0}\eta^{\dagger}_{\mathbf{k}\sigma}|\Psi_{0},0_{\mathbf{k}\sigma}\rangle,\nonumber\\
a_{0}|\Psi_{0},0_{\mathbf{k}\sigma}\rangle &=& a_{1}\eta_{\mathbf{k}\sigma}|\Psi_{1},1_{\mathbf{k}\sigma}\rangle~,\label{transition-eqns}
\end{eqnarray}
where the operators $\hat{\omega}$, $\eta^{\dagger}_{\mathbf{k}\sigma}$ and $\eta_{\mathbf{k}\sigma}$ are defined as
\begin{eqnarray}
&&\hspace*{-0.5cm}\hat{\omega}=H^{'D}+H^{'X}_{\bar{\mathbf{k}\sigma}}-H^{X}_{\bar{\mathbf{k}\sigma}},\label{quantum fluctuation scale}\\
&&\hspace*{-0.5cm}\eta^{\dagger}_{\mathbf{k}\sigma}=\frac{1}{\hat{\omega}-Tr_{\mathbf{k}\sigma}(H^{D}\hat{n}_{\mathbf{k}\sigma})\hat{n}_{\mathbf{k}\sigma}}c^{\dagger}_{\mathbf{k}\sigma}Tr_{\mathbf{k}\sigma}(Hc_{\mathbf{k}\sigma}),\label{eh-transition-1} \\
&&\hspace*{-0.5cm}\eta_{\mathbf{k}\sigma}=\frac{1}{\hat{\omega}-Tr_{\mathbf{k}\sigma}(H^{D}(1-\hat{n}_{\mathbf{k}\sigma}))(1-\hat{n}_{\mathbf{k}\sigma})}Tr_{\mathbf{k}\sigma}(c^{\dagger}_{\mathbf{k}\sigma}H)c_{\mathbf{k}\sigma}.~~~~~\label{eh-transition-2}
\end{eqnarray} 
Here, $Tr_{\mathbf{k}\sigma}(\cdot)$ stands for a partial trace in the Fock space, where the usual fermion anti-commutation rules are followed. 
$Tr_{\mathbf{k}\sigma}(c^{\dagger}_{\mathbf{k}\sigma}H)c_{\mathbf{k}\sigma}$ and its conjugate are obtained from $H^{X}_{\mathbf{k}\sigma}$ as 
\begin{equation}
Tr_{\mathbf{k}\sigma}(c^{\dagger}_{\mathbf{k}\sigma}H)c_{\mathbf{k}\sigma} = (1-\hat{n}_{\mathbf{k}\sigma})H^{X}_{\mathbf{k}\sigma}\hat{n}_{\mathbf{k}\sigma}~.
\end{equation}
Using eqns.\eqref{transition-eqns}, we arrive at
\begin{eqnarray}
\eta^{\dagger}_{\mathbf{k}\sigma}\eta_{\mathbf{k}\sigma}|\Psi_{1},1_{\mathbf{k}\sigma}\rangle &=& \hat{n}_{\mathbf{k}\sigma}|\Psi_{1},1_{\mathbf{k}\sigma}\rangle =|\Psi_{1},1_{\mathbf{k}\sigma}\rangle~,\nonumber\\ 
\eta_{\mathbf{k}\sigma}\eta^{\dagger}_{\mathbf{k}\sigma}|\Psi_{0},0_{\mathbf{k}\sigma}\rangle &=& (1-\hat{n}_{\mathbf{k}\sigma})|\Psi_{0},0_{\mathbf{k}\sigma}\rangle = |\Psi_{0},0_{\mathbf{k}\sigma}\rangle~.~~\label{relation-set1}
\end{eqnarray}
Further, from the definitions of $\eta_{\mathbf{k}\sigma}$ and $\eta^{\dagger}_{\mathbf{k}\sigma}$ (eq.\eqref{eh-transition-2}), we obtain the relations
\begin{eqnarray}
&&\eta_{\mathbf{k}\sigma}|\Psi_{0},0_{\mathbf{k}\sigma}\rangle = 0\implies \eta^{\dagger}_{\mathbf{k}\sigma}\eta_{\mathbf{k}\sigma}|\Psi_{0},0_{\mathbf{k}\sigma}\rangle = 0~,\nonumber\\
&&\eta^{\dagger}_{\mathbf{k}\sigma}|\Psi_{1},1_{\mathbf{k}\sigma}\rangle = 0\implies \eta_{\mathbf{k}\sigma}\eta^{\dagger}_{\mathbf{k}\sigma}|\Psi_{0},0_{\mathbf{k}\sigma}\rangle = 0~.
\label{relation-set2}
\end{eqnarray}
Combining eqs.\eqref{relation-set1} and\eqref{relation-set2}, we arrive at the following algebraic relations for $\eta_{\mathbf{k}\sigma}$ and $\eta^{\dagger}_{\mathbf{k}\sigma}$
\begin{eqnarray}
\lbrace\eta^{\dagger}_{\mathbf{k}\sigma},\eta_{\mathbf{k}\sigma}\rbrace =1~~,~~ [\eta^{\dagger}_{\mathbf{k}\sigma},\eta_{\mathbf{k}\sigma}]=2\hat{n}_{\mathbf{k}\sigma}-1~.\label{eta-algebra}
\end{eqnarray}
Also, we note that using eqs.\eqref{transition-eqns} together with the form of the state $|\Psi\rangle$ (eq.\eqref{superposition}), we obtain a similarity transformation between $|\Psi_{1},1_{\mathbf{k}\sigma}\rangle$ and $|\Psi\rangle$
\begin{eqnarray}
a_{0}|\Psi_{0},0_{\mathbf{k}\sigma}\rangle &=& \mathcal{N}^{-1}\eta_{\mathbf{k}\sigma}|\Psi_{1},1_{\mathbf{k}\sigma}\rangle\nonumber\\
|\Psi\rangle &=&a_{1}(1+\eta_{\mathbf{k}\sigma})|\Psi_{1},1_{\mathbf{k}\sigma}\rangle \nonumber\\
&=& a_{1}\exp(\eta_{\mathbf{k}\sigma})|\Psi_{1},1_{\mathbf{k}\sigma}\rangle~,\nonumber\\
\implies |\Psi_{1},1_{\mathbf{k}\sigma}\rangle &=& a_{1}^{-1}\exp(-\eta_{\mathbf{k}\sigma})|\Psi\rangle~,
\end{eqnarray}
where $\mathcal{N}=a_{1}$. Importantly, in the many-body state $|\Psi_{1},1_{\mathbf{k}\sigma}\rangle$, the single electronic state labelled $\mathbf{k}\sigma$ is disentangled. Thus, the operator $\exp(-\eta_{\mathbf{k}\sigma})$ removes the many-body entanglement content between the state $|\mathbf{k}\sigma\rangle$ and the rest. From this similarity transformation, we can construct the unitary operator~\cite{shavitt1980quasidegenerate,suzuki1982construction}
\begin{eqnarray}
U_{\mathbf{k}\sigma}&=&\exp\arctan\text{h}(\eta^{\dagger}_{\mathbf{k}\sigma}-\eta_{\mathbf{k}\sigma})\nonumber\\
&=&\exp\frac{\pi}{4}(\eta^{\dagger}_{\mathbf{k}\sigma}-\eta_{\mathbf{k}\sigma})=\frac{1}{\sqrt{2}}(1+\eta^{\dagger}_{\mathbf{k}\sigma}-\eta_{\mathbf{k}\sigma})~,\label{unitary operator}
\end{eqnarray}
that transforms $|\Psi\rangle$ to $|\Psi_{1},1_{\mathbf{k}\sigma}\rangle = U|\Psi\rangle$. The unitarity property $U_{\mathbf{k}\sigma}U^{\dagger}_{\mathbf{k}\sigma}=U^{\dagger}_{\mathbf{k}\sigma}U_{\mathbf{k}\sigma}=I$ can be verified using the algebra of $\eta_{\mathbf{k}\sigma}$ and $\eta^{\dagger}_{\mathbf{k}\sigma}$ operators given in eq.\eqref{eta-algebra}. Thus, $U_{\mathbf{k}\sigma}$ can be interpreted as a disentangling transformation. 
\subsection{Derivation for the rotated Hamiltonian $U_{\mathbf{k}\sigma}HU^{\dagger}_{\mathbf{k}\sigma}$}\label{derive-rot-H}
Having obtained the unitary operation that carries out the disentanglement of single-particle states, we will now compute the form of the rotated Hamiltonian. We note that the rotated Hamiltonian should be purely diagonal in the occupation-number basis states $1_{\mathbf{k}\sigma}$ and $0_{\mathbf{k}\sigma}$. In order to verify this, we decompose the rotated Hamiltonian into a diagonal and an off-diagonal component
\begin{eqnarray}
U_{\mathbf{k}\sigma}HU_{\mathbf{k}\sigma}^{\dagger} &=& H_{1}+H_{2},\nonumber\\
H_{1}&=&\frac{1}{2}\bigg[H+[\eta^{\dagger}_{\mathbf{k}\sigma}-\eta_{\mathbf{k}\sigma},H]\nonumber\\
&+&\eta_{\mathbf{k}\sigma}H\eta^{\dagger}_{\mathbf{k}\sigma}+\eta^{\dagger}_{\mathbf{k}\sigma}H\eta_{\mathbf{k}\sigma}\bigg]~,\nonumber\\
H_{2}&=&\frac{1}{2}\bigg[H^{X}_{\mathbf{k}\sigma}-\eta^{\dagger}_{\mathbf{k}\sigma}Tr_{\mathbf{k}\sigma}(c^{\dagger}_{\mathbf{k}\sigma}H)c_{\mathbf{k}\sigma}\eta^{\dagger}_{\mathbf{k}\sigma}\nonumber\\
&-&\eta_{\mathbf{k}\sigma}c^{\dagger}_{\mathbf{k}\sigma}Tr_{\mathbf{k}\sigma}(Hc_{\mathbf{k}\sigma})\eta_{\mathbf{k}\sigma}\bigg]~,
\end{eqnarray}
where the off-diagonal component $H_{2}$ must vanish. To show that, we first set up the preliminaries
\begin{eqnarray}
\eta^{\dagger}_{\mathbf{k}\sigma}\eta_{\mathbf{k}\sigma}&=&\hat{n}_{\mathbf{k}\sigma}~,\nonumber\\
\implies \hat{\omega}-Tr_{\mathbf{k}\sigma}(H^{D}\hat{n}_{\mathbf{k}\sigma})\hat{n}_{\mathbf{k}\sigma}&=&c^{\dagger}_{\mathbf{k}\sigma}Tr_{\mathbf{k}\sigma}(Hc_{\mathbf{k}\sigma})\eta_{\mathbf{k}\sigma}~,\nonumber\\
\implies\eta_{\mathbf{k}\sigma}c^{\dagger}_{\mathbf{k}\sigma}Tr_{\mathbf{k}\sigma}(Hc_{\mathbf{k}\sigma})
\eta_{\mathbf{k}\sigma} &=& c^{\dagger}_{\mathbf{k}\sigma}Tr_{\mathbf{k}\sigma}(Hc_{\mathbf{k}\sigma})~.\label{vanishing-offdiag}
\end{eqnarray}
The definition of $H^{X}_{\mathbf{k}\sigma}=c^{\dagger}_{\mathbf{k}\sigma}Tr_{\mathbf{k}\sigma}(Hc_{\mathbf{k}\sigma})+h.c.$, along with eq.\eqref{vanishing-offdiag}, then implies that $H_{2}=0$. In the other component, $H_{1}$, we first  unravel the terms $\eta_{\mathbf{k}\sigma}H\eta^{\dagger}_{\mathbf{k}\sigma}$ and $\eta^{\dagger}_{\mathbf{k}\sigma}H\eta_{\mathbf{k}\sigma}$. Using eq.\eqref{quantum fluctuation scale}, eq\eqref{eh-transition-1} and eq\eqref{eh-transition-2}, we obtain
\begin{eqnarray}
&&\frac{1}{\hat{H}'-Tr_{\mathbf{k}\sigma}(H\hat{n}_{\mathbf{k}\sigma})\hat{n}_{\mathbf{k}\sigma}}c^{\dagger}_{\mathbf{k}\sigma}Tr_{\mathbf{k}\sigma}(Hc_{\mathbf{k}\sigma})\nonumber\\
&=&c^{\dagger}_{\mathbf{k}\sigma}Tr_{\mathbf{k}\sigma}(Hc_{\mathbf{k}\sigma})\frac{1}{\hat{H}'-Tr_{\mathbf{k}\sigma}(H(1-\hat{n}_{\mathbf{k}\sigma}))(1-\hat{n}_{\mathbf{k}\sigma})}~,~~\nonumber\\
\implies &&Tr_{\mathbf{k}\sigma}(H\hat{n}_{\mathbf{k}\sigma})\hat{n}_{\mathbf{k}\sigma}c^{\dagger}_{\mathbf{k}\sigma}Tr_{\mathbf{k}\sigma}(Hc_{\mathbf{k}\sigma})\nonumber\\
&=&c^{\dagger}_{\mathbf{k}\sigma}Tr_{\mathbf{k}\sigma}(Hc_{\mathbf{k}\sigma})Tr_{\mathbf{k}\sigma}(H(1-\hat{n}_{\mathbf{k}\sigma}))(1-\hat{n}_{\mathbf{k}\sigma})~.\label{electron-hole-transition}
\end{eqnarray}
The above relation 
then allows us to simplify $\eta_{\mathbf{k}\sigma}H\eta^{\dagger}_{\mathbf{k}\sigma}$ and $\eta^{\dagger}_{\mathbf{k}\sigma}H\eta_{\mathbf{k}\sigma}$ as follows
\begin{eqnarray}
\eta_{\mathbf{k}\sigma}H\eta^{\dagger}_{\mathbf{k}\sigma} &=& Tr_{\mathbf{k}\sigma}(H(1-\hat{n}_{\mathbf{k}\sigma}))(1-\hat{n}_{\mathbf{k}\sigma})~,\nonumber\\
\eta^{\dagger}_{\mathbf{k}\sigma}H\eta_{\mathbf{k}\sigma} &=& Tr_{\mathbf{k}\sigma}(H\hat{n}_{\mathbf{k}\sigma})\hat{n}_{\mathbf{k}\sigma}~.\label{Piece 1 of H1}
\end{eqnarray}
Next, we deduce $[\eta^{\dagger}_{\mathbf{k}\sigma}-\eta_{\mathbf{k}\sigma},H]$, i.e., the renormalization of the Hamiltonian using the relations obtained above
\begin{eqnarray}
[\eta^{\dagger}_{\mathbf{k}\sigma}-\eta_{\mathbf{k}\sigma},H]=2\tau_{\mathbf{k}\sigma}\lbrace c^{\dagger}_{\mathbf{k}\sigma}Tr_{\mathbf{k}\sigma}(Hc_{\mathbf{k}\sigma}),\eta_{\mathbf{k}\sigma}\rbrace~.\label{ren_H}
\end{eqnarray} 
Finally, by combining the result $H_{2}=0$ together with eqs.\eqref{Piece 1 of H1} and \eqref{ren_H}, we obtain the form of the rotated $H$
\begin{eqnarray}
U_{\mathbf{k}\sigma}HU^{\dagger}_{\mathbf{k}\sigma}&=& \frac{1}{2}Tr_{\mathbf{k}\sigma}(H)+\tau_{\mathbf{k}\sigma}Tr_{\mathbf{k}\sigma}(H\tau_{\mathbf{k}\sigma})
\nonumber\\
&+&\tau_{\mathbf{k}\sigma}\lbrace c^{\dagger}_{\mathbf{k}\sigma}Tr_{\mathbf{k}\sigma}(Hc_{\mathbf{k}\sigma}),\eta_{\mathbf{k}\sigma}\rbrace~.\label{rot_ham}
\end{eqnarray}
One can easily check that the rotated Hamiltonian $[U_{\mathbf{k}\sigma}HU^{\dagger}_{\mathbf{k}\sigma},\hat{\tau}_{\mathbf{k}\sigma}]=0$, i.e., $\tau_{\mathbf{k}\sigma}$ is an integral of motion. Turning to the quantum fluctuation operator $\omega$ (eq.\eqref{quantum fluctuation scale}), we note that its eigenvalues represent energy scales for the fluctuations in the occupation number of state $|\mathbf{k}\sigma\rangle$. 
\par
We can now put our unitary disentangling tranformation in context with the canonical transformations used in various other RG methods, including continuous unitary transformation (CUT) RG~\cite{glazek1993,glazek1994,wegner1994,savitz2017stable}, strong disorder RG~\cite{rademaker2016explicit,monthus2016flow}                                                                                                                                                                        and spectrum bifurcation RG~\cite{you2016entanglement}. We recall that CUT RG schemes aim, via the iterative application of unitary transformations, to remove off-diagonal entries coupling various energy configurations using a variety of choices for the RG flow generator. The goal is, in this way, to make the Hamiltonian matrix more band-diagonal. Nevertheless, this implementation of the RG in terms of unitary transformations eventually becomes perturbative in nature, as at any given RG step, the rotated Hamiltonian cannot be computed exactly owing to the appearance of an infinite series expansion in the couplings. Instead, an effective Hamiltonian is obtained perturbatively through a truncation of the coupling expansion. This is also true of the recently developed entanglement-CUT RG scheme~\cite{sahin2017entanglement}, where the RG flow of the entanglement content between operators is studied using tensor networks. 
Similarly, in various recent strong disorder RG schemes~\cite{rademaker2016explicit,monthus2016flow}, the generator of transformations is chosen such that certain terms in the Hamiltonian can be dropped. As with the CUT RG, this leads to only the partial disentanglement of a single electronic degree of freedom at any given RG step. Finally, in the spectrum bifurcation RG scheme~\cite{you2016entanglement}, the Hamiltonian is made progressively block diagonal at each RG step via the iterative application of local unitary rotations along with coarse-graining transformations that are perturbative in nature.
\par\noindent
This should be contrasted with the non-local nature of the unitary operations employed in our RG scheme (eq.\eqref{unitary operator}), that implement non-perturbative coarse-graining transformations through the precise disentanglement of one electronic state at every step.
Further, unlike the RG schemes discussed above, we obtain close-form analytic expressions for the rotated Hamiltonian at every step of the RG transformations.
Finally, our Hamiltonian RG flow evolves across multiple quantum-fluctuation scales, the eigenvalues ($\omega$) of eq.\eqref{quantum fluctuation scale}. This helps obtain effective theories for various subparts of the many-body spectrum. 
\par
This brings us to an important outcome of our RG transformation scheme $H\to U_{\mathbf{k}\sigma}HU^{\dagger}_{\mathbf{k}\sigma}$~:-\emph{if along the RG flow, one of the energy eigenvalues of $\hat{\omega}$ operator matches with an eigenvalue of the diagonal operator $H^{D}$, we obtain a stable fixed point of the RG transformations that is signalled via the vanishing of the off-diagonal blocks in the occupation basis of the electronic state being disentangled at that step.} 
This can be seen by starting from equation eq.\eqref{relation-set1}, with $\eta_{\mathbf{k}\sigma}$ acting on any one of  the eigenstates of the $\hat{\omega}$ operator (say $|\Phi_{1},1_{\mathbf{k}\sigma}\rangle$) with eigenvalue $\omega$  
\begin{eqnarray}
(\omega-Tr_{\mathbf{k}\sigma}(H^{D}\hat{n}_{\mathbf{k}\sigma})|\Phi_{1},1_{\mathbf{k}\sigma}\rangle &=& c^{\dagger}_{\mathbf{k}\sigma}Tr(Hc_{\mathbf{k}\sigma})\eta_{\mathbf{k}\sigma}|\Phi_{1},1_{\mathbf{k}\sigma}\rangle\nonumber\\
\textrm{Det}(\omega-Tr_{\mathbf{k}\sigma}(H^{D}\hat{n}_{\mathbf{k}\sigma}))=0 &\implies & H^{X}_{\mathbf{k}\sigma}|\Phi_{1},1_{\mathbf{k}\sigma}\rangle =0~.
\label{quantum_fluc_switch_off}
\end{eqnarray}
This shows that if an eigenvalue of $H^{D}$ becomes equal to $\omega$, a stable fixed point is reached due to a vanishing off-diagonal block~\cite{Glazek2004}.
\par
\begin{figure}[!h]
\includegraphics[width=0.5\textwidth]{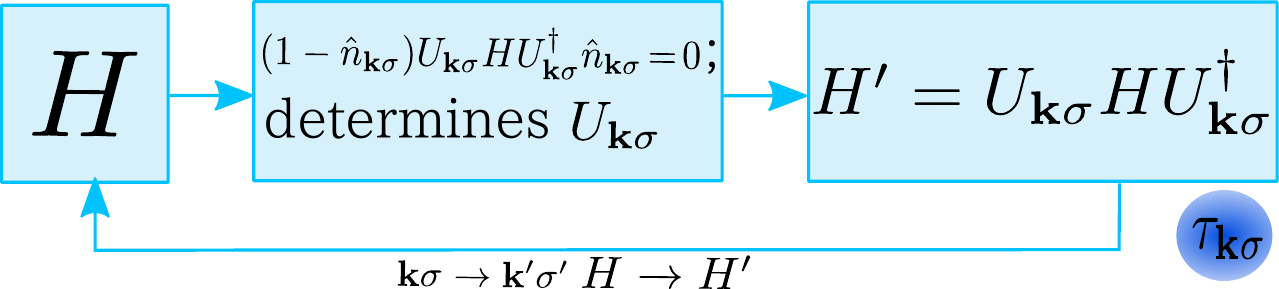}
\caption{Schematic diagram of the RG procedure. The feedback loop shows the replacement of $H\to H'$ and $|\mathbf{k}\sigma\rangle \to |\mathbf{k}'\sigma'\rangle$. $H'$ and $U_{\mathbf{k}'\sigma'}$ satisfy the condition shown in the middle block, thereby generating the Hamiltonian at the next step. At each step, a  commuting operator $\tau_{\mathbf{k}\sigma}$ (represented by the blue circle) is generated via the many-body rotation, such that its eigenvalues are integrals of motion.}\label{RG-scheme}
\end{figure}

In the next section, by using results from the above, we will implement the unitary transformations $U_{\mathbf{k}\sigma}'s$ iteratively on the Hubbard Hamiltonian (eq.\eqref{Hubbard Hamiltonian}) by progressively decoupling the highest energy state $\epsilon_{\mathbf{k}\sigma}$ and scaling gradually towards the Fermi energy $E_{F}$. This will allow us to set up a momentum-space Hamiltonian RG theory~\cite{pal2019} for the 2D Hubbard model as 
visualized in Fig.\ref{RG-scheme}.
\par 
\subsection{RG via the decoupling of single-particle occupation states}\label{decoupling single-particle states}
\begin{figure}
\includegraphics[width=0.54\textwidth]{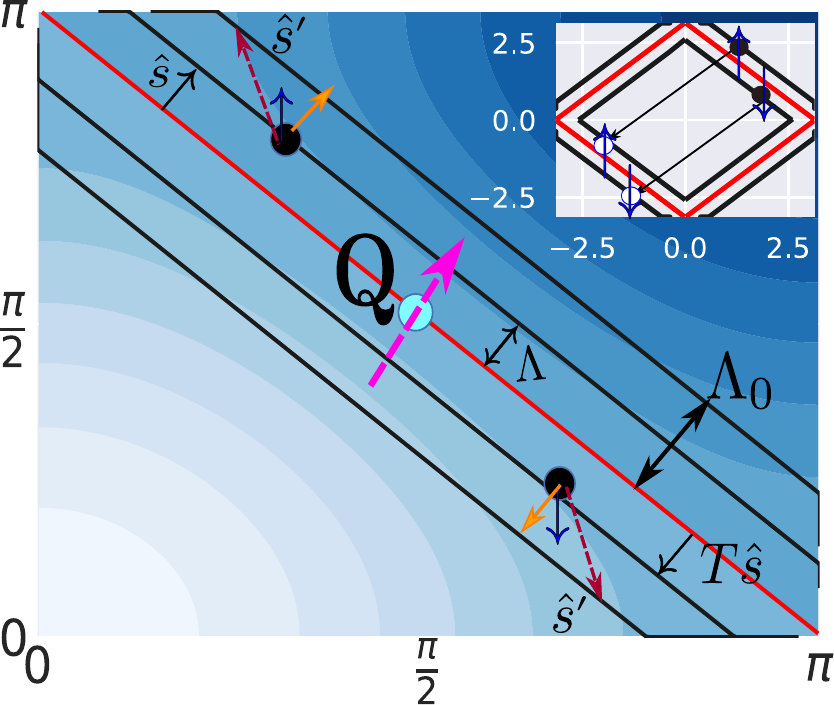} 
\caption{(Colour Online) (a) Schematic representation of  shells (black lines parallel to and formed around the FS (red line) with spacing $\Lambda$) of states that are integrated out from first quadrant of Brillouin zone (BZ):~$0<k_{x},k_{y}<\pi$. Inset (top right): Umklapp scattering of electron pairs. The pair of black dots represents electron-electron/electron-hole pair with opposite spins for charge/spin excitations formed around the node (cyan dot); white dots represents the hole-hole/hole-electron pair. $\hat{s}$ represents the direction normal to F, the orange arrows represent the forward scattering process. $T\hat{s}=(s_{y}, s_{x})$ represents the orientation vector symmetrically placed about the nodal vector $\mathbf{Q}$. The pair of brown dashed lines shows  tangential scattering from $\hat{s},T\hat{s}\to \hat{s}',T\hat{s}'$.
 (b) The arrows (i) represents the resonant two electron scattering process across FS via states (black dots) located symmetrically at $-\Lambda, \Lambda$ distances, arrows (ii) represents the off-resonant two electron scattering process where the two electrons (green dots) are placed at $\Lambda +\delta$, $-\Lambda$.}\label{Fermi_surface_scatt}
\end{figure}

In this section, we design the RG scheme that implements the algorithm shown in Fig.~\ref{RG-scheme} for decoupling single-particle Fock states. We will define shells that are isogeometric to the non-interacting Fermi surface (see Fig.~\ref{Fermi_surface_scatt}). This involves identifying the Fermi surface of the half-filled tight-binding model on the 2d square lattice at $E_{F}=\mu^{0}_{eff}=0$. The Fermi surface (FS) is then defined as a collection of unit normal wave-vectors $\hat{s}=\nabla\epsilon_{\mathbf{k}}/|\nabla\epsilon_{\mathbf{k}}||_{\epsilon_{\mathbf{k}}=E_{F}}$. The $C_{4}$ symmetric square FS  also has four van-Hove singularities along the antinodal (AN) directions: two along $\mathbf{Q}_{y}=(0, \pi)$ and another two along $\mathbf{Q}_{x}=(\pi, 0)$ (Fig.~\ref{Fermi_surface_scatt}). The nodal (N) directions are given by the bisectors: $\mathbf{Q}=\mathbf{Q_{x}}+\mathbf{Q_{y}}$ and $\mathbf{Q}_{\perp}\perp \mathbf{Q}$. The normal vectors are defined as $\hat{s}=\mathbf{Q}/|\mathbf{Q}|$ on one quadrant of the square Fermi surface, which on crossing the van-Hove to the other arm becomes orthogonally oriented to $\hat{s}$: $\hat{s}_{\perp}=\mathbf{Q}_{\perp}/|\mathbf{Q}_{\perp}|~,~\hat{s}_{\perp}\cdot\hat{s}=0$.
\par 
The normal translations of the Fermi surface wave-vectors $\mathbf{k}_{\Lambda,\hat{s}}=\mathbf{k}_{F}(\hat{\mathbf{s}})+\Lambda\hat{\mathbf{s}}$ represent \textit{isogeometric curves} displaced parallel by distance $\Lambda$ from the FS (i.e., the black lines parallel to the FS shown in Fig.~\ref{Fermi_surface_scatt}(a)). Importantly, the anisotropy of the dispersion term with $\hat{s}$ on the square Fermi surface ($k_{Fx}+k_{Fy}=\pi$), together with the non-commutativity of the dispersion and Hubbard $U$ terms (eq.\eqref{non-commutativity}), leads to a variety of quantum fluctuation scales across F ranging from the anti-nodes (AN: $k_{Fx}=0$) to the nodes (N: $k_{Fx}=\pi/2$) 
\begin{eqnarray}
\epsilon_{\mathbf{k}_{\Lambda\hat{s}}} = -2t\sin\frac{\Lambda}{\sqrt{2}}\sin k_{Fx}~,\label{dispersion_differentiation}
\end{eqnarray}
where we have set $|\hat{s}_{x}|=|\hat{s}_{y}|=\sqrt{2^{-1}}$.
The momentum-space representation of the Hubbard term contains four-fermionic off-diagonal scattering pieces coupling states between isogeometric curves (longitudinal scattering), and between normal directions $\hat{s}$ (tangential scattering). Thus, the renormalization group (RG) flow takes place via the decoupling of an isogeometric curve ($\Lambda_{j}$) farthest from the FS at every step by using a product of unitary operations ($U_{(j)}$), itself a product of unitary operators $U_{(j, l)}$ that decouple individual states $(j, l)\equiv (\mathbf{k}_{\Lambda_{j}\hat{s}}, \sigma)$ along a given normal $\hat{s}$ 
\begin{eqnarray}
U_{(j)} &=& \prod_{l=(\hat{s}, \sigma)}U_{(j, l)}~,~\label{Unitary_operator_prod_decoupling_curve}\\
U_{(j, l)}&=&\frac{1}{\sqrt{2}}[1+\eta^{\dagger}_{(j, l)}-\eta_{(j, l)}]\label{unitary_op_def}~.
\end{eqnarray}
In the above expression, $\eta^{\dagger}_{(j, l)}$ is an operator that causes transitions from hole-occupied many-body configurations to electron-occupied configurations, while $\eta_{(j, l)}$ does the reverse. They have the following properties
\begin{eqnarray}
(1-\hat{n}_{j, l})\eta_{(j, l)}\hat{n}_{j, l}&=&\eta_{(j, l)}~,~\hat{n}_{j, l}\eta_{(j, l)}(1-\hat{n}_{j, l})=0~,\nonumber\\
\eta_{(j, l)}^{2}&=&0~,~\left[\eta^{\dagger}_{(j, l)}, \eta_{(j, l)}\right]=2\hat{n}_{j, l}-1~,\nonumber\\
\lbrace\eta^{\dagger}_{(j, l)}, \eta_{(j, l)}\rbrace &=&1~.\label{eta-operator-rel}
\end{eqnarray}
The operator $\eta^{\dagger}_{(j, l)}$ is defined as a sum over 
projections of various eigen-configurations of the renormalised Hamiltonian at the RG step $j$ ($H_{(j)}$)
\begin{eqnarray}
\eta^{\dagger}_{(j, l)}&=& \frac{1}{\hat{\omega}-Tr_{j,l}(H^{D}_{(j,l)}\hat{n}_{j,l})}c^{\dagger}_{j,l}Tr_{j,l}(H_{(j,l)}c_{j,l})\nonumber\\
&=&\sum_{i}[\eta^{\dagger}_{(j, l)}\hat{O}_{(j, l)}](\omega_{(j), i})~,~\label{e-h transition operator}\\
\eta^{\dagger}_{(j, l)}(\omega_{(j), i}) &=& \frac{1}{\omega_{(j), i} -Tr_{j}(H^{D}_{(j,l)}\hat{n}_{j, l})\hat{n}_{j, l}}c^{\dagger}_{j, l}Tr_{j, l}(H_{(j,l)}c_{j, l})~,\nonumber
\end{eqnarray}
where $c^{\dagger}_{\mathbf{k}_{\Lambda_{j}, \hat{s}}, \sigma}=c^{\dagger}_{j, l}$, $\hat{n}_{\mathbf{k}_{\Lambda_{j}\hat{s}}\sigma}=\hat{n}_{j, l}$, $(\hat{O}(\omega_{i}))_{(j, l)}$ are the projection operators involved in projecting onto various many-body configurations of $H_{(j,l)}$ and 
$H^{D}_{(j,l)}$ represents the renormalized diagonal piece of the Hamiltonian in the occupation number basis. The additional index $l$ in $H_{(j,l)}$ denotes its intermediate evolution along a given isogeometric curve through a successive set of unitaries $U_{(j,l)}$. The $\omega$'s are eigenvalues of $\hat{\omega}$ (eq.\eqref{quantum fluctuation scale}).

The flow equation for the Hamiltonian  is then given by 
\begin{eqnarray}
H_{(j-1)} &=& U_{(j)}H_{(j)}U_{(j)}^{\dagger}~,	
\label{isogeometric_curve_RG_flow}
\end{eqnarray}
where the count of RG step $j$ involves a countdown from $N$ (the number of isogeometric curves from the BZ boundaries to FS), such that the bare Hamiltonian $H\equiv H_{(N)}$. In a later section, we show the method for obtaining the vertex RG flow equations using the form of the rotated Hamiltonian like that obtained in  eq.\eqref{rot_ham}.
\subsection{Correspondence between $\omega$ and an emergent thermal scale $T$}\label{meaning_of_omega}
In the above RG scheme the renormalized Hamiltonian can be decomposed over all fluctuation scale as follows $H_{(j)}=\sum_{\omega}(H(\omega)\hat{O}(\omega))_{(j)}$. Here the Hamiltonian $H_{(j)}(\omega_{i})$ can include the effect of many body correlations differently compared to $H_{(j)}(\omega_{l})$. This would mean that the nature of the low energy excitations for the different $\omega$ dependent Hamiltonians might also be different. This provides the perfect setting to ask the following question: Can the effects of $\omega$ dependent many body correlations on the non-interacting Fermi gas at 0K be manifested in a finite temperature T scale? If the answer is affirmative then this will allow us to classify the different phases of the Hamiltonian across varying temperature scales? 
\par\noindent
We introduce the imaginary time evolution operator of the renormalized Hamiltonian at RG step $j$, $U_{(j)}(\tau) =\exp(-\tau H_{(j)})$, where time $t\to i\tau$, $\tau\in[0, \beta]$. Here, $\beta=(k_{B}T)^{-1}$ is the imaginary time period for the evolution operator $U_{(j)}(\tau)=-U_{(j)}(\tau+\beta)$, and the (-) sign is present as we are dealing with fermions~\cite{haag1967}. We now proceed to find the relation between the thermal energy $k_{B}T$ and energy broadening of the quasiparticle. Till step $j$, we have decoupled $N-j$ single particle states labelled $j<l\leq N$. Thus, the time evolution operator attains the form
\begin{eqnarray}
U_{(j)}(\tau) = U_{<j+1}(\tau)\otimes U_{>j}(\tau)~,
\end{eqnarray}  
where $U_{<j+1}(i\tau)=Tr_{N, \ldots, j+1}(U_{(j)}(\tau))$ is the time evolution operator for the coupled states, $U_{>j}=Tr_{1, \ldots, j}(U_{(j)}(\tau))=\exp(-\tau H_{>j})$ is the same for the subspace of decoupled states and $H_{>j}=Tr_{1,\ldots, j}(H_{(j)})$ is the number diagonal Hamiltonian of the decoupled subspace.
\par\noindent
A decomposition of the Hamiltonian among various fluctuation scales, $H_{>j}=\sum_{i}(H_{>j}(\omega_{i})\hat{O}(\omega_{i}))_{(j)}$, then allows us to extract the effective Hamiltonian at a scale $\omega_{i}$ 
\begin{eqnarray}
H_{>j}(\omega_{i}) &=& \sum_{j}\epsilon_{j}(\omega_{i})\hat{n}_{j}+\sum_{jj'}f_{jj'}(\omega_{i})\hat{n}_{j}\hat{n}_{j'}\nonumber\\
&+&\sum_{jj'j''}f_{jj'l'}(\omega_{i})\hat{n}_{j}\hat{n}_{j'}(1-\hat{n}_{l'})\nonumber\\
&+&\sum_{jj'l'l''}f_{jj'j''}(\omega_{i})\hat{n}_{j}\hat{n}_{j'}\hat{n}_{l'}(1-n_{l''})+\ldots~. 
\end{eqnarray}
From $H_{>j}(\omega_{i})$, we can form the evolution operator $U_{>j}(\omega_{i},\tau) = \exp(-\tau H_{>j}(\omega_{i}))$. 
Using the mapping $\hat{n}_{j}\to \tau_{j}=\hat{n}_{j}-\frac{1}{2}$ the Hamiltonian can be decomposed into a irreducible sum of 1-particle, 2-particle, 3-particle etc. self/correlation energies as follows
\begin{eqnarray}
H_{>j}(\omega_{i}) &=& \sum_{j}\tilde{\epsilon}_{j}(\omega_{i})\tau_{j}+\sum_{jj'}\Gamma^{4}_{jj'}(\omega_{i})\tau_{j}\tau_{j'}\nonumber\\&+&\sum_{jj'j''}\Gamma^{6}_{jj'l'}(\omega_{i})\tau_{j}\tau_{j'}\tau_{l'}+\ldots~,\label{effnonintmetal}
\end{eqnarray}
where $\tilde{\epsilon}_{j}(\omega_{i}) =\epsilon_{j}+\Sigma_{j}(\omega_{i})$ with $\Sigma_{j}$ composed of all higher order correlations. $U_{>j}(\omega_{i})$, therefore, can also be decomposed into a product of a evolution operators for 1-particle, 2-particle, 3-particle etc. Hamiltonians (as all terms commute with each other)
\begin{eqnarray}
U_{>j}(\tau, \omega_{i}) =U^{1}_{>j}(\tau, \omega_{i})U^{2}_{>j}(\tau, \omega_{i})\ldots U^{j}_{>j}(\tau, \omega_{i})~,
\end{eqnarray}
where $U^{n}_{>j}(i\beta, \omega) = \exp(-\beta\sum_{j_{1}\ldots j_{n}}\Gamma^{2n}_{j_{1}\ldots  j_{n}}(\omega)\tau_{j_{1}}\ldots \tau_{j_{n}})$ with $n=1$ $\Gamma^{2n}_{j_{1}\ldots j_{n}}=\tilde{\epsilon}_{j}$. For instance, the imaginary time evolution operator $U^{1}_{>j}(\tau,\omega_{i})$ is for the effective single particle Hamiltonian $H^{1}_{>j}(\omega_{i})= \sum_{l=j+1}^{N}\tilde{\epsilon}_{l}(\omega_{i})\tau_{l}$.
\par\noindent
Given that we are decoupling precisely one single-particle state at every RG step using a unitary operation, a thermal scale arises by limiting our perspective to many-body correlations within the single-particle Hamiltonian $H^{1}_{>j}(\omega_{i})$. We can now use the Kubo-Martin-Schwinger condition~\cite{haag1967}
$U^{1}_{>j}(\tau, \omega)=-U^{1}_{>j}(\tau+\beta, \omega)$ to attain a Matsubara spectral representation (where $\tilde{\omega}_{m}=\frac{\pi (2m+1)}{\beta}$ are the harmonics)
\begin{eqnarray}
U^{1}_{>j}(
i\tilde{\omega}_{m}, \omega)& =& \sum_{\beta} e^{\tau\tilde{\omega}_{m}}\tilde{U}^{1}_{>j}(\tau, \omega)\nonumber\\
& = &\sum_{l=j+1}^{N}\frac{|l\rangle\langle l|}{i\tilde{\omega}_{m}-\epsilon_{l}\tau_{l}-\Sigma_{l}(\omega)\tau_{l}}~.
\end{eqnarray}
Here, $\Sigma_{l}(\omega)$ is the frequency-dependent renormalized self-energy and $\epsilon_{l}$ is the bare electronic dispersion. We can define a complex self-energy, $\bar{\Sigma}_{j+1}(\omega) =\Sigma_{j+1}(\omega)-i\tilde{\omega}_{m}$, where $\tilde{\omega}_{m}$ is the Matsubara frequency. As the single-particle states are quantum mechanically decoupled from the rest, any mixedness in the quantum state of the effective non-interacting metal obtained (eq.\ref{effnonintmetal}) can be attributed to a thermal scale $\tilde{\omega}_{0}=2\pi\beta^{-1}$. The Matsubara frequencies $\tilde{\omega}_{m}=\pi(2m+1)  \beta^{-1}$ are defined as the $m$th harmonics of $\beta =1/k_{B}T$. Here we choose $m=0$, i.e., $\tilde{\omega}_{0}=\pi \beta^{-1}$ in order to find the largest temperature scale $T$ upto which the poles will persist. By writing the imaginary part of the self energy as a Kramers-Kronig partner of the real self-energy, we obtain an equivalent temperature scale
\begin{eqnarray}
\frac{\hbar}{\tau}=\frac{1}{\pi}\mathcal{P}\int_{-\infty}^{\infty}\frac{\Sigma_{j+1}(\bar{\omega})}{\bar{\omega} - \tilde{\omega}_{0}}d\bar{\omega}\equiv\hbar\tilde{\omega}_{0}\label{equivalent_Thermal scale}~.~~~~~
\end{eqnarray}
\par\noindent
This temperature scale provides the highest thermal scale upto which the one-particle excitations can survive. Beyond it, they are replaced by 2e-1h composite excitations. The above relation shows the finite lifetime $\tau$ of the single-particle states can be viewed as an effective temperature scale arising out of the unitary disentanglement. A temperature scale for emergent gapped states of matter can be obtained similarly, and will be presented in Sec.~\ref{mottliquid}.
\subsection{Fermi surface instabilities}\label{FSinstab}
As we will now see, the perfect nesting of the square FS (Fig.\ref{Fermi_surface_scatt}) indicates a putative instability of the FS via Umklapp back-scattering, i.e., scattering processes connecting states across the FS via a multiple of the reciprocal lattice vector ($2\mathbf{Q}$). In order to identify the dominant low-energy subspace~\cite{anderson1958random} where the Umklapp \textit{back scattering} processes contribute maximally, we first choose a pair of states, one from distance $\Lambda$ above FS ($\mathbf{k}_{\Lambda\hat{s}}$) and another at a distance $-\Lambda+\delta$ below FS ($\mathbf{k}_{-\Lambda +\delta, \hat{s}'}$). 
The net momentum for such a pair of states is given by
 \begin{eqnarray}
 \mathbf{k}_{\Lambda,\hat{s}}+\mathbf{k}_{-\Lambda +\delta, \hat{s}'} = \mathbf{k}_{F\hat{s}}+\mathbf{k}_{F\hat{s}'}+\Lambda\hat{s}-\Lambda\hat{s}'+\delta\hat{s}'.
\end{eqnarray}   
By summing over the elements of the energy ($\Omega$) dependent transition matrix ($T (\Omega)$) for the Umklapp back-scattering processes $(\hat{s}, \hat{s}')\to (-\hat{s}, -\hat{s}')$, we obtain the net second order $T$-matrix element connecting states on nested surfaces and low energies as
\begin{eqnarray}
\lim_{\Omega\to 0} T^{(2)}_{\hat{s}, \hat{s}'\to -\hat{s}, -\hat{s}'}(\Omega) &=& \frac{1}{vol^{2}}\lim_{\Omega\to 0}\sum^{W}_{\Delta\epsilon^{pair}_{\Lambda, \delta}
(\hat{s})}\frac{U_{0}^{2}}{\Omega - \Delta\epsilon^{pair}_{\Lambda, \delta}
(\hat{s}, \hat{s}')}, \nonumber\\
&=&\frac{U_{0}^{2}}{(vol)^{2}W}\ln\frac{W}{\Delta\epsilon^{pair}_{\Lambda, \delta}
(\hat{s}, \hat{s}')}, ~\label{T-matrix_argument}
\end{eqnarray} 
where $\epsilon_{\Lambda, \hat{s}}=\epsilon_{\mathbf{k}_{\Lambda\hat{s}}}$, $W$ is the bandwidth, $vol$ is the system volume and the energy difference between scattering pairs is denoted by
\begin{equation}
\Delta\epsilon^{pair}_{\Lambda, \delta}
(\hat{s}, \hat{s}') = (\epsilon_{-\Lambda, \hat{s}}+\epsilon_{\Lambda +\delta,  \hat{s}'})-(\epsilon_{-\Lambda, -\hat{s}}+\epsilon_{\Lambda-\delta, -\hat{s}'})~. 
\end{equation}
For pairs positioned symmetrically about the nodal vector $\mathbf{Q}$, $\hat{s}'=T\hat{s}=(s_{y}, s_{x})$, the green lines in Fig.~(\ref{Fermi_surface_scatt}(b)) connect the filled and unfilled green circles with total momenta along $\mathbf{Q}$ given by 
\begin{equation}
(\mathbf{k}_{\Lambda+\delta\Lambda, \hat{s}}+\mathbf{k}_{-\Lambda, T\hat{s}})\cdot\mathbf{Q} = 2\pi +\delta~.
\end{equation}
\par
For such pairs, the $T$-matrix has a leading order logarithmic divergence with the branch cut located along the line $\delta =0$ with $\lim_{\delta\to 0}\Delta\epsilon^{pair}_{\Lambda,\delta}\to 0$. This indicates that the \textit{resonant pairs} (i.e., with wavevector $\delta =0$, placed symmetrically above and below the FS) are more susceptible to the Umklapp back scattering instability compared to their off-resonant ($\delta \neq 0$) counterparts, and will therefore dominate the physics of the Mott insulating state at low energies ($\Omega \to 0$). As in the Kondo problem \cite{anderson1970poor}, such a logarithmic divergence of the $T$-matrix signals the need for a RG treatment of the FS instability. 
\par
A similar instability can be shown due to the \textit{spin backscattering} process of opposite spin electron pairs ($\uparrow$)-hole ($\downarrow$) across the FS. The pair of states labelled by $(\mathbf{k}_{\Lambda+\delta, \hat{s}}, \mathbf{k}_{-\Lambda, T\hat{s}})$ in the electron/hole configurations at distances $\Lambda+\delta, -\Lambda$ (i.e. above/below FS) along normals at $\hat{s}$ and $T\hat{s}$ possess a net pair-dispersion along $\hat{s}$ given by 
\begin{eqnarray}
\Delta E_{\Lambda, \hat{s}, \delta}&=&4t\sin k_{Fx}\sin\frac{\Lambda}{\sqrt{2}}\left[\cos\frac{\delta}{\sqrt{2}}-1\right]~,\label{Fermi_surface_geometry}
\end{eqnarray}
where $s_{x}=s_{y}=\cos\frac{\pi}{4}$, and the square shape of FS gives $k_{Fx}+k_{Fy}=\frac{\pi}{2}$. For every normal direction $\hat{s}$, the bare e-h opposite spin pairs with energy difference $\Delta E_{\Lambda, \hat{s}, \delta}$ undergo backscattering (eq.\eqref{Hubbard Hamiltonian}) across the FS due to the term $-U_{0}c^{\dagger}_{\Lambda, \hat{s}, \sigma}c_{-\Lambda +\delta , T\hat{s}, -\sigma}c^{\dagger}_{\Lambda, -\hat{s}, -\sigma}c_{\Lambda+\delta-2\Lambda, -T\hat{s}, \sigma}$.  At the level of second order perturbation level theory, the energy cost ($E_{corr}(\hat{s})$) associated with such scattering is given by 
\begin{eqnarray}
E_{corr}(\hat{s}, \Lambda, \delta) = \frac{U_{0}^{2}}{\Delta E_{\Lambda, \hat{s}, \delta}}~.\label{spin_backscattering_Corr}
\end{eqnarray}
Summing $E_{corr}(\hat{s},\Lambda,\delta)$ over all $0\leq\Lambda\leq W$ 
gives the associated $T$-matrix contribution with a similar logarithmic singularity for $\delta=0$ pairs as observed above in eq.\eqref{T-matrix_argument}. It is important to note that the expressions for the $T$-matrix elements arising from Umklapp charge backscattering (eq.\eqref{T-matrix_argument}) and spin backscattering (eq.\eqref{spin_backscattering_Corr}) contain information of the Fermi surface geometry (eq.\eqref{Fermi_surface_geometry}). This results in {\it electronic differentiation}: a range of quantum fluctuation scales associated with the instabilities across the FS (i.e., from the AN to the N), one for every $\hat{s}$ normal to Fermi surface. In the next section, we will treat these instabilities via the Hamiltonian Renormalization group procedure eq.\eqref{isogeometric_curve_RG_flow}, as well as identify the parent interacting metallic state of the Mott problem at $1/2$-filling. We will also see that electronic differentiation leads to the nodal-antinodal dichotomy at the heart of the pseudogap phenomenon observed in doped Mott insulators ~\cite{tallon2001doping,imada2013,keimer2015quantum}.
 \subsection{Renormalization group flow equations for Longitudinal and Tangential scattering processes}\label{RG_flow_long_tan}
Here, we treat the instabilities arising from the two-particle scattering processes discussed earlier section via the unitary operator based Hamiltonian RG formalism.
From the discussion leading upto eq.\eqref{isogeometric_curve_RG_flow}, it follows that the operator RG equations for forward ($V_{l}^{(j)}(\delta)$) and backward ($K_{l}^{(j)}(\delta)$) scattering vertices (orange and green arrows in Fig.\ref{Fermi_surface_scatt}(a) and (b) respectively) are given by (details are provided in Appendix \ref{derv_RG}) 
\begin{eqnarray}
\Delta V^{(j)}_{l}(\delta)&=&\frac{4(V^{(j)}_{l}(\delta))^{2}\hat{\tau}_{j, l}\hat{\tau}_{j, l'}}{[G_{j, l}]^{-1}-V^{(j)}_{l}(\delta)\hat{\tau}_{j, l}\hat{\tau}_{j, l'}},\nonumber\\
\Delta K^{(j)}_{l}(\delta)&=&\frac{4V^{(j)}_{l}(\delta)K^{(j)}_{l}(\delta)\hat{\tau}_{j, l}\hat{\tau}_{j, l'}}{[G_{j, l}]^{-1}-V_{l}^{(j)}(\delta)\hat{\tau}_{j, l}\hat{\tau}_{j, l'}},\nonumber\\
\left[G_{j, l}\right]^{-1}&=&\hat{\omega}-\tilde{\epsilon}_{j, l}\hat{\tau}_{j, l}-\tilde{\epsilon}_{j, l'}\hat{\tau}_{j, l'}~,\label{RG_flow_eqn_FS_BS}
\end{eqnarray}
where $(j, l')\equiv(\mathbf{k}_{-\Lambda_{j}+\delta, \hat{s}'}, -\sigma$) and $\hat{\tau}_{j, l}=\hat{n}_{j, l}-\frac{1}{2}$, and $\tilde{\epsilon}_{j, l}=\epsilon_{j, l}-\Delta\mu_{eff}$ is the electronic dispersion measured with respect to the effective chemical potential measured with respect to $1/2$-filling ($\Delta\mu_{eff}$). 
From the RG eqns.\eqref{RG_flow_eqn_FS_BS}, we obtain a RG invariant $C$ that characterises the flows
\begin{equation}
CK_{l}^{(j)}(\delta)=V_{l}^{(j)}(\delta)~.
\end{equation}
The uniform magnitude of the bare scattering vertices in the Hubbard Hamiltonian (eq.\eqref{Hubbard Hamiltonian}), $K_{l}^{(N)}(\delta)=V_{l}^{(N)}(\delta)=\bar{U}_{0}$~($\bar{U}_{0}\equiv U_{0}/\sqrt{vol}$), fixes the RG invariant to $C=1$. 
Fig.~\ref{three_scattering_pathways} (a) and Fig.~\ref{three_scattering_pathways}(b) represent the renormalization contributions of the two-particle scattering vertices in the electron-electron ($\hat{\tau}_{i}$'s with eigenvalue $\frac{1}{2}$, eq.\eqref{RG_flow_eqn_FS_BS}) and electron-hole ($\hat{\tau}_{i}$'s with eigenvalue $\pm\frac{1}{2}$, eq.\eqref{RG_flow_eqn_TS}) intermediate configuration channels respectively. 
\par\noindent
Interaction vertices involving tangential scattering are denoted as $L^{(j)}(\delta)$ (brown arrows in Fig.~\ref{Fermi_surface_scatt}(a)). For tangential scattering processes, the intermediate state configuration necessarily involves electronic states on the entire isogeometric curve, i.e., the various many-body configurations obtained for a collective density operator $L^{z}_{j}= 2^{-1}\sum_{l}(\hat{n}_{j, l}+\hat{n}_{j, l'}-1)$. The scattering of a collective configuration of electronic states on the isogeometric curve is described by pairwise electron raising lowering operators $L^{+}_{j}=\sum_{l}c^{\dagger}_{j, l}c^{\dagger}_{j, l'}$ and $L^{-}_{j}$ operators. Following the appendix \ref{derv_RG}, we can write down the operator RG equations for the tangential scattering processes as
\begin{eqnarray}
\Delta L^{(j)}   = \frac{(L^{(j)})^{2}(L^{2}_{j}-L^{z2}_{j}-L^{z}_{j})}{\hat{\omega}-\tilde{\epsilon}^{c}_{j,avg}L^{z}_{j}-L^{(j)}L^{z2}_{j}}.\label{RG_flow_eqn_TS}
\end{eqnarray}
where 
$\tilde{\epsilon}^{c}_{j, avg}=N_{j}^{-1}\sum_{l}(\epsilon_{j, l}+\epsilon_{j, l'}-2\Delta\mu_{eff})$ is the mean kinetic energy of the occupied pair of electrons along a isogeometric curve. $N_{j}$ is the number of electronic states on an isogeometric curve.
\subsection{Mixing between e-e and e-h configurations in the RG procedure}\label{mixing}
\begin{figure}
\includegraphics[scale=0.4]{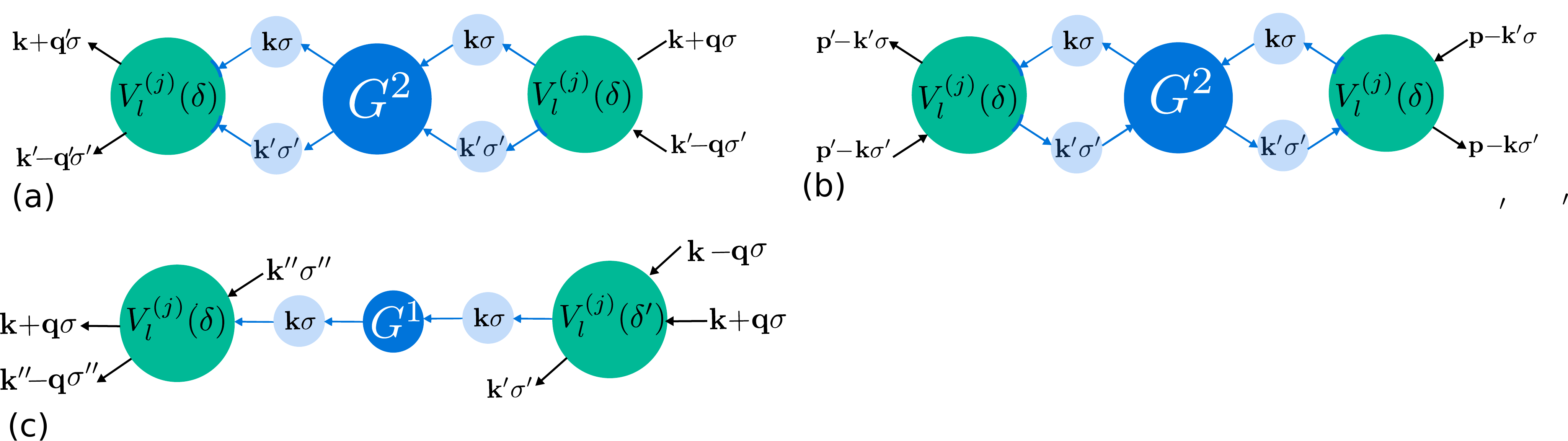} 
\caption{(Colour Online) Figures
(a) and (b) Renormalization of the forward scattering processes in the (ee/hh- grey bordered dumbbell) and (eh/he- red bordered dumbbell ) configurations respectively of the states $\mathbf{k}\sigma$ and $\mathbf{k}'\sigma'$. The intermediate 4-point diagonal propagator containing the diagonal interaction piece (eq.\eqref{RG_flow_eqn_FS_BS}) is represented by the blue circles. (c) 6-point vertices generated out of two different off-resonant pair scattering terms ($\mathbf{k}\sigma$, $\mathbf{k}'\sigma'$), ($\mathbf{k}\sigma, \mathbf{k}''\sigma'')$ sharing a common electronic state.
}\label{three_scattering_pathways}
\end{figure}

In Fig.~\ref{three_scattering_pathways}(c), we observe that mixing between various electron-electron and electron-hole scattering terms (i.e., the ee/hh and eh/he pairs shown in Fig.~\ref{three_scattering_pathways}(a, b)) leads to three-particle (or six-point) scattering vertices. This is an outcome of the non-commutativity between the composite electron creation operator $(1-\hat{n}_{\mathbf{k}\sigma})c^{\dagger}_{\mathbf{k}'\sigma'}$ and the ee/eh pair creation operators\cite{anderson1958random} $c^{\dagger}_{\mathbf{k}\sigma}c^{\dagger}_{\mathbf{k}'\sigma'}$ and $c^{\dagger}_{\mathbf{k}\sigma}c_{\mathbf{k}'\sigma'}$ operators. The operator RG flow equation for the three-particle scattering vertex ($R_{l, \delta\delta'}^{(j)}$) is given by (see also 
eq.\eqref{Three-particle-vertex-derv-RG}) 
\begin{eqnarray}
\Delta R_{l, \delta\delta'}^{(j)} &=& \frac{V^{(j)}_{l}(\delta)V^{(j)}_{l}(\delta')}{\hat{\omega} - \tilde{\epsilon}^{(j)}_{j, l}\hat{\tau}_{j, l}}+\frac{K^{(j)}_{l}(\delta)K^{(j)}_{l}(\delta')}{\hat{\omega} - \tilde{\epsilon}^{(j)}_{j, l}\hat{\tau}_{j, l}}\nonumber\\
&+&\sum_{\Lambda'<\Lambda_{j}}\frac{8R_{l, \delta\delta''}^{(j)}R_{l,\delta''\delta'}^{(j)}}{\left[G_{j, l, 3}\right]^{-1}-R_{l, \delta''\delta''}^{(j)}\prod_{i=1}^{3}\hat{\tau}_{i}}~,\nonumber\\
\left[G_{j, l, 3}\right]^{-1}&=&\hat{\omega}-\tilde{\epsilon}_{j, l}\hat{\tau}_{j, l}-\tilde{\epsilon}_{j, l''}\hat{\tau}_{j, l''}-\tilde{\epsilon}_{j', l},
\hat{\tau}_{j', l}~,\hspace*{-1cm}\label{six-point flow}
\end{eqnarray}
where the collective index $i$ in the e-h imbalance operator $\hat{\tau}_{i}$ given by $i=1:(\Lambda_{j},\hat{s},\sigma$), $i=2:(-\Lambda_{j}+\delta'', T\hat{s},-\sigma$) and $i=3:(\Lambda',\hat{s},\sigma$).
We note that eq.\eqref{six-point flow} contains contributions from longitudinal forward/backward scattering terms, while that from tangential scattering is absent. This is because 
contributions from the latter to the renormalization of three-particle terms is subdominant, owing to the maximal entanglement present in the intermediate configuration ($m=0$ in eq\eqref{tang_configurations}). 
\par\noindent
In order to see the effects of the mixing of ee/hh (charge channel) and eh/he (spin channel) terms on the RG eqns.\eqref{RG_flow_eqn_FS_BS} for longitudinal scattering,  
we perform a $\omega$-dependent rotation, $\tan^{-1}(\sqrt{\frac{1-p}{p}})$, in the space of the electron/hole configurations of $(\mathbf{k}', -\sigma)=(j, l')$, where $\mathbf{k}'= \mathbf{k}_{-\Lambda_{j} +\delta, T\hat{s}}$
\begin{eqnarray}
|\Uparrow\rangle := |1_{\mathbf{k}\sigma}\psi_{\mathbf{k}' -\sigma}\rangle =\sqrt{p}|1_{\mathbf{k}\sigma}1_{\mathbf{k}'-\sigma}\rangle+\sqrt{1-p}|1_{\mathbf{k}\sigma} 0_{\mathbf{k}'-\sigma}\rangle~,~~\label{ee_eh_mixed_configuration}\\
|\Downarrow\rangle := |0_{\mathbf{k}\sigma}\psi^{\perp}_{\mathbf{k}' -\sigma}\rangle =\sqrt{1-p}|1_{\mathbf{k}\sigma} 0_{\mathbf{k}'-\sigma}\rangle -\sqrt{p}|0_{\mathbf{k}\sigma}0_{\mathbf{k}'-\sigma}\rangle ~.~~
\label{hh_he_mixed_configuration}
\end{eqnarray}
the spin-charge mixing parameter $0\leq p\leq 1$
is determined as a function of $(\omega,\Delta\mu_{eff})$ by maximizing the two-particle Greens function $G_{j, l}$'s matrix element in the ee/eh mixed pair configuration eq.\eqref{ee_eh_mixed_configuration} for all energies $(\tilde{\epsilon}_{j, l},\tilde{\epsilon}_{j, l'})$
\begin{eqnarray}
G_{j, l}^{p', \Uparrow}&=&\langle 1_{j, l}\psi_{j, l'}|G_{j, l}|1_{j, l}\psi_{j, l'}\rangle\nonumber\\
& = &[\omega - p\left(\frac{\epsilon_{j, l}+\epsilon_{j, l'}}{2}-\Delta\mu_{eff}\right)-(1-p)\frac{\tilde{\epsilon}_{j, l}-\tilde{\epsilon}_{j, l'}}{2}]^{-1}\label{Green_func_long},\nonumber\\
p:=p'&&\text{~s.t.~~~}G_{j, l}^{p, \Uparrow} =\smash{\displaystyle\max_{p'}}~G_{j, l}^{p', \Uparrow}\label{determining p}~.~
\end{eqnarray}
This special value of the parameter $p$ in eq\eqref{determining p} in turn causes the maximization of the 2-particle vertex RG flows, ensuring their domination over the 3-particle off-diagonal vertex RG flows.
\par 
In the (ee \& eh)/(hh \& he) opposite-spin pair configurations $|1_{j, l}\psi_{j, l'}\rangle$ (see eq.\eqref{ee_eh_mixed_configuration}), the RG flow equations for longitudinal forward/backward scattering vertices (eq.\eqref{RG_flow_eqn_FS_BS}) of the charge ($(V_{c}/K_{c})_{l}^{(j)}$) and spin ($(V_{s}/K_{s})^{(j)}_{l}$) kinds in the presence of spin-charge mixing are given by 
\begin{eqnarray}
\frac{\Delta V_{c, l}^{(j)}(\delta)}{\Delta \log\frac{\Lambda_{j}}{\Lambda_{0}}}&=& \frac{p(V_{c, l}^{(j)}(\delta))^{2}}{e^{i\gamma_{l}^{\Downarrow}}|G^{p, \Downarrow}_{j, l}|^{-1}-\frac{V_{p, l}^{(j)}(\delta)}{4}}, \nonumber\\
\frac{\Delta K_{c, l}^{(j)}(\delta)}{\Delta \log\frac{\Lambda_{j}}{\Lambda_{0}}}&=& \frac{p(K_{c, l}^{(j)}(\delta))^{2}}{e^{i\gamma_{l}^{\Uparrow}}|G_{j, l}^{p, \Uparrow}|^{-1}-\frac{K_{p, l}^{(j)}(\delta)}{4}}, \nonumber\\
\frac{\Delta V_{s, l}^{(j)}(\delta)}{\Delta \log\frac{\Lambda_{j}}{\Lambda_{0}}}&=& -\frac{(1-p)(V_{s, l}^{(j)}(\delta))^{2}}{e^{i\gamma_{l}^{\Downarrow}}|G_{j, l}^{p, \Downarrow}|^{-1}-\frac{V_{p, l}^{(j)}(\delta)}{4}}, \nonumber\\
\frac{\Delta K_{s, l}^{(j)}(\delta)}{\Delta \log\frac{\Lambda_{j}}{\Lambda_{0}}}&=& -\frac{(1-p)(K_{s, l}^{(j)}(\delta))^{2}}{e^{i\gamma_{l}^{\Uparrow}}|G_{j, l}^{p, \Uparrow}|^{-1}-\frac{K_{p, l}^{(j)}(\delta)}{4}}~,
\label{Long RG equations with doping}
\end{eqnarray}
where $V_{p,l}$ and $K_{p,l}$ are strengths for forward and backward interaction couplings respectively with spin-charge mixing 
\begin{eqnarray}
V_{p, l}^{(j)}(\delta) &=& pV^{(j)}_{c, l}(\delta)-(1-p)V^{(j)}_{s, l}(\delta)~,\nonumber\\
K_{p, l}^{(j)}(\delta) &=& pK^{(j)}_{c, l}(\delta)-(1-p)K^{(j)}_{s, l}(\delta)~,
\label{spin_charge_interaction_strength}
\end{eqnarray}  
and $\Lambda_{j}=\Lambda_{0}e^{-j}$~,~ $\Delta\ln\frac{\Lambda_{j}}{\Lambda_{0}} =1$. 
Further, $\gamma^{(\Uparrow, \Downarrow)}_{\hat{s}}(\omega)$ is the topological phase of the Green function eq.\eqref{Green_func_long} 
\begin{eqnarray}
\gamma^{\Uparrow, \Downarrow}_{l}(\omega):=e^{i\pi(N^{\Uparrow, \Downarrow}_{l}(\omega)+1)},\label{top_phase}
\end{eqnarray}
with the topological invariant~\cite{volovik2009universe} 
\begin{equation}
N^{\Uparrow, \Downarrow}_{l}(\omega)=\oint dz [\mathcal{G}_{\Uparrow/\Downarrow}^{(j, l)}]^{-1}\partial_{z}\mathcal{G}_{\Uparrow/\Downarrow}^{(j, l)}~,~ \label{Top_term}
\end{equation}
where $[\mathcal{G}_{\Uparrow/\Downarrow}^{(j, l)}]^{-1} =z-[\hat{G}_{p, (\Uparrow/\Downarrow)}^{(j, l)}]^{-1}$. These topological invariants are constrained by the relation 
\begin{equation}
N^{\Uparrow}_{l}(\omega)+N^{\Downarrow}_{l}(\omega)=1~,\label{constraint}
\end{equation}
such that a RG relevant forward scattering coupling ensures an RG irrelevant backward scattering coupling, and vice versa.  
\par\noindent
We now discuss briefly the choices made for the appearance of certain Greens functions above in the RG equations eq\eqref{Long RG equations with doping}. The pair of electronic states $|j, l\rangle : =|\mathbf{k}_{\Lambda_{j}, \hat{s}}\sigma\rangle$ and $|j, l'\rangle:=|\mathbf{k}_{-\Lambda_{j}+\delta, \hat{s}}-\sigma\rangle$ (for $\delta>0$) in the $|\Uparrow\rangle$ configuration (eq.\eqref{ee_eh_mixed_configuration}) have a net spin-charge hybridized energy: $p(\epsilon_{j, l}+\epsilon_{j, l'})+(1-p)(\epsilon_{j, l}-\epsilon_{j, l'})>0$, as $\epsilon_{j, l}+\epsilon_{j, l'}>0$ for $\epsilon_{j, l}>0, \epsilon_{j, l'}<0$. The poles of the Greens function for $\Uparrow$ are situated along the $\omega>0$, and act as intermediate channels for the renormalisation of spin and charge backscattering vertices. On the other hand, for the forward scattering channel along a given $\hat{s}$, the intermediate configuration ($\Downarrow$) is chosen with a net energy below $E_{F}$, such that the pole of the Greens function lies along $-\omega$. These channels have been chosen in such a way that a relevant renormalisation of the backscattering vertices ($K_{c,l}^{(j)}(\delta),K_{s,l}^{(j)}(\delta)$) in intermediate configuration $\Uparrow$ is associated with an irrelevant renormalisation of the forward scattering vertices ($V_{c,l}^{(j)}(\delta),V_{s,l}^{(j)}(\delta)^{(j)}$) in intermediate configuration $\Downarrow$, and vice-versa. The constraint on the topological invariants in eq.\eqref{constraint} is simply a manifestation of this choice of the intermediate channels.
\par\noindent
From the longitudinal scattering vertex flow eq.\eqref{Long RG equations with doping}, we get the RG invariant ($C$)
\begin{eqnarray}
\int \frac{dV^{(j)}_{s, l}(\delta)}{(V^{(j)}_{s, l}(\delta))^{2}} &=&-\frac{1-p}{p}\int \frac{dV^{(j)}_{c, l}(\delta)}{(V^{(j)}_{c, l}(\delta))^{2}},\nonumber\\
C&=& [pV^{(j)}_{s, l}(\delta)]^{-1}+[(1-p)V^{(j)}_{c, l}(\delta)]^{-1}\nonumber\\
\int \frac{dK^{(j)}_{s, l}(\delta)}{(K^{(j)}_{s, l}(\delta))^{2}} &=&-\frac{1-p}{p}\int \frac{dK^{(j)}_{c, l}(\delta)}{(K^{(j)}_{c, l}(\delta))^{2}},\nonumber\\
C&=& [pK^{(j)}_{s, l}(\delta)]^{-1}+[(1-p)K^{(j)}_{c, l}(\delta)]^{-1},~~~
\label{RG_invariant_spin_charge_long_scatt}
\end{eqnarray}
Given the form of the Hubbard interaction in eq.\eqref{Hubbard Hamiltonian}, the bare values of various couplings are: $V^{(N)}_{s, l}=V^{(N)}_{c, l}=K^{(N)}_{s, l}=V^{(N)}_{c, l}=\bar{U}_{0}$. We obtain, therefore, the value of the RG invariant as $C^{-1}=p(1-p)\bar{U}_{0}$. 
\par\noindent
The tangential scattering processes can involve the following class of intermediate state configurations
\begin{eqnarray}
|L_{j} =m-\frac{1}{2}\rangle &=& \sum_{\mathcal{C}}\frac{1}{\sqrt{\binom{2N_{j}}{N_{j}-m}}}\prod_{i=1}^{N_{j}-m}|0_{l_{i}}0_{l'_{i}}\rangle_{j}\nonumber\\
&&\prod_{i=N_{j}-m+1}^{2N_{j}}|1_{l_{i}}1_{l'_{i}}\rangle_{j}~,\label{tang_configurations}
\end{eqnarray}
where the quantum number $m$ indicates the number of electronic states on the isogeometric curve. These states are coupled by tangential scattering, such that the lower the magnitude of $|m|$, the more highly entangled is the state $|L_{j} =m-\frac{1}{2}\rangle$. This can be seen from the fact that a larger number of configurations enter into superposition with a decreasing magnitude of $m$, i.e., $\binom{2N_{j}}{N_{j}}>\binom{2N_{j}}{N_{j}-m}$ for all $m\neq 0$.  
\par\noindent 
The RG flow equation of the tangential scattering vertices can be found from the operator equation eq.\eqref{RG_flow_eqn_TS} for the configuration given above in eq.\eqref{tang_configurations} 
\begin{eqnarray}
\Delta L^{(j)}&=&\frac{(N_{j}+m)(N_{j}-m+1)(L^{(j)})^{2}}{\omega +W\text{sgn}(\Delta\mu_{eff}) +\tilde{\epsilon}^{c}_{\Lambda_{j}, avg}-\frac{1}{4}L^{(j)}}~,\label{tang_scatt_RG_flows}
\end{eqnarray}
where $\tilde{\epsilon}^{c}_{j,avg}=N_{j}^{-1}\sum_{l}(\epsilon_{j,l}+\epsilon_{j,l'}-2\Delta\mu_{eff})$ is the average kinetic energy of the electrons on the high-energy isogeometric curve. The eigenvalues of $L^{2}_{j}$ and $L^{z2}_{j}$ are $=N_{j}(N_{j}+1)$ and $m^{2}$ respectively. We observe that the highly entangled $m=0$ configuration maximizes the RG flow rate in eq.\eqref{RG_flow_eqn_TS}. This indicates that due to the rich entanglement structure of the state with $m=0$, the breaking of an electronic configuration with off-resonance 
pairs is unfavourable under RG.
The value of the fluctuation operator scale $\omega$ is given by $\omega +W\text{sgn}(\Delta\mu_{eff})$, where $W=8t$ is the single-particle bandwidth. This can be argued as follows. For $\Delta\mu=0$, beyond a minimum value of $\Delta\mu^{min}_{eff}=-\Delta U^{max}_{0}/2=-W$, the tight-binding band has only holes with a Fermi surface shifted to the BZ center $\mathbf{k}=(0, 0)$. As the low energy off-diagonal tangential scattering processes ($L^{(j)}(\delta)$) cause fluctuations of the minimum hole energy $E^{min}_{hole}=-(W/2)\times 2=-\Delta\mu^{min}_{eff}$, the correct energy scale for quantum fluctuations is now given by $\omega +W\text{sgn}(\Delta\mu_{eff})$.
\par\noindent
Gapless parts of the FS neighbourhood are characterised by back-scattering ($Q_{1}=\pi$ in eq.\eqref{Long RG equations with doping}) being RG irrelevant, but with forward scattering ($Q_{1}=0$ in eq\eqref{Long RG equations with doping}) being RG relevant. The low lying excitations on such gapless stretches of the FS are strongly influenced by the RG flow equations of the three-particle scattering vertices in eq.\eqref{six-point flow}. We choose the 2 electron-1 hole intermediate configuration for the three states in the neighbourhood of the FS as follows 
\begin{eqnarray}
&&1:(0_{j, l}, 0_{j, l'})~~,~~2:(0_{j, l},1_{j, l''})~~,~~3:0_{0, l}~,\label{three particle intermediate config}  
\end{eqnarray}  
where $(\mathbf{k}_{\Lambda_{j}\hat{s}}, \sigma)\equiv j, l$, $(\mathbf{k}_{-\Lambda_{j}+\delta'', T\hat{s}}, -\sigma)\equiv j, l''$, $\mathbf{k}_{0\hat{s}\sigma}\equiv 0, l$, i.e., states labelled $1$ and $2$ are in the hh/he mixed configuration with energies below $E_{F}$ (see eq.\eqref{hh_he_mixed_configuration}), and the occupied state labelled $3$ is precisely at $E_{F}=0$, such that $\Lambda'=0$ in eq.\eqref{six-point flow}). For such an intermediate configuration, we obtain the flow equation for the three-particle scattering vertices as
\begin{eqnarray}
\Delta R^{(j)}_{\hat{s}, \delta\delta'} &=& \frac{V^{(j)}_{l}(\delta)V^{(j)}_{l}(\delta')}{\omega - \tilde{\epsilon}^{(j)}_{j, l}}+\frac{K^{(j)}_{l}(\delta)K^{(j)}_{l}(\delta')}{\omega - \tilde{\epsilon}^{(j)}_{j, l}}\nonumber\\
&+&\frac{R^{(j)}_{l, \delta\delta''}R^{(j)}_{l, \delta''\delta'}}{\left[G_{j, l, 3}\right]^{-1}+\frac{1}{8}R^{(j)}_{l, \delta''\delta''}},~\nonumber\\
\left[G_{j, l, 3}\right]^{-1}&=&\frac{1}{2}(p\frac{\tilde{\epsilon}_{j, l}+\tilde{\epsilon}_{j, l'}}{2}+(1-p)\frac{\tilde{\epsilon}_{j, l}-\tilde{\epsilon}_{j, l'}}{2})-\omega~.~~\label{six_point_RG_flows}
\end{eqnarray}
It is easily seen that the choice of $\Lambda'=0$, together with the extremal choice of the mixing parameter $p$ in eq.\eqref{determining p}, maximises the 2e-1h contribution to the above RG equation. 
\par
We now mention some other salient features of this RG formulation. First, the effective Hamiltonian at a given RG step can be formulated, with contributions from longitudinal (forward and backscattering, eq.\eqref{Long RG equations with doping}), tangential (eq.\eqref{tang_scatt_RG_flows}) and three-particle diagonal and off-diagonal scattering vertices (eq\eqref{six_point_RG_flows}). The detailed form of the effective Hamiltonian is shown in Appendix \ref{renm_H}. Next, the configuration energy for an e-h/e-e intermediate pair (see discussion below eq.\eqref{Green_func_long}) 
\begin{equation}
\tilde{\epsilon}_{j,l}-\tilde{\epsilon}_{j, l'} = 2\sin k_{Fx}\left[\sin\frac{\Lambda+\delta}{\sqrt{2}}+\sin\frac{\Lambda}{\sqrt{2}}\right],~\delta>0
\end{equation}
is minimum for \textit{resonant pairs} ($\delta =0$). This leads to the propagator for $\delta =0$ resonant pairs ($|\hat{G}^{j, l}_{p, \Downarrow}|$,) having the highest magnitude in the RG equations for longitudinal scattering (eq.\eqref{Long RG equations with doping}). In turn, this leads to the smallest denominators in these RG relations, ensuring that the resonant pairs dominate the RG flows for longitudinal scattering vertices. 
\par
Further, fixed points of the RG flows equations for longitudinal, tangential and three-particle vertices (eqns.\eqref{RG_flow_eqn_FS_BS}, \eqref{RG_flow_eqn_TS} and \eqref{six_point_RG_flows} respectively) are associated with the vanishing of their respective denominators: attaining a stable fixed point is related to the vanishing of quantum fluctuations such that no further decoupling of states can be carried out under the RG transformations (see eq.\eqref{quantum_fluc_switch_off})~\cite{Glazek2004}. Given that the resonant pairs dominate the RG flows, the spectral weight (characterised by the final distance from the FS, $\Lambda^{*}(\delta, \hat{s}, \omega)$) is also the highest at a RG fixed point for such pairs
\begin{eqnarray}
\Lambda^{*}(\delta = 0, \hat{s}, \omega)>\Lambda^{*}(\delta > 0, \hat{s}, \omega).\label{highest_spectral_weight}
\end{eqnarray} 
As can be seen from eqns.\eqref{RG_flow_eqn_FS_BS}, the resonant pairs carry the highest spectral weight in the backscattering process ($Q_{1}=\pi$). 
In this way, we have provided a RG-based justification for the backscattering T-matrix argument given earlier (eq.\eqref{T-matrix_argument}). Finally, the RG equations can be solved numerically in an iterative manner on a two-dimensional  momentum-space grid, leading to fixed point values of various couplings, spectral weights and gaps (the details of the algorithm for which are presented in Appendix-\ref{AlgoSimRG}). From these, we can draw a RG phase diagram, as well as compute several physical observables. In the following sections, we adopt this procedure in unveiling the physics of the $T=0$ Mott-Hubbard transitions at and away from $1/2$-filling.
\begin{figure}[h!]
\hspace*{-0.5cm}\includegraphics[width=0.5\textwidth]{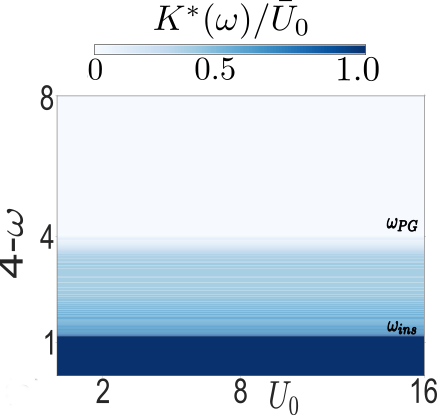}
\caption{(Colour Online) (a) Renormalisation group phase diagram at $1/2$-filling in the quantum fluctuation energyscale ($4-\omega$)-bare repulsion ($U_{0}$) plane. Colourbar represents ratio of renormalized coupling to bare coupling $K^{*}/\bar{U}_{0}$. Transition from non-Fermi liquid (NFL, white) to Mott liquid (ML, dark blue) insulator is through a pseudogap (PG, shaded blue) for all $U_{0}>0$. $\omega_{PG}$ and $\omega_{ins}$ are energy scales for Lifshitz transitions that initiate and end the PG respectively. (b) Schematic representation of  shells of states (black lines parallel to, and formed around, the FS (red line) with spacing $\Lambda$) that are integrated out from first quadrant of Brillouin zone (BZ):~$0<k_{x}, k_{y}<\pi$. Inset (top right): Umklapp scattering of electron pairs. Inset (bottom left): variation in density of states from antinode (AN) to node (N). The pair of black dots represents electron pair pseudospin for charge/spin excitations in the up orientation formed around the node (cyan dot); white dots represents the opposite orientation. $\hat{s}$ represents the direction normal to FS.}
\label{half_filled_phase_diag} 
\end{figure}
\begin{figure}
\includegraphics[width=0.45\textwidth]{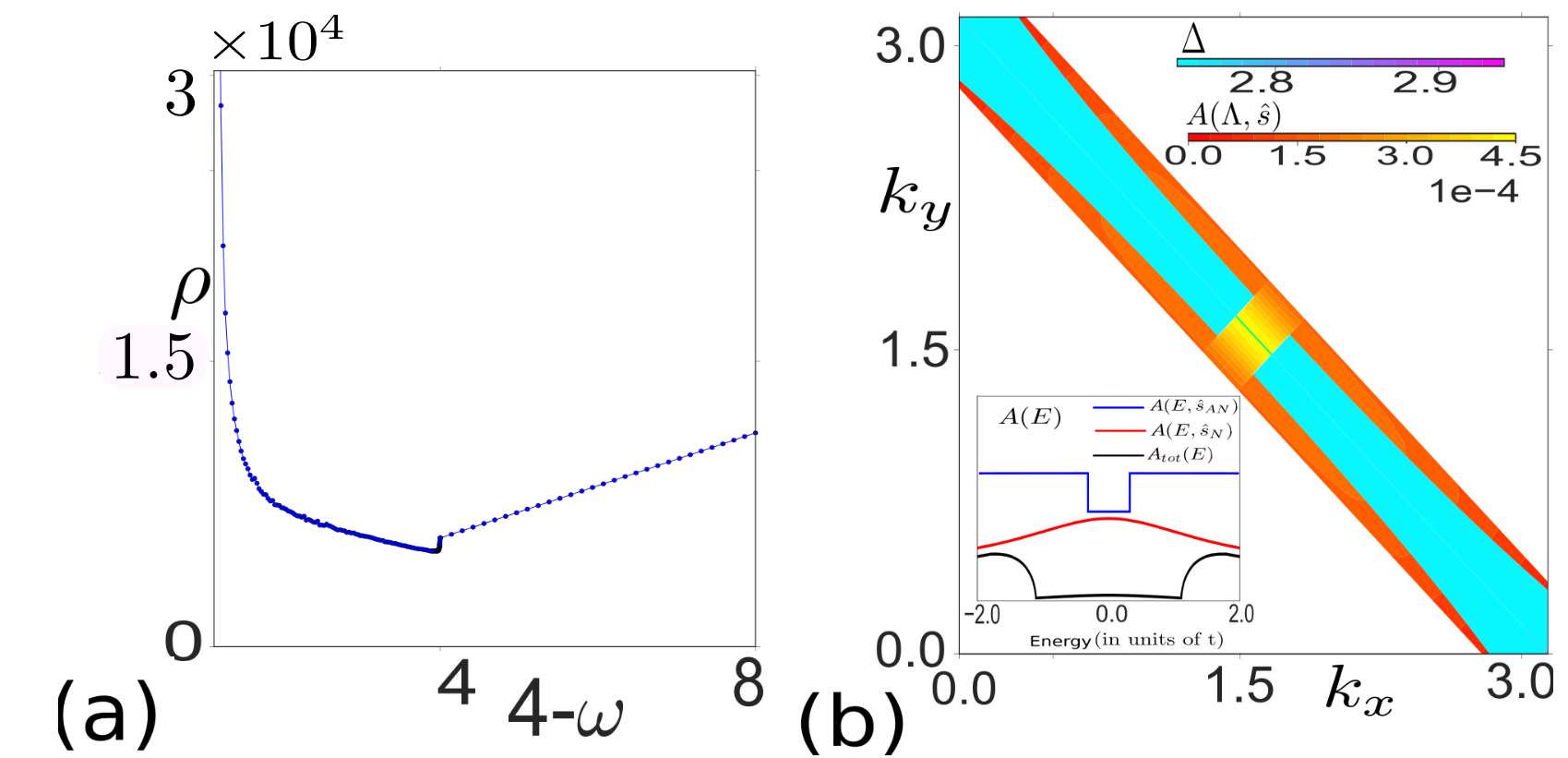}
\caption{(Colour Online)(a) Linear variation of resistivity ($\rho$) with $\omega<\omega_{PG}$ in NFL crosses over through PG ($\omega_{PG}<\omega<\omega_{ins}$) into the ML (diverging $\rho$ for $\omega>\omega_{ins}$).(b) $A(E)$ (colourbar: red to yellow) and $\Delta$ (colourbar: cyan to violet) in first quadrant of BZ with changing $\omega=2.5$. Inset: Quasiparticle (qp) spectral function ($A(E)$) for $\omega=2.5$, showing NFL at N (red curve) and gap at AN (blue curve) and averaged over FS ($A_{tot}$, black curve).}\label{passage_through_half_fill_ph_diag}
\end{figure}
\section{Mott MIT at $1/2$-filling}\label{NFLSection} 
The $T=0$ phase diagram obtained by integrating the RG equations set out in the previous section at $1/2$-filling ($\mu^{0}_{eff}=0$) is shown in Fig.~\ref{half_filled_phase_diag}. Prior to the detailed discussions that will follow, we outline the key aspects displayed in the RG phase diagram. First, an explanation of the axes: the $y$-axis represents the energy scale for quantum fluctuations discussed earlier, i.e., the eigenvalues of fluctuation operator $\omega$ (eq.\eqref{quantum fluctuation scale}) 
\begin{equation}
\frac{W}{2}-\omega\in [0, W=8t]~~(t=1)~,\label{fluctuation_scale_span}
\end{equation}
and the x-axis represents the bare value of the on-site Hubbard coupling ranging from weak to strong coupling ($0<U_{0}\leq 16=2W$). A striking observation is that the Mott metal-insulator transition (MIT) involves the passage from a gapless metallic normal state at high $\omega$ to a gapped insulating Mott liquid (ML) groud state at low $\omega$, but through a pseudogapped (PG) state of matter (at intermediate values of $\omega$) arising from a differentiation of electrons based on the monotonic variation of their kinetic energy (see, e.g., eq.\eqref{spin_backscattering_Corr}) from node (N) to antinode (AN) ~\cite{imada2010unconventional,*imada2010unconventional0}.
\par 
The PG phase is described by partial gap in the neighbourhood of AN, with a gapless stretch centered around N. The gapping process is initiated at the ANs as an FS topology-changing Lifshitz transition of the normal phase at fluctuation scale $\omega=\omega_{PG}\equiv 0.034t$, and proceeds until the Ns are gapped out in a Mott liquid state via a second Lifshitz transition at $\omega = \omega_{ins}$ (Fig.~\ref{passage_through_half_fill_ph_diag}(b), Video S1). We develop the RG fixed point theory for the normal phase in this section. The next section is devoted to the PG phase and the Lifshitz transition leading to it. 
We complete our discussion for the Mott MIT at $1/2$-filling by by focusing on the Mott liquid in detail in a subsequent section. It is also worth noting the flatness of the phase boundaries: this indicates the absence of a critical $(U/t)_{c}$ for the metal-Mott insulator transition for the 1/2-filled Hubbard model on the 2d square lattice with only nearest neighbour hopping, and results from the perfectly nested FS~\cite{anderson-advphys-1997, moukouri2001absence}. This is consistent with recent D$\Gamma$A and quantum Monte Carlo simulations of the unfrustrated Hubbard model by Schafer et. al.~\cite{schafer2015}. We anticipate the presence of a critical $(U/t)_{c}$ in the generalised Hubbard model with a frustrating additional next nearest neighbour hopping, as has been demonstrated in dynamical mean-field theory (DMFT) studies~\cite{rozenberg1994,georges1996}. 
\par\noindent
The $U_{0}$ independent gapping of the antinodes can be anticipated from the divergence of the second order T-matrix element~eq.\eqref{T-matrix_argument} for both resonant ($\delta =0$) as well as off-resonant ($\delta\neq 0$). This results from the vanishing of the energy transfer at the antinodes $k_{x},k_{y}=\pi,0$ and $0,\pi$ (as can be seen from eq.\eqref{Fermi_surface_geometry}) 
at $\Omega=0$ and the existence of van Hove singularities of the DOS at the antinodes. Thus, this event marks the onset energy scale of the pseudogap ($\omega_{PG}$). On the other hand, along the nodal direction $k_{x}=k_{y}=\pi/2$, the energy transfer is $\Delta E = 4t\sin\frac{\Lambda}{\sqrt{2}}\left[\cos\frac{\delta}{\sqrt{2}}-1\right]\neq 0$ (eq.\eqref{Fermi_surface_geometry}). This lowers considerably the gapping of the nodal points on the FS, and therefore the onset energy scale for the Mott insulator ($\omega_{ins}$). This is indicated by the fact that only the resonant ($\delta= 0$) scattering events
along the nodal direction contribute to the T-matrix element in a divergent manner at $\Omega\to 0$ (eq.\eqref{T-matrix_argument}). Further, this divergence is again $U_{0}$ independent.
These arguments show that the Fermi surface topology-changing events, i.e., the disconnection of the antinodes at $\omega_{PG}$ and the vanishing of the nodal arcs at $\omega_{ins}$, are both $U_{0}$ independent and are only related to the geometry of the underlying lattice.
\subsection{Normal state for the Mott insulator}\label{MottMetal}
In charting the physics of the normal state from which the gapped Mott insulator arises, we will carry out the RG analysis in two parts. The first part of RG involves revealing the normal state lying farthest away from the FS of the non-interacting tight-binding problem. For this, we note that in the quantum fluctuation range
\begin{eqnarray}
\frac{W}{2}-\omega\in \left[\left(\frac{W}{2}-\omega_{PG}=\frac{W}{2}\right), W\right]~, \label{NFL_cond}
\end{eqnarray}
all the backscattering vertices that can lead to instabilities of the FS (eq.\eqref{T-matrix_argument}) are found to be RG irrelevant. This arises from the global topological index, $N^{\Downarrow}_{\hat{s}}(\omega)=1~\forall \hat{s}$, near $E_{F}$ in eq.\eqref{Long RG equations with doping}. The forward scattering coupling is, on the other hand, RG relevant. 
Further, the tangential scattering coupling (whose flow equation is shown in eq.\eqref{tang_scatt_RG_flows}) is also found to be RG irrelevant, as the denominator has an overall negative signature ($\omega+\tilde{\epsilon}^{c}_{\Lambda_{j},avg}<0$). Thus, this leads to a fixed point effective Hamiltonian with different pairs (with charge 2e) marked by $\delta$ (eq.\eqref{hh_he_mixed_configuration}) are involved in forward scattering 
\begin{eqnarray}
\hat{H}^{*}(\omega)&=&\sum_{k, l}\epsilon_{k, l}\tau_{k, l}+\sum_{\delta >0}\hat{H}^{1, *}_{\delta}(\omega)+H^{*}_{dec}(\omega)~,~~\label{full_Hamiltonian_Metal}\\
\hat{H}^{1, *}_{\delta}(\omega)&=&\sum_{k, k', l}V^{*}_{l}(\omega, \delta)c^{\dagger}_{k, l}c^{\dagger}_{k, l'}c_{k', l'}c_{k', l}~,\nonumber
\end{eqnarray}
where $(k, l)\equiv (\mathbf{k}_{\Lambda, \hat{s}}, \uparrow)$, $(k, l') \equiv (\mathbf{k}_{-\Lambda+\delta, T\hat{s}}, \downarrow)$ and $(k', l') \equiv (\mathbf{k}_{-\Lambda'+\delta, T\hat{s}}, \downarrow)$. The fixed point value of the coupling $V^{*}_{l}(\omega, \delta)$ is given by vanishing of the denominator in eq.\eqref{RG_flow_eqn_FS_BS}~\cite{Glazek2004}
\begin{eqnarray}
V^{*}_{l}(\omega, \delta) = [\frac{1}{2}(\epsilon_{k^{*}, l}+\epsilon_{k^{*}, l'})-\omega]~,\nonumber\\
\epsilon_{k^{*}, l} = \epsilon_{\mathbf{k}_{\Lambda^{*}\hat{s}}}~,~\epsilon_{k^{*}, l'} = \epsilon_{\mathbf{k}_{-\Lambda^{*}+\delta T\hat{s}}}~.
\label{forward_scattering_magnitude}
\end{eqnarray}
The momentum state $\mathbf{k}_{\Lambda, \hat{s}}$ marked by the pair of indices $(k, l)$ is summed over the range $0<\Lambda < \Lambda^{*}$ for every direction $\hat{s}$ normal to the FS. The window $\Lambda^{*}$ partitions the Hamiltonian eq.\eqref{full_Hamiltonian_Metal} into two subparts: one involving the decoupled degrees of freedom ($\mathbf{k}_{\Lambda, \hat{s}}\sigma$) given by
\begin{eqnarray}
H^{*}_{dec}(\omega) &=& \hspace*{-0.2cm}\sum_{\substack{k, l, l', \\ \Lambda_{k}<\Lambda^{*}_{l}}}\hspace*{-0.2cm}\left[(\epsilon_{k, l}-\Delta\mu_{eff})\tau_{k, l}+V^{(k)}_{l}(\omega,\delta)\hat{n}_{k, l}\hat{n}_{k, l'}\right]~,\label{FL_Ham}
\end{eqnarray} 
and another for the coupled degrees of freedom involved in forward scattering: $\hat{H}^{*}(\omega)-H^{*}_{dec}(\omega)= \sum_{k, l}\epsilon_{k, l}\tau_{k, l}+\sum_{\delta >0}\hat{H}^{1, *}_{\delta}(\omega)$. $H^{*}_{dec}(\omega)$ clearly describes a Fermi liquid-like gapless metallic state of matter. This Fermi liquid is positioned farthest away from the non-interacting FS in energy as well as in $k$-space. As we shall now see from the second part of the RG analysis, it undergoing a gradual crossover to a very different gapless metallic state of matter in the immediate neighbourhood of the FS.
\par
\begin{figure}
\includegraphics[scale=0.5]{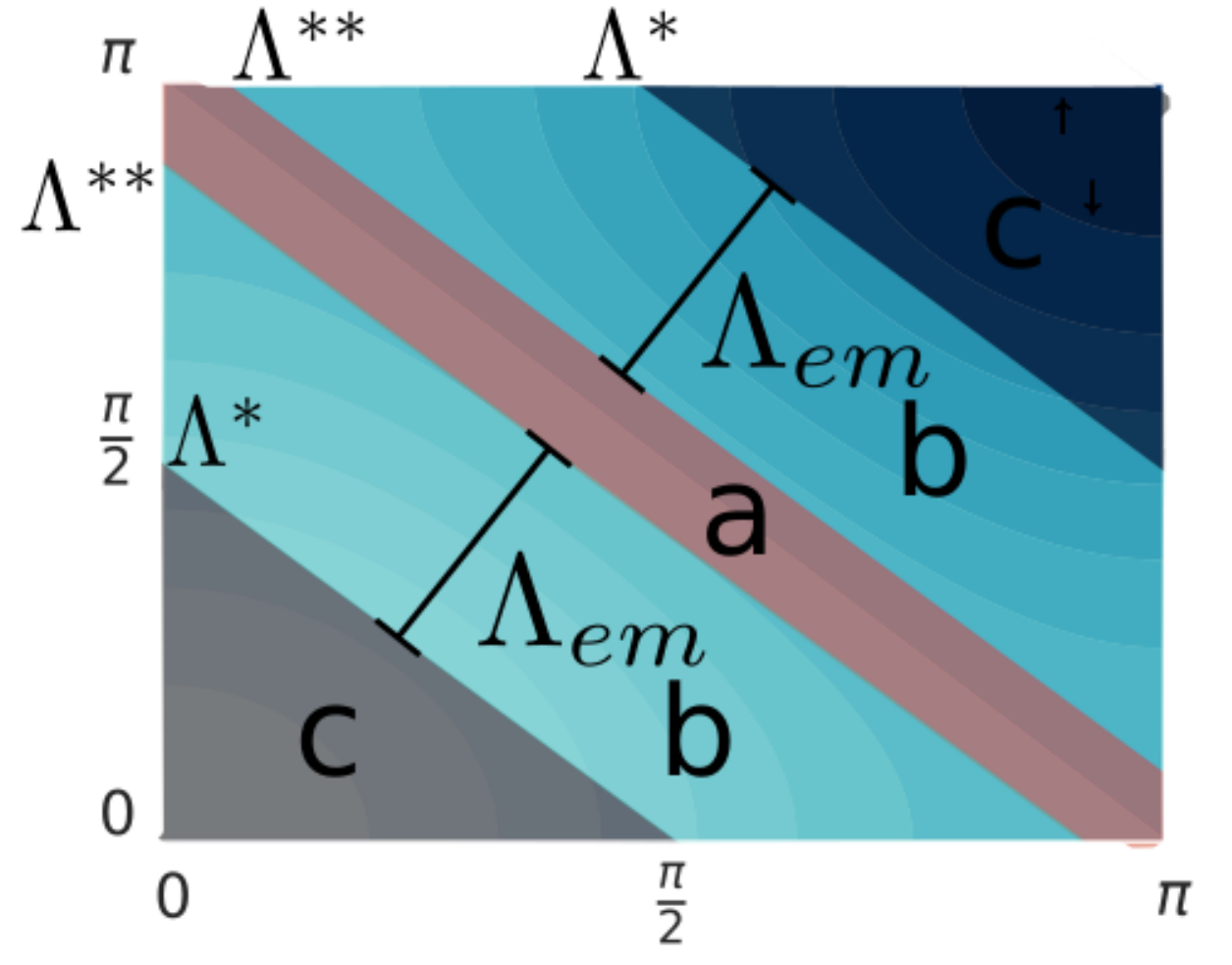}
\caption{(Colour Online) Momentum-space map partitioned into regions with different Hamiltonian structures for the normal state. (a) represents the Marginal Fermi liquid (MFL), (b) represents a correlated Fermi liquid, i.e., containing both MFL and Fermi liquid (FL), (c) represents a FL.}\label{metallic-state}
\end{figure}

\noindent
The value of $\Lambda^{*}$ is determined from numerical solution for the vanishing denominator in eq.\eqref{Long RG equations with doping}
\begin{eqnarray}
e^{i\gamma^{(\Downarrow)}_{\hat{s}}(\omega)}|\hat{G}_{j^{*}, l}(\pm\omega)|^{-1}=\frac{V_{p}^{*})}{4}\vert_{p=1}~,\label{fp_condition_MFL}
\end{eqnarray}
where we determined $p=1$ from eq.\eqref{determining p} for $\omega$ lying in the range given in eq.\eqref{NFL_cond}. 
As discussed above eq.\eqref{six-point flow}, the non-commutativity between different $\delta$-pair momenta scattering Hamiltonians, $[\hat{H}^{1, *}_{\delta}(\omega), \hat{H}^{1, *}_{\delta'}(\omega)]\neq 0$, leads to the effective three-particle (2-electron and 1 hole) scattering terms in Fig.\ref{three_scattering_pathways}. The net dispersion energy for the 2e-1h intermediate configuration (eq.\eqref{three particle intermediate config}) lies in energy range 
\begin{eqnarray}
[0, \frac{\epsilon_{j, l}+\epsilon_{j, l'}}{2}]\subset[0, W]~.~~ \label{energy_range_2+1}
\end{eqnarray} 
For such 2e-1h composites, an electron-hole pair is fixed on the FS while another electronic state $\mathbf{k}_{\Lambda, \hat{s}}$ is taken from within FS. 
These quantum fluctuations, when treated via RG at the lowest fluctuation scale 
\begin{eqnarray}
\max\limits_{\delta, \hat{s}}\frac{\epsilon_{j*, l}+\epsilon_{j*, l'}}{2} = \bar{\omega}\label{lowest_energy_end_of_spectrum}
\end{eqnarray} 
residing in the low energy spectrum of the fixed point Hamiltonian $\hat{H}^{*}(\omega)$ (eq.\eqref{full_Hamiltonian_Metal}) leads to flow equations (eq.\eqref{Long RG equations with doping}, eq.\eqref{six_point_RG_flows}, with $\omega$ replaced by $\bar{\omega}$)
\begin{eqnarray}
\frac{\Delta V^{(j)}_{c, l}}{\Delta\ln\frac{\Lambda_{j}}{\Lambda_{0}}} &=& -\frac{(V^{(j)}_{c, l})^{2}}{\left|\frac{1}{2}(\epsilon_{j, l}+\epsilon_{j, l'})-\bar{\omega}\right |+\frac{V^{(j)}_{c, l}}{4}}~,\label{fwd_scattering_near_Fermi_energy}~~\\
\Delta R^{(j)}_{l\delta\delta'} &=& \frac{V^{(j)}_{c, l}V^{(j)}_{c, l'}}{\bar{\omega} - \frac{1}{2}\epsilon^{(j)}_{\Lambda_{j}\hat{s}}}-\sum_{\delta''}\frac{R^{(j)}_{l\delta\delta''}R^{(j)}_{l\delta''\delta'}}{[G_{j, l, 3}]^{-1}-\frac{1}{8}R^{(j)}_{l\delta''}}~,~~\label{2e1h_scattering_near_Fermi_energy}\\
G_{j, l, 3}&=&\frac{1}{\frac{1}{2}(\epsilon_{j, l}+\epsilon_{j, l'})-\bar{\omega}}~.\nonumber
\end{eqnarray}
The number diagonal 2e-1h terms are represented as $R^{(j)}_{l\delta\delta}\equiv R^{(j)}_{l\delta}$.
The lowest fluctuation scale is attained by maximizing over the multiple scales im eq.\eqref{lowest_energy_end_of_spectrum} for every $\hat{s}$ and $\delta$, such that the 2e-1h scattering vertices have the dominant renormalization. 
From the negative sign in the denominator in eq.\eqref{fwd_scattering_near_Fermi_energy}, we find that the forward scattering RG flow equations at the lowest energy end of the spectrum $\bar{\omega}$ (eq.\eqref{lowest_energy_end_of_spectrum}) are irrelevant and flow towards vanishing coupling, $V^{*, l}_{c} = 0$. On the other hand, we find that the flow of the 2e-1h off-diagonal scattering terms 
generated from the forward scattering processes (first term in eq.\eqref{2e1h_scattering_near_Fermi_energy}) are initially RG relevant. However, as the 2e-1h scattering processes becomes bigger in magnitude, the second term in eq.\eqref{2e1h_scattering_near_Fermi_energy} eventually cuts off their growth and leads to 
a nonzero fixed point at 
\begin{eqnarray}
\frac{R^{*}_{l\delta}}{8} = &=& \bar{\omega}-\frac{1}{2}(\epsilon_{k*, l}+\epsilon_{k*, l'})~,~~\label{fixed_point_three_particle}
\end{eqnarray}  
where $(k^{*}, l)\equiv \mathbf{k}_{\Lambda^{**}\hat{s}}~,~ (k^{*}, l')\equiv \mathbf{k}_{-\Lambda^{**}+\delta T\hat{s}}$.
At this fixed point, 
the effective Hamiltonian is given by $H^{3*}(\bar{\omega}) = \sum_{\hat{s}}H^{3*}(\bar{\omega}, \hat{s})$
\begin{eqnarray}
\hspace*{-1cm}
H^{3*}(\bar{\omega}, \hat{s}) &=&\sum_{\Lambda<\max\limits_{\delta}\Lambda^{**}_{\hat{s}, \delta}}\epsilon_{k, l}\hat{n}_{k, l}+H'^{*}_{dec}(\bar{\omega})\nonumber\\
&+&\sum_{\substack{\Lambda, \Lambda'<\max\limits_{\delta}\Lambda^{**}_{\hat{s},\delta}, \\ \delta>0}}R^{*}_{l\delta}\hat{n}_{k, l}\hat{n}_{k, l'}(1-\hat{n}_{k', l'})\label{fixed point Hamiltonian 2nd step}~.
\end{eqnarray}
The first and third terms of eq.\eqref{fixed point Hamiltonian 2nd step} represent the Hamiltonian for the degrees of freedom lying closest to the FS of the non-interacting tight-binding problem. It is easily seen that this gapless, metallic state of matter is composed of composite 2e-1h degrees of freedom. On the other hand, the intermediate range of energies described by the window 
\begin{equation}
\Lambda\in [\Lambda_{em}]=[\max\limits_{\delta}{\Lambda^{*}_{\hat{s}, \delta}},\max\limits_{\delta}{\Lambda^{**}_{\hat{s}, \delta}}],\label{momentum space window_FL_MFL_admix}
\end{equation} 
contains two-particle as well as three-particle vertices, leading to 
\begin{eqnarray}
H'^{*}_{dec}(\bar{\omega}) &=& \sum_{\hat{s},\Lambda\in [\Lambda_{em}]}(\epsilon_{\Lambda\hat{s}}-\Delta\mu_{eff})\left(\hat{n}_{\Lambda, \hat{s}, \sigma}-\frac{1}{2}\right)\label{decoupled_states_Hamiltonian_2nd_level}\\
&+&\sum_{\hat{s}, \delta, \Lambda_{k}\in [\Lambda_{em}]}V^{(j)}_{k, l}(\omega,\delta)\hat{n}_{k, l}\hat{n}_{k, l'}\nonumber\\
&+&\sum_{\substack{\hat{s}, \delta>0\\ \Lambda_{j}\in [\Lambda_{em}]}}R^{(j)}_{l\delta}\hat{n}_{k, l}\hat{n}_{-k, l'}(1-\hat{n}_{k', l'})~.\nonumber
\end{eqnarray}
Thus, the intermediate window involves a gradual crossover from a Fermi liquid to another metallic state of matter (see Fig.\ref{metallic-state}) which we  characterise below.
\subsubsection{Marginal Fermi liquid in the IR}
We will now see that the gapless state of matter lying at lowest energies possesses properties ascribed phenomenologically to the Marginal Fermi liquid~\cite{varma-PhysRevLett.63.1996}. Of primary importance is the renormalisation of the 1-particle self-energy arising from 2e-1h off-diagonal scattering terms in the neighbourhood of the fixed point (eq.\eqref{fixed_point_three_particle}).
This renormalisation arises intermediate scattering configurations involving two electrons and three holes. 
The 2e-3h configuration energy is obtained from off-resonant pairs $(j, l):(\Lambda_{j}, \hat{s}, \sigma)$, $(j, l'):(-\Lambda_{j}+\delta, T\hat{s}, -\sigma)$ in electron-occupied configuration, along with three holes present at Fermi surface. We then obtain the leading contribution to the 1-e self energy RG equation
\begin{eqnarray}
\Delta\Sigma^{Re, (j)}_{\Lambda, \hat{s}}(\bar{\omega}) &=& \sum_{\Lambda<\Lambda_{j}}\frac{(R^{(j)}_{X})^{2}}{\bar{\omega} -\frac{1}{2}\epsilon^{c}_{j, l}+\frac{1}{8}R^{(j)}_{D}}~.\label{1-particle_self_energy_RG}\nonumber\\
&=&N(\Lambda_{j},\hat{s})\frac{(R^{(j)}_{X})^{2}}{R^{(j)}_{D}}\int^{\Lambda_{j}}_{0, \delta\to 0} \frac{d\epsilon_{\Lambda_{j}}}{\bar{\omega}-\frac{1}{2}\epsilon^{c}_{j, l}+\frac{1}{8}R^{(j)}_{D}} \nonumber\\
	&=&	N(\Lambda_{j},\hat{s})\frac{(R^{(j)}_{X})^{2}}{R^{(j)}_{D}}\ln\frac{\bar{\omega}}{\bar{\omega}-\frac{1}{2}\epsilon^{c}_{j, l}+\frac{1}{8}R^{(j)}_{D}}~	,	\label{self_energy_flow_3_p}
\end{eqnarray}
where $R^{(j)}_{X} =R^{(j)}_{l\delta\delta'}$, $R^{(j)}_{D} = R^{(j)}_{l\delta}$ and $\epsilon^{c}_{j,l}=\epsilon_{j,l}+\epsilon_{j,l'}$ is the dispersion for the 2e-1h composite with the momentum indices locked by the choice of the 2e-3h state described earlier. In the second step of the above set of equations, we have introduced the total number of states along a given $\hat{s}$, $N(\Lambda,\hat{s})$, upon replacing the summation by a integration. Near the FS, using eq.\eqref{2e1h_scattering_near_Fermi_energy} and in the vicinity of the fixed point eq.\eqref{fixed_point_three_particle}, the flow equations attain the form
\begin{eqnarray}
\frac{\Delta R^{(j)}_{X}}{\Delta\log_{b}
\frac{\Lambda_{j}}{\Lambda_{0}}}=\frac{\Delta R^{(j)}_{D}}{\Delta\log_{b}
\frac{\Lambda_{j}}{\Lambda_{0}}} = -\frac{(R^{(j)}_{X})^{2}}{\bar{\omega}-\frac{1}{2}\epsilon^{c}_{j, l}+\frac{1}{8}R^{(j)}_{D}}~.~~~\label{simple_form_3}
\end{eqnarray} 	
Using the RG invariant $C=R^{(j)}_{X}-R^{(j)}_{D}$ (eq.\eqref{simple_form_3}) with $C=0$, 
the self energy has the form
\begin{eqnarray}
\Delta\Sigma^{Re, (j)}_{\Lambda_{j}, \hat{s}}(\bar{\omega}_{\hat{s}}) = N(\hat{s},\Lambda_{j})\left(\bar{\omega}-\frac{1}{2}\max\limits_{\delta}\epsilon^{c}_{j, l}\right)\ln\left\vert \frac{\bar{\omega}}{\bar{\omega}-\frac{1}{2}\max\limits_{\delta}\epsilon^{c}_{j,l}}\right\vert~.~~~~~~ 
\end{eqnarray}
As the self energy renormalization has a branch-cut log singularity at the FS, we may approximate 
$\Sigma^{Re, (j)}_{\mathbf{k}_{\Lambda\hat{s}}}(\omega)\approx \Delta\Sigma^{Re, (j)}_{\mathbf{k}_{\Lambda\hat{s}}}(\omega)+O(\omega)$. From these relations, we obtain the self energy $\Sigma$ and the quasiparticle residue $Z_{1}$ as
\begin{eqnarray}
\Sigma(\tilde{\omega}_{\hat{s}}) &=& \tilde{\omega}_{\hat{s}}\ln\left\vert\frac{N^{*}(\hat{s},\omega)\bar{\omega}}{\tilde{\omega}_{\hat{s}}}\right\vert~,~ Z_{1}(\tilde{\omega}_{s})=\frac{1}{2-\ln\left\vert\frac{\tilde{\omega}_{\hat{s}}}{N(\hat{s},\Lambda^{**})\bar{\omega}}\right\vert}~, \label{1-p self energy}
\end{eqnarray}
where $\tilde{\omega}_{\hat{s}} = N^{*}(\hat{s},\omega)(\bar{\omega}-\frac{1}{2}\epsilon^{c}_{\Lambda^{**},\hat{s}})$ and $\bar{\omega}$ is defined in eq\eqref{lowest_energy_end_of_spectrum}. The quasiparticle residue $Z_{1}(\tilde{\omega}_{s})$ vanishes as $\tilde{\omega}_{\hat{s}}\to 0$, indicating breakdown of Landau's quasiparticle picture. These well-known expressions for the Marginal Fermi liquid have been proposed on phenomenological grounds towards understanding the strange metal phase encountered in the hole-doped cuprates~\cite{varma-PhysRevLett.63.1996}. While $\Sigma$  
has the same structure as proposed in \cite{varma-PhysRevLett.63.1996}, it is worth noting that the Marginal Fermi liquid we find arises from singular longitudinal scattering along directions normal to the FS ($\hat{s}$).
\par\noindent
The imaginary part of the self energy can be computed from the real part of the self energy using the Kramers-Kronig relations
\begin{eqnarray}
\Sigma^{Im, (j)}_{\Lambda_{j}, \hat{s}}(\bar{\omega}_{\hat{s}})= \frac{1}{\pi}\mathcal{P}\int_{-\infty}^{\infty}\frac{\Sigma^{Re, (j)}_{\Lambda_{j}, \hat{s}}(\omega)}{\omega-\tilde{\omega}_{\hat{s}}} = \tilde{\omega}_{\hat{s}}~.
\end{eqnarray}
From here, we obtain the quasiparticle lifetime as $\tau = 2\pi\tilde{\omega}_{\hat{s}}^{-1}$, in keeping with the proposed relation for the marginal Fermi liquid. Further using the equivalence relation between the quantum fluctuations assisted broadening $\tilde{\omega}_{\hat{s}}$ and thermal broadening (eq.\eqref{equivalent_Thermal scale}), we can obtain the largest  temperature scale ($T$) upto which the single particle description is well defined
\begin{eqnarray}
k_{B}T = \hbar \max_{\hat{s}}\tilde{\omega}_{\hat{s}}~.\label{Thermal scale}
\end{eqnarray}
The inverse lifetime $\tau^{-1}$ is thus associated with a linear-in-$T$ Drude resistivity, $\rho\propto T$, arising from the excitations of the gapless Fermi surface. This temperature scale will appear in a later section as the source of ``Planckian dissipation" when a quantum critical point is reached upon doping away from $1/2$-filling. 
\par\noindent
The 1-particle self energy is singular leading to $Z_{1}(\tilde{\omega}_{\hat{s}})\to 0$,but from the form of the Hamiltonian eq.~\eqref{fixed point Hamiltonian 2nd step} its clear that the 2-e 1-h self energy is well defined and given by 
\begin{eqnarray}
\Sigma^{3*}_{\Lambda\hat{s}}(\omega) &=& R^{*}_{\hat{s}, \delta}, Z_{3, \Lambda\hat{s}}(\omega) = \frac{1}{1-\frac{\partial\Sigma^{3*}_{\Lambda\hat{s}}(\omega)}{\partial\omega}}.~~~~~~\label{2-p 1-h self energy}
\end{eqnarray}
As $\omega \to 0$ and $\Lambda^{**}_{\hat{s}, \delta''}\to 0$, the 2e-1h dispersion vanishes $\Sigma^{3*}_{\Lambda\hat{s}}(\omega)\to 0$ and $Z_{3, \Lambda\hat{s}}(\omega)\to 1$, making the composite degree of freedom well defined. This is exhibited by the fixed point Hamiltonian eq.\eqref{fixed point Hamiltonian 2nd step}.
\subsubsection{Luttinger sum and the f-sum rule}
\begin{figure}\includegraphics[scale=0.25]{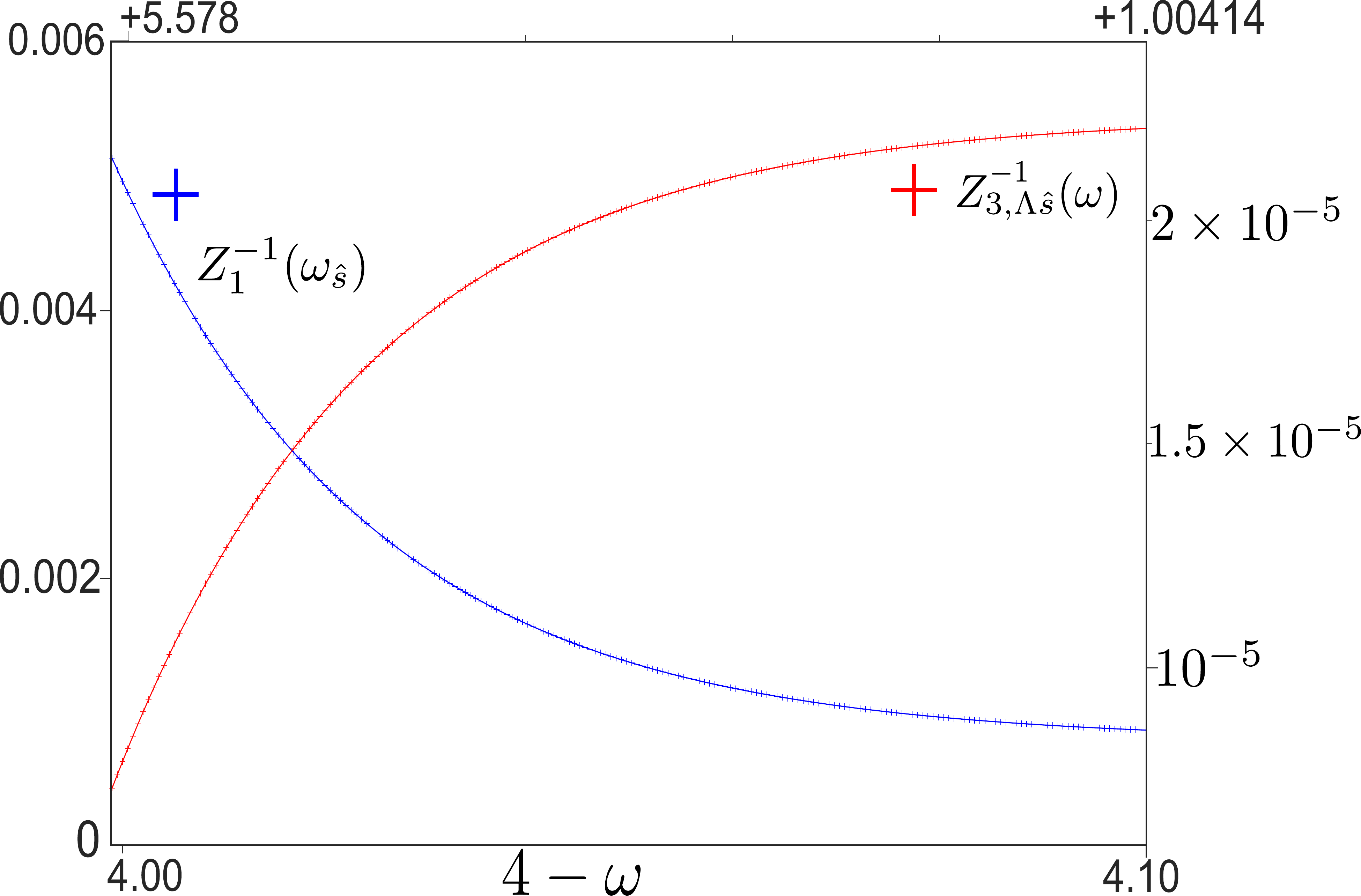}
\caption{(Colour Online) Inverse quasiparticle residue averaged over a gapless connected Fermi surface $Z^{-1}_{avg}$ at energies above the entry into the pseudogap phase shows a logarithmic growth as $4-\omega$ is lowered toward the Fermi energy ($\omega\to 4-$). Vanishing of the quasiparticle residue (as the Fermi energy is approached) is an important characteristic of the destruction of the Fermi liquid. Instead, $Z^{-1}_{3,\Lambda\hat{s}}$ (indicating a marginal Fermi liquid of composite 2e-1h quasiparticles) gradually rises to unity near the Fermi energy.}
\label{qp-residue}
\end{figure}

As observed in eq.~\eqref{1-p self energy} and eq.~\eqref{Thermal scale}, the metallic state within the window from $[0, \Lambda^{**}_{\hat{s}, \delta}]$ displays properties of the marginal Fermi liquid arising from a coherence for the 2e-1h composite degrees of freedom near the Fermi surface (eq.\eqref{2-p 1-h self energy}). Outside the window $\Lambda\in[\Lambda_{em}]$ resides a correlated Fermi liquid with an effective Hamiltonian eq.\eqref{decoupled_states_Hamiltonian_2nd_level}, and with the real part of the 1-electron ($\Sigma^{1, em}_{re, \Lambda_{k}, \hat{s}}$) and 2e-1h ($ \Sigma^{3, em}_{re, \Lambda_{k}, \hat{s}}$) self energies given by
\begin{eqnarray}
 \Sigma^{1, em}_{re, \Lambda_{k}, \hat{s}}&=&\frac{(V_{k, l}^{(k)})^{2}}{\bar{\omega} - \frac{1}{2}\epsilon_{\Lambda_{k}, \hat{s}}}+R^{(k)}_{\hat{s},\delta}\ln\frac{\bar{\omega}-\frac{1}{2}\epsilon^{c}_{\Lambda_{k}, \hat{s},\delta}+\frac{1}{8}R^{(k)}_{\hat{s}, \delta}}{\bar{\omega}}\nonumber\\
 \Sigma^{3, em}_{re, \Lambda_{k}, \hat{s}}&=&R^{(k)}_{\hat{s}, \delta}~.\label{self_energy_window_1}
\end{eqnarray}
Further outside $\Lambda>\Lambda^{*}_{\hat{s}, \delta}$ resides the Fermi liquid (effective Hamiltonian in eq.\eqref{FL_Ham}) with 1-electron self energy given by
\begin{eqnarray}
\Sigma^{1, FL}_{re, \Lambda_{k}, \hat{s}}=\sum_{\delta}V^{(k)}_{\hat{s}}(\omega,\delta)~.\label{self_energy_window_2}
\end{eqnarray}
From these relations, we find the net 1-electron ($\Sigma^{1}_{\Lambda, \hat{s}}(\omega)$) and composite 2e-1h ($\Sigma^{3}_{re, \Lambda\hat{s}, \delta}(\omega)$) self energies to be
\begin{eqnarray}
\Sigma^{1}_{\Lambda, \hat{s}}(\omega) &=& \Sigma(\tilde{\omega}_{\hat{s}})\theta(\Lambda^{**}_{\hat{s}}-\Lambda)+\Sigma^{1, em}_{re, \Lambda\hat{s}}\theta(\Lambda-\Lambda^{**}_{\hat{s}})\theta(\Lambda^{*}_{\hat{s}}-\Lambda)\nonumber\\
&+&\Sigma^{1, FL}_{re, \Lambda\hat{s}}\theta(\Lambda-\Lambda^{*}_{\hat{s}}),\nonumber\\
\Sigma^{3}_{re, \Lambda\hat{s}, \delta}(\omega)&=&\Sigma_{re, \Lambda, \hat{s}}^{*}(\omega)\theta(\Lambda^{**}_{\hat{s}}-\Lambda)\nonumber\\
&+&\Sigma^{3, em}_{re, \Lambda, \hat{s}, \delta}\theta(\Lambda-\Lambda^{**}_{\hat{s}})\theta(\Lambda^{*}_{\hat{s}}-\Lambda)~,\label{gapless_self}
\end{eqnarray}
\par\noindent
from which we obtain the 1-e and 2e-1h Green functions as
\begin{eqnarray}
G^{1}_{\Lambda ,\hat{s},\sigma}(\omega) &=& \frac{1}{\omega -\epsilon_{\Lambda,\hat{s}}-\Sigma^{1}_{\Lambda,\hat{s}}(\omega)-i\Sigma^{Im,1}_{\Lambda,\hat{s}}(\omega)},\nonumber\\
G^{3}_{\Lambda,\hat{s},\sigma}(\omega)&=&\frac{1}{\omega -\frac{1}{2}\epsilon^{c}_{\Lambda,\hat{s}}-\Sigma^{3}_{re,\Lambda,\hat{s}}(\omega)-i\Sigma^{3,Im}_{re,\Lambda,\hat{s}}(\omega)}\label{self_energy_imaginary}~,
\end{eqnarray}
where $\Sigma^{3, Im}_{re, \Lambda, \hat{s}}(\omega)$ and $\Sigma^{Im, 1}_{\Lambda, \hat{s}}(\omega)$ are the imaginary parts of the 2e-1h and 1e self-energies obtained from Kramers-Kronig relations. The spectral weights/residues for 1e ($Z_{1}(\omega_{\hat{s}})$) and 2e-1h composite ($Z_{3, \Lambda\hat{s}}(\omega)\to 1$) are computed numerically from the above Greens functions, and shown in Fig.~\ref{qp-residue}. The figure shows that as the FS is approached ($\omega\to 4$), a vanishing $Z_{1}(\omega_{\hat{s}})$ is compensated by a $Z_{3,\Lambda\hat{s}}\to 1$. Then, from the poles of these two Greens functions, we find that the Luttinger sum is preserved 
\begin{equation}
N=\sum_{\omega >0, \Lambda, \hat{s}}G^{3}_{\Lambda, \hat{s}, \sigma}(\omega)+\sum_{\omega >0, \Lambda>\Lambda^{*}, \hat{s}}G^{1}_{\Lambda, \hat{s}, \sigma}(\omega)~, \label{modified-Luttinger-sum}
\end{equation}
i.e., the net spectral weight equals the total number of electrons ($N$). 
\par\noindent
Further, this metallic phase is characterised by a \textit{global topological index} $N^{\Downarrow}_{\hat{s}}(\omega) =1, \forall\hat{s}$ normal to the FS. From the constraint eq.\eqref{constraint}, it is clear that backscattering is RG irrelevant everywhere on FS because of the topological phase $e^{i\pi(N^{\Uparrow}_{\hat{s}}(\omega)+1)}=-1, \forall \hat{s}$. This manifests in the topological protection for the metallic phase. Similarly, the net lifetime ($\tau$) arising from 1-e,  2e-1h excitation scattering processes 
$\tau(\omega , \Lambda, \hat{s})=(\Sigma^{Im, 1}_{\Lambda, \hat{s}})^{-1}+(\Sigma^{3, Im}_{\Lambda, \hat{s}})^{-1}$ can be shown to satisfy the conductivity sum-rule (or the f-sum rule) 
\begin{eqnarray}
f=\frac{e^{2}}{m}\sum_{\Lambda\hat{s}}\nu_{\hat{s}}\int d\omega~\tau(\omega ,\Lambda\hat{s})~,\label{f_sum_rule}
\end{eqnarray}
where $\nu_{\hat{s}}=\sum_{\Lambda}\theta(E_{F}-\epsilon_{\Lambda\hat{s}})$ is the partial Luttinger sum along $\hat{s}$.
In the following sections, we describe the destablization of the marginal Fermi liquid metallic state leading to various other exotic phases at lower fluctuation scales.
\section{RG flow through the Pseudogap}\label{PGSection}
\begin{figure}
\hspace*{-0.5cm}
\includegraphics[width=0.55\textwidth]{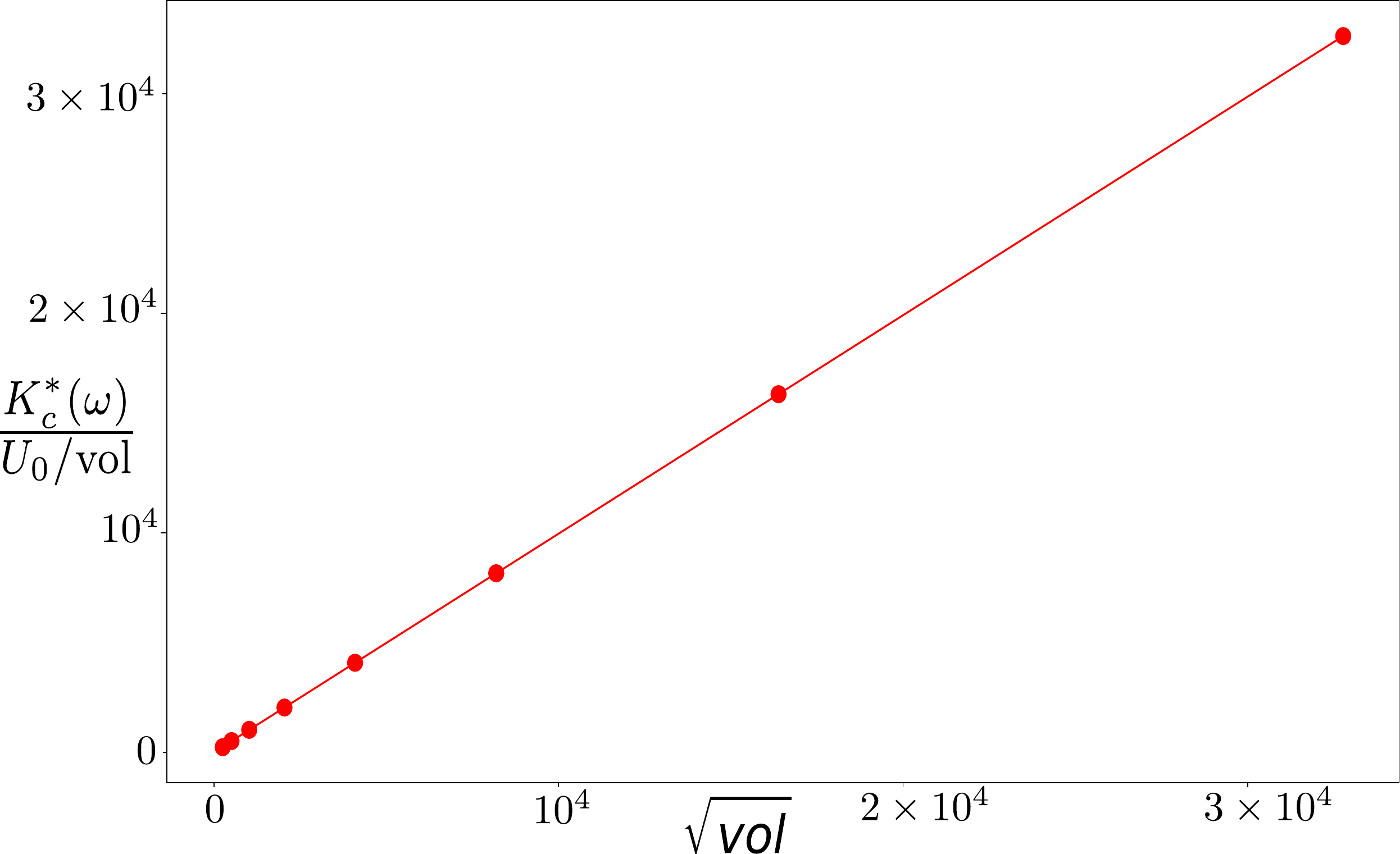} 
\caption{(Colour Online) Ratio of renormalized coupling to bare coupling ($K_{c}^{*}/(U_{0}/vol)$) showing a linear scaling with $\sqrt{\textrm{volume}}$.}\label{scaling_vertex}
\end{figure}

In the energy range,
\begin{eqnarray}
\frac{W}{2}-\omega \in \left[\frac{W}{2}-\omega_{ins}, \frac{W}{2}-\omega_{PG}\right], \label{pseudogap_energy_range}
\end{eqnarray}
the pseudogap is initiated in the form of a FS topology-changing Lifshitz transition that disconnects the connected FS via a gaping of the antinodes for $\omega<\omega_{PG}\equiv 0$, and proceeds until the nodes are gapped via a second Lifshitz transition at $\omega = \omega_{ins}$ (Fig.~1(d), Video S1). This is an outcome of electronic differentiation arising out of a monotonic variation of the electronic dispersion along the Fermi surface~\cite{imada2010unconventional} which influences the ee or eh paring energy (see eq.\eqref{Fermi_surface_geometry}). 
While the resistivity shows a linear behaviour with $W/2-\omega$ in the metallic phase (i.e., in the range $\omega_{PG}\leq W/2-\omega\leq W$), in the PG phase (Fig.\ref{half_filled_phase_diag}(c)), the resistivity shows a crossover into a insulator phase, i.e., an increase with lowering $\frac{W}{2}-\omega$ beyond $\omega =\omega_{PG}$. At a given fluctuation scale within energy range given by eq.\eqref{pseudogap_energy_range}, the fixed point Hamiltonian of the gapped parts of the FS is described by $\delta =0$ resonant pairs carrying the highest spectral weight (eq.\eqref{highest_spectral_weight})
\begin{eqnarray}
\hat{H}^{*}(\omega)&=&\sum_{\hat{s}}N^{\Uparrow}_{\hat{s}}(\omega)\bigg(K_{c,\hat{s}}^{*}(\omega))\mathbf{A}_{*, \hat{s}}\cdot\mathbf{A}_{*, -\hat{s}}\nonumber\\
&&\hspace*{1.5cm}-K_{s,\hat{s}}^{*}(\omega)\mathbf{S}_{*, \hat{s}}\cdot\mathbf{S}_{*, -\hat{s}}\bigg) \nonumber\\    
&+&\sum_{\hat{s}}N^{\Downarrow}_{\hat{s}}(\omega)H^{3*}(\omega,\hat{s})+H^{'*}_{dec}(\omega)~,~~~~\label{Hamiltonian_pseudogap}
\end{eqnarray}
where the charge ($\mathbf{A}$) and spin ($\mathbf{S}$) pseudospin operators are defined as
\begin{eqnarray}
\mathbf{A}_{*,\hat{s}} &=& \sum_{\Lambda<\Lambda^{*}_{\hat{s},\omega}}\mathbf{A}_{\Lambda,\hat{s}}~,~\mathbf{S}_{*,\hat{s}}=\sum_{\substack{\Lambda<\Lambda^{*}_{\hat{s},\omega}}}\mathbf{S}_{\Lambda,\hat{s}}~,~\nonumber\\
\mathbf{A}_{\Lambda,\hat{s}}&=&f^{c;\dagger}_{\Lambda,\hat{s}}\frac{\boldsymbol{\sigma}}{2}f^{c}_{\Lambda,\hat{s}}~~,~~\mathbf{S}_{\Lambda,\hat{s}}=f^{s;\dagger}_{\Lambda,\hat{s}}\frac{\boldsymbol{\sigma}}{2}f^{s}_{\Lambda,\hat{s}}~,\nonumber\\ 
f^{c;\dagger}_{\Lambda,\hat{s}} &=&\left[c^{\dagger}_{\Lambda,\hat{s},\sigma} ~~c_{-\Lambda,T\hat{s},-\sigma}\right]~,~\nonumber\\
f^{s;\dagger}_{\Lambda,\hat{s}} &=& \left[c^{\dagger}_{\Lambda,\hat{s},\sigma}~~ c^{\dagger}_{\Lambda - 2\Lambda^{*}_{\hat{s}},T\hat{s},-\sigma}\right]~,\label{pseudospins}
\end{eqnarray}
where $(\Lambda, \hat{s})$ are as defined earlier and $\Lambda^{*}_{\hat{s}}$ is the window width along $\hat{s}$ at the fixed point. This Hamiltonian is easily seen as a sum of mutually commuting Hamiltonian of one-dimensional systems, each involving a distinct pair of normal directions $(\hat{s}, T\hat{s})$. This results from the fact that tangential scattering between different $\hat{s}$ directions is RG irrelevant. Further, as each of these 1D Hamiltonians involves scattering across all ranges in momentum-space, the scattering vertices $K_{c,\hat{s}}^{*}(\omega))$ and $K_{s,\hat{s}}^{*}(\omega))$ are inversely proportional to the number of states along a given $\hat{s}$: $K_{c,\hat{s}}^{*}(\omega))\sim U_{0}/\sqrt{Vol}$~,~$K_{s,\hat{s}}^{*}(\omega))\sim U_{0}/\sqrt{Vol}$ where $Vol$ indicates the number of lattice sites (see Fig.\ref{scaling_vertex}).
\par\noindent 
The charge/spin pseudospin flip scattering terms $(A^{+}_{*, \hat{s}}A^{-}_{*, -\hat{s}}+h.c.)$~/~$(S^{+}_{*, \hat{s}}S^{-}_{*, -\hat{s}}+h.c.)$ present in the fixed point Hamiltonian eq.\eqref{Hamiltonian_pseudogap} comprise the charge/spin backscattering processes for resonant pairs ($\delta=0$). We had earlier shown in eq.\eqref{T-matrix_argument} and eq.\eqref{spin_backscattering_Corr} the appearance of log-divergences in the 2nd order corrections of the T-matrix arising from scattering in the resonant-pair subspaces. Here, through the RG flow, we show the condensation of the pseudospins in these subspaces
\begin{eqnarray}
\hat{n}_{\mathbf{k}_{\Lambda_{j}\hat{s}}\sigma} &=& \hat{n}_{\mathbf{k}_{-\Lambda_{j}T\hat{s}}-\sigma}\to (\mathbf{A}_{\Lambda\hat{s}})^{2}=\frac{3}{4}~,\label{charge_subspace}\\
\hat{n}_{\mathbf{k}_{\Lambda_{j}\hat{s}}\sigma} &=& 1-\hat{n}_{\mathbf{k}_{\Lambda_{j}-2\Lambda^{*}_{\hat{s}, \omega}
T\hat{s}}-\sigma}\to (\mathbf{S}_{\Lambda\hat{s}})^{2}=\frac{3}{4}~.\label{spin_subspace}
\end{eqnarray} 
The fixed point values of the backscattering couplings are given by
\begin{eqnarray}
[\omega - p\frac{\epsilon^{*}_{l}+\epsilon^{*}_{l'}}{2}-(1-p)\frac{\epsilon^{*}_{l}-\epsilon^{*}_{l'}}{2}]=\frac{K^{*}_{p,\hat{s}}}{4}~,\label{fixed_point_2_particle}
\end{eqnarray}
where the form of $V^{*, l}_{p}$ is given in eq\eqref{spin_charge_interaction_strength} in terms of the charge and spin backscattering strengths and the spin charge mixing parameter $p$. As seen in the previous section, the decoupled degrees of freedom residing outside the window $\Lambda^{*}_{\hat{s}}$ have a Hamiltonian $H^{'*}_{dec}(\omega)$ given by eq.\eqref{decoupled_states_Hamiltonian_2nd_level}. 
\par\noindent
The topological indices $N^{\Uparrow}_{\hat{s}}(\omega)$ and $N^{\Downarrow}_{\hat{s}}(\omega)$ appearing in eq.\eqref{Hamiltonian_pseudogap} characterise the pseudogap phase as follows
\begin{eqnarray}
\text{I}:N^{\Uparrow}_{\hat{s}} (\omega)&=&1-N^{\Downarrow}_{\hat{s}} (\omega)=1~\forall~\hat{s}\in [\hat{s}_{AN}, \hat{s}'],\nonumber\\
\text{II}:N^{\Downarrow}_{\hat{s}}(\omega)&=&1-N^{\Uparrow}_{\hat{s}}(\omega)=1~\forall~\hat{s}\in [\hat{s}', \hat{s}_{N}].\label{region_PG}
\end{eqnarray}
In this way, the first term in Hamiltonian $H^{*}(\omega)$ describes the gapped parts of the FS (I in eq.\eqref{region_PG}), 
while the second term $H^{3*}(\omega,\hat{s})$ (of the form eq.\eqref{fixed point Hamiltonian 2nd step}) describes the gapless terms of the FS (II in eq\eqref{region_PG}) in terms of composites of 2e 1h degrees of freedom.
At the second Lifshitz transition involving the gapping of the FS at the nodes, $\hat{s}'=\hat{s}_{N}$ 
and the resulting Mott liquid (discussed in more detail in Section \ref{mottliquid}) is described by the global topological invariant $N^{\Uparrow}_{\hat{s}}(\omega)=1~\forall \hat{s}$. The pseudogap phase is thus a coexistence of gapped and gapless parts of the FS, and can also be characterised by a different global topological invariant
\begin{eqnarray}
N_{\hat{s}}^{PG} = N_{R\hat{s}} + N_{T\hat{s}}~,
\label{Glob-PG}
\end{eqnarray} 
where $N_{R/T\hat{s}} = |N^{\Uparrow}_{\hat{s}}(\omega) - N^{\Uparrow}_{\hat{s}+(R/T)\hat{s}}(\omega)|$, and the parity operation $R\hat{s}:\hat{s}_{x} \leftrightarrow \hat{s}_{x}, :\hat{s}_{y} \leftrightarrow -\hat{s}_{y}$ (or vice versa). It is easily seen that $N^{PG}_{\hat{s}}=1 ~\forall ~\hat{s}$ in the PG phase, and vanishes in the metallic and insulating phases. These non-local order parameters ensure that the two $T=0$ Fermi surface topology-changing Lifshitz transitions at the passage into and out of the pseudogap phase do not belong to the Ginzburg-Landau-Wilson paradigm~\cite{imada2010unconventional}.
\subsection{Pole-to-zero conversion and the f-sum rule}\label{pole-zero-conv}
In the normal metallic state discussed in the previous section, low energy scattering processes were found to renormalise the 1e (eq.\eqref{1-p self energy}) and 2e-1h (eq.\eqref{2-p 1-h self energy}) self energies. This analysis can be extended to the gapped region centered around the AN's by looking at the renormalization of the 2e-1h, 1e and 2e self energies via the backscattering vertex 
\begin{eqnarray}
\Sigma^{*}_{\Lambda\hat{s}}(\omega_{1}) &=& \frac{p\left(K_{c}^{*}\right)^{2}+(1-p)\left(K_{s}^{*}\right)^{2}}{\omega_{1} -\omega -\frac{1}{2}\epsilon_{\Lambda, \hat{s}}},\label{1-electron_self_energy_backscattering}\\
\Sigma^{2, *}_{\Lambda\hat{s}}(\omega) &=& [pK^{*}_{c, \hat{s}}(\omega)+(1-p)K^{*}_{s, \hat{s}}(\omega)]~,\label{2-electron_self_energy}\\
\Sigma^{3, *}_{\Lambda\hat{s}, \delta}(\omega)&=&\sum_{\delta'}\frac{-(R^{*}_{l,\delta\delta'})^{2}}{\omega - \frac{1}{2}(\epsilon_{\Lambda^{*}\hat{s}}+\epsilon_{-\Lambda^{*}+\delta'\hat{s}})+\frac{1}{2}\epsilon_{-\Lambda\hat{s}}-\frac{1}{8}R^{*}_{\hat{s}, \delta}},\label{2-electron_1-hole self energy}~~~~~\\
R^{*}_{\hat{s}, \delta, \delta'}&=&\frac{p(K^{*}_{c, \hat{s}, \delta}K^{*}_{c, \hat{s}, \delta'})+p_{s}(K^{*}_{s, \hat{s}, \delta}K^{*}_{s, \hat{s},\delta'})}{\bar{\omega}-\frac{1}{2}\epsilon_{\Lambda^{*}\hat{s}}}~,\nonumber
\end{eqnarray}
where ($\epsilon^{c/s}_{\Lambda^{*}, \hat{s}}=\epsilon_{\Lambda^{*}, \hat{s}}
\pm\epsilon_{-\Lambda^{*}, T\hat{s}}$). In the 1e and 2e self energy renormalization, the dominant contribution from the resonant pairs is obtained from the fixed point values $K^{*}_{c, \hat{s}, \delta=0}=K^{*}_{c, \hat{s}}$, $K^{*}_{s, \hat{s}, \delta=0}=K^{*}_{s, \hat{s}}$. In parallel with an earlier discussion of the marginal Fermi liquid, the 2e-1h self energy in eq.\eqref{2-electron_1-hole self energy} is observed to be an outcome of the three-particle scattering processes that arise out of non-commutativity between resonant and off-resonant pair {\it backscattering} vertices. 
From the Kramers-Kronig relation, we obtain the imaginary parts of the 1-e (eq.\eqref{1-electron_self_energy_backscattering}), 2-e (eq.\eqref{2-electron_self_energy}) and 2e-1h (eq.\eqref{2-electron_1-hole self energy}) self energies
\begin{eqnarray}
\Sigma^{Im*}_{\Lambda, \hat{s}}(\omega_{1})&=&\frac{1}{\pi}\mathcal{P}\int_{-\infty}^{\infty} d\omega'\frac{\Sigma^{*}_{\Lambda, \hat{s}}(\omega')}{\omega'-\omega_{1}}=\Sigma^{*}_{\Lambda, \hat{s}}(\omega_{1}), \nonumber\\
		 &=& \frac{p\left(K^{*}_{c, \hat{s}}\right)^{2}+(1-p)\left(K^{*}_{s, \hat{s}}\right)^{2}}{\omega_{1} - \omega -\frac{1}{2}\epsilon_{\Lambda,\hat{s}}},~~~\nonumber\\
\Sigma^{2, Im*}_{\Lambda, \hat{s}}(\omega_{1})&=&\frac{1}{\pi}\mathcal{P}\int_{-\infty}^{\infty}d\omega' \frac{\Sigma^{2, *}_{\Lambda, \hat{s}}(\omega')}{\omega'-\omega_{1}}=\Sigma^{2, *}_{\Lambda, \hat{s}}(\omega_{1}), \label{imaginary_self_2}\nonumber\\		 
\Sigma^{3, Im*}_{\Lambda, \hat{s}, \delta}(\omega_{1})&=&\frac{1}{\pi}\mathcal{P}\int_{-\infty}^{\infty} d\omega'\frac{\Sigma^{3, *}_{\Lambda, \hat{s}, \delta}(\omega')}{\omega'-\omega_{1}}=\Sigma^{3, *}_{\Lambda, \hat{s}, \delta}(\omega_{1}), \nonumber\\
				&=& -\frac{\sum_{\delta'}(R^{*}_{\hat{s}, \delta\delta'})^{2}}{\omega_{1} -\omega-\frac{1}{2}\epsilon_{\Lambda, \hat{s}}}~.\label{imaginary_self_3}
\end{eqnarray}
The 1e, 2e and 2e-1h fixed point Greens functions for the gapped regions can then be written as
\begin{eqnarray}
&&\hspace*{-0.5cm}G^{1}_{\Lambda,\hat{s}, \sigma}(\omega_{1})=\frac{1}{\omega_{1}-\epsilon_{\Lambda\hat{s}}-\Sigma^{*}_{\Lambda, \hat{s}}-i\Sigma^{Im}_{\Lambda\hat{s}}}~,\label{1-electron green}\\
&&\hspace*{-0.5cm}G^{2}_{\Lambda, \hat{s}, \sigma}(\omega_{1})=\frac{1}{\omega_{1} - \frac{\tilde{\epsilon}}{2}
-\Sigma^{2, *}_{\Lambda, \hat{s}}-i\Sigma^{2, Im }_{\Lambda, \hat{s}}},~~~\label{2-electron green}~~~\\
&&\hspace*{-0.5cm}G^{3}_{\Lambda\hat{s}, \sigma, \delta}(\omega_{1})=\frac{1}{\omega_{1}-\frac{1}{2}\epsilon^{3}_{\Lambda, \hat{s}, \delta}-\Sigma^{3, *}_{\Lambda, \hat{s}, \delta}-i\Sigma^{3, Im}_{\Lambda, \hat{s}, \delta}}
\label{2-electron 1-hole green function}~~~
\end{eqnarray}
where $\tilde{\epsilon}=p_{c}\tilde{\epsilon}^{c}_{\Lambda, \hat{s}, \delta}+p_{s}\tilde{\epsilon}^{s}_{\Lambda, \hat{s}, \delta}$ is the spin-charge admixed pair-dispersion and $\epsilon^{3}_{\Lambda, \hat{s}, \delta}=\epsilon_{\Lambda, \hat{s}}+\epsilon_{-\Lambda, T\hat{s}}-\epsilon_{\Lambda+\delta, \hat{s}}$ is the net dispersion of the 2e-1h composite.
\par\noindent
From the above imaginary self energy expressions for the 1e and 2e-1h composite excitations, we observe zeros in their respective Greens functions (eq.\eqref{1-electron green}, eq.\eqref{2-electron 1-hole green function}) near the FS ($\Lambda\to 0$) as $\omega_{1}\to \omega$, i.e., $\Sigma^{Im}_{\Lambda\hat{s},\delta}\to \infty$ and $\Sigma^{3, Im}_{\Lambda\hat{s}, \delta}\to \infty$ in this limit. Crucially, these zeros of the Greens functions are concomitant with the appearance of poles in the ee/eh spin-charge hybridized pseudospin Greens function (eq.\eqref{2-electron green})~ \cite{martin-PhysRevLett.48.362}. This signals the condensation of emergent bound states discussed earlier. Further, the excitations residing outside the gapped directions $\hat{s}'s$ describes a correlated Fermi liquid: they are described by the decoupled Hamiltonian $H^{'*}_{dec}(\omega)$ (eq.\eqref{decoupled_states_Hamiltonian_2nd_level}), and composed of 1e ($\Sigma^{1, em}_{re, \Lambda\hat{s}}$) and 2e-1h composites. 
The complete self-energies are obtained as 
\begin{eqnarray}
\Sigma^{1}_{\Lambda, \hat{s}}(\omega) &=& \Sigma^{*}_{\Lambda, \hat{s}}(\omega)\theta(\Lambda^{*}_{\hat{s}}-\Lambda)+\Sigma^{1, em}_{re, \Lambda, \hat{s}}\theta(\Lambda-\Lambda^{*}_{\hat{s}}),\nonumber\\
\Sigma^{3}_{re, \Lambda, \hat{s}, \delta}(\omega)&=&\Sigma_{re, \Lambda, \hat{s}}^{*}(\omega)\theta(\Lambda^{*}_{\hat{s}}-\Lambda)+\Sigma^{3, em}_{re, \Lambda, \hat{s}}\theta(\Lambda-\Lambda^{*}_{\hat{s}}), \nonumber\\
\Sigma^{2}_{\Lambda, \hat{s}}(\omega) &=&\Sigma^{2*}_{\Lambda\hat{s}} (\omega).\label{gapped_self}
\end{eqnarray}
We obtain the net lifetime by uniting the spectral weights of the excitations outside the window with that for the emergent degrees of freedom lying within the window 
\begin{eqnarray}
\tau(\omega, \Lambda, \hat{s})=(\Sigma^{Im, 1}_{\Lambda, \hat{s}})^{-1}+(\Sigma^{2}_{\Lambda, \hat{s}})^{-1}+(\Sigma^{3, Im}_{\Lambda, \hat{s}})^{-1}~,~~~~~
\end{eqnarray} 
Importantly, we can now use the lifetime given above in formulating the f-sum rule (eq.\eqref{f_sum_rule}) for the PG phase. Indeed, this is analogous to Cooper's demonstration of the formation of bound pairs of electrons leading to an instability of the Fermi surface as being responsible for the onset of superconductivity~\cite{cooper1956}. 
\section{Properties of the Mott liquid}\label{mottliquid}
In the energy range
$\frac{W}{2}>\omega >\omega_{ins}$,
the spin-charge interplay parameter is found to be $p=0$ from eq.\eqref{determining p}. In turn, this makes the RG flow for the Umklapp process marginal, such that we obtain the fixed point coupling $K_{c}^{*}(\omega)=\bar{U}_{0}$. Then, using the RG invariant relation eq.\eqref{RG_invariant_spin_charge_long_scatt}, we obtain the fixed point value for the spin backscattering coupling as  
\begin{eqnarray}
K_{s}^{*}(\omega) = \frac{\bar{U}_{0}(1-p)K_{c}^{*}(\omega)}{K_{c}^{*}(\omega)-p\bar{U}_{0}}|_{p=0, K_{c}^{*}(\omega)}=\bar{U}_{0} .
\end{eqnarray}
Thus, we obtain the fixed point Hamiltonian for the Mott liquid state as
\begin{eqnarray}
\hat{H}^{*}(\omega)&=&\sum_{\hat{s}}\bar{U}_{0}\bigg[\mathbf{A}_{*, \hat{s}}\cdot\mathbf{A}_{*, -\hat{s}} - \mathbf{S}_{*, \hat{s}}\cdot\mathbf{S}_{*, -\hat{s}}\bigg]~,\label{fixed_point_Ham_mott}
\end{eqnarray}
where the psuedospins $\mathbf{A}_{*, \hat{s}}$ and $\mathbf{S}_{*, \hat{s}}$ are as defined in earlier sections. As mentioned in an earlier section, this Hamiltonian is again a collection of 1D Hamiltonians, and the renormalised coupling $\bar{U}_{0}$ is therefore given by $\bar{U}_{0}=U_{0}/\sqrt{Vol}$ (see Fig.\ref{scaling_vertex}). The Hamiltonian thus obtained has antiferromagnetic and ferromagnetic exchange interactions in the charge-type pseudospin and spin-type pseudospin sectors respectively. For this theory, the 1e (eq.\eqref{1-electron green}) and 2e-1h (eq.\eqref{2-electron 1-hole green function}) Greens function has a zero for all $\hat{s}$, i.e., everywhere on the FS  (as $\omega_{1}\to \omega$, $\Lambda\to 0$). This is a Luttinger surface of zeros. A temperature scale $T_{ML}$ associated with the formation of the gapped Mott liquid can be obtained by using the connection between a quantum fluctuation scale and a thermal scale (eq.\ref{equivalent_Thermal scale}) as follows. From the form of the self energy for the decoupled degrees of freedom (eq.\eqref{gapped_self}), we employ the relations given in eqs.\eqref{forward_scattering_magnitude}, \eqref{lowest_energy_end_of_spectrum}, \eqref{fixed_point_three_particle} and \eqref{1-p self energy}, we determine $T_{ML}$ as 
\begin{eqnarray}
T_{ML} &=& \frac{\hbar}{k_{B}}\max_{\hat{s}}\tilde{\omega}_{\hat{s}}\nonumber\\
&=& \frac{\hbar N^{*}(\hat{s}_{1},0)}{2k_{B}}\left(\epsilon_{\Lambda_{0},\hat{s}_{N}}-\epsilon_{\Lambda^{**}(\omega_{ins}),\hat{s}_{N}}\right)~,\label{frequency_ML}
\end{eqnarray}
where the normal $\hat{s}_{1}=\left(1-\frac{\Lambda_{0}}{\sqrt{2}\pi}\right)\hat{s}_{AN}$ is defined in the immediate vicinity of $\hat{s}_{AN}$ in order to avoid the discontinuity at the van Hove points (see Appendix \ref{AlgoSimRG}). The normal distance $\Lambda^{**}(\omega_{ins})$ is obtained from the eq.\eqref{fixed_point_three_particle}.
\subsection{Low energy eigenstates of the Mott liquid}\label{eigMottLiquid}
\par\noindent 
Some of the many-body eigenstates for $H$ (eq.\eqref{fixed_point_Ham_mott}) are obtained by entangling every charge pseudospin with a unique pair of occupied electronic states residing outside the window, such that the spin pseudospins have vanishing magnitude ($(\mathbf{S}_{\Lambda\hat{s}})^{2}=0$). We now lay out the construction of such states. The vacuum state along a pair of normal $\hat{s}$ is given by
\begin{eqnarray}
|0\rangle_{d} = |[0_{1'}0_{1}0_{-1}0_{-1'}]..[0_{1'}0_{1}0_{-1}0_{-1'}]..[0_{n'}0_{n}0_{-n}0_{-n'}]\rangle_{d}~,
\end{eqnarray}
where the labels $(j,d):=\mathbf{k}_{\Lambda_{j}\hat{s}},\sigma$ and $(-j,d):=\mathbf{k}_{-\Lambda_{j},T\hat{s}},-\sigma$ represent momentum vectors \textit{within} the emergent window $\Lambda_{j}<\Lambda^{*}_{\hat{s}}$. On the other hand, the labels $(j',d): = \mathbf{k}_{\Lambda_{j}-
2\Lambda^{*}_{\hat{s}},\hat{s}},\sigma$, $(-j',d)=\mathbf{k}_{\Lambda_{j}-
2\Lambda^{*}_{\hat{s}},\hat{s}},-\sigma$ represent the momentum vectors residing \textit{outside} the emergent window $\Lambda^{*}_{\hat{s}}<|\Lambda_{j}-
2\Lambda^{*}_{\hat{s}}|$. The boundary states of the emergent window are represented by $(n,d)=\mathbf{k}_{\Lambda_
{j^{*}}\hat{s}},\sigma$ and $(-n,d)=\mathbf{k}_{-\Lambda_{j^{*}}T\hat{s}},-\sigma$.
\par\noindent
We denote the configurations of a single charge pseudospin and its associated pair of electronic states by
\begin{eqnarray}
 &&\hspace*{-0.1cm}|\uparrow_{j}\rangle_{d}:= |1_{\mathbf{k}_{\Lambda_{j}\hat{s}}\sigma}1_{\mathbf{k}_{-\Lambda_{j}, T\hat{s}}, -\sigma}\rangle , |\downarrow_{j}\rangle_{d}:= |0_{\mathbf{k}_{\Lambda_{j}, \hat{s}}, \sigma}0_{\mathbf{k}_{-\Lambda_{j}, T\hat{s}}, -\sigma}\rangle ~.~~~\label{states_of_1}
\end{eqnarray}
The vacuum can then be rewritten in the charge pseudospin basis
\begin{eqnarray}
|0\rangle_{d} &=& |[0_{1'}\downarrow_{1}0_{-1'}]..[0_{2'}\downarrow_{2}0_{-1'}]..[0_{n'}\downarrow_{n}0_{-n'}]\rangle_{d}\nonumber\\
&=&|A_{*,\hat{s}}=N^{*}_{\hat{s}},A^{z}_{*,\hat{s}}=-N^{*}_{\hat{s}}~,\rangle
\end{eqnarray}
where $2N^{*}_{\hat{s}}$ is the number of charge pseudospins along $\hat{s}$. The application of operator $M^{+}_{k,d} = c^{\dagger}_{j,d}A^{+}_{j',d}c^{\dagger}_{-j',d}$ flips a charge pseudospin in the $\uparrow$  configuration along with creating electrons in two associated states: $M^{+}_{j,d}|0\rangle = |[1_{j'}\uparrow_{j}1_{-j'}]\rangle_{d}$. From the above operation, it can be easily seen that for every charge pseudospin-flip operation, a pair of electrons created outside the window get entangled. Further, we find that the spin pseudospin raising and lowering operators annihilate this composite space
\begin{eqnarray}
S^{\pm}|[1_{j'}\uparrow_{j}1_{-j'}]\rangle = 0~,~S^{\pm}|[0_{j'}\downarrow_{j}0_{-j'}]\rangle = 0~.
\end{eqnarray}
The $z$-component can also be shown to vanish in a similar fashion. With these constraints, we can now determine eigenstates of the total charge pseudospin angular momentum operator along $\hat{s}$ $\mathbf{A}_{*,\hat{s}}=\sum_{\Lambda =0}^{\Lambda^{*}_{\hat{s}}}\mathbf{A}_{\Lambda\hat{s}}$ 
\begin{eqnarray}
\hspace*{-0.6cm}|A_{*,\hat{s}}=N=A^{z}_{*,\hat{s}}\rangle 
&=&\hspace*{-0.2cm}\sum_{j_{1},..,j_{l}}\hspace*{-0.1cm}\begin{vmatrix} e^{ij_{1}q_{1}}&\ldots &e^{ij_{1}q_{l}}\\
\vdots & \ddots  &\vdots\\
e^{ij_{l}q_{1}}&\ldots &e^{ij_{l}q_{l}}\end{vmatrix}
\prod_{i}M^{+}_{j_{i},d}|0\rangle_{d}
\end{eqnarray}
where the Slater determinant involves states with wavevectors $q_{s}=\frac{s\pi }{2N^{*}_{\hat{s}}}\in[-\pi,\pi)$, $l=N+N^{*}_{\hat{s}}$ . The eigenvalue of $\mathbf{A}_{*,\hat{s}}^{2}$ for the state $|A_{*,\hat{s}}=N=A^{z}_{*,\hat{s}}\rangle$ is given by $A_{*,\hat{s}}^{2} = N(N+1)$. This can be obtained by noting that
\begin{eqnarray}
\mathbf{A}_{*,\hat{s}}^{2} = \frac{3}{4}(2N^{*}_{\hat{s}})+\sum_{i<j=1}^{2N^{*}_{\hat{s}}}P_{ij}-\frac{1}{2}\binom{2N^{*}_{\hat{s}}}{2}
\end{eqnarray}
where $P_{ij}=(2\mathbf{A}_{\Lambda_{i}\hat{s}}\cdot\mathbf{A}_{\Lambda_{j}\hat{s}}+\frac{1}{2})$ is the permutation operator that exchanges the $i$ and $j$ charge pseudospin configurations. 
\par\noindent
One class of eigenstates for the Hamiltonian eq.\eqref{fixed_point_Ham_mott}
can now be obtained by entangling states $|A_{*,\hat{s}}=N,A^{z}_{*,\hat{s}}=p\rangle$ from one side of Fermi surface with $|A_{*,-\hat{s}}=N,A^{z}_{*,-\hat{s}}=p\rangle$ from the diametrically opposite side
\begin{eqnarray}
&&|A_{*} = m,A^{z}_{*} = p, A_{*,\hat{s}}=N_{1},A_{*,-\hat{s}}=N_{2}\rangle \\
&=&\sum C^{m,p}_{N_{1},p_{1};N_{2},p_{2}}
\nonumber\\
&&\times |A_{*,\hat{s}}=N_{1},A^{z}_{*,\hat{s}}=p_{1};A_{*,-\hat{s}}=N_{2},A^{z}_{*,-\hat{s}}=p_{2}\rangle\nonumber
\end{eqnarray} 
where $p_{1}=p-p_{2}$ and $C^{m,p}_{N_{1},p_{1};N_{2},p_{2}}$ are Clebsch-Gordon coefficients~\cite{edmond1985}. The energy eigenvalues for these states are then obtained as
\begin{equation}
E=\sum_{\hat{s}}\frac{\bar{U}_{0}}{2}[m(m+1)-N_{1}(N_{1}+1)-N_{2}(N_{2}+1)]~. 
\end{equation}
From the expression, we observe that low-lying eigenstates of the spectrum reside in the space of states  
\begin{eqnarray}
|\Psi_{\hat{s}}, m\rangle = |A_{*}=m,A_{*,\hat{s}}=A_{*,-\hat{s}} = N^{*}_{\hat{s}},S_{\Lambda\hat{s}}=0\rangle\label{class_1_wvfn}  
~,~~\end{eqnarray}
where the charge pseudospin angular momentum of the two normal directions $\hat{s}$ and $-\hat{s}$ have magnitude $N^{*}_{\hat{s}}(N^{*}_{\hat{s}}+1)$. sThe eigenvalue of the Hamiltonian $H^{*}(\omega)|\Psi_{\hat{s},m}\rangle = E^{c}_{m}|\Psi_{\hat{s},m}\rangle$ associated with the eigenstates eq.\eqref{class_1_wvfn} are
\begin{eqnarray}
E^{c}_{m} = \sum_{\hat{s}}\bar{U}_{0}\frac{1}{2}\left(m(m+1)-2N^{*}_{\hat{s}}\left(N^{*}_{\hat{s}}+1\right)\right)~,\label{energy-1}
\end{eqnarray}
where $N^{*}_{\hat{s}}$ is the number of states in the window $\left[0, \Lambda^{*}_{\hat{s}}\right]$ 
and $0\leq m \leq 2N^{*}_{\hat{s}}$. The lowest lying eigenstate $|\Psi_{\hat{s}}, 0\rangle$ is obtained for $m=0$, i.e., a charge  pseudospin singlet with energy $E_{1}=-\bar{U}_{0}\sum_{\hat{s}}N^{*}_{\hat{s}}\left(N^{*}_{\hat{s}}+1\right)$.
\par\noindent 
Another family of eigenstates 
\begin{equation}
|\Phi_{\hat{s}},m\rangle = |S_{*}=m,S_{*,\hat{s}}=S_{*,-\hat{s}} = N^{*}_{\hat{s}},A_{\Lambda\hat{s}}=0\rangle \label{class_2_wvfn}
\end{equation} are obtained from the spin pseudospin subspace $\mathbf{S}_{\Lambda\hat{s}}^{2}=\frac{3}{4}$, and where all the individual
charge pseudospins $\mathbf{A}_{\Lambda\hat{s}}=0$.
The energy spectrum for this class of wave functions, $H^{*}(\omega)|\Phi_{\hat{s}, m}\rangle = E^{s}_{m}|\Phi_{\hat{s}, m}\rangle$, is given by 
\begin{eqnarray}
E^{s}_{m} = -\sum_{\hat{s}}\bar{U}_{0}\frac{1}{2}\left(m(m+1)-2N^{*}_{\hat{s}}\left(N^{*}_{\hat{s}}+1\right)\right)~,\label{energy-2}
\end{eqnarray}
where $m\in (0, 2N^{*}_{\hat{s}})$ once again. 
For the lowest lying eigenstate $m=2N^{*}_{\hat{s}}$ configuration here, the energy is given by $E_{2}=-\bar{U}_{0}\sum_{\hat{s}}N^{*2}_{\hat{s}}$.
\par\noindent
But have we accounted for all the states within the window $\Lambda^{*}_{\hat{s}}$? An answer can be found by first computing the size of the Hilbert space for the Hamiltonian $H^{*}(\omega)$. By choosing every pair of two states $(\Lambda, \hat{s}),(-\Lambda, T\hat{s})$, where $\Lambda<\Lambda^{*}_{\hat{s}}$, there are two more electronic states that are chosen accordingly, i.e., $(\Lambda-2\Lambda^{*}_{\hat{s}}, \hat{s})$, $(-\Lambda-2\Lambda^{*}_{T\hat{s}}, T\hat{s})$. Thus, the total size of the Hilbert space within the window $\Lambda^{*}_{\hat{s}}$ is given by $2^{2N^{*}_{\hat{s}}}$. 
On the other hand, the total number of eigenstates of type $|\Psi_{\hat{s}},m\rangle$ is $2^{N^{*}_{\hat{s}}}$, as there are $2$ choices per composite subspace (eq.\eqref{states_of_1}) and there are $N^{*}_{\hat{s}}$ such subspaces. An identical count of $2^{N^{*}_{\hat{s}}}$ is obtained similarly for the Hilbert space of the states $|\Phi_{\hat{s}}, m\rangle$. Therefore, these two sets of eigenstates do not exhaust the entire eigenspectrum; there are other exotic combinations of entangled states possible, whose count is $2^{2N^{*}_{\hat{s}}}-2^{N^{*}_{\hat{s}}+1}$. 
\par\noindent
The state spaces eqs.\eqref{class_1_wvfn} and \eqref{class_2_wvfn} have been chosen such that they possess either $\langle(\Delta S^{x, y, z}_{\Lambda\hat{s}})^{2}\rangle= 0$ or $\langle(\Delta A^{x, y, z}_{\Lambda\hat{s}})^{2}\rangle= 0$. This sets them out as excellent candidate members of the low energy spectrum: any locally spin-charge combination must, on the other hand, involve non-zero uncertainties $\langle(\Delta S^{x, y, z}_{\Lambda\hat{s}})^{2}\rangle\neq 0$ as well as $\langle(\Delta A^{x, y, z}_{\Lambda\hat{s}})^{2}\rangle\neq 0$, raising their energy. 
We can now conclude on the ground states of $H^{*} (\omega)$. Clearly, $|\Psi_{\hat{s}}, m=0\rangle$ and  $|\Phi_{\hat{s}}, m=2N^{*}_{\hat{s}}\rangle$ are the lowest energy states (with energies $E_{1}$ and $E_{2}$ respectively. In order for them to be degenerate ground states, we need to take the thermodynamic limit $N^{*}_{\hat{s}}>>1$, such that  $N^{*2}_{\hat{s}}>N^{*}_{\hat{s}}$ leading to $E_{1}=E_{2}\equiv E_{g}$. Further, in this limit, the sum and difference combinations of the states 
eqs.\eqref{class_1_wvfn} and \eqref{class_2_wvfn}
\begin{eqnarray}
|\Gamma_{\pm}, m\rangle = \prod_{\hat{s}}\frac{1}{\sqrt{2}}\left[|\Psi_{\hat{s}}, m\rangle\pm |\Phi_{\hat{s}}, 2N^{*}_{\hat{s}}-m\rangle\right]
\end{eqnarray}  
are also eigenstates of $H^{*}(\omega)$. 
\par\noindent
One of the lowest excitations lying above these ground states can be written by noting that the magnitude of the pseudospins $A_{*}^{\hat{s}}$ and $A_{*}^{-\hat{s}}$ are lowered from $N_{\hat{s}}^{*}$ to $N_{\hat{s}}^{*}-1$, such that the charge pseudospin singlet formed from these two has the form 
\begin{eqnarray}
| A_{*}=0, A_{*,\hat{s}}\equiv A_{*,-\hat{s}}=N^{*}_{\hat{s}}-1,S_{\Lambda\hat{s}}=0\rangle~.\label{exc-1}
\end{eqnarray}
Similarly, another lowest lying excitation is given by 
\begin{eqnarray}
|A_{\Lambda\hat{s}}=0, S_{*}=2N^{*}_{\hat{s}}-2, S_{*,\hat{s}}\equiv S_{*,-\hat{s}}=2N^{*}_{\hat{s}}-2\rangle~.\label{exc-2}
\end{eqnarray}
The energy cost to reach the lowest lying excited state (i.e., the spectral gap) in the thermodynamic limit ($N^{*}_{\hat{s}}>>1$) is also easily obtained as $\Delta E=\bar{U}_{0}\sum_{\hat{s}}N^{*}_{\hat{s}}$.
\subsubsection{Topological features of the Mott liquid}
By putting periodic boundary conditions on the momentum space window, we can construct a set of nonlocal operators $W_{m}$
\begin{eqnarray}
W_{m}&=&\exp\left[i\frac{\pi}{2}(|\Gamma_{+},m\rangle\langle\Gamma_{+},m|-|\Gamma_{-},m\rangle\langle\Gamma_{-},m|]-1)\right]~,\nonumber
\end{eqnarray}
such that $W_{m}$ commutes with the $SU(2)\times SU(2)$ pseudospin rotational invariant Hamiltonian in the projected subspace of $P_{c}+P_{s}$
\begin{equation}
[(P_{c}+P_{s})H^{*}(P_{c}+P_{s}), W_{m}]=0~ \forall~m~,
\end{equation}
where the projection operators are given by $P_{c}=\sum_{m}|\Psi_{\hat{s}}, m\rangle\langle\Psi_{\hat{s}}, m|$ and $P_{s}=\sum_{m}|\Phi_{\hat{s}}, m\rangle\langle\Phi_{\hat{s}}, m|$. 
Further, it is clear that $W_{m}|\Gamma_{\pm},m\rangle = \pm |\Gamma_{\pm},m\rangle$~.
Thus, the simultaneous degenerate ground state eigenfunctions of the Hamiltonian $H^{*}$ of the Mott liquid and the operator $W_{m=0}$ can be written as
\begin{eqnarray}
|\Gamma_{\pm},0\rangle = \prod_{\hat{s}}\left[\frac{1}{\sqrt{2}}\left(|\Psi_{\hat{s}}, 0\rangle \pm |\Phi_{\hat{s}}, 2N_{\hat{s}}^{*}\rangle\right)\right]~.\label{ground_state}
\end{eqnarray}
These two degenerate ground states are connected via a twist operator/ nonlocal gauge transformation $\hat{O}$
\begin{eqnarray}
\hat{O} = \exp\left[4\pi i \sum_{\Lambda\hat{s}}\frac{(S^{z}_{\Lambda\hat{s}})^{2}}{N^{*}_{\hat{s}}}\right]~,~\hat{O}|\Gamma_{\pm},0\rangle = |\Gamma_{\mp},0\rangle~.
\end{eqnarray}
As these two states are protected by the many body gap $\Delta E = \bar{U}_{0}\sum_{\hat{s}}N^{*}_{\hat{s}}$ (where $\bar{U}_{0}=U_{0}/\text{vol}$) given above, adiabatic passage between these degenerate ground states via the application of the twist operator $\hat{O}$ involve the creation of charge-$1/2$ excitations~\cite{wen1990,oshikawa2006,chen2010}. This can be seen from the anticommutation relation
\begin{eqnarray}
\lbrace \hat{O},W_{0}\rbrace =0~.
\label{nontriv-anticommute}
\end{eqnarray}
The above relation, along with $[\hat{O}^{2},W_{0}]=0$, allows us to conclude that the ground state manifold is {\it topologically} degenerate in the thermodynamic limit.
\subsection{Benchmarking against existing numerical results}
\begin{figure}[h!]
\hspace*{-1.6cm}
\includegraphics[width=0.59\textwidth,height=7cm]{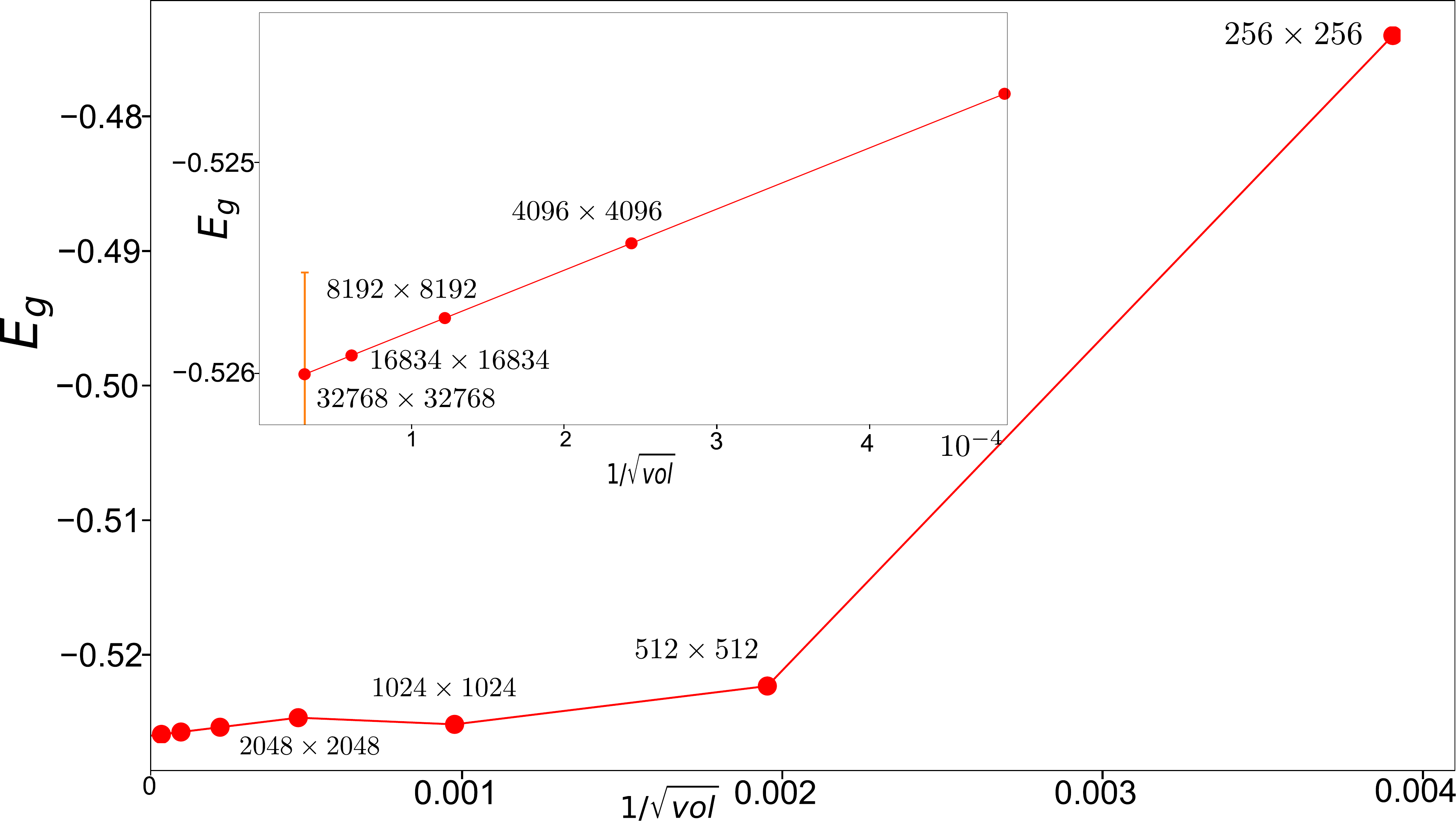}
\caption{(Colour Online) Finite-size scaling of the saturation value for $E_{g}\equiv E_{gs}$ with $1/\sqrt{\text{Volume}}$ with increasing $k$-space grid size from $2^{8}\times 2^{8}$ to $2^{15}\times 2^{15}$. The saturation $E_{gs}$ for the largest grid is observed to be $-0.526$(for $t=1$). The error bar for all data points is $\sim \text{O}(10^{-}4t)$. Inset: Zoomed view of finite-size scaling plot for lattice sizes $2^{11}\times 2^{11}$ to $2^{15}\times 2^{15}$.}
\label{groundstateenergy}
\end{figure}

\begin{figure}[h!]
\includegraphics[scale=0.45]{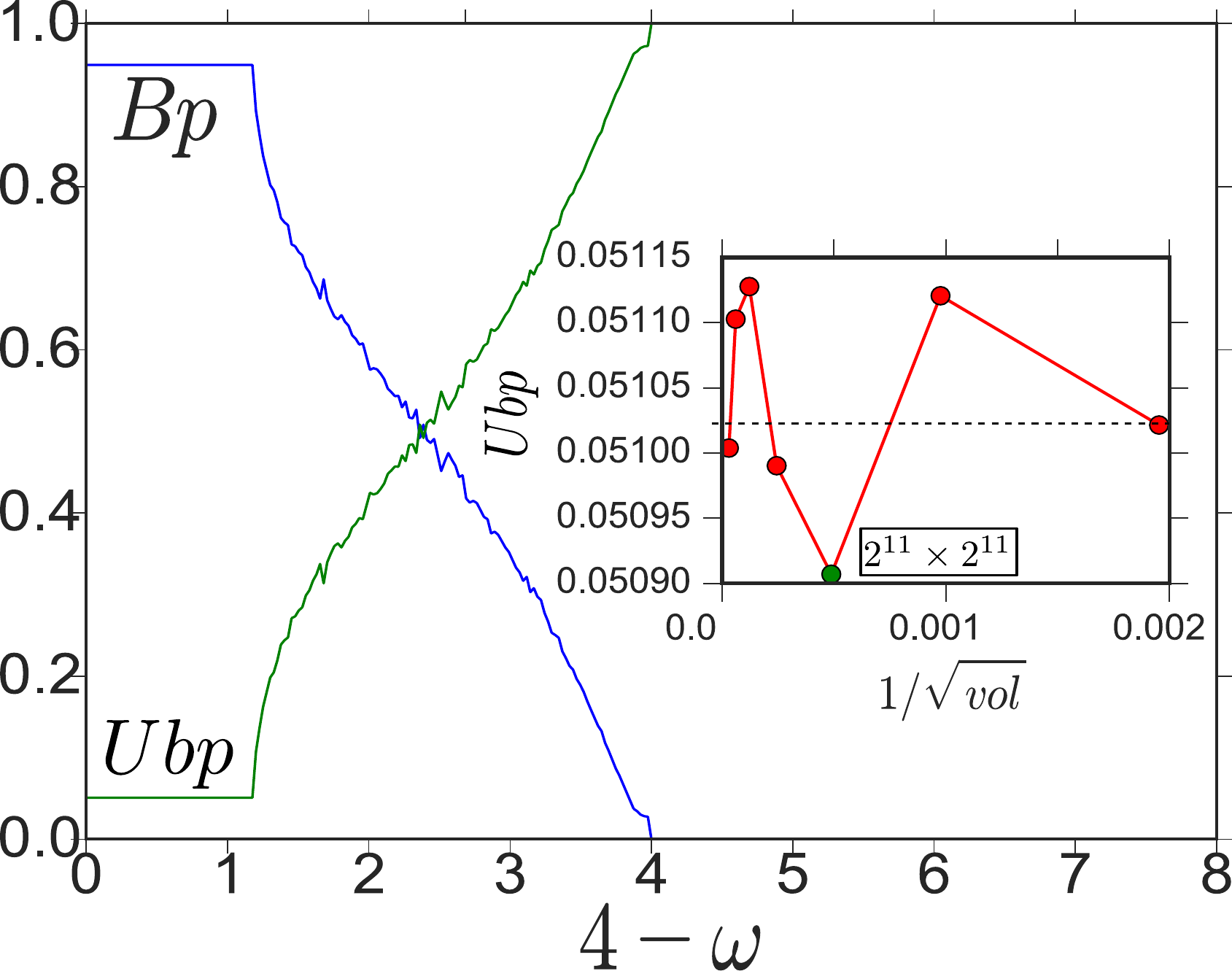}
\caption{(Colour Online) Blue curve: Growth in fraction of bound pairs ($Bp$) of electronic states (holon-doublon/two-spinon) within the low-energy window at half-filling and $U=8t$ during the passage from metal ($\omega <0$) to the Mott liquid ($\omega >2.8$) through the pseudogap ($0\leq \omega\leq 2.8$). Green curve: Concomitant decay in fraction of unbound pairs ($Ubp$). Calculations are done for a $k$-space grid of $2^{11}\times 2^{11}$. Inset: Finite-size scaling of the saturation value for the fraction of unbound pairs $(Ubp^{*})$ in the Mott liquid with $1/\sqrt{\text{Volume}}$ with increasing $k$-space grid size from $2^{9}\times 2^{9}$ to $2^{15}\times 2^{15}$. The saturation fraction display small fluctuations about an average value $Ubp^{*}\sim 0.051$. Green point is the $k$-space grid used in the main figure. The error bar for all data points is $\sim \text{O}(10^{-5})$~.}
\label{windowwidth}
\end{figure}

In order to benchmark the results obtained from the effective low-energy Hamiltonian and wavefunctions given above against those found from existing numerical methods applied to the 2D Hubbard model on the square lattice~\cite{leblanc2015solutions,ehlers2017hybridDMRG,dagotto1992}, we present results for the ground state energy per particle $E_{g}$ and the fraction of bound pairs ($Bp$) in the gapped Mott liquid ground state. The analytic forms of $E_{g}$ and $Bp$ are computed from the spin and charge backscattering parts of the effective Hamiltonian given above ($H^{*}(\omega)$) and are found to be
\begin{eqnarray}
E_{g} &=& \frac{E_{2}}{N_{e}} = -\sum_{\hat{s}}\frac{\bar{U}_{0}(N^{*}_{\hat{s}})^{2}}{N_{e}}~,\nonumber\\
Bp &=& \frac{\sum_{\hat{s}}N^{*}_{\hat{s}}}{N_{e}}~,~Ubp = 1 - Bp~,
\end{eqnarray} 
where the fraction of unbound pairs is denoted by $Ubp$.
The plots for $E_{g}$ and $(Bp, Ubp)$ versus the probe energy scale $\omega$ are shown in Fig.~\ref{groundstateenergy} and Fig.~\ref{windowwidth} respectively. A comparison of the saturation value of the ground state energy for the largest $k$-space grid used in our simulations ($2^{15}\times 2^{15}$), $E_{g}^{*}=-0.526t$ (shown in Fig.~(\ref{groundstateenergy})), is well within the range $-0.51t<E_{gs}<-0.53t$ obtained in the thermodynamic limit from several state-of-the-art numerical methods applied to the half-filled 2D Hubbard model at $U=8t$~\cite{leblanc2015solutions,ehlers2017hybridDMRG,dagotto1992}. Similarly, the average saturation value for the fraction of unbound pairs ($Ubp^{*}$) in the Mott liquid obtained from a 
finite-size scaling analysis (inset of Fig.~(\ref{windowwidth})), $Ubp^{*}\sim 0.051$, is also well within the range $0.0535 < Ubp <0.0545$ obtained in Ref.\cite{leblanc2015solutions}. The code used for the numerical computation of the ground state energy shown here is made available electronically~\cite{anirbanhubbard2019code}. 
As an added note, we point the reader to further benchmarking exercises presented in Appendix-\ref{further benchmarking} for the cases of $U/t=2,4,6,10$ at half-filling. We continue to find excellent quantitative agreement with exact diagonalization studies from Ref.\cite{dagotto1992} and other numerical methods reported in Refs.\cite{leblanc2015solutions,ehlers2017hybridDMRG}. We stress that this offers confidence in the effective Hamiltonian and ground state wavefunction we have obtained for the half-filled Mott liquid.  
\newpage
\subsection{Symmetry breaking of the Mott Liquid}\label{symmbreakMottliquid}
Our RG analysis can also be extended to show that the topologically ordered Mott liquid ground state is replaced, under renormalisation, by a chosen ordered state with an order parameter corresponding to a defined broken symmetry. As RG transformations are, by definition, meant to preserve the symmetries of the Hamiltonian they act upon, such symmetry-broken states can only be reached from our RG analysis by first explicitly including an order parameter-field term in the Hamiltonian and then proceeding with the RG transformations\cite{salmhofer2004}. In this way, for instance, we can reach a $(\pi,\pi)$ charge density wave (CDW) broken symmetry ground state in the presence of a staggered chemical potential, as well as  
a $(\pi, \pi)$ spin density wave (SDW) Ne\'{e}l ground state in the presence of a staggered magnetic field. We demonstrate the case of the SDW below. First, we write the Hamiltonian
\begin{eqnarray}
\hat{H}^{*}_{SB} = \sum_{\mathbf{r}}(-1)^{x+y}\hat{n}_{\mathbf{r}\sigma}+H^{*}(\omega)~,
\end{eqnarray}
where $H^{*}(\omega)$ is the fixed point Hamiltonian of the topologically ordered Mott liquid obtained earlier. Let us now restart a renormalization procedure with the channels $N_{\Lambda_{n}}, m$ where tangential scattering is present alongside the effects of a staggering field given by
\begin{eqnarray}
H_{stag} = h\sum_{\mathbf{r}}e^{i\pi(i+j)}S^{z}_{\mathbf{r}}~,~ \mathbf{r} = i\hat{x}+j\hat{y}
\end{eqnarray}
This staggering field in momentum space can be represented by $H_{stagg} = \sum_{\hat{s}}H_{stag,\hat{s}}$, where
\begin{eqnarray}
H_{stagg, \hat{s}} &=& h\sum_{\mathbf{k}, \mathbf{Q}}(c^{\dagger}_{\mathbf{k}_{\Lambda}(\hat{s})\sigma}c_{\mathbf{k}_{\Lambda}(-T\hat{s})\sigma}-c^{\dagger}_{\mathbf{k}_{\Lambda}(\hat{s})-\sigma}c_{\mathbf{k}_{\Lambda}(-T\hat{s})-\sigma})~~~~~
\end{eqnarray}
The effects of tangential scattering terms, when included along with forward and backward scattering gives, are given by the Hamiltonian
\begin{eqnarray}
H_{comp} = V\sum_{\Lambda, \Lambda', \hat{s}, \hat{s}'}c^{\dagger}_{\Lambda\hat{s}\sigma}c_{\Lambda' -T\hat{s}\sigma}c^{\dagger}_{\Lambda'T\hat{s}'-\sigma}c_{\Lambda -\hat{s}'-\sigma}~.
\end{eqnarray}
Let us consider the staggering field as the larger energy scale in comparison to the kinetic energy scale, and confine our attention to the participation of the electrons close to the AN points. The latter is justified by the presence of the van Hove singularities in the single-particle density of states in the half-filled tight-binding problem on the square lattice. The basis states in the high energy sector taking part in the RG are completely entangled via tangential scattering and essentially correspond to the following channels $|N_{\Lambda_{n}}, m\rangle$. Here, $N_{\Lambda_{n}}$ gives the electronic density of states at a normal deviation $\Lambda_{n}$ from the AN point of the Fermi surface which takes part in the RG process ($k=\sqrt{k_{x}^{2}+k_{y}^{2}}$),
$N_{\Lambda_{n}} = \ln\left|\frac{\Lambda_{0}}{\Lambda_{n}}\right|$~.
Using the Hamiltonian renormalization equation given earlier, the RG equation for the tangential Umklapp scattering processes near the antinodes is given in terms of the dimensionless coupling $V'_{\Lambda}= \frac{V_{\Lambda}}{h_{\Lambda}}$
\begin{eqnarray}
\frac{d(V'_{\Lambda})}{d\ln\frac{\Lambda}{\Lambda_{0}}}&=&N_{\Lambda}\frac{V^{'2}_{\Lambda}}{1-V^{'}_{\Lambda}N_{\Lambda}(N_{\Lambda}+1)}. 
\end{eqnarray}
For weak coupling
\begin{eqnarray}
V'_{\Lambda}<<\frac{1}{N_{\Lambda}(N_{\Lambda}+1)} = \left(1/\left(\frac{N_{e}}{P_{FS}}\frac{\Lambda^{*}(\omega)}{\omega}\right)\right)^{2}~,
\end{eqnarray}
where $P_{FS}=\sum_{\hat{s}}1$, the RG equation for $V'_{\Lambda}$ reduces to
\begin{eqnarray}
\frac{dV'_{\Lambda}}{d\ln\frac{\Lambda}{\Lambda_{0}}} &=& 
N_{\Lambda}V^{'2}_{\Lambda}\to  V^{'}_{\Lambda}=\frac{V^{'}_{0}}{1-V^{'}_{0}\ln^{2}\left(\frac{\Lambda}{\Lambda_{0}}\right)}~,~\nonumber\\
V^{'}_{0}&=&\frac{U_{0}}{h_{0}}~,~\frac{\Lambda}{\Lambda_{0}} = \exp\left(-\frac{1}{\sqrt{U_{0}}}\right)
\end{eqnarray}
In this way, we find the the well-known form for the gap function $\Lambda$ corresponding to the Neel SDW state on the 2D square lattice at $1/2$-filling, and as obtained from a mean-field analysis\cite{fradkin2013field}.
\subsection{Cooper pair fluctutations within Mott liquid}
\label{cooperflucMOtt}
The pseudospin flip interaction $-\bar{U}_{0}(S^{+}_{*, \hat{s}}S^{-}_{*, -\hat{s}}+h.c.)$ present within the Mott liquid can be rewritten in terms of interactions between finite-momentum Cooper pairing terms as follows:
\begin{eqnarray}
\sum_{\Lambda\Lambda'}\left[S^{+}_{\Lambda\hat{s}}S^{-}_{\Lambda'\hat{s}}+h.c.\right]
= \sum_{\mathbf{k}, \mathbf{p}}c^{\dagger}_{\mathbf{k}\uparrow}c^{\dagger}_{\mathbf{p}-\mathbf{k}\downarrow} c_{\mathbf{p}-\mathbf{k}'\downarrow}c_{\mathbf{k}'\uparrow}~,
\end{eqnarray}
where $\mathbf{k}=\mathbf{k}_{\Lambda, \hat{s}}$ and $\mathbf{k}' = \mathbf{k}_{2\Lambda^{*}_{\hat{s}}-\Lambda, \hat{s}}$. Note that the wave vector $\mathbf{p}-\mathbf{k}$ is centered around the opposite normal $-\hat{s}$, as $\mathbf{p}-\mathbf{k} = \mathbf{k}_{\Lambda' -\hat{s}}$. The condition of choosing electronic states from diametrically opposite parts of the Fermi surface is satisfied with the pair momentum $0\leq \mathbf{p}\leq2\Lambda^{*}_{\hat{s},\omega}\hat{s}$. The presence of entanglement between the charge/spin type pseudospins in eq.\eqref{ground_state}, along with mixing between different $\mathbf{p}$-momentum pairs, is captured in the off-diagonal long range order (ODLRO)
\begin{eqnarray}
\rho(\mathbf{r}-\mathbf{r'}) &=& \langle \Psi_{1}|\psi^{\dagger}_{\mathbf{r}\uparrow}\psi^{\dagger}_{\mathbf{r}\downarrow}\psi_{\mathbf{r}'\downarrow}\psi^{\dagger}_{\mathbf{r}'\uparrow}|\Psi_{1}\rangle \nonumber\\
&=& \frac{1}{2}\sum^{2\Lambda^{*}_{\hat{s}}}_{\hat{s}, \mathbf{p}=0}\frac{N_{\Lambda^{*}_{\hat{s}}}}{L^{2}}\cos(\mathbf{p}\cdot(\mathbf{r}-\mathbf{r}'))~.
\end{eqnarray}
In the calculation above, we have chosen the quantization axis of $\mathbf{S}_{\Lambda\hat{s}}$ along the x-direction. From this expression, we observe that the ODLRO decays for $|(\mathbf{r}-\mathbf{r}')\cdot\hat{s}|>(2\Lambda^{*}_{\hat{s}})^{-1}$. The prefactor of $1/2$ in the expression for the ODLRO arises from the superposition of the two types of many body states (see eq\eqref{ground_state}), while its anisotropic form results from the geometry of the Fermi surface in the tight-binding problem. Finally, $N_{\Lambda^{*}_{\hat{s}}}$ is the number of composite objects along the $\hat{s}$ direction.
\par\noindent 
We saw earlier the susceptibility of the Mott liquid towards a symmetry broken Neel antiferromagnet. From the analysis presented for the ODLRO above, we find that the global spin-charge entanglement prevents the condensation of zero-momentum Cooper pairs. This leads us to conclude that antiferromagnetism is clearly favored over a $U(1)$-symmetry broken superconducting state as an instability of the Mott liquid. This appears to be consistent with the finding of subdominant superconducting correlations in the insulating phase at half filling from variational Monte Carlo studies of the $1/2$-filled Hubbard model~\cite{tocchio2016}. We will return to the investigation of superconducting fluctuations upon considering the case of the doped Mott liquid in a later section. Finally, we can also conclude that an instability towards a $(\pi, \pi)$ CDW symmetry-broken ordered state is clearly subdominant to the Neel SDW, as the former must overcome the entanglement of the charge pseudospin singlet state $|\Psi_{\hat{s}, m}\rangle$ of the Mott liquid. 
\section{Mottness Collapse and quantum criticality with hole doping}\label{MottCollapseSection}
\begin{figure}[h!]
\hspace*{-1.5cm}
\includegraphics[width=0.6\textwidth]{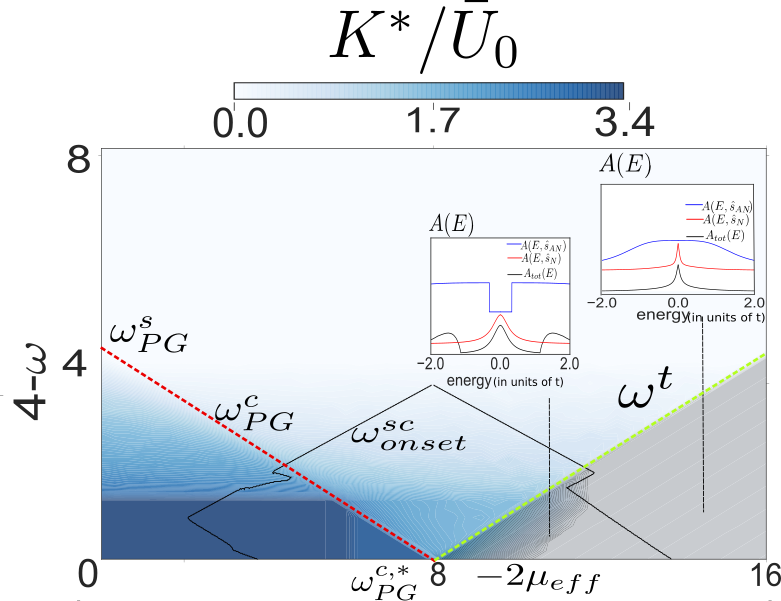}
\caption{(Colour Online) RG phase diagram with hole doping showing quantum critical point (QCP) at $-\Delta\mu_{eff}=4=\omega$ and its wedge extending to higher energies. All phases (NFL, PG, ML, correlated Fermi liquid (CFL), PG-CFL and QCP-wedge) and related energy scales are shown in the colour bars, and discussed in detail in the text. Dashed line shows highest energy scale for superconducting fluctuations. Insets: N, AN and FS-averaged spectral function ($A(E)$) for the QCP and gapless CFL (grey region: $\omega=3.2,\Delta\mu_{eff}=-7.5$).}\label{Phase_diagram_with_doping-1}
\end{figure}

~We present the RG phase diagram for the 2D Hubbard model with doping away from 1/2-filling in Fig.~\ref{Phase_diagram_with_doping-1}. This phase diagram results from longitudinal and tangential scattering RG equations (Sec.\ref{RG_flow_long_tan}) upon including effects of doping $\Delta\mu_{eff}\neq 0$ and Cooper pair scattering processes across the Fermi surface. 
\begin{figure}[h!]
\hspace*{-0.5cm}
\includegraphics[width=0.45\textwidth, height=0.95\textwidth]{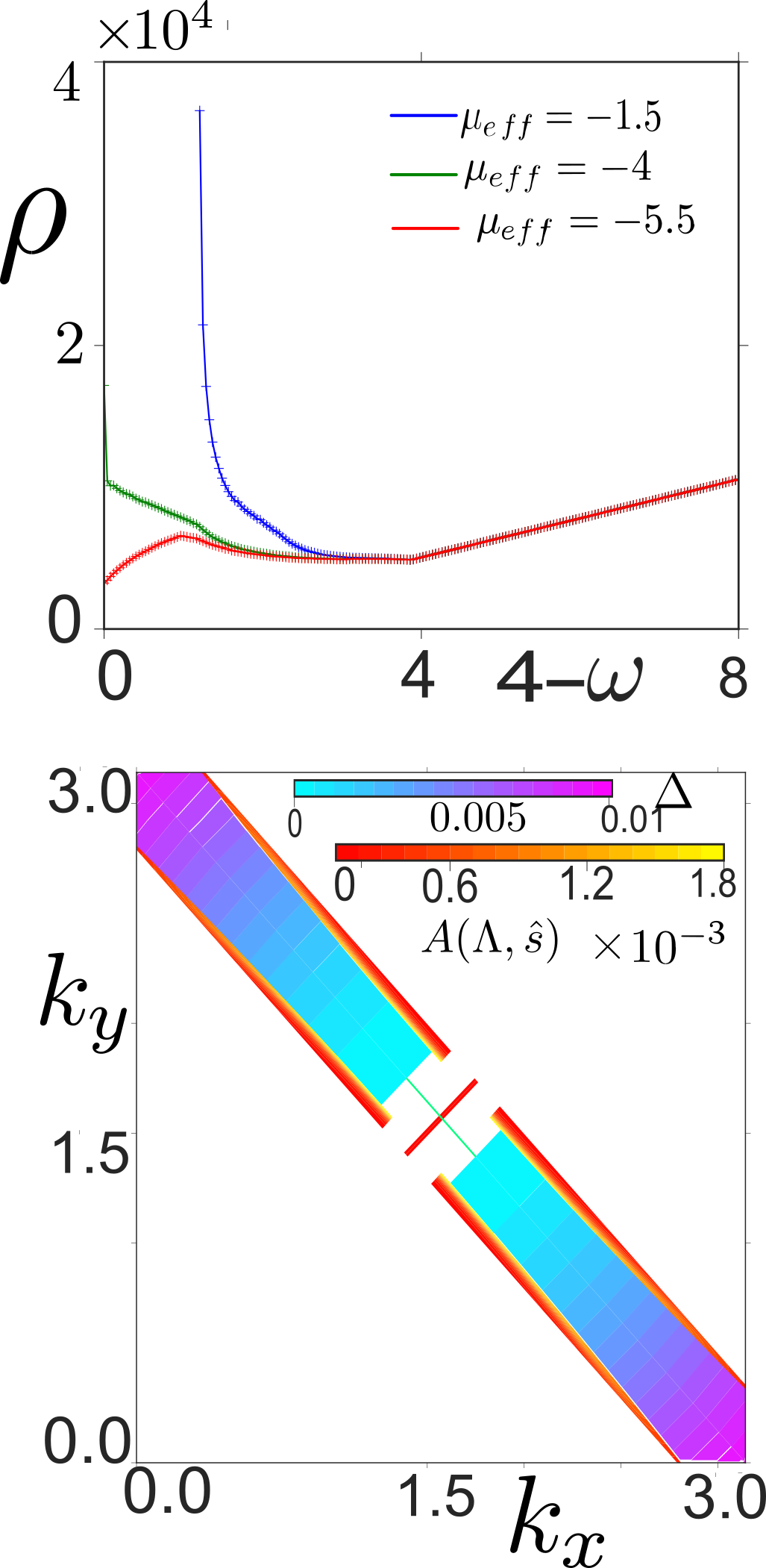}
\caption{(Colour Online) Upper Panel: Resistivity ($\rho$) vs. $4-\omega$ for various $\Delta\mu_{eff}$, showing passage from NFL into (blue) ML, (green) the QCP and (red) PG-CFL. Lower Panel: Map of $A(E)$ at the QCP.}\label{Phase_diagram_with_doping-2}
\end{figure}
Upon doping the Mott insulator via an effective chemical potential, we observe a marginal Fermi liquid metal (whitish-blue region in Fig.~\ref{Phase_diagram_with_doping-1}) in the energy scale $\frac{W}{2}<\frac{W}{2}-\omega<W$. This phase is analytically continued from that observed at $1/2$-filling case ($\Delta\mu_{eff}=0$) with a 2e-1h dispersion contained in the Hamiltonian eq.\eqref{fixed point Hamiltonian 2nd step}. On lowering the energy scale, the marginal Fermi liquid undergoes once again a Lifshitz transition into the pseudogap phase (misty blue region in Fig.~\ref{Phase_diagram_with_doping-1}) with spin gapping at the antinodes (AN). The onset energy scale for spin fluctuations that gap the AN is $\omega>\omega_{PG}\equiv \omega^{s}_{PG}$. 
\par\noindent
Further lowering the energy scale gaps the charge fluctuations at the AN, i.e., creates a charge pseudogap at AN. The energy scale for the gapping of charge fluctuations at the AN (dotted red line in Fig.~\ref{Phase_diagram_with_doping-1} is delayed, falling linearly with increasing doping: $\omega>\omega^{c}_{PG}=-\Delta\mu_{eff}$. This delayed entry into the charge pseudogap can be seen from the dependence of the sign of the Greens function  $sgn(G^{j,l}_{p,\Uparrow})$ on $\Delta\mu_{eff}$ (eq.\eqref{Green_func_long}). Upon yet lowering the energy scale for fluctuations causes the exit from the pseudogap via a second Lifshitz transition into a Mott liquid state (solid blue region), with a Luttinger surface of zeros observed in the 1e and 2e-1h Greens functions (eqs.\eqref{1-electron green} and \eqref{2-electron 1-hole green function}). The Mott liquid insulator comprises the following region in the phase diagram ((Fig.\ref{Phase_diagram_with_doping-1}, Video S4)): ($\frac{W}{2}-\omega<1.2t$) with $\Delta\mu_{eff}>-2.8t$, and 
($\frac{W}{2}-\omega<-2\Delta\mu_{eff}$) with $\Delta\mu_{eff}^{*}<\Delta\mu_{eff}<-2.8t$ and $\Delta\mu_{eff}^{*}=-\frac{W}{2}$. Importantly, the nature of the spin-charge hybridization parameter $p(\omega,\Delta\mu_{eff})$ (eq.\eqref{determining p})
is now doping dependent, and changes the nature of the Mott liquid from spin-like $p=0$ for $\Delta\mu_{eff}>-2.8t$ to charge-like $p=1$ for $\Delta\mu_{eff}^{*}<\Delta\mu_{eff}<-2.8t$. 
\par\noindent
This variation of $p$ with doping $\Delta\mu_{eff}$ leads to a separation of the charge gap-dominated and spin gap-dominated parts of the pseudogap within the phase diagram Fig.\ref{Phase_diagram_with_doping-1}, such that the boundary of the charge gap-dominated pseudogap (dotted red line in Fig.\ref{Phase_diagram_with_doping-1}) is observed to fall with increasing doping straight into a quantum critical point (QCP) at $\omega^{c, *}_{PG}= W/2=-\Delta\mu_{eff}^{*}$. This is called the collapse of Mottness~\cite{phillips2011mottness, zaanen2011mottness}.  
The QCP is associated with point-like Fermi surfaces at the four nodal points, and spin-gapping that increases monotonically from near the nodes to the antinodes (Fig.\ref{Phase_diagram_with_doping-2}(b), Videos S2, S3). The resulting nodal metal supports a 2e-1h dispersion that is analytically continued from $\frac{W}{2}-\omega = W$ at $\Delta\mu_{eff}=0$ down to $\frac{W}{2}-\omega = 0$, and given by
$\epsilon_{\mathbf{k}_{\Lambda\hat{s}_{N}}} = -\sqrt{2}t\Lambda$. This linear dispersion signals the Lorentz invariance emergent precisely at the QCP, with 
chiral $(SU(2)_{charge}\times SU(2)_{spin})_{R/L}$ symmetries. 
Thus, the $U(1)$ symmetry of the charge fluctuations is promoted to $SU(2)$ at the QCP due to the emergent particle-hole symmetry of the gapless nodal Dirac electrons.  
Recall that a similar $(SU(2)_{charge}\times SU(2)_{spin})_{R/L}$ symmetries exist for the particle-hole symmetric 1/2-filled Hubbard model. Indeed, these findings for the QCP are in striking agreement with the works of Phillips and co-workers \cite{phillips-PhysRevB.82.214503}.
\par\noindent
For $\omega > \omega^{c,*}_{PG}$, to the left of the QCP, lies the Mott liquid  discussed earlier. For $\omega > \omega^{t} =\Delta\mu_{eff}-2\Delta\mu_{eff}^{*}$, to the right of the QCP, lies a correlated Fermi liquid (CFL) arising from RG relevant tangential scattering (Fig.~2(a) inset). The CFL is associated with well-defined electronic quasiparticles coexisting with the 2e-1h composites of the marginal Fermi liquid on different parts of the gapless FS. For $\Delta\mu_{eff}^{*}<\Delta\mu_{eff}<\Delta\mu_{eff}^{S}$, the FS forms four gapless stretches centred around the nodes, and with spin-gapped regions at the antinodes. Here, the FL quasiparticles are present in regions around the nodes, while the MFL 2e-1h composites are positioned at the ends of the gapless stretches. For $\Delta\mu_{eff}>\Delta\mu_{eff}^{S}$, the FS reconnects, and the MFL is confined to the antinodes. 
\par\noindent
\subsection{Low energy eigenstates of the doped Mott liquid and the QCP}
\begin{figure} 
\hspace*{-0.5cm}
\includegraphics[width=0.59\textwidth,height=7cm]{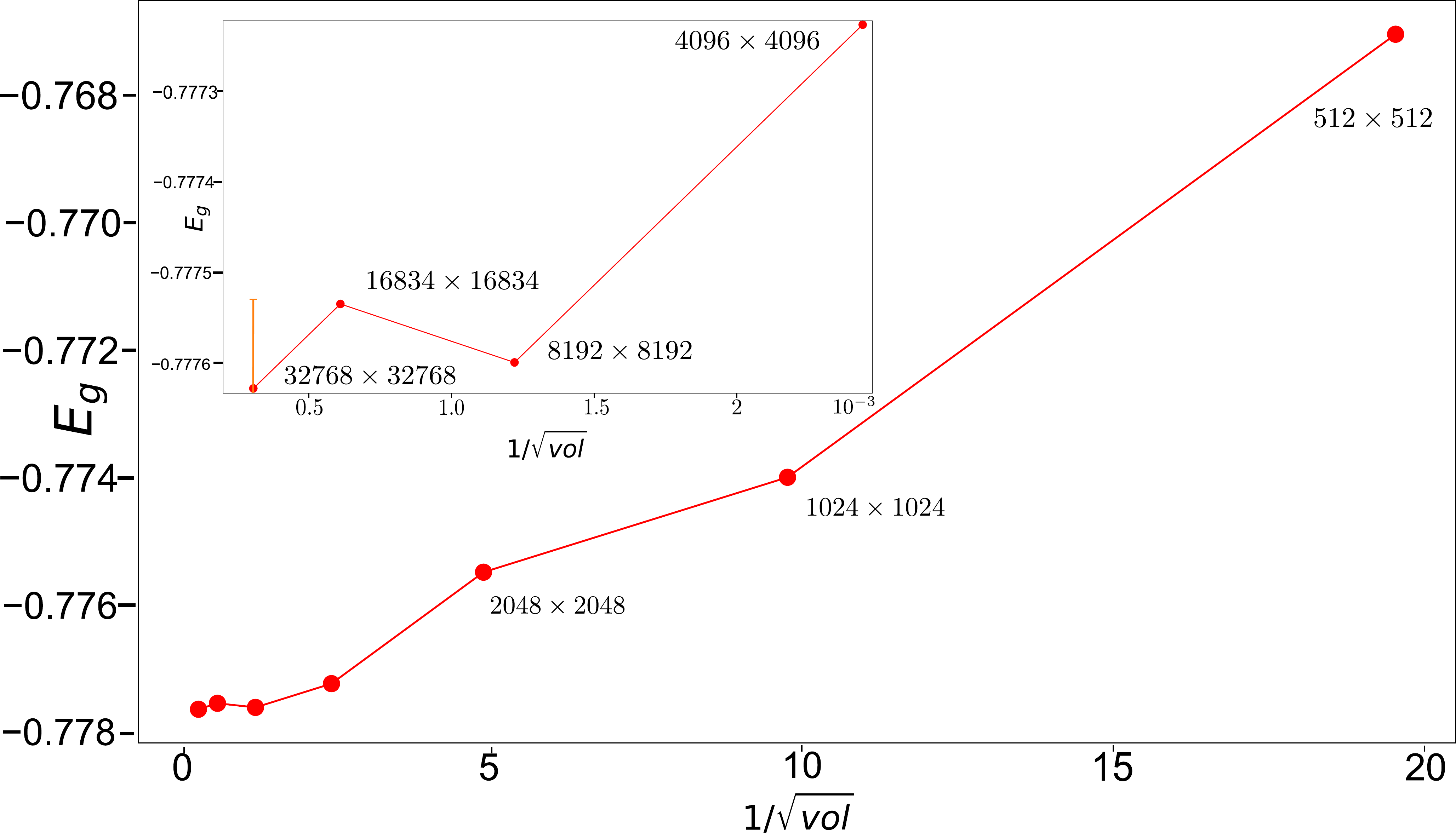}
\caption{(Colour Online) Variation of ground state energy per particle $E_{g}$ of the doped Mott liquid (for $U_{0}=8t$ and hole doping $f_{h}=0.125$) with inverse square root of system size ($1/\sqrt{vol}$, ranging from $512\times 512$- to $32768\times 32768$-sized $k$-space grids) and showing saturation at $-0.776t$.}\label{GSdopedML} 
\end{figure}

Using the RG invariant relation eq.\eqref{RG_invariant_spin_charge_long_scatt} at finite $\Delta\mu_{eff}$, we obtain the
relation between the fixed point spin/charge backscattering coupling
\begin{eqnarray}
K^{*}_{s}(\omega) &=& \frac{\bar{U}_{0}(1-p)K^{*}_{c}(\omega)}{K^{*}_{c}(\omega)-\bar{U}_{0}p}~,
\end{eqnarray}
where the parameter $p$ will be determined by a non-zero hole doping $\Delta\mu_{eff}$. For instance, at $\omega = \frac{W}{2}$, we obtain from the extremization condition (eq.\eqref{determining p}) a value of $p=0$ within the chemical potential ranges $2.9>-\Delta\mu_{eff}>0$, and $p=1$ within the range $W/2>-\Delta\mu_{eff}>2.9$. For these two cases,  $K^{*}_{s}(\omega,\Delta\mu_{eff})=K^{*}_{c}(\omega,\Delta\mu_{eff})=\bar{U}_{0}$, such that the fixed point Hamiltonian is given by 
\begin{eqnarray}
H^{*}(\Delta\mu_{eff})&=&
\sum_{\hat{s}}\bar{U}_{0}\left(\mathbf{A}^{*}_{\hat{s}}\cdot\mathbf{A}^{*}_{-\hat{s}}-\mathbf{S}^{*}_{\hat{s}}\cdot\mathbf{S}^{*}_{-\hat{s}}\right)\nonumber\\
&+&\Delta\mu_{eff}\sum_{\hat{s}}A^{z}_{\hat{s}}~.\label{dopedMottLiquidHam}
\end{eqnarray}
\begin{widetext}
The low-lying eigenstates and eigenenergies of the doped Mott liquid can now be sought. We begin by writing the vacuum within the emergent window as
\begin{eqnarray}
|0\rangle &=& \prod_{\hat{s}}|[0_{1'}0_{1}0_{-1}0_{-1'}]..[0_{j'}0_{j}0_{-j}0_{-j'}]..[0_{n'}0_{n}0_{-n}0_{-n'}]\rangle_{d}= |[0_{1'}\downarrow _{1} 0_{-1'}]\ldots [0_{1'}\downarrow _{j} 0_{-1'}]\ldots [0_{1'}\downarrow _{n} 0_{-1'}]\rangle_{d}~,\nonumber\\
&=&\prod_{\hat{s}}|A=N^{*},A^{z}=-N^{*}\rangle_{d}~,
\end{eqnarray}
where the labels denote (as earlier) $(j,d):=\mathbf{k}_{\Lambda, \hat{s}}, \uparrow$, $(-j,d):=\mathbf{k}_{-\Lambda, T\hat{s}}, \downarrow$, $(j',d):=\mathbf{k}_{2\Lambda^{*}_{\hat{s}}-\Lambda, T\hat{s}}, \downarrow$, $(-j',d):=\mathbf{k}_{-2\Lambda^{*}_{\hat{s}}-\Lambda, \hat{s}}, \uparrow$. The states $|\downarrow_{j}\rangle =|0_{1}0_{-1}\rangle$ and $|\uparrow\rangle_{j}=|1_{1}1_{-1}\rangle$ are eigenstates of the $A^{z}_{j}$ operator, i.e., $A^{z}_{j}|\downarrow\rangle_{j} = -2^{-1}|\downarrow_{j}\rangle$ and $A^{z}_{j}|\uparrow_{j}\rangle = 2^{-1}|\uparrow_{j}\rangle$. 
We recall from section \ref{eigMottLiquid} that one of the ground states of the Mott liquid Hamiltonian eq\eqref{fixed_point_Ham_mott} at half-filling ($\Delta\mu_{eff}=0$)  is given by (eq.\eqref{class_2_wvfn})
\begin{eqnarray}
\prod_{\hat{s}}|S=N^{*},S^{z}=-N^{*}\rangle_{d} &=& \prod_{d}|[1_{1'}0_{1}1_{-1}0_{-1'}]..[1_{j'}0_{j}1_{-j}0_{-j'}]..[1_{n'}0_{n}1_{-n}0_{-n'}]\rangle_{d}\nonumber\\
&=& \prod_{i=1}^{2N^{*}_{\hat{s}}}S^{-}_{i}c^{\dagger}_{i}|0\rangle =  \prod_{\hat{s}}|[\Downarrow_{1}\Downarrow_{-1}]..[\Downarrow_{j}\Downarrow_{-j}]..[\Downarrow_{n}\Downarrow_{-n}]\rangle_{d}~,
\end{eqnarray} 
with energy $-\bar{U}_{0}\sum_{\hat{s}}(N^{*}_{\hat{s}})^{2}$ and where $S^{z}_{j}|\Uparrow_{j}\rangle = 2^{-1}|\Uparrow_{j}\rangle~,~S^{z}_{j}|\Downarrow_{j}\rangle = -2^{-1}|\Downarrow_{j}\rangle $. In the presence of a chemical potential term $\Delta\mu_{eff}(A^{z}_{*,\hat{s}}+A^{z}_{*,-\hat{s}})$, some of the spin-type pseudospin configurations (say $2K$ in number) are broken and replaced by charge-type pseudospin configurations. This results in eigenstates of the form
\begin{eqnarray}
&&|S_{*,\hat{s}}=N_{\hat{s}}^{*}-K, S_{*,\hat{s}}^{z}=K-N_{\hat{s}}^{*},A_{*,\hat{s}}=K,A_{*,\hat{s}}^{z}=-K\rangle  \nonumber\\
&=& \frac{1}{\sqrt{\binom{N^{*}_{\hat{s}}}{N^{*}_{\hat{s}}-K}}}\sum_{\alpha}\prod_{i=1}^{N^{*}_{\hat{s}}-K}S^{-}_{\alpha_{i}}c^{\dagger}_{\alpha_{i},d}|0\rangle\nonumber\\
 &=& \frac{1}{\sqrt{\binom{N^{*}_{\hat{s}}}{N^{*}_{\hat{s}}-K}}}\sum_{\alpha}|(1)\ldots[0_{\alpha_{1}'-1}0_{\alpha_{1}-1}0_{-\alpha_{1}+1}0_{-\alpha_{1}'+1}][1_{\alpha_{1}'}0_{\alpha_{1}}0_{-\alpha_{1}}0_{-\alpha_{1}'}][0_{\alpha_{1}'+1}0_{\alpha_{1}+1}0_{-\alpha_{1}-1}0_{-\alpha_{1}'-1}]\ldots(\alpha_{2N^{*}_{\hat{s}}-2K})\ldots(2N^{*}_{\hat{s}})\rangle\nonumber\\
& = &\frac{1}{\sqrt{\binom{N^{*}_{\hat{s}}}{N^{*}_{\hat{s}}-K}}}\sum_{\alpha}|[_{1}]\ldots[0_{\alpha'_{1}-1}\downarrow _{\alpha_{1}}0_{-\alpha_{1}'+1}][\Downarrow_{\alpha_{1}}0_{-\alpha_{1}}0_{-\alpha_{1}'}][0_{\alpha'_{1}+1}\downarrow_{\alpha} 0_{-\alpha_{1}'-1}]\ldots (\alpha_{2N^{*}_{\hat{s}}-2K})...(2N_{\hat{s}}^{*})\rangle\label{basal_wvfn}
\end{eqnarray} 
where $S_{*,\hat{s}},S^{z}_{*,\hat{s}}$ are quantum numbers associated with the operator $\mathbf{S}_{*,\hat{s}}=\sum_{i=1,\hat{s}}^{N_{\hat{s}}^{*}}\mathbf{S}_{i,\hat{s}}$. By applying spin- and charge -pseudospin raising operators on the above wavefunction, we can obtain the general class of wavefunctions given by
\begin{eqnarray}
&&|\Psi\rangle = \prod_{\hat{s}}|S_{*,\hat{s}}=S_{*,-\hat{s}}=N^{*}-K,S = l, S^{z}=p, A_{*,\hat{s}}= A_{*,-\hat{s}}= K, A=n, A^{z}=m\rangle\nonumber\\
&=&\prod_{\hat{s}}\sum_{m_{1}}C^{n,m}_{K,m_{1};K,m_{2}}C^{l,p}_{K,m_{1};K,m_{2}}\times\nonumber\\
&&\times|S_{*,\hat{s}}=S_{*,-\hat{s}}=N_{\hat{s}}^{*}-K, S^{z}_{*,\hat{s}}=m_{1},S^{z}_{*,-\hat{s}}=p-m_{1},A_{*,\hat{s}}=A_{*,-\hat{s}}=K,A^{z}_{*,\hat{s}}=m_{1},A^{z}_{*,-\hat{s}}=m-m_{1}\rangle~,~~~~\label{manyBodyState}
\end{eqnarray} 
where $C^{n,m}_{K,m_{1};K,m-m_{1}}$ and $C^{l,p}_{K,m_{1};K,p-m_{1}}$ are the Clebch Gordon coefficients~\cite{edmond1985} for charge/spin like large pseudospins respectively.
 \end{widetext}
Angular momentum algebra constrains the integers $K,l,n,m$ as follows
\begin{eqnarray}
&& 0\leq K\leq N^{*}_{\hat{s}}~,~0\leq l\leq 2N^{*}_{\hat{s}}-2K~,~ -l\leq p\leq l\nonumber\\
&& 0\leq n\leq 2K~,~ -n<m\leq n~.
\end{eqnarray} 
The energy eigenvalue $E(K,l,n,m)$ of the state $|\Psi\rangle$ (eq.\eqref{manyBodyState}) is then obtained as
\begin{eqnarray}
\hspace*{-1cm}
&&\hspace*{-1cm}E(K,l,n,m) = -\sum_{\hat{s}}\frac{\bar{U}_{0}}{2}\bigg[l(l+1)-2(N^{*}_{\hat{s}}-K)(N^{*}_{\hat{s}}-K+1)\bigg]\nonumber\\
&& +\frac{1}{2}\sum_{\hat{s}}\bar{U}_{0}\left[n(n+1)-2K(K+1)\right]-\Delta\mu_{eff}m~.
\label{eigenValue}
\end{eqnarray} 
For $l=2N^{*}-2K$ and $m=-n$, the low-lying energy states of the spectrum can be accessed by respecting the attractive nature of the spin pseudospin interactions and the repulsive nature of the charge pseudospin interactions in the Hamiltonian eq.\eqref{dopedMottLiquidHam}
\begin{eqnarray}
E(K,n) &=& -\bar{U}_{0}\sum_{\hat{s}}(N^{*}_{\hat{s}}-K)^{2}\nonumber\\
&+&\frac{1}{2}\sum_{\hat{s}}\bar{U}_{0}\left[n(n+1)-2K(K+1)\right]\nonumber\\
&+&\Delta\mu_{eff}n~.\label{minimizeE}
\end{eqnarray}
The ground state energy $E_{g}(\Delta\mu_{eff})= E(K^{*},n^{*})$, and hence $n^{*},K^{*}$, can then be obtained from a numerical search of the lowest eigenvallue of the energy function E(K,n) eq\eqref{minimizeE}. 
From this, we can obtain the hole-doping fraction $f_{h}$ as
\begin{eqnarray}
f_{h}(\Delta\mu_{eff}) = \frac{n^{*}}{2N}~,
\end{eqnarray}
where the factor $1/2$ prevents double counting of the number of normal $\hat{s}$'s in the Hamiltonian eq.\eqref{dopedMottLiquidHam}. We conducted a test of these results by comparing the ground state energy obtained for $U=8t$ and at a hole-doping fraction of $f_{h}=0.125$ (i.e., $12.5\%$ hole doping) against that obtained from various other numerical methods as given in Ref.\cite{leblanc2015solutions}. Fig.\ref{GSdopedML} shows a finite-size scaling plot of $E_{g}(\Delta\mu_{eff})$ per particle at  $12.5\%$ hole doping and system sizes (i.e., number of sites) varying between $32\times 32$ to $32768\times 32768$. The value of $E_{g}(\Delta\mu_{eff})$ per particle in the thermdodynamic limit from this analysis appears to be converging towards $-0.776$ (in units of the hopping amplitude $t$), and compares well with the range of $-0.74>(E_{g}/\textrm{per particle})>-0.77$ obtained from Ref.\cite{leblanc2015solutions,ehlers2017hybridDMRG}. We also point the reader to further benchmarking exercises presented in Appendix-\ref{further benchmarking} with $U/t=2,4,6,10$ at $12.5\%$  doping. Once more, we find excellent agreement with the results obtained from the numerical methods employed in Refs.\cite{dagotto1992} and \cite{leblanc2015solutions,ehlers2017hybridDMRG}, offering confidence on the effective Hamiltonian and ground state wavefunction derived for the doped Mott liquid. 
\par
\begin{figure}
\hspace*{-1.5cm}
\includegraphics[width=0.59\textwidth]{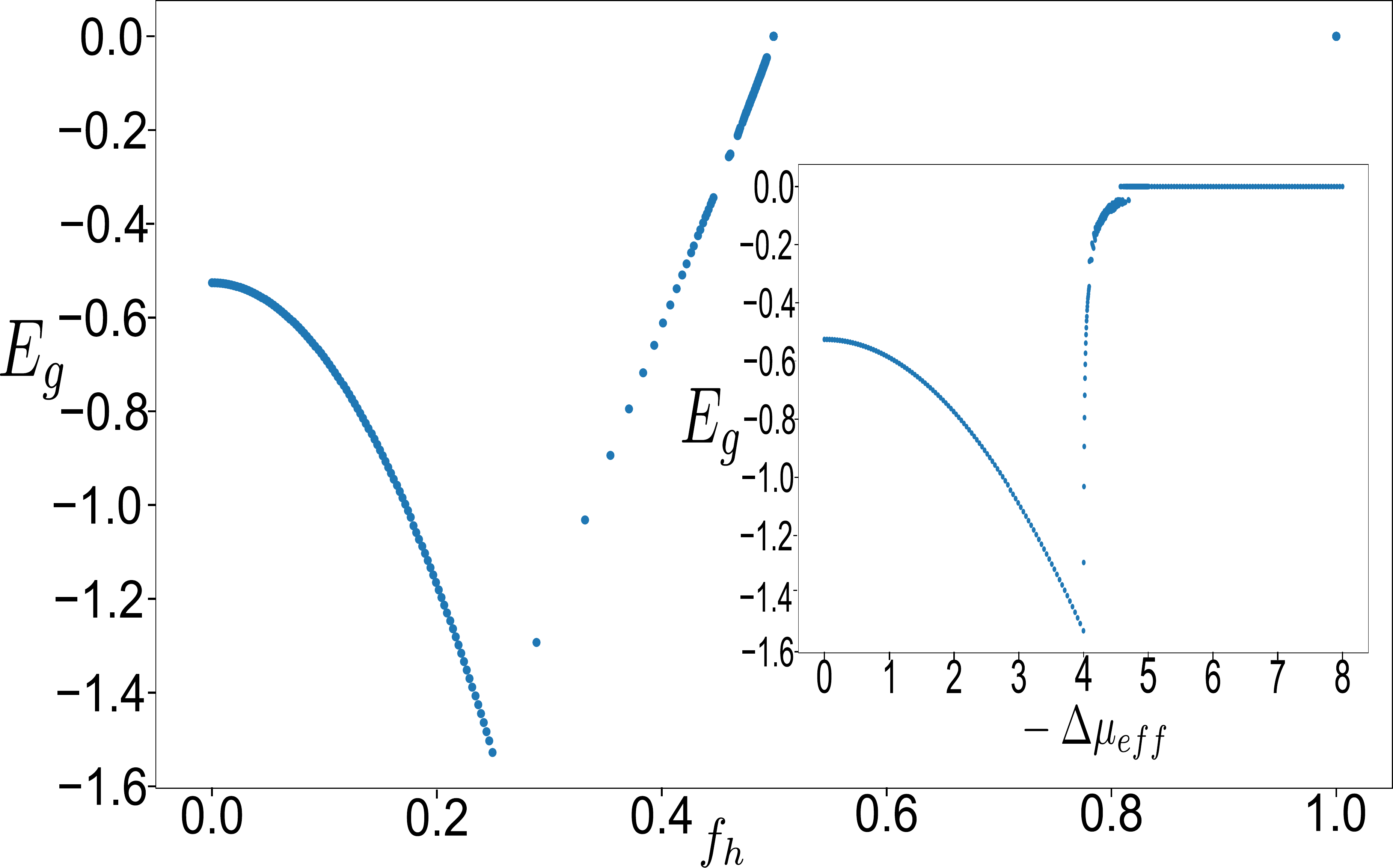}
\caption{(Colour Online) Variation of ground state energy $E_{g}$ with hole doping $f_{h}$ for lattice size $1024\times 1024$ and $U_{0}=8t$. Inset: Variation of ground state energy $E_{g}$ with $-\Delta\mu_{eff}$.}\label{GSwithDoping} 
\end{figure}

\begin{figure}
\includegraphics[width=0.5\textwidth]{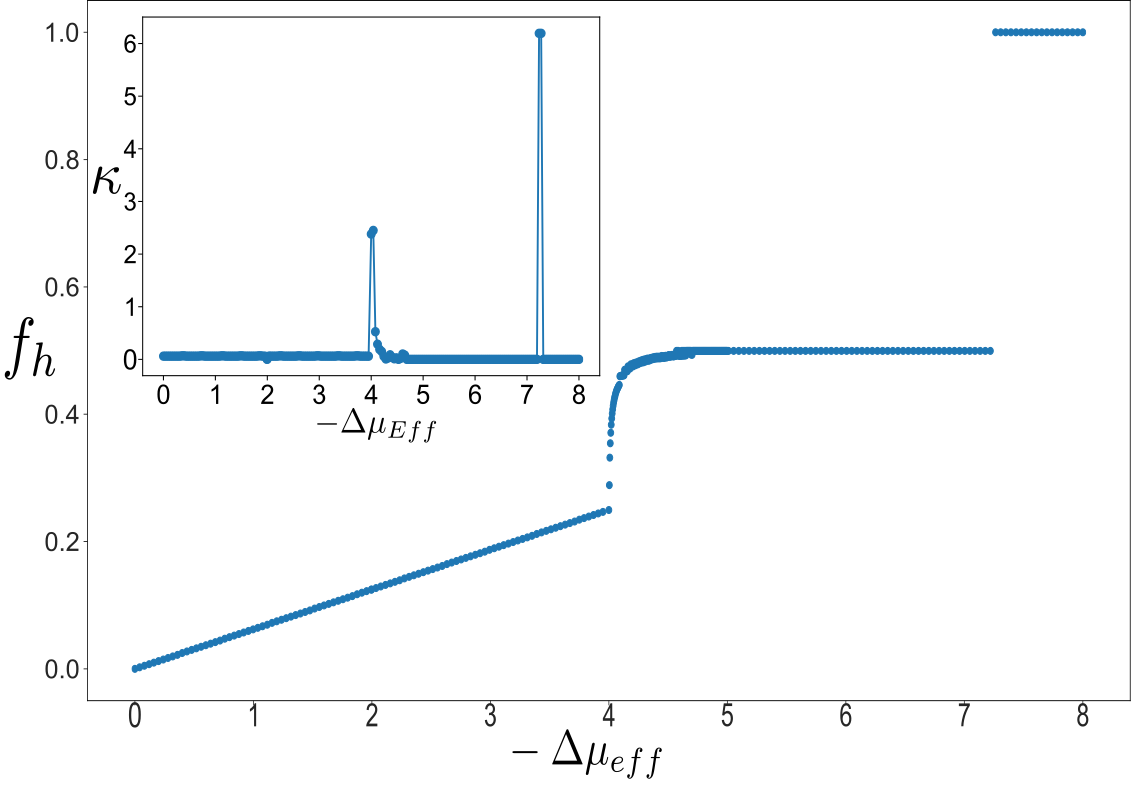} 
\caption{(Colour Online) Variation of hole doping $f_{h}$ with chemical potential $\Delta\mu_{eff}$ for lattice size $1024\times 1024$. Jumps in $f_{h}$ are clearly visible as the chemical potential is tuned first through the QCP ($\Delta\mu_{eff}=-4$) and then through the topological reconstruction of a fully connected Fermi surface ($\Delta\mu_{eff}\sim -7.2t$). Inset: Number compressibility $\kappa$ as a function of chemical potential $\Delta\mu_{eff}$. Spikes in $\kappa$ are visible at $-\Delta\mu_{eff}=4$ and at $-\Delta\mu_{eff} \sim 7.2t$.}\label{DopingwithChemPot}
\end{figure}

\begin{figure} 
\hspace*{-0.75cm}
\includegraphics[width=0.5\textwidth]{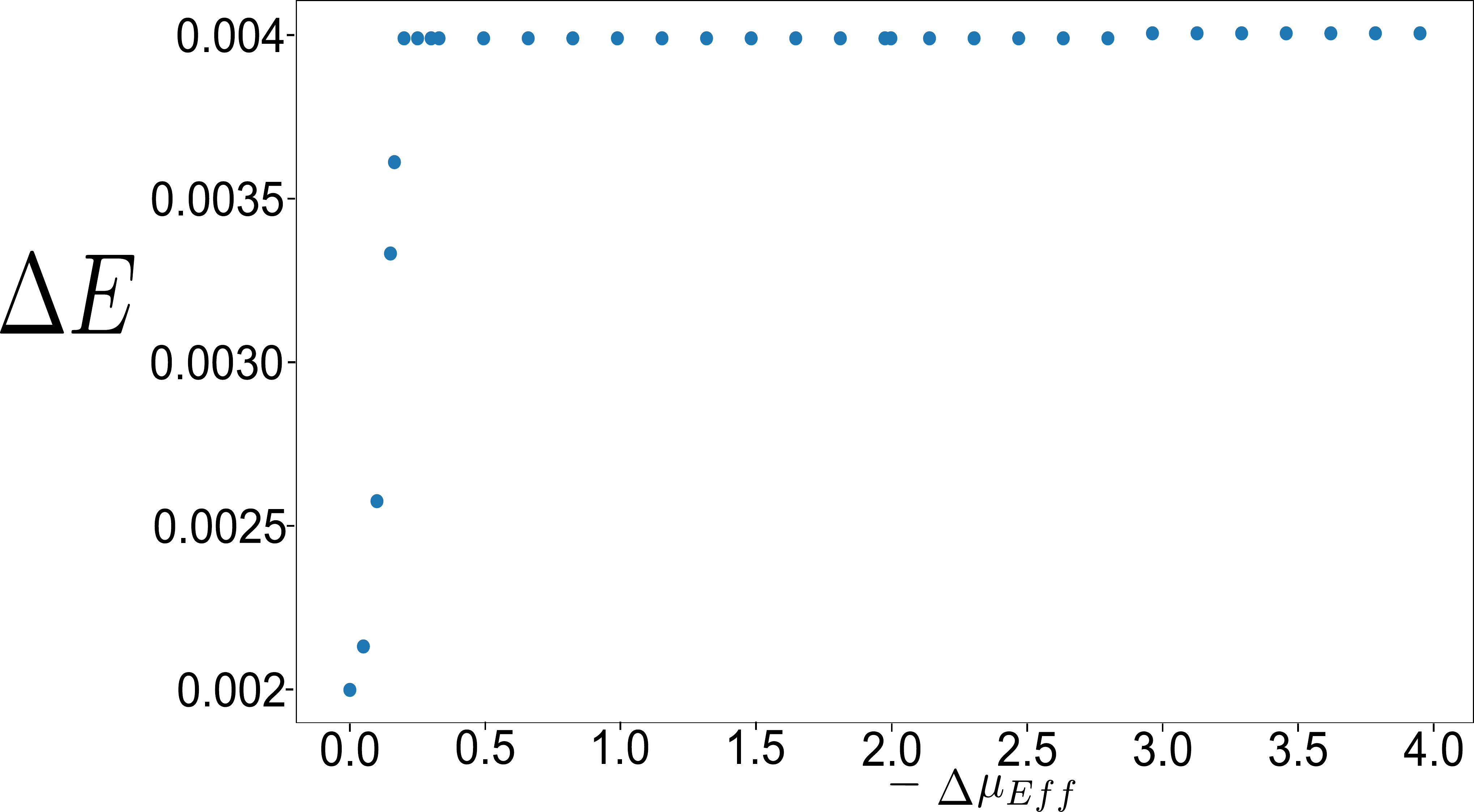}
\caption{(Colour Online) Variation of the energy gap $\Delta E$ per particle of the doped Mott Liquid along the nodal direction with the chemical potential $\Delta\mu_{eff}$. $\Delta E$ is seen to rise steadily from $0.002t$ upon doping the half-filled Mott liquid with holes and saturating to a finite value ($0.004t$). The plot ends just before the QCP is reached in $\Delta\mu_{eff}$, where $\Delta E\to 0$ abruptly (not shown).}\label{GapwithChemPot} 
\end{figure}

\begin{figure}
\hspace*{-1.0cm}
\includegraphics[scale=0.2]{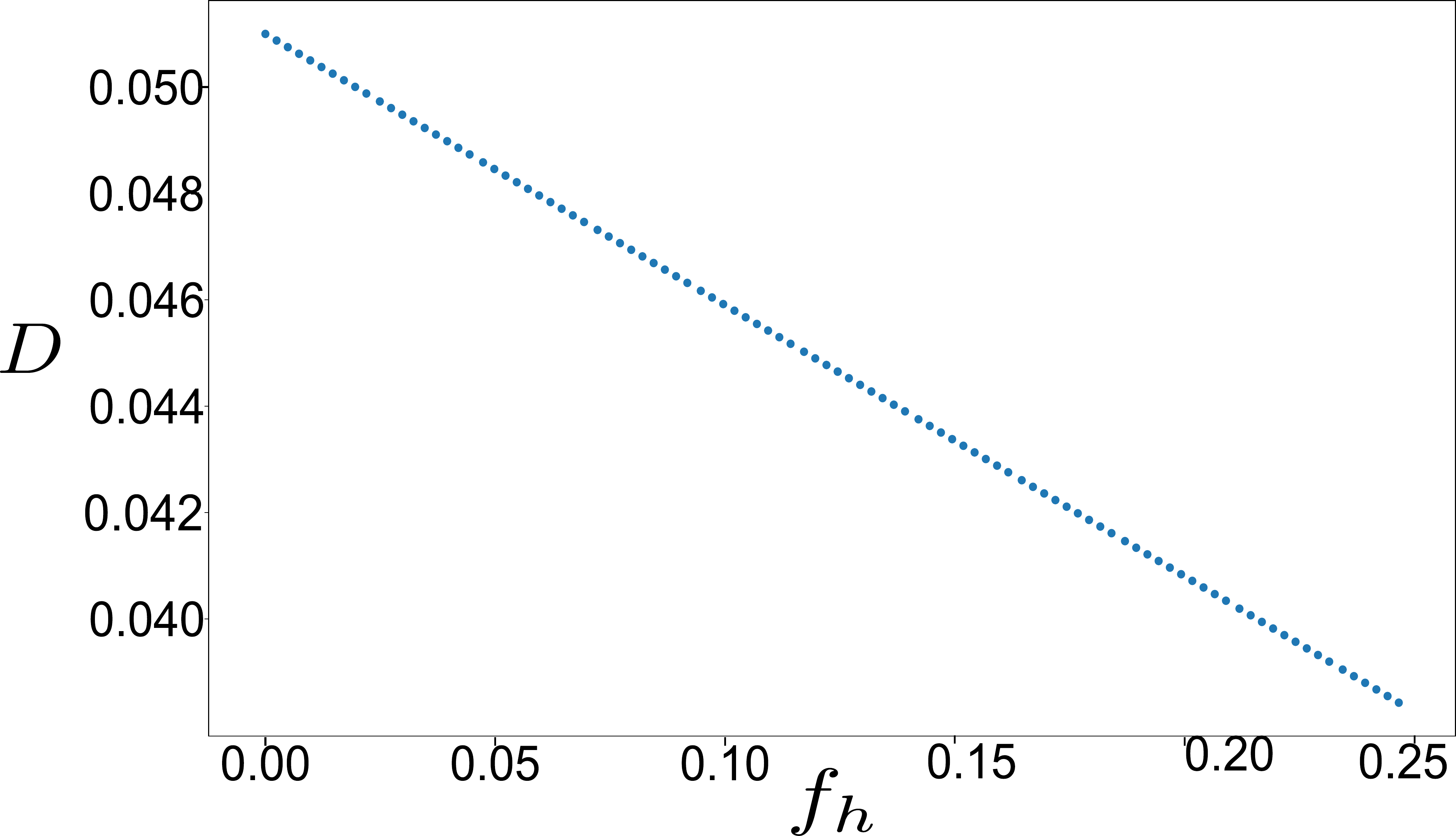}
\caption{(Colour Online) Variation of the Double occupancy $D$ of the doped Mott liquid with the hole fraction $f_{h}$. The plot shows a steady decline of $D$ as $f_{h}$ is increased towards the QCP.}\label{doublonwithdoping}
\end{figure}

\noindent
A plot of $E_{g}$ as a function of $f_{h}$ (Fig.\ref{GSwithDoping}) clearly shows cusp-like behaviour at $f_{h}=0.25$, signalling a quantum critical point (QCP). The discontinuity in the first derivative is also seen in a plot of $E_{g}$ vs. $-\Delta\mu_{eff}$ (inset of Fig.\ref{GSwithDoping}), as well as a plot of $f_{h}$ vs. $-\Delta\mu_{eff}$ (Fig.\ref{DopingwithChemPot}). The step-like discontinuity observed in $f_{h}$ at $-\Delta\mu_{eff}\simeq 7.2$ indicates the topological reconstruction of a fully connected Fermi surface as the spin pseudogapped parts of the FS vanish at this value of the chemical potential. This is reinforced by a plot of the number compressibility $\kappa$ with the chemical potential $\Delta\mu_{eff}$ in the inset of Fig.\ref{DopingwithChemPot}): the first spike indicates the appearance of point-like nodal FS at the QCP, while the second denotes the reconstruction of a fully connected FS.
In order to understand better the nature of the QCP, we present a plot of the energy gap $\Delta E$ per particle along the nodal direction in the doped Mott liquid with varying $-\Delta\mu_{eff}$ in Fig.\ref{GapwithChemPot}. The plot shows that $\Delta E$ per particle rises steadily from a value of $0.002t$ and saturates at $0.004t$ well before the QCP is approached from the underdoped side. This saturated value of $\Delta E$ appears to be robust very close to the QCP, and collapses abruptly to $0$ at the QCP. This signals the nucleation of gapless nodal Fermi points precisely at the QCP, and corresponds to a topology change in the Luttinger surface of zeros for the gapped Mott liquid. The QCP is an example of a Lifshitz transition driven by electronic correlations~\cite{imada2010unconventional}. We can also compute the doublon occupancy from the hole-doping fraction $f_{h}$,$D=D_{0}(1-f_{h})$, where $D_{0}$ is the doublon fraction at half-filling. This is shown in Fig.\ref{doublonwithdoping}, and shows a steady decline in the doublon occupancy of the doped Mott liquid from half-filling till the QCP. We also note that the value of $0.045$ is obtained for the doublon occupancy $D$ at a hole doping of $12.5\%$, and compares well with the range $0.04< D <0.045$ obtained from various numerics in Ref.\cite{leblanc2015solutions}.
\par\noindent
The eigenenergies for the gapped parts of the FS at and across the QCP $-W\leq\Delta\mu_{eff}\leq-\frac{W}{2}$ can be obtained from the fixed point Hamiltonians at $\omega = \frac{W}{2}$ (eq.\eqref{effHamforQCPregion}) and are given by
\begin{eqnarray}
E &=& -\sum_{\hat{s}=\hat{s}_{AN}}^{\hat{s}^{*}}\frac{K_{s,0,\hat{s}}^{*}(\frac{W}{2},\Delta\mu_{eff})}{2}\bigg[l(l+1)\nonumber\\
&-&2(N^{*}_{\hat{s}}-K)(N^{*}_{\hat{s}}-K+1)\bigg]\nonumber\\
& +&\frac{1}{2}\sum_{\hat{s}_{AN}}^{\hat{s}^{*}}K^{*}_{c}(\frac{W}{2},\Delta\mu_{eff})\left[n(n+1)-2K(K+1)\right]\nonumber\\
&-&\Delta\mu_{eff}m~.
\end{eqnarray}
For the gapless regions, the lowest excitations appear near zero energy. Taken together, this allows for a determination of the ground state energy ($E_{g}$) and hole-doping fraction ($f_{h}$) as a function of chemical potential $\Delta\mu_{eff}$ for a range of chemical potential that crosses the QCP. The results are shown in Fig.\ref{GSwithDoping} and \ref{DopingwithChemPot}. Specifically, we find that $E_{g}$ rises steadily with $\Delta\mu_{eff}$ from its value at the QCP to zero as the chemical potential leads to the reconstruction of a fully connected FS (inset of Fig.\ref{GSwithDoping}). Interestingly, $f_{h}$ is seen to attain a plateau upon increasing $\Delta\mu_{eff}$ away from the QCP, transitioning abruptly at the FS reconstruction. 
These findings are consistent with results obtained from the dynamical cluster quantum Monte Carlo method applied to the 2D Hubbard model with doping away from 1/2-filling~\cite{vidhyadhiraja2009quantum,zaanenjarrell-PRL-2011}, and unveil the mechanism responsible for the experimentally observed topological reconstruction of the Fermi surface near critical doping~\cite{he2014, fujita2014, balakirev2003}. We will further discuss these results below, as we study the interplay of spin, charge, Cooper and tangential scattering processes in the vicinity of the QCP. The codes used for the numerical computations of the ground state energy and hole fraction shown here are made available electronically~\cite{anirbanhubbard2019code}. 
\subsection{Signature of QCP in ground state fidelity susceptibility with doping}
Recently fidelity computed as an overlap between many body wavefunctions with parameter tuning has emerged as a possible candidate for tracking quantum phase transition(QPT) \cite{wang2015fidelity,you2007fidelity}. In this section we will use fidelity to observe for signatures of QCP seen earlier from cusp in ground state energy density Fig.\ref{GSwithDoping}(b), bulk compressiblity (inset of Fig.\ref{DopingwithChemPot}). Fidelity is computed from the overlap between  ground states at $\frac{W}{2}-\omega=0$ obtained for infinitesmal variation of $\Delta\mu_{eff}$ the effective chemical potential($\epsilon>0$),
\begin{eqnarray}
F(\Delta\mu_{eff},\epsilon) = |\langle\Psi_{g}(\Delta\mu_{eff})|\Psi_{g}(\Delta\mu_{eff}-\epsilon)\rangle|~.
\end{eqnarray}The ground state wavefunctions for $\Delta\mu_{eff}>-\frac{W}{2}$ can be represented purely in terms of pseudospins eq\ref{manyBodyState}, where $n(\Delta\mu_{eff})$ is determined by minimizing the energy density eq\ref{minimizeE}. For $-\Delta\mu_{eff}<\frac{W}{2}$ the fidelity is obtained as, $F(\Delta\mu_{eff},\epsilon)=\delta_{n(\Delta\mu_{eff}),n(\Delta\mu_{eff}-\epsilon)}$. 
\par\noindent
For $-\Delta\mu_{eff}>W/2$ a gapless node is formed at $N=(\pi/2,\pi/2)$ that streches into an arc in momentum space with gapped AN regions. The states in AN regions can still be described in terms of the pseudospin wavefunctions eq.(\ref{manyBodyState}), however the  the gapless regions are described by separable eigenstates of the Hamiltonian eq.(\ref{fixed point Hamiltonian 2nd step}) in the occupation number basis of $|\mathbf{k}\sigma\rangle$. The ground states describing this journey across the QCP is written as $|\Psi_{g}(\Delta\mu_{eff})\rangle = |\Psi_{gapped}\rangle|\Psi_{gapless}\rangle$ given by eq.\ref{metEig}, eq.\ref{manyBodyState} and the overlap 
%{\color{blue}\begin{eqnarray}
%|\Psi_{g}(\Delta\mu_{eff})\rangle &=& |\Psi_{gapped}\rangle|\Psi_{gapless}\rangle\nonumber\\
%|\Psi_{gapped}\rangle &=& \prod^{\hat{s}_{AN}}_{\hat{s}(\Delta\mu_{eff})}|A=-A^{z}=n, A^{*}_{\hat{s}}=A^{*}_{-\hat{s}}=N^{*}_{\hat{s}},\nonumber\\&&~~~~~~~~~S^{*}_{\hat{s}}=S^{*}_{-\hat{s}}=0\rangle\nonumber\\
%|\Psi_{gapless}\rangle &=& \prod_{\hat{s}_{N}}^{\hat{s}(\Delta\mu_{eff})}\prod_{\Lambda\Lambda'\delta}|1_{\Lambda\hat{s}}1_{-\Lambda+\delta T\hat{s}}0_{-\Lambda'\hat{s}}\rangle,
%\end{eqnarray}}
%i.e. a antisymmetrized product of composites formed out of pair of electronic states one above and another below the Fermi surface and a hole within the Fermi surface. 
between neighboring ground states for $-\Delta\mu_{eff}>W/2$ axis is then obtained as \begin{equation}
\langle\Psi_{g}(\Delta\mu_{eff})|\Psi_{g}(\Delta\mu_{eff}-\epsilon)\rangle = {2N^{*}\choose n(\Delta\mu_{eff}-\epsilon)+N^{*}}^{-l(\Delta\mu_{eff},\epsilon)}
\end{equation}
where $l(\Delta\mu_{eff},\epsilon)$ is the increment of the gapless stretch, $2N^{*}$ is the number of electronic states that have transformed into $N^{*}$ bound pairs in the momentum space window. 
\par
\begin{figure}
\includegraphics[height=0.4\textwidth, width=0.6\textwidth]{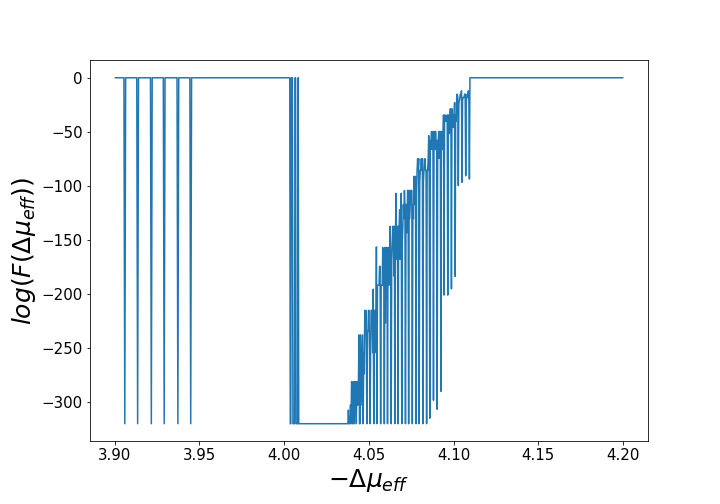}
\includegraphics[height=0.4\textwidth, width=0.6\textwidth]{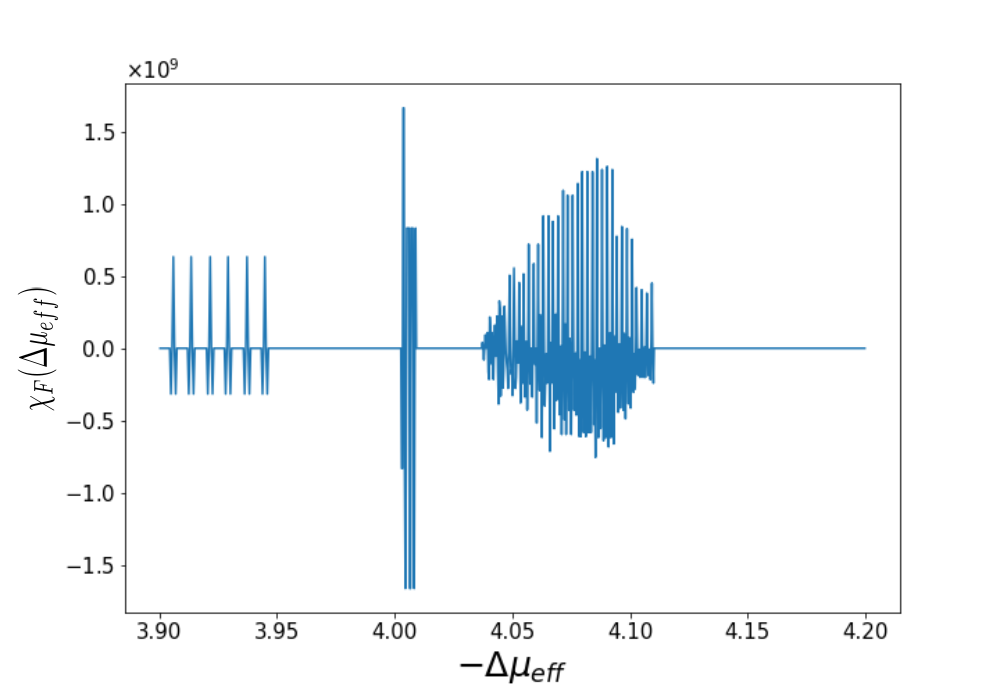}
\caption{(a)$\log(F(\Delta\mu_{eff}))$ Logarithm of fidelity, (b)$\chi(\Delta\mu_{eff})$ fidelity susceptibility plotted against effective chemical potential($\Delta\mu_{eff}$).}\label{fidelity}
\end{figure}

\noindent
The $\log(F(\Delta\mu_{eff}))$ plot Fig.\ref{fidelity}(a) for $-\Delta\mu_{eff}<W/2$ shows oscillations between $0$ and $1$ due to integer increment in the number of holes($n$) being added to the ground state by changing $\Delta\mu_{eff}$. The oscillation period is of the order $10^{-3}t$. In Fig. \ref{DopingwithChemPot}($f_{h}$ vs. $\Delta\mu_{eff}$ curve) we see this as a linear growth because the step size$>10^{-3}t$. To avoid the log singularities in the numerics associated with $F(\Delta\mu_{eff})=0$ we add a cutoff of $10^{-323}$. For $-\Delta\mu_{eff}>W/2$ the log of fidelity jumps as the nodal points become gapless. In the intervening pseudogap phase the fidelity rises with increase in the gapless stretch saturating to 1 as the Fermi surface is recreated.
Around the QCP $4<-\Delta\mu_{eff}<4.02$ a huge amount of fidelity oscillations in Fig.\ref{fidelity}(a,b) could be arising out of spin gapping fluctuations competing with marginal Fermi liquid fluctuations. Again the flat stretch of vanishing fidelity for $4.00<-\Delta\mu_{eff}<4.05$ and its susceptibility means the many body wavefunction is facing a cascade of orthogonality catastrophies. Following this there is a noisy rise of the fidelity in the intervening pseudogap phase which could be resulting from the interplay between tangential scattering process and spin backscattering processes  Similarly the fidelity susceptibility $\chi_{F}( \Delta\mu_{eff})=d^{2}\log(F(\Delta\mu_{eff}))/d\Delta\mu_{eff}^{2}$ shows a spike in the curve at $\Delta\mu_{eff}=-\frac{W}{2}$ giving obvious evidence of the QCP. Even though the critical features of this plot is well behaved the noisiness could be additionally coming from convergence issues at fixed point coming from finite number of RG steps. 

% 
%, its mathematically defined as,
%\begin{eqnarray}
%l=\sum_{\hat{s}_{N}}^{\hat{s}(\Delta\mu_{eff}-\epsilon)}1-\sum_{\hat{s}_{N}}^{\hat{s}(\Delta\mu_{eff})}1~.
%\end{eqnarray}
\subsection{Theory for the vicinity of the QCP}\label{QCPtheory}
Doping via an effective chemical potential $\Delta\mu_{eff}=-\frac{\Delta U_{0}}{2}$ generates a doublon-holon imbalance in the $\mathbf{Q}_{1}=(\mathbf{k}_{\Lambda\hat{s}}+\mathbf{k}_{-\Lambda T\hat{s}})$ pair-momentum resonant-pair subspace. This doublon-holon imbalance is associated with a field-like term for the charge pseudospins ($-\Delta\mu_{eff}(A^{z}_{*, \hat{s}}+A^{z}_{*, -\hat{s}})$). As mentioned earlier, this lowers the symmetry of the many-body state from $SU(2)_{\text{charge}}\to U(1)_{\text{charge}}$ within the gapped parts of the FS in the pseudogap and the Mott liquid phases. We also briefly recall the twofold reason for emphasizing the physics of the resonant-pair subspace: it contains the log-divergences arising from Umklapp scattering, as well as carries the highest spectral weight. Due to doublon-holon imbalance, the contribution of Umklapp scattering processes ($A^{+}_{*, \hat{s}}A^{-}_{*, -\hat{s}} + h.c.$) to gap formation is reduced compared to that due to spin backscattering ($S^{+}_{*, \hat{s}}S^{-}_{*, -\hat{s}}+ h.c.$) in eq.\eqref{Hamiltonian_pseudogap}. This doublon-holon imbalance 
also disfavours the sub-dominant  off-resonant pair ($\delta\neq 0$) Umklapp scattering terms.
\par\noindent
On the other hand, this uniform field 
favours spin-backscattering of all different Anderson pseudospin/Cooper pairs $(c^{\dagger}_{\mathbf{k}_{\Lambda\hat{s}}\uparrow}c^{\dagger}_{\mathbf{p}-\mathbf{k}_{\Lambda\hat{s}}\downarrow})$ with $\mathbf{p}$ pair-momentum centered about $\mathbf{p}=0$. 
Indeed, there exists a direct equivalence between the set of resonant/off-resonant spin-backscattering terms and the set of  $\mathbf{p}$ pair momentum terms centered around $\mathbf{p}=0$, i.e.,
\begin{eqnarray}
\hspace*{-0.5cm}&&\sum_{\mathbf{p},\Lambda, \delta, \hat{s}}K_{s,\mathbf{p}, \hat{s}}(\delta)c^{\dagger}_{\mathbf{k}_{\Lambda, \hat{s}}, \uparrow}c^{\dagger}_{\mathbf{p}-\mathbf{k}_{\Lambda, \hat{s}}, \downarrow}c_{\mathbf{k}_{(-2\Lambda^{*}+\Lambda+\delta), T\hat{s}, \downarrow}}c_{\mathbf{p}-\mathbf{k}_{(-2\Lambda^{*}+\Lambda+\delta), T\hat{s}}, \uparrow}\nonumber\\
&&\hspace*{-.75cm}= -\sum_{\mathbf{p}, \Lambda, \delta, \hat{s}}K_{s,\mathbf{p}, \hat{s}}(\delta)c^{\dagger}_{\mathbf{k}_{\Lambda, \hat{s}}, \uparrow}c_{\mathbf{k}_{(-2\Lambda^{*}+\Lambda+\delta), T\hat{s}}, \downarrow}c^{\dagger}_{\mathbf{k}_{\Lambda', -\hat{s}},\downarrow}c_{\mathbf{k}_{(-2\Lambda^{*}+\Lambda'+\delta), -T\hat{s}}, \uparrow}~,~\nonumber\\
&&\hspace*{-0.5cm}=-\sum_{\mathbf{p}, \Lambda, \delta, \hat{s}}K_{s,\mathbf{p}, \hat{s}}(\delta)(S^{+}_{\Lambda,\delta,\hat{s}}S^{-}_{\Lambda',\delta,-\hat{s}}+h.c.)~,\
\end{eqnarray}
where $\mathbf{k}_{\Lambda',-\hat{s}}= \mathbf{p}-\mathbf{k}_{\Lambda\hat{s}}$ and the spin pseudospins $\mathbf{S}_{\Lambda, \hat{s}, \delta}$ are defined as
\begin{eqnarray}
\mathbf{S}_{\Lambda, \hat{s}, \delta}&=& f^{s;\dagger}_{\Lambda, \delta, \hat{s}}\frac{\mathbf{\sigma}}{2}f^{s}_{\Lambda, \delta, \hat{s}}\nonumber\\
f^{s; \dagger}_{\Lambda, \delta\hat{s}} &=& \left[c^{\dagger}_{\Lambda, \hat{s},\sigma}~,~ c^{\dagger}_{(-2\Lambda^{*}+\Lambda+\delta), T\hat{s}, -\sigma}\right]~.
\end{eqnarray}
\par\noindent
In accounting for all possible pair-scattering processes in the vicinity of the QCP, we include the contribution of Cooper and tangential scattering channels in the spin-backscattering RG equation ($\Delta K^{(j)}_{s, \mathbf{p}, \hat{s}}(\delta)$) in eqs.\eqref{Long RG equations with doping}
\begin{eqnarray}
&&\Delta K^{(j)}_{s, \mathbf{p}, \hat{s}}(\delta) = \frac{-(1-p)(K^{(j)}_{s, \mathbf{p}, \hat{s}}(\delta))^{2}}{e^{i\gamma_{l}^{\Uparrow}}|G^{p, \Uparrow}_{j, l}|^{-1}-\frac{K_{p, \mathbf{p}, \hat{s}}^{(j)}(\delta)}{4}}\nonumber\\
&+&\frac{(K^{(j)}_{s, \mathbf{p}, \hat{s}}(\delta))^{2}}{\omega -\frac{1}{2}(\epsilon_{\mathbf{k}_{\Lambda\hat{s}}}+\epsilon_{\mathbf{p}-\mathbf{k}_{\Lambda\hat{s}}})-|\Delta\mu_{eff}-\Delta\mu^{*}_{eff}|-\frac{K^{(j)}_{p, \mathbf{p},\hat{s}}(\delta)}{4}}\nonumber\\
&-& \frac{N_{j}^{2}(L^{(j)}_{\delta})^{2}}{\omega +sgn(\Delta\mu_{eff})W+(\epsilon^{s}_{j, avg}-\Delta\mu_{eff})-L^{(j)}(\delta)}~,\label{spin_cooper_RG}
\end{eqnarray}
where $\Delta\mu_{eff}-\Delta\mu_{eff}^{*}$ is the effective chemical potential for the marginal Fermi liquid centered around the node, and renormalized by the fluctuation scale $\omega^{c, *}_{PG}=\Delta\mu^{*}_{eff}$. In the above RG equation, the spin-charge hybridized backscattering strength (with Cooper channel contribution) is given by $K^{(j)}_{p, \mathbf{p}, \hat{s}}(\delta) =p_{c}K^{(j)}_{c, \mathbf{p},\hat{s}}(\delta)+(1-p_{s})K^{(j)}_{s, \mathbf{p}, \hat{s}}(\delta)$. The negative sign of the first and third terms in the above RG equation originates from the interchanging of electron and hole creation operators in eq.\eqref{spin_cooper_RG}. The second term contains contributions to spin-backscattering from the Cooper channel due to Cooper pairs with zero as well as non-zero pair-momentum. The third term accounts for the influence of tangential scattering on the spin-backscattering. 
\par\noindent
On either side of the QCP, spin-backscattering with $\mathbf{p}=0$ pair-momentum is suppressed by tangential scattering ($\omega>\omega_{t}$) and $\pi$-momentum e-h pair-backscattering ($\omega>\omega^{c}_{PG}$) processes, leading to the establishing of the CFL and Mott liquid phases respectively. Remarkably, right above the QCP (i.e., $\omega^{c}_{PG}<\omega<\omega_{t}$), we observe that the antinodal spin gap (composed of $\mathbf{p}=0$ pair-momentum Cooper pairs) carries the highest spectral weight. This can be seen from the smallest magnitude of the inverse Green function for $\mathbf{p}=0$ pair-momentum: $G_{\Lambda\hat{s}}^{-1}(\omega, \mathbf{p})= (\omega - (\epsilon_{\mathbf{k}_{\Lambda\hat{s}}}+\epsilon_{\mathbf{p}-\mathbf{k}_{\Lambda\hat{s}}})/2)$. We also note that for $\Delta\mu_{eff}=0$, only the first term has leading contribution in the above RG equation, thus analytically continuing to the $\Delta\mu_{eff}=0$ (i.e., $1/2$-filled) case (eq.\eqref{Long RG equations with doping}). 
\par\noindent
The fixed point Hamiltonian  obtained by investigating the RG relation eq.(\eqref{spin_cooper_RG}) for the gapped parts and the 2e-1hole RG relation for the gapless parts of the FS (eq.\eqref{2e1h_scattering_near_Fermi_energy})
\begin{widetext}
\begin{eqnarray}
&&\hspace*{-1cm} H^{*}(\omega,\Delta\mu_{eff}) = \sum^{\hat{s}(\omega, \Delta\mu_{eff})}_{\Lambda,\hat{s}=\hat{s}_{N}, \delta}R^{*}_{\hat{s}, \delta}\hat{n}_{\mathbf{k}_{\Lambda, \hat{s}}, \uparrow}\hat{n}_{\mathbf{k}_{-\Lambda +\delta T, \hat{s}}, \downarrow}(1-\hat{n}_{\mathbf{k}_{\Lambda', \hat{s}}, \uparrow})
+\sum_{\hat{s}(\omega, \Delta\mu_{eff})}^{\hat{s}=\hat{s}_{AN}}K^{*, l}_{c}(\omega,\Delta\mu_{eff})\mathbf{A}_{*, \hat{s}}\cdot\mathbf{A}_{*, -\hat{s}}\nonumber\\
&+&\frac{1}{2}\sum_{\hat{s}=\hat{s}_{AN},\Lambda,\delta}^{\hat{s}_{\omega}}K^{*, l}_{s, 0,\hat{s}}(\omega,\Delta\mu_{eff}, \delta)(B^{+}_{\Lambda,\hat{s}}B^{-}_{-2\Lambda^{*}+\Lambda+\delta,T\hat{s}}+h.c.)
-\sum_{\delta, \hat{s}, \hat{s}', \Lambda}K^{*, l}_{s, 0,\hat{s}}(\omega,\Delta\mu_{eff}, \delta)S^{z}_{\Lambda,\hat{s}}S^{z}_{-2\Lambda^{*}+\Lambda+\delta,-\hat{s}}\nonumber\\
&+&\sum_{\delta, \hat{s}, \hat{s}', \Lambda}L^{*}_{\delta}\hat{n}_{\mathbf{k}_{\Lambda,\hat{s}}, \uparrow}\hat{n}_{\mathbf{k}_{-\Lambda+\delta , T\hat{s}\downarrow}}+\sum^{\hat{s}(\omega,\Delta\mu_{eff})}_{\Lambda,\hat{s}=\hat{s}_{N}}(\epsilon_{\mathbf{k}}-\Delta\mu_{eff})\hat{n}_{\mathbf{k}_{\Lambda, \hat{s}}}
+\sum_{\hat{s}}(\Delta\mu_{eff}-\Delta\mu^{*}_{eff})B^{z}_{\Lambda, \hat{s}} +\sum_{\hat{s}}\Delta\mu_{eff}^{*}A^{z}_{\Lambda,\hat{s}}\nonumber\\
&-& \sum_{\hat{s}\neq\hat{s}', \delta}T^{*}\left(B^{+}_{*, \Lambda, \hat{s}}B^{-}_{*, \Lambda', T\hat{s}'}+h.c.\right)
+\frac{1}{2}\sum_{\hat{s}=\hat{s}_{AN},\Lambda,\delta}^{\hat{s}_{\omega}}K^{*, l}_{s, \mathbf{p},\hat{s}}(\omega,\Delta\mu_{eff}, \delta)(B^{+}_{\mathbf{p},\Lambda,\hat{s}}B^{-}_{\mathbf{p},-2\Lambda^{*}+\Lambda +\delta,T\hat{s}}
+ h.c.)\nonumber\\
&-&\sum_{\delta, \hat{s}, \hat{s}', \Lambda}K^{*, l}_{s, \mathbf{p},\hat{s}}(\omega,\Delta\mu_{eff}, \delta)S^{z}_{\Lambda,\hat{s}}S^{z}_{-2\Lambda^{*}+\Lambda'+\delta,-\hat{s}}~,
\label{effHamforQCPregion}
\end{eqnarray}
\end{widetext}
where the charge ($\mathbf{A}$) fluctuation pseudospins has been defined earlier, and the zero momentum Cooper pair/Anderson pseudospins are given by 
\begin{equation}
\mathbf{B}_{\Lambda\hat{s}}=2^{-1}f^{cp\dagger}_{\Lambda, \hat{s}}\boldsymbol{\sigma}f^{cp}_{\Lambda, \hat{s}}~,~f^{cp}_{\Lambda, \hat{s}}=(c^{\dagger}_{\mathbf{k}_{\Lambda\hat{s}}\uparrow}~c_{-\mathbf{k}_{\Lambda\hat{s}}\downarrow})~,
\end{equation}
operating in the subspace
\begin{eqnarray}
\hat{n}_{\mathbf{k}_{\Lambda\hat{s}}\uparrow} = \hat{n}_{-\mathbf{k}_{\Lambda\hat{s}}\downarrow}~~,~~ \Lambda <\Lambda^{*}~.\label{anderson_pseudo}
\end{eqnarray} 
The finite momentum Cooper pairs are given by $B^{+}_{\mathbf{p},\Lambda,\hat{s}}=c^{\dagger}_{\mathbf{k}_{\Lambda\hat{s}}\sigma}c^{\dagger}_{\mathbf{p}-\mathbf{k}_{\Lambda\hat{s}},-\sigma}$~.
The  $\mathbf{p}=0$ pair-momentum Cooper pair pseudospin scattering terms can, as discussed earlier, be recast in terms of the electron-hole pseudospins 
\begin{eqnarray}
\hspace*{-1cm}-(S^{+}_{\Lambda, \hat{s}, \delta}S^{-}_{-\Lambda, -\hat{s},\delta}+h.c.) = (B^{+}_{\Lambda,\hat{s}}B^{-}_{-2\Lambda^{*}+\Lambda+\delta,T\hat{s}}+h.c.)~.\label{transmutation-1}
\end{eqnarray}
This confirms that even as we study a model of repulsive electronic correlations, the effective attractive nature of spin pseudospin backscattering is responsible for the formation of preformed Cooper pairs in the gapped antinodal regions of the FS~\cite{anderson1987,emery1995}. Thus, the first term in eq.\eqref{effHamforQCPregion} corresponds to the 2e-1h nodal Marginal Fermi liquid metal. The second and third terms govern the gapping mechanisms of the Fermi surface by charge and Anderson pseudospins respectively.
The fourth term is associated with tangential scattering in the gapless stretches of the FS, while the fifth term (proportional to $B^{z}_{\Lambda,\hat{s}}$) reflects the doublon-holon disparity with hole doping. Finally, the sixth term 
(i.e., involving $T^{*}_{\hat{s}}$) represents Josephson Cooper-pair tunnelling between the collection of 1D gapless MFL regions at and near the N and the collection of 1D spin gapped regions at and near the AN. 
\par\noindent
The Hamiltonian $H^{*}(\omega,\Delta\mu_{eff})$ naturally encompasses the doped Mott liquid Hamiltonian with insulating ground states in the region $0>-\Delta\mu_{eff}>-\Delta\mu_{eff}^{*}$ and $\omega =\frac{W}{2}$. This can be seen as follows: $R^{*}_{\hat{s},\delta}=0$ for all $\hat{s}$, $L^{*}_{\delta}=0$ from eq\eqref{Long RG equations with doping}, eq\eqref{tang_scatt_RG_flows}. Then, $K^{*,l}_{s,\mathbf{p},\hat{s}}(\omega,\Delta\mu_{eff},\delta)=K^{*,l}_{s,\mathbf{p},\hat{s}}(\omega,\Delta\mu_{eff})$ for $\delta=0$ and zero otherwise                                                                                                                                                                                                                                     eq\eqref{spin_cooper_RG}. Further, the z-component of the total Cooper-pair pseudospin is equal to the z-component of the total charge pseudospin: $\sum_{\Lambda,\hat{s}}B^{z}_{\Lambda,\hat{s}}=\sum_{\Lambda,\hat{s}}A^{z}_{\Lambda,\hat{s}}$. This leads to the Hamiltonian for the insulating Mott liquid eq\eqref{dopedMottLiquidHam} centered about the QCP
\begin{eqnarray}
H^{*}(\omega,\Delta\mu_{eff}) &=& \sum_{\hat{s}(\omega, \Delta\mu_{eff})}\bar{U}_{0}(\mathbf{A}_{*, \hat{s}}\cdot\mathbf{A}_{*, -\hat{s}}-\mathbf{S}_{*, \hat{s}}\cdot\mathbf{S}_{*, -\hat{s}})\nonumber\\
&+&\sum_{\hat{s}}\Delta\mu_{eff}^{*}A^{z}_{*,\hat{s}}\nonumber\\
&+&\sum_{\hat{s}}(\Delta\mu_{eff}-\Delta\mu^{*}_{eff})B^{z}_{*, \hat{s}}.\label{dopedMlQCP}
\end{eqnarray}
\par
\begin{figure}
\includegraphics[width=0.45\textwidth]{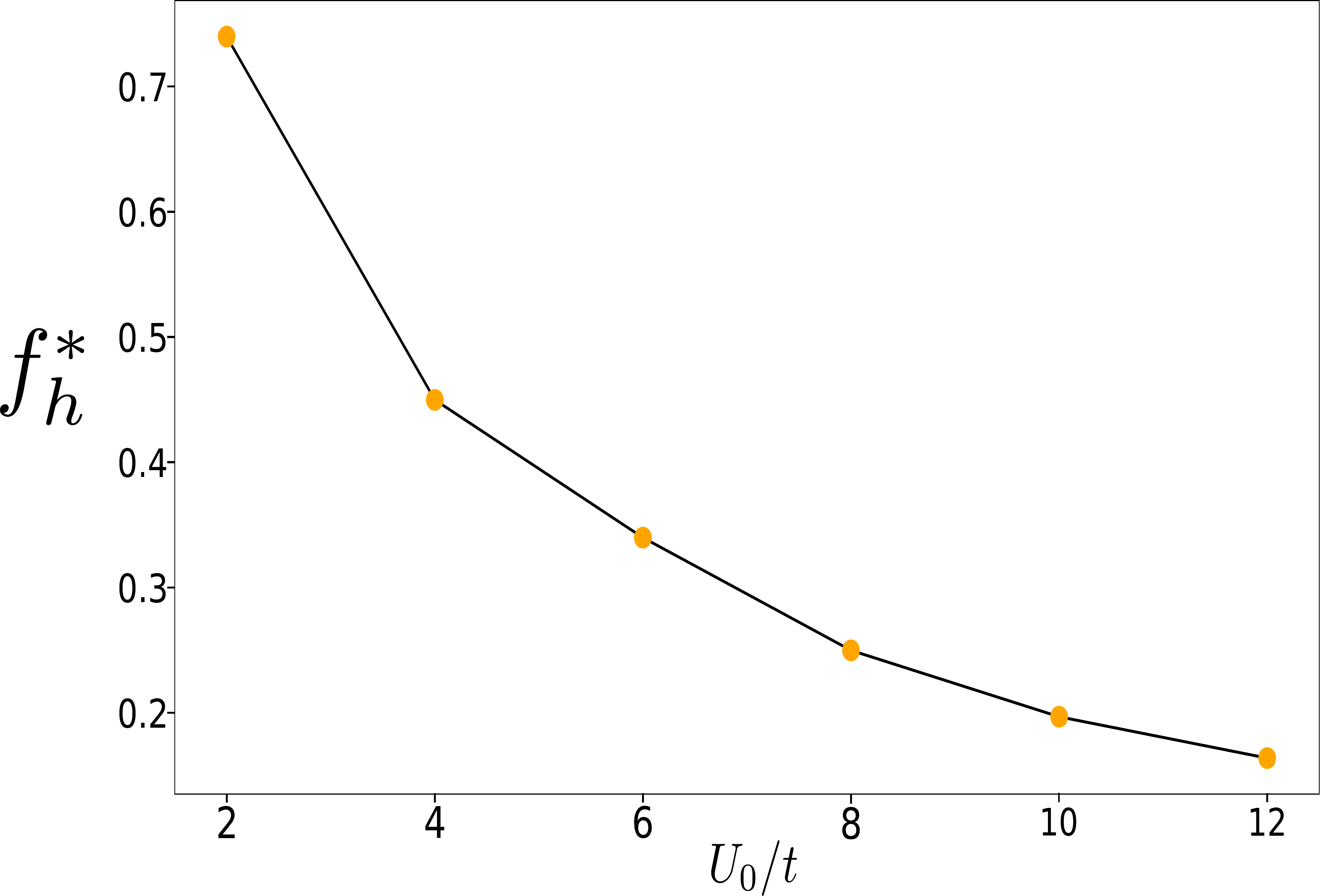}
\caption{Figure shows variation of the critical doping($F_{h}^{*}$) value against the value of bare Hubbard coupling($U_{0}$). Numerically plotted for $U=12t,10t,8t,6t,4t,2t$ with $t=1$, we find the critical doping $f^{*}_{h}$ varying from $16.4\%-74\%$. This has been done for lattice size $2048\times 2048$}~.
\label{critvariationwithU}
\end{figure}

\noindent
The existence of competing gapping instabilities of some parts of the FS, as well as tangential scattering of Landau qausiparticles and preformed Cooper pairs between the gapless nodal regions and the gapped antinodes signals a drastic change in the nature of many-particle entanglement across the QCP~\cite{zaanen2011mottness}. This can be seen simply from the nature of the state of the nodal points on the FS. The QCP at optimal doping is associated with a sudden appearance of the nodal Marginal Fermi liquid's excitations which can be written in terms of a separable state, while on either side the state at the node is strongly entangled through longitudinal and tangential  scattering in the Mott liquid (underdoped) and CFL (overdoped) phases respectively. Thus, the nodal state is an eigenstate of the Mott liquid Hamiltonian (involving the charge pseudospins $\mathbf{A}$) upon underdoping away from the QCP as discussed above, and possesses Landau quasiparticles with $0< Z_{1}<1$ upon overdoping away from the QCP. Finally, as discussed earlier in subsection \ref{MottMetal}, the nodal excitations precisely at the QCP show $Z_{1}\to 0$ and $Z_{3}\to 1$. The phase precisely above the QCP contains preformed Cooper pairs in spin-gapped antinodal regions coupling with the MFL's excitations in the gapless nodal stretches, extending into the entire conical region lying above the QCP in the phase diagram shown in Fig.\ref{Phase_diagram_with_doping-1}. This feature of the phase diagram is reminiscent of the \textit{quantum critical cone} typically observed at finite temperatures above a QCP, but with temperature (i.e., the energy scale for thermal fluctuations) here replaced by $\omega$ (the scale for the quantum fluctuations). We will turn to a discussion of this in the next subsection. We end this subsection with an interesting observation on the variation of critical doping for the QCP ($f^{*}_{h}$) against the Hubbard repulsion $U$ (shown in Fig.\ref{critvariationwithU}).
The plot shows that the value of critical doping lowers with increasing $U_{0}$: $f^{*}_{h} = 74\%$ for $U_{0}=2t$, while $f^{*}_{h} = 16.4\%$ for $U=12t$. We stress that we have computed  $f^{*}_{h}$ only for those values of $U_{0}/t$ at which we have benchmarked our ground state energy against other numerical methods (see Appendix~\ref{further benchmarking}). We have also checked that the qualitative feature of the phase diagram presented in Fig. \ref{Phase_diagram_with_doping-1} for $U_{0}/t=8$ remains unchanged for $2\leq U_{0}/t \leq 12$. We also note that for this entire range, the QCP always appears at a fixed chemical potential independent of $U_{0}$: $\Delta\mu_{eff}=-\frac{W}{2}$. The variation of $f^{*}_{h}$ with $U/t$ certainly deserves further investigation.
\subsection{$T=0$ origin of Homes Law: Planckian dissipation and preformed Cooper pairs}
The V-shaped region bounded by the lines $\omega^{c}_{PG}<\omega<\omega^{t}$ in the phase diagram Fig.~\ref{Phase_diagram_with_doping-1} has a simpler fixed point Hamiltonian compared to eq.\eqref{effHamforQCPregion} owing to the RG irrelevance of charge backscattering (with coupling $K_{c}^{*}$) and tangential scattering (with coupling $L^{*}_{\delta}$) terms
\begin{eqnarray}
&&\hspace*{-0.5cm}H^{*}(\omega,\Delta\mu_{eff})= \sum^{\hat{s}(\omega, \Delta\mu_{eff})}_{\hat{s}=\hat{s}_{N}, \delta}R^{*}_{\hat{s}, \delta}\hat{n}_{\mathbf{k}_{\Lambda, \hat{s}},\uparrow}\hat{n}_{\mathbf{k}_{-\Lambda +\delta, T\hat{s}}, \downarrow}(1-\hat{n}_{\mathbf{k}_{\Lambda', \hat{s}}, \uparrow})\nonumber\\
&+&\sum_{\hat{s}=\hat{s}_{AN},\Lambda,\delta}^{\hat{s}_{N}}K^{*, l}_{s, 0,\hat{s}}(\omega,\Delta\mu_{eff}, \delta)(B^{+}_{\Lambda,\hat{s}}B^{-}_{-2\Lambda^{*}+\Lambda+\delta,T\hat{s}}+h.c.)\nonumber\\
&+&\sum_{\mathbf{k}}(\epsilon_{\mathbf{k}}-\Delta\mu_{eff})\hat{n}_{\mathbf{k}_{\Lambda\hat{s}}}
+\sum_{\hat{s}}(\Delta\mu_{eff}-\Delta\mu^{*}_{eff})B^{z}_{\Lambda,\hat{s}}~.
\label{H_V}
\end{eqnarray}
Further, we have neglected the tangential (Josephson) scattering of Cooper pairs between the gapless and gapped parts (with coupling $T^{*}$) are they are subdominant. 
We will now focus our attention on the physics of the AN and N points on the FS. The nodal points are important as the QCP is realized with four nodal gapless MFL metals. On the other hand, the antinodes possess the largest spin gap (Fig.~\ref{Phase_diagram_with_doping-2}(b)): the highest single-particle spectral weight (due to the van Hove singularities of the electronic dispersion) are converted to that of Cooper pairs at the antinodes.
\par\noindent
The spin-backscattering RG equations for the N and AN points at fluctuation scales ($\tilde{\omega}_{c}, \Delta\mu_{eff}$) and ($\tilde{\omega}=0, \Delta\mu_{eff}=\Delta\mu^{*}_{eff}$) within the `V' shaped region in the vicinity of the QCP are given by
 \begin{widetext}
\begin{eqnarray}
\Delta K^{(j)}_{s, 0, \hat{s}_{AN}} &=& \frac{-p_{s}(K^{(j)}_{s, 0, \hat{s}_{AN}})^{2}}{\omega_{c} - p_{s}\epsilon^{s}_{\Lambda_{j}\hat{s}_{AN}}-(1-p_{s})\epsilon^{c}_{\Lambda_{j}\hat{s}_{AN}}-\frac{K^{(j)}_{s, 0, \hat{s}_{AN}}}{4}}+\frac{(K^{(j)}_{s, 0, \hat{s}_{AN}})^{2}}{\omega_{c} -\frac{1}{2}(\epsilon_{\mathbf{k}_{\Lambda_{j}\hat{s}_{AN}}}+\epsilon_{-\mathbf{k}_{\Lambda_{j}\hat{s}_{AN}}})-|\Delta\mu_{eff}-\Delta\mu_{eff}^{*}|-\frac{K^{(j)}_{s, 0, \hat{s}_{AN}}}{4}}\nonumber\\
\Delta K^{(j)}_{s, 0,\hat{s}_{N}} &=& \bigg(\frac{-(K^{(j)}_{s, 0, \hat{s}_{N}})^{2}}{\frac{W}{2}-\frac{1}{2}(\epsilon_{\mathbf{k}_{\Lambda\hat{s}_{N}}}-\epsilon_{\mathbf{k}_{-\Lambda T\hat{s}_{N}}})-\frac{K^{(j)}_{s, 0, \hat{s}_{N}}}{4}}+\frac{(K^{(j)}_{s, 0, \hat{s}_{N}})^{2}}{\frac{W}{2} -\frac{1}{2}(\epsilon_{\mathbf{k}_{\Lambda\hat{s}_{N}}}+\epsilon_{-\mathbf{k}_{\Lambda\hat{s}_{N}}})-\frac{K^{(j)}_{s, 0, \hat{s}_{N}}}{4}}\bigg)~\nonumber\\
&=&\frac{(K^{(j)}_{s, 0, \hat{s}_{N}})^{2}}{\frac{W}{2} -\frac{1}{2}(\epsilon_{\mathbf{k}_{\Lambda\hat{s}_{N}}}+\epsilon_{-\mathbf{k}_{\Lambda\hat{s}_{N}}})}-\frac{(K^{(j)}_{s, 0, \hat{s}_{N}})^{2}}{\frac{W}{2}-\frac{1}{2}(\epsilon_{\mathbf{k}_{\Lambda\hat{s}_{N}}}-\epsilon_{\mathbf{k}_{-\Lambda T\hat{s}_{N}}})}~=~0~.\label{RG_equations_AN_N}
\end{eqnarray}
\end{widetext}
From the second of these RG relations, we find that the Cooper instability is marginal along the nodal directions. This results from the fact that the Cooper pair and e-h pair along $\hat{s}\equiv$ N has the same dispersion energy: $(\epsilon_{\mathbf{k}_{\Lambda\hat{s}_{N}}}+\epsilon_{-\mathbf{k}_{\Lambda\hat{s}_{N}}})= (\epsilon_{\mathbf{k}_{\Lambda\hat{s}_{N}}}-\epsilon_{-\mathbf{k}_{\Lambda\hat{s}_{N}}}) \approx v_{F\hat{s}_{N}}\Lambda\hat{s}_{N}$. 
This then leads to the exact cancellation of the two terms in the RG relation for $\Delta K^{(j)}_{s, 0,\hat{s}_{N}}$. This leads to protection of the gapless nodal points from gapping via a Cooper instability. Therefore, the nodal Hamiltonian precisely at the QCP is composed of 2e-1h degrees of freedom described by the above Hamiltonian (eq.\eqref{H_V}), but with $K^{*, l}_{s, 0}(\omega, \Delta\mu_{eff}, \delta)=0$ and $B^{z}_{\Lambda\hat{s}}=0$ 
\begin{eqnarray}
&&\hspace*{-1.4cm}H^{*}_{QCP}(\omega, \Delta\mu_{eff}^{*})= \sum^{\hat{s}(\omega, \Delta\mu_{eff}^{*})}_{\hat{s}=\hat{s}_{N}, \delta}R^{*}_{\hat{s}, \delta}\hat{n}_{\mathbf{k}_{\Lambda, \hat{s}},\uparrow}\hat{n}_{\mathbf{k}_{-\Lambda +\delta, T\hat{s}}, \downarrow}(1-\hat{n}_{\mathbf{k}_{\Lambda', \hat{s}}, \uparrow})\nonumber\\
&+& \sum_{\mathbf{k}}(\epsilon_{\mathbf{k}}-\Delta\mu_{eff}^{*})\hat{n}_{\mathbf{k}_{\Lambda\hat{s}}}~.\label{nodal_MFL}
\end{eqnarray}
The nodal liquid state described here possesses 2e-1h composite excitations of the MFL with a gapless Dirac dispersion, i.e., the dynamical exponent of the excitations is $z=1$. Signatures of a nodal liquid state appear to have been observed in ARPES measurements carried out within the PG phase of the slightly underdoped cuprate Bi2212 above the superconducting dome \cite{kanigel2006evolution}, as well as in transport measurements~\cite{legros2019}. The direct experimental evidence for the 2e-1h composite excitations is, however, desirable.
\par\noindent
At the AN points, however, the Cooper-channel scattering RG relation is relevant, with a final fixed point window ($\Lambda^{*}_{\hat{s}_{AN}}$) determined from the criterion
\begin{equation}
\omega^{sc}_{onset} - |\Delta\mu_{eff}-\Delta\mu_{eff}^{*}|-\frac{1}{2}(\epsilon_{\Lambda^{*}\hat{s}_{AN}}+\epsilon_{-\Lambda^{*}\hat{s}_{AN}})=\frac{K_{s, 0, \hat{s}}^{(j^{*})}}{4}~.
\label{optcriterion}
\end{equation}
The frequency scale ($\omega^{sc}_{onset}$) lying above the QCP in the phase diagram Fig.~\ref{Phase_diagram_with_doping-1} at which superconducting fluctuations are able to condense into preformed pairs is a function of $\Delta\mu_{eff}$. A non-zero spectral weight for Cooper pairs along the AN direction at optimal doping $\rho=\Lambda^{*}_{\hat{s}_{AN}}(\omega^{sc}_{onset},\Delta\mu_{eff}=\Delta\mu_{eff}^{*})$ is then obtained from eq.\eqref{optcriterion} for $\omega^{sc}_{onset}\to 0+$. Therefore, the fluctuation scale $\frac{W}{2}-\omega^{sc}_{onset}$ is largest at optimal doping $\Delta\mu_{eff} = \Delta\mu_{eff}^{*}$, and falls away from optimality. 
This explains the dashed line in Fig.~\ref{Phase_diagram_with_doping-1} displaying the onset of a gapped state at the AN regions involving the formation of Cooper pairs. We stress that this state comprises of a fixed number of Cooper pairs, and therefore lacks the off-diagonal long-ranged order (ODLRO) associated with phase stiffness characteristic of superconductivity. Thus, such a state should display large superconducting phase fluctuations. Indeed, signatures of large superconducting phase fluctuations at temperatures much higher than the superconducting $T_{c}$ have been observed in careful Nernst effect measurements on the cuprates~\cite{ong-PhysRevB.73.024510}. Remarkably, the experiments reveal a ``dome" associated with the onset of phase fluctuations enveloping the dome of true d-wave superconductivity, similar to that observed in the RG phase diagram Fig.\ref{Phase_diagram_with_doping-1}. 
\par\noindent
We have just established that the net spectral weight of the spin gapped regions around the AN is converted into that for preformed Cooper pairs within the low energy subspace (eq.\eqref{anderson_pseudo}) in the vicinity of the QCP. Further, the nodal marginal Fermi liquid at the QCP can be shown to adiabatically continue to the region lying outside the gapped parts of the FS, and described by Hamiltonian eq.\eqref{fixed point Hamiltonian 2nd step}. This marginal Fermi liquid metal has earlier been shown to follow the Planckian dissipation law (eq.\eqref{Thermal scale}). Indeed, Planckian dissipation has received experimental confirmation as the origin of the linear-in-$T$ resistivity in several members of the cuprate family of materials recently~\cite{legros2019}. Thus, the coexistence of preformed pairs at the AN alongwith a Planckian dissipator at the N calls for an investigation of the experimetally observed Homes law~\cite{homes2004}. This is an empirical relation between the normal state Drude conductivity $\sigma (T_{c})$, the superconducting critical temperature $T_{c}$ and the superfluid weight $\rho_{s}$ observed for the d-wave superconductivity in the cuprates: $A\sigma_{DC}(T_{c})T_{c}=\rho_{s}$, where $A$ is a universal constant. We search for the $T=0$ origin of this relation in the theory of the state residing within the conical-shaped part of the phase diagram lying above the QCP. Thus, we first compute the antinodal (AN) superfluid weight from the Ferrel-Glover-Tinkham (FGT) sum rule
\begin{equation}
\int_{0}^{\infty} d\tilde{\omega}~Re[\sigma_{n, \hat{s}_{AN}}(\tilde{\omega}) -\sigma_{s, \hat{s}_{AN}}(\tilde{\omega})] =\frac{\pi}{2}\rho_{s}~, \label{integral_FGT}
\end{equation}
where $\sigma_{n, \hat{s}_{AN}}(\tilde{\omega})$ is the normal state conductivity for the marginal Fermi liquid metal present initially at the AN before the instability. On the other hand, $\sigma_{s, \hat{s}_{AN}}(\tilde{\omega})$ is the conductivity of the many-body state that contains preformed Cooper pairs, and we have labelled $\tilde{\omega}=\frac{W}{2}-\omega$. We then decompose the integral over $\tilde{\omega}$ into one over $0<\tilde{\omega}<\tilde{\omega}^{sc}_{onset}$ and another over 
$\tilde{\omega}^{sc}_{onset}<\tilde{\omega}<\infty$
\begin{eqnarray}
&&\int_{0}^{\infty} d\tilde{\omega} Re[\sigma_{n,\hat{s}_{AN}}(\tilde{\omega}) -\sigma_{s, \hat{s}_{AN}}(\tilde{\omega})] \nonumber\\
&=& \int_{0}^{\tilde{\omega}^{sc}_{onset}} d\tilde{\omega} Re[\sigma_{n, \hat{s}_{AN}}(\tilde{\omega}) -\sigma_{s, \hat{s}_{AN}}(\tilde{\omega})]\nonumber\\
&+&\int_{\tilde{\omega}^{sc}_{onset}}^{\infty} d\tilde{\omega} Re[\sigma_{n, \hat{s}_{AN}}(\tilde{\omega}) -\sigma_{s, \hat{s}_{AN}}(\tilde{\omega})]~.
\end{eqnarray}
The second integral vanishes because the integrated spectral weight of the 2e-1h continuum in the normal and gapped state of preformed Cooper pairs is equal. Further, in the first integral, there is no contribution from $\sigma_{s,\hat{s}_{AN}}(\tilde{\omega})$ as the 2e-1h degrees of freedom are not present within the momentum space window along $\hat{s}_{AN}$ ($[-\Lambda^{*}_{\hat{s}_{AN}}, \Lambda^{*}_{\hat{s}_{AN}}]$). 
For $\tilde{\omega}<\tilde{\omega}^{sc}_{onset}$, denoting the antinodal conductivity as $\sigma_{n, \hat{s}_{AN}}(\tilde{\omega}^{sc}_{onset})=\sigma_{n, \hat{s}_{AN}}(\tilde{\omega})$, the FGT sum rule amounts to
\begin{eqnarray}
&&\int_{0}^{\omega^{sc}_{onset}} d\tilde{\omega} Re[\sigma_{n, \hat{s}_{AN}}(\tilde{\omega}^{sc}_{onset})] = \tilde{\omega}^{sc}_{onset}\sigma_{n, \hat{s}_{AN}}(\tilde{\omega}^{sc}_{onset})\nonumber\\
&=&\frac{e^{2}\Lambda_{\hat{s}_{AN}}^{*}(\Delta\mu_{eff})}{m}=\frac{\pi}{2}\rho_{s}~,\label{FGT_superfluid}
\end{eqnarray}
where we have used the relation $n=\Lambda_{\hat{s}_{AN}}^{*}(\Delta\mu_{eff})/2\pi$ for the state number density $n$ in the Drude relation for the conductivity. Using the Planckian dissipation law eq.\eqref{Thermal scale}, along with the FGT relation for the superfluid weight given above, we obtain a relation analogous to Homes law for 
the onset scale for superconducting fluctuations at the antinodes ($T_{ons}=\hbar \tilde{\omega}^{sc}_{onset}/k_{B}$)
\begin{eqnarray}
\rho_{s}(\Delta\mu_{eff},0)= \frac{4k_{B}}{h}\sigma_{n,\hat{s}_{AN}}(T_{ons})~T_{ons}(\Delta\mu_{eff})~.\label{onset_superconductivity}
\end{eqnarray}
The frequency scale $\tilde{\omega}^{sc}_{onset}$ is itself obtained from eqs.\eqref{forward_scattering_magnitude},\eqref{lowest_energy_end_of_spectrum},\eqref{fixed_point_three_particle} and \eqref{1-p self energy}, and by picking up the normal $\hat{s}_{1}=\left(1-\frac{\Lambda_{0}}{\sqrt{2}\pi}\right)\hat{s}_{AN}$ in the vicinity of $\hat{s}_{AN}$ to avoid the discontinuity at the antinodal van Hove singularities (see Appendix\ref{AlgoSimRG})
\begin{eqnarray}
\tilde{\omega}^{sc}_{onset} &=& N^{*}(\hat{s}_{1},0)(\frac{1}{2}\max_{\delta,\hat{s}}(\epsilon_{\Lambda^{*},\hat{s}}+\epsilon_{-\Lambda^{*}+\delta,\hat{s}})\nonumber\\
&-&\frac{1}{2}\max_{\delta}(\epsilon_{\Lambda^{**},\hat{s}_{1}}+\epsilon_{-\Lambda^{**}+\delta,\hat{s}_{1}}))\nonumber\\
&=& \frac{N^{*}(\hat{s}_{1},0)}{2}\left(\epsilon_{\Lambda_{0},\hat{s}_{N}}-\epsilon_{\Lambda^{**},\hat{s}_{1}}\right)~.\label{frequency_onset_sc}
\end{eqnarray}
From the geometry of the square Fermi surface the number of states within the gapped window along normal $\hat{s}$ is determined to be $N^{*}(\hat{s}_{1},0) = \frac{\Lambda^{**}}{\Lambda_{0}}N\left(\frac{1}{2\sqrt{2}}-\frac{\Lambda_{0}}{4\pi}\right)$.
In the above equation $\Lambda^{*}$, $\Lambda^{**}$ are determined from eqs.\eqref{forward_scattering_magnitude} and \eqref{fixed_point_three_particle} respectively and $N^{*}(\hat{s}_{1},0)$ is the total number of gapped states at $\omega=0$. In keeping with our earlier discussion, $T_{ons}$ is largest at optimality ($\Delta\mu^{*}_{eff}$) and falls off with doping away from optimality on either side. 
In the next section, these fluctuations will be seen to interplay with the spin-gap in leading to d-wave superconductivity~ \cite{zaanenjarrell-PRL-2011}. Further, we have shown in eq.\eqref{critOns} (see Appendix \ref{symmbreakAppendix} for a detailed derivation) that the superconducting critical temperature $T_{C}$ is linearly related to $T_{ons}$, with a proportionality constant related to the extent of the pseudogap (seen in terms of the difference between the electronic dispersion at the antinodes and the nodes). In this way, we offer insight into the $T=0$ origin of Homes law~\cite{homes2004}.
\subsection{Mixed optical conductivity of the Correlated Fermi liquid}\label{mixed_optical}
The fluctuation scale $\omega_{t}:\omega-\frac{\Delta\mu_{eff}}{2}+Wsgn(\Delta\mu_{eff})>0$ marks the boundary across which the tangential scattering processes become RG relevant (eq.\eqref{tang_scatt_RG_flows}). Starting from the QCP ($\frac{W}{2}-\omega=0, \Delta\mu_{eff}=\Delta\mu_{eff}^{*}$), and proceeding into the region $\omega>\omega_{t}$, the gapless stretch of the FS increases steadily. This enhances the spectral weight of states taking part in the tangential scattering process, $N_{j}(\omega, \Delta\mu_{eff})=\sum^{\hat{s}_{\omega}}_{\hat{s}=\hat{s}_{N}}1$. As can be seen from the numerator of eq\eqref{tang_scatt_RG_flows}, a growing spectral weight gradually enhances the RG flow rate for tangential scattering processes.
It is also important to note that on these gapless stretches, the effect of forward scattering is still RG relevant (eq.\eqref{fwd_scattering_near_Fermi_energy}) leading to  2e-1h composite excitations of the marginal Fermi liquid. The tangential scattering processes, on the other hand, enhance the quasi-particle excitations of the Fermi liquid. The outcome of these competing tendencies is captured by the effective 1-particle self energy
\begin{eqnarray}
\Sigma_{\Lambda\hat{s}}(\omega) &=& L^{*}(\delta, \omega)\theta(\Lambda^{*}_{1}(\omega)-\Lambda)\nonumber\\
&+&R^{*}_{\hat{s}, \delta}\theta(\Lambda^{*}_{2}(\omega)-\Lambda)\ln\vert \frac{\bar{\omega}+\Delta\mu -\epsilon_{\Lambda\hat{s}}}{\bar{\omega}}\vert~,
\end{eqnarray}
where $\bar{\omega}=\textrm{max}_{\delta,\hat{s}}\frac{\epsilon_{j^{*}, l}+\epsilon_{j^{*}, l'}}{2}$, and $j^{*}$ is obtained from the fixed point condition eq.\eqref{fp_condition_MFL}. The fixed point values of the couplings are given by
\begin{eqnarray}
L^{*} &=& \omega +W\text{sgn}(\Delta\mu_{eff}) +\epsilon^{c}_{\Lambda^{*}_{1}, avg}-\Delta\mu_{eff}~, \label{tan_scatt_coup}\nonumber\\
\frac{R^{*}_{l\delta}}{8}  &=& \bar{\omega}-\frac{1}{2}(\epsilon_{\Lambda^{*}_{2}, l}+\epsilon_{\Lambda^{*}_{2}, l'})+\Delta\mu_{eff}~.\label{MFL_coup}
\end{eqnarray}  
The above equations for $L^{*}$ and $R^{*}_{l\delta}$ are satisfied within the windows $\Lambda^{*}_{1}$ and $\Lambda^{*}_{2}$ respectively. Near the QCP, $\Lambda^{*}_{1}<\Lambda^{*}_{2}$, such that the MFL dominates over the FL within the gapless regions of the FS. Upon gradually increasing the doping away from the QCP leads to $\Lambda^{*}_{1}>\Lambda^{*}_{2}$, such that the enhanced tangential scattering on the enlarged gapless strecthes leads to the Fermi liquid quasiparticles dominating over the composite excitations of the MFL. From the detailed discussions of subsection \ref{MottMetal}, the imaginary part of the single particle self-energy obtained from the Kramers-Kronig relation then allow us to obtain the inverse quasiparticle lifetime as a mixture of MFL and FL forms 
\begin{equation}
(\tau (\Delta\mu_{eff}))^{-1} = \alpha(\Delta\mu_{eff})\omega + \beta(\Delta\mu_{eff})\omega^{2}~,\label{mixed_lifetime}
\end{equation}
where the coefficients $\alpha$ and $\beta$ are functions of the chemical potential $\Delta\mu_{eff}$, and undergo a gradual evolution from QCP to the overdoped regions of the phase diagram. This gives rise to a mixed nature of the optical conductivity as a function of $\Delta\mu_{eff}$. This appears to be consistent with the observations of van der Marel et al.~\cite{van2003}, where a form of the optical conductivity that departs from the expected MFL form was attributed to quantum critical scaling in the neighbourhood of a QCP at optimal doping.  
\section{Symmetry breaking Orders and superconductivity}\label{SymmetryBreakingSection}
\begin{figure}
\hspace*{-1cm}
\includegraphics[width=0.6\textwidth]{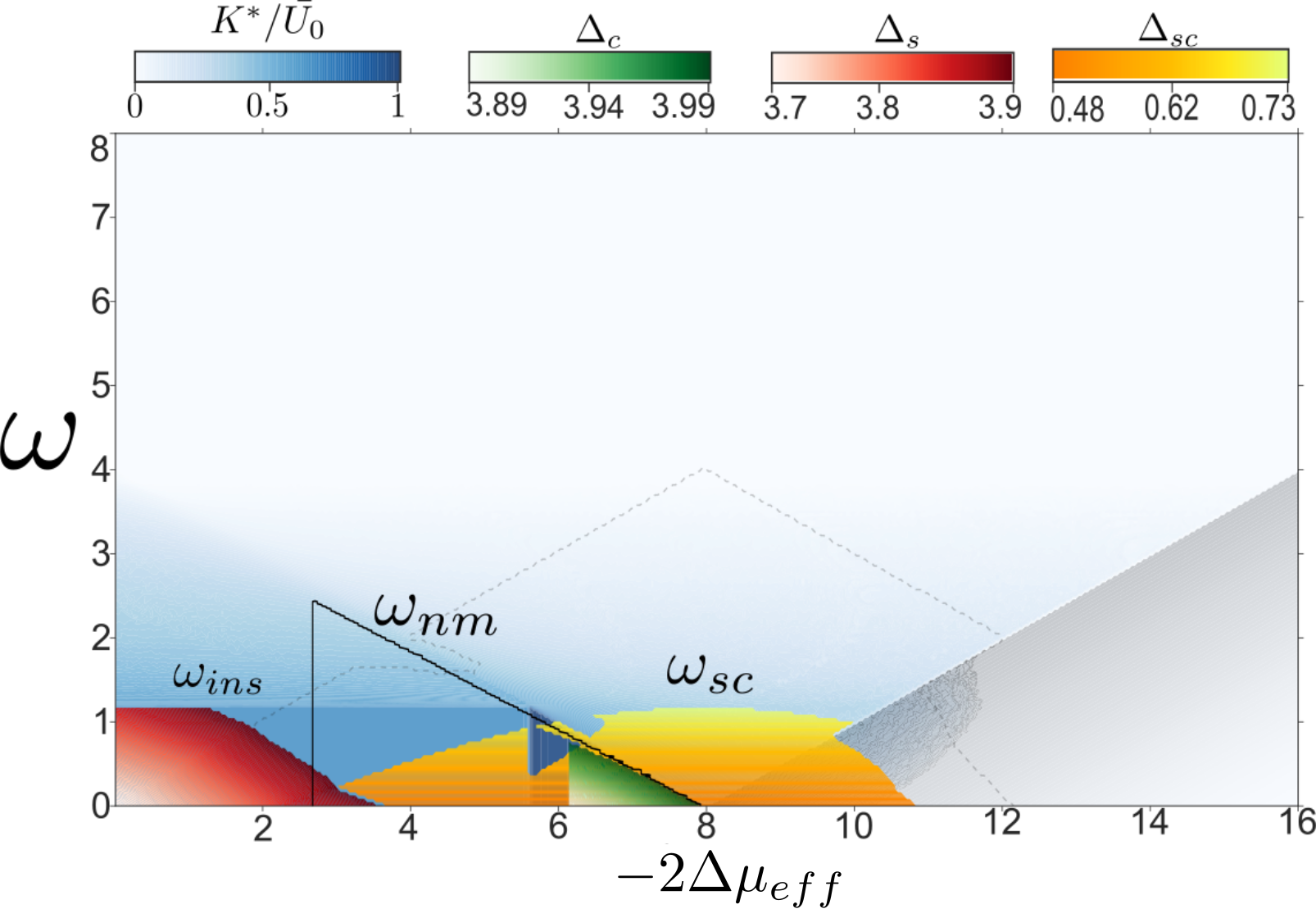}
\caption{(Colour Online) RG phase diagram for 2D Hubbard model with doping and $(\pi, \pi)$ CDW  (green), $(\pi, \pi)$ SDW (red) and d-wave superconducting (SC) orders (yellow) included. $K^{*}/\bar{U}_{0}$ in the white-deep blue colourbar represents the ratio of renormalized coupling to bare coupling symmetry unbroken PG and ML states. Gap scales for various symmetry broken phases are shown in the colour bars. Origin of superconductivity from spin-PG nodal NFL with superconducting fluctuations is described in text. The SC ``dome" is centered about the QCP (optimality) and falls away on either side due to competition with insulating orders (underdoped) and gapless CFL (overdoped). The solid black line indicates the chemical potential dependent energy scale for the onset of nematic fluctuations, while the thin dashed line denotes the onset of superconducting fluctuations.}\label{phasediagwithsymmbreak}
\end{figure}

We encountered in subsection \ref{symmbreakMottliquid} the analysis of symmetry-broken states of matter that are obtained from the $1/2$-filled Mott liquid under RG. Here, we repeat the investigation for the Hubbard model upon doping away from $1/2$-filling. 
As before, this involves a renormalization group analysis in the background of symmetry breaking fields: staggered chemical potential $\sum_{i, j}(-1)^{i+j}\hat{n}_{\mathbf{r}}$ (for the $\pi, \pi$ charge density wave), staggered magnetic field $\sum_{i, j}(-1)^{i+j}S^{z}_{\mathbf{r}}$ (for the $\pi, \pi$ spin density wave), the spin-nematic order field $Q_{\alpha\beta}^{0}f(\mathbf{r}_{1}-\mathbf{r}_{2})S^{\alpha}_{\mathbf{r}_{1}}S^{\beta}_{\mathbf{r}_{2}}$, and finally a $U(1)$ symmetry-breaking field $\sum_{\mathbf{k}\sigma}c^{\dagger}_{\mathbf{k}\sigma}c^{\dagger}_{-\mathbf{k}-\sigma}$ for superconductivity. The corresponding RG equations are given in Appendix \ref{symmbreakAppendix}. The full RG phase diagram away from 1/2-filling, and with $(\pi, \pi)$ CDW, $(\pi, \pi)$ SDW, spin-nematic and d-wave superconducting broken symmetry orders, is shown in Fig.\ref{phasediagwithsymmbreak}. The effective Hamiltonians and gap functions familiar for these symmetry broken states of matter are easily obtained from the dominant symmetry breaking coupling at the RG fixed point (e.g., the discussion for the SDW presented in subsection \ref{symmbreakMottliquid}). Results for symmetry-broken orders at finite-temperature can then be obtained via standard mean-field methods. Below, we confine ourselves to a discussion of the findings shown in the phase diagram Fig.\ref{phasediagwithsymmbreak}.
\par
\begin{figure}
(a)\includegraphics[scale=0.22]{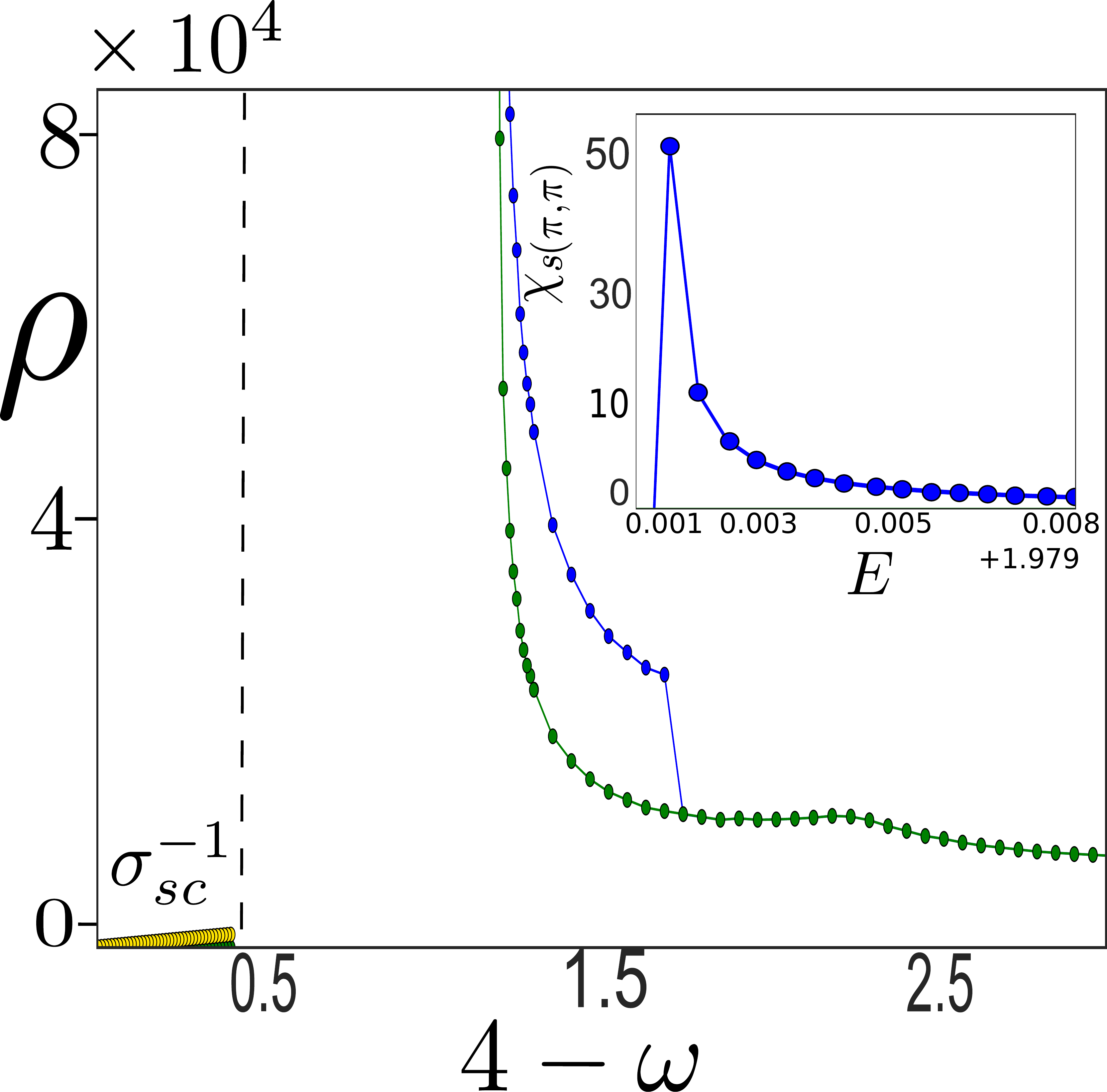} 
(b)\includegraphics[scale=0.24]{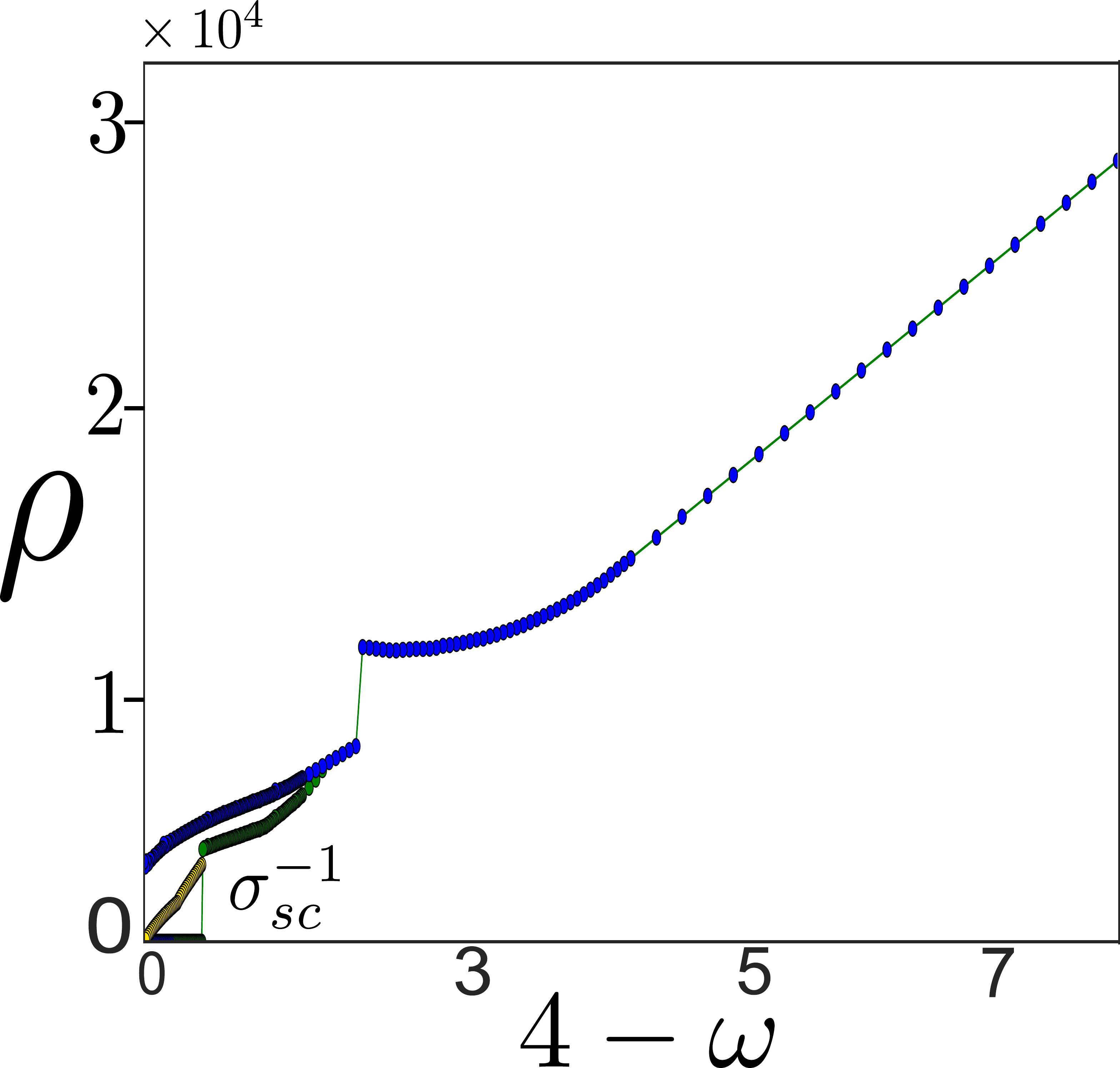}
\caption{(Colour Online)Figures (a) and (b) represents resistivity with (green) and without any form of symmetry breaking (blue), and inverse superfluid stiffness ($\sigma^{-1}_{sc}$, yellow) at various dopings. (a) $\Delta\mu_{eff}=-1.75$: passage from PG to ML (blue), from PG to SC through spin-gap dominated ML. Inset: Peak in spin susceptibility within SC region $\Delta\mu_{eff}=-1.75$. (b)$\Delta\mu_{eff}=-5.5$: passage from NFL to PG-CFL (blue), from NFL to SC through PG-CFL (green).}\label{CFLAndSDW}
\end{figure}

\begin{figure}
(a)\includegraphics[scale=0.26]{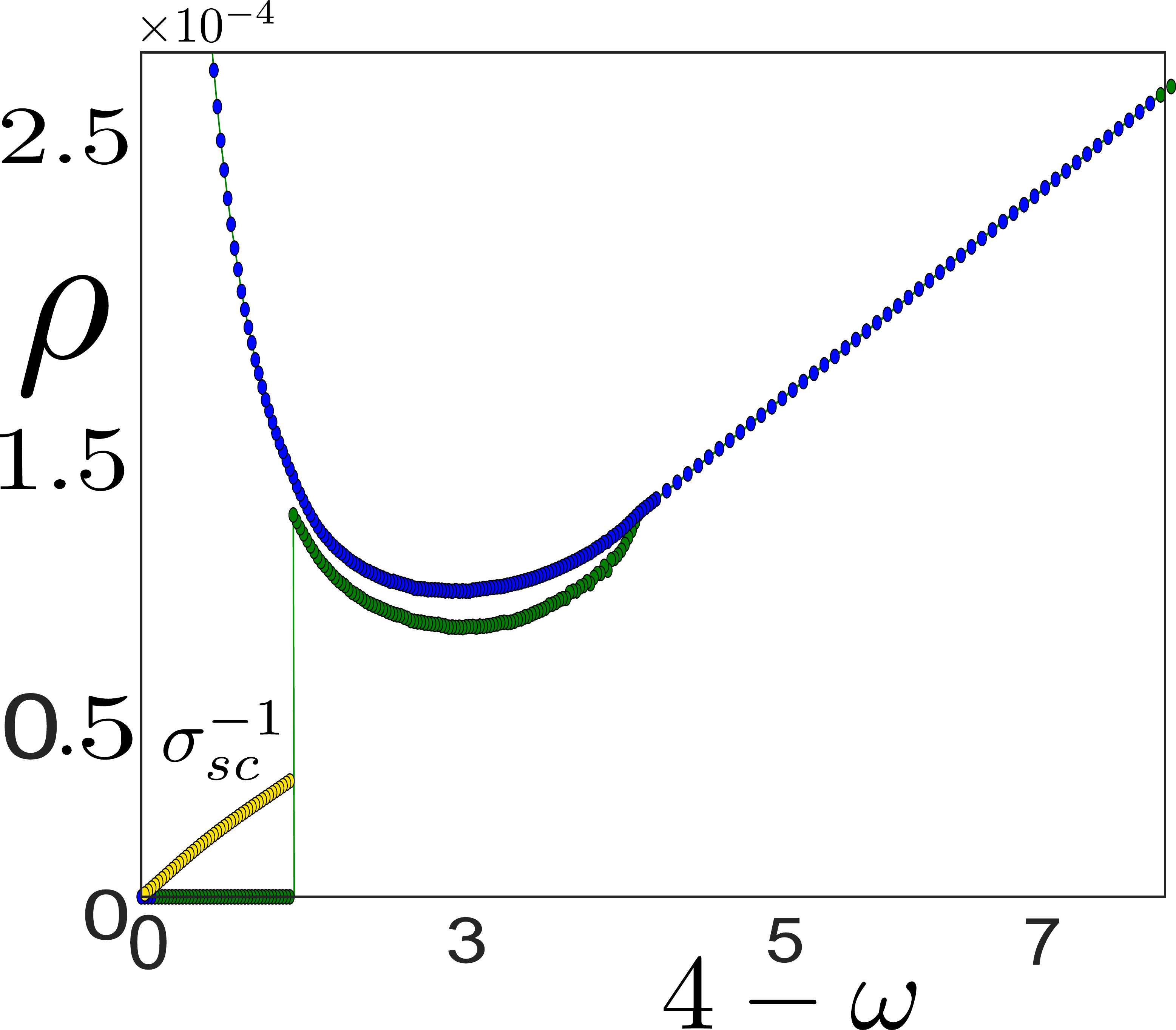} 
(b)\includegraphics[scale=0.23]{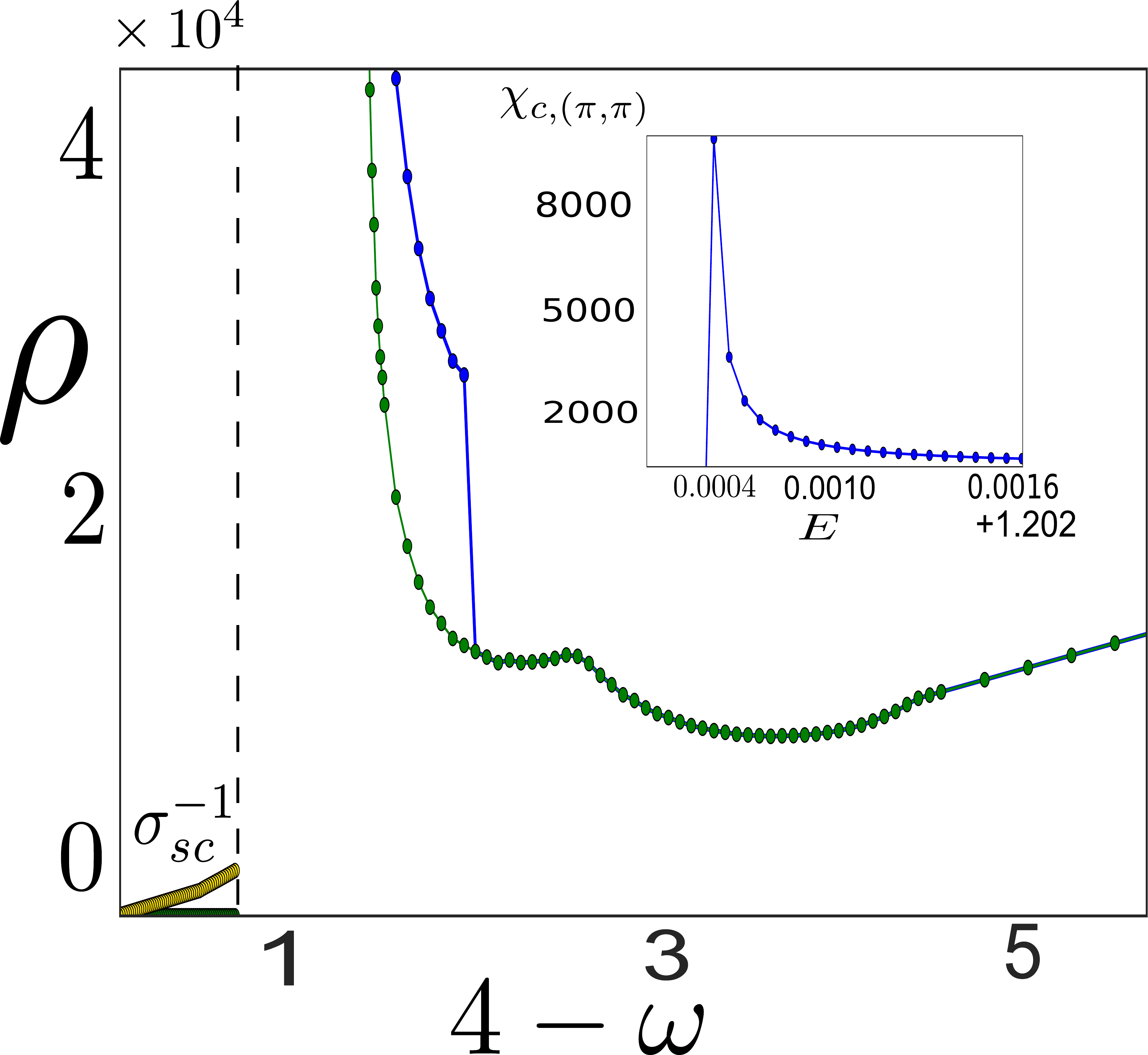}
\caption{(Colour Online) Figures (a) and (b) show resistivity ($\rho$) with (green) and without any form of symmetry-breaking (blue), and inverse superfluid stiffness ($\sigma_{SC}^{-1}$, yellow) at various dopings. (a) $\Delta\mu_{eff}=-4$: passage from PG to ML (blue), from PG to SC through charge-gap dominated ML. Inset: Peak in charge susceptibility within SC region. (b) $\Delta\mu_{eff}=-3.25$: passage from NFL with connected FS to spin-PG nodal NFL at QCP (blue), from NFL to SC through spin-PG nodal NFL (green).}\label{SCdomeAndCDW}
\end{figure}

\noindent
We find the $(\pi, \pi)$ SDW Neel antiferromagnet for energies $\omega \geq \omega_{ins}=2.8$ and doping $0>\Delta\mu_{eff}\geq -1.75$ (deep underdoping), and the $(\pi, \pi)$ CDW for energies $\omega\geq 3.2$ and doping $-3.05\geq \Delta\mu_{eff}\geq -4$ (moderately underdoped). A d-wave superconducting dome is found to extend between a doping range of $-1.5\geq \Delta\mu_{eff}\geq -6$ and has an optimal gap scale ($\Delta_{SC}$) at the critical doping $\Delta\mu_{eff}=-4$ corresponding to the QCP. We have already presented earlier our finding for the variation of the superfluid density $\rho_{s}$ with chemical potential, with a discussion on why $\rho_{s}$ is maximal at optimal doping. In keeping with this, the highest $\Delta_{SC}$ arises from a maximal density of bound Cooper pairs in the gapped regions of the Fermi surface interplaying with the critical fluctuations of the gapless regions for a chemical potential tuned to the QCP~\cite{pushp2009} (see green curve in Fig.\ref{arc_to_point_spinon_holon_Arc}(b)). The optimal quantum fluctuation energy scale for the onset of superconductivity is the kinetic energy of a pair of electrons from diametrically opposite nodal points on the FS: $W/2-\omega_{SC}=W/2-2\epsilon_{\Lambda^{*}\hat{s}_{node}}$, where $\Lambda^{*}/\Lambda_{0}$ is the dimensionless spectral weight for the marginal nodal electronic quasiparticles obtained from the RG. Deep in the underdoped regime ($-1.5\geq \Delta\mu_{eff}\geq -1.75$), a coexistence of $(\pi, \pi)$ SDW and d-wave superconducting orders appears likely~\cite{sebastian2010} (see inset of Fig.~\ref{CFLAndSDW}(a)). Similarly, within the lightly underdoped region, a coexistence of $(\pi, \pi)$ CDW and d-wave SC orders appears likely (see inset of Fig.~\ref{SCdomeAndCDW}(b)). 
\begin{figure}
\includegraphics[width=0.5\textwidth, height=0.35\textwidth]{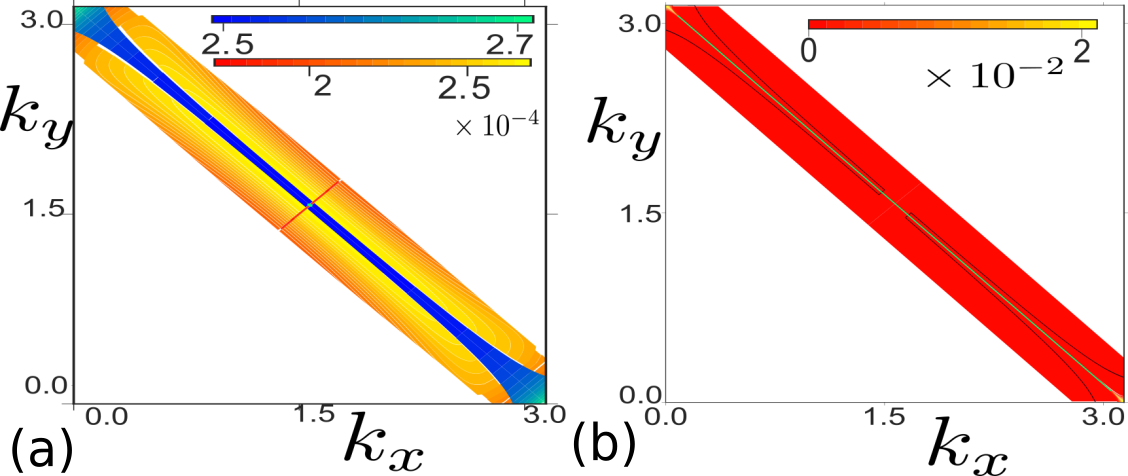}
\caption{(Colour Online)Figures (a) and (b) shows the full momentum space resolved spectral function $A(\mathbf{k}_{\Lambda\hat{s}})$, where the black curves show extent of SC fluctuations around Fermi surface. (a): to the right of SC dome (CFL). (b): within d-wave SC dome, with nodal NFL.} \label{MapsInCFLandSC}
\end{figure}
\begin{figure}
\includegraphics[width=0.5\textwidth, height=0.35\textwidth]{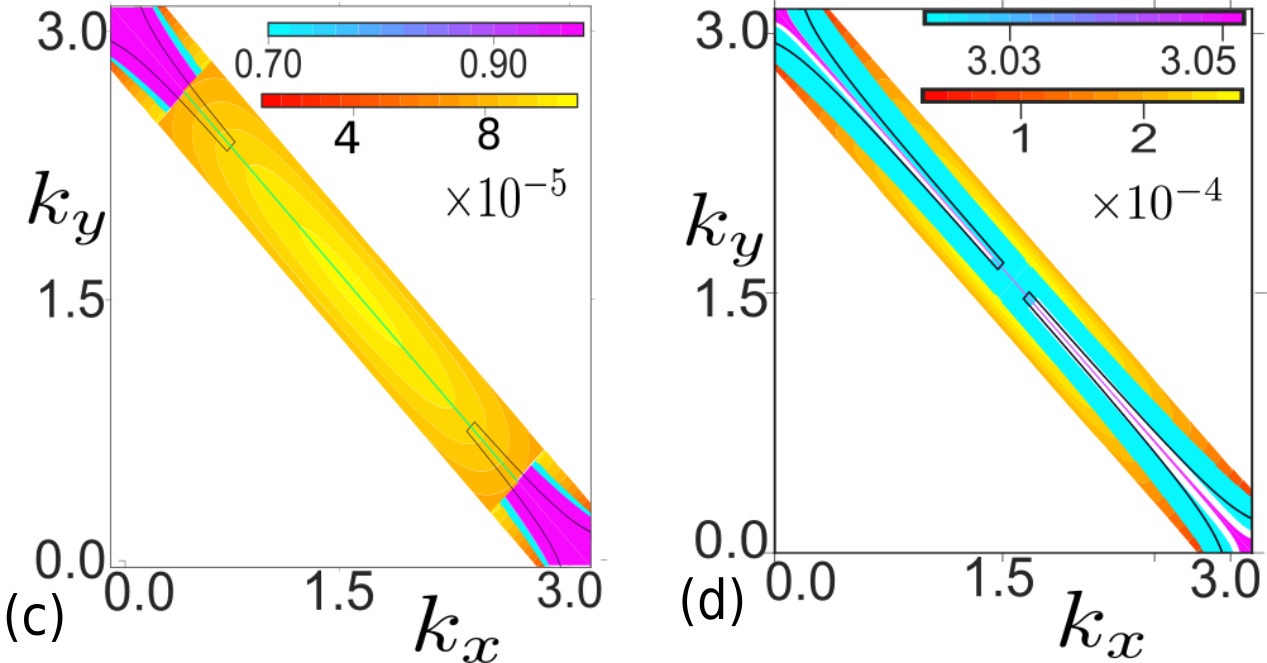}
\caption{(Colour Online)Figures (a) and (b) shows the full momentum space resolved spectral function $A(\mathbf{k}_{\Lambda\hat{s}})$, where the black curves show extent of SC fluctuations around Fermi surface (a): to the left of SC dome (ML), (b):  vertically above SC dome at optimal doping.} \label{MapsInMLandAboveSC}
\end{figure}
\par
\begin{figure}[h]
\hspace*{-1cm}
\includegraphics[scale=0.35]{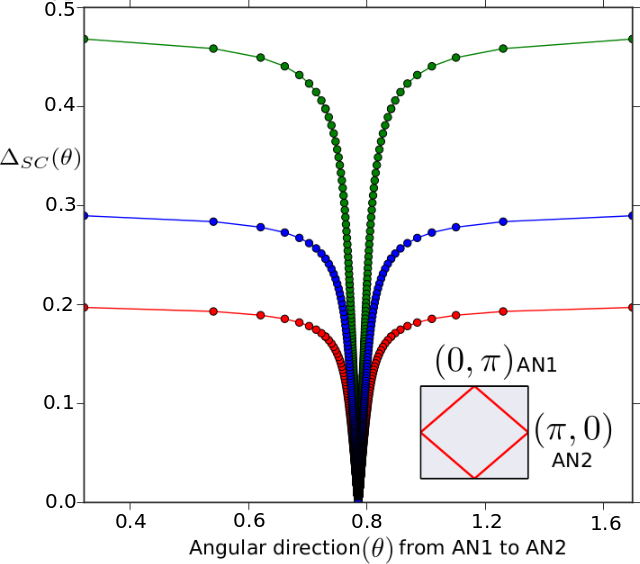}
\caption{(Colour Online) Gap $\Delta_{SC}(\theta)$ vs the angular coordinate $\theta$ for chemical potential and $4-\omega$ values given by $\mu_{eff}=-2.5$,$4-\omega_{c1} = 0.8$ (red curve) in the underdoped regime. At optimal doping $\Delta\mu_{eff}=-4.0$,$4-\omega_{c2} = 1.2$ (green curve). At overdoping $\Delta\mu_{eff}=-5.5$,$4-\omega_{c3} = 0.5$ (blue curve).}\label{doping nodal structure}
\end{figure}

\begin{figure*}
(a)\includegraphics[scale=0.9]{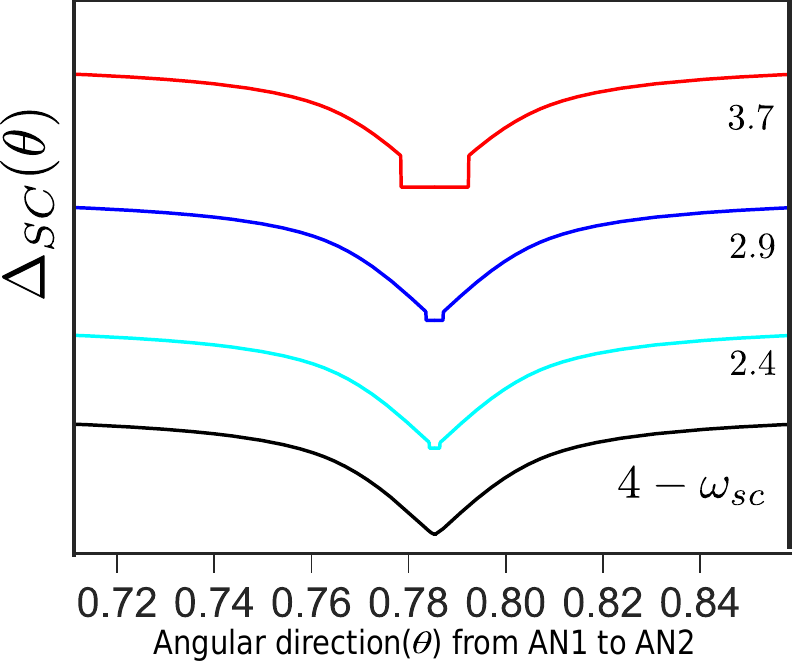}
(b)\includegraphics[scale=0.36]{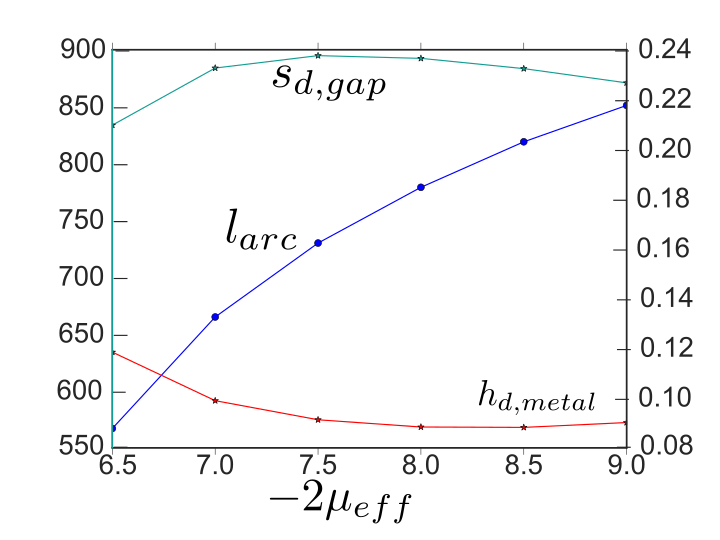}
\caption{(a)(Colour Online) Gap $\Delta(\theta)$ vs. the angular coordinate $\theta$ along the $\Delta\mu_{eff}=-4.0$ optimal doping axis for different values of $4-\omega_{sc}$ lying above the dome. The different curves have been staggered for purposes of clarity. The gap $\Delta(\theta)$ vs. the angular coordinate upon just entering the superconducting dome is given by the black curve. (b) Left vertical axis: Blue curve represents the gapless arc length of the Fermi surface at the brink of the superconducting transition for six different $\omega^{cric}_{sc}$ and $-2\Delta\mu_{eff}$ values i.e.,($2.922,6.5$),($2.869,7.0$),($2.837,7.5$),($2.827,8.0$),($2.839,8.5$), and ($2.87,9.0$). Right vertical axis: At those same values, the green curve represents the Cooper pair density in the gapped regions in the neighbourhood of the antinodes, and the red curve represents the holon densities in the gapless parts neighboring the nodes.}
\label{arc_to_point_spinon_holon_Arc}
\end{figure*}

\noindent
As a clear demonstration of the d-wave nature of the superconducting order parameter, we show in Fig.\ref{doping nodal structure} the variation of the superconducting(SC) gap from maximum at one antinode(AN1=$(0,\pi)$) to a node at the nodal point $k_{x},k_{y}=\left(\pm \pi/2,\pm\pi/2\right)$ and finally maximizing again at (AN2=$(\pi,0)$). The three curves are made at the brink of entry into SC dome from underdoping (blue), optimal (green) doping, and overdoping (red) sides. At optimality, the connected Fermi surface becomes an arc with lowering quantum fluctuation scale $4-\omega$ and collapses to a point on entering the superconducting dome. The journey from above the superconducting dome to the final entry into superconducting dome at optimal doping can be seen from the variation of the gap from antinodes to the nodes in Fig.\ref{arc_to_point_spinon_holon_Arc}(a).
\par\noindent
The experiments in Ref.(\cite{pushp2009}) appear to show that optimal superconductivity is associated with the largest Fermi surface volume at the brink of entering SC dome as optimality is approached from the underdoped side. We find evidence for this, as shown in Fig.\ref{arc_to_point_spinon_holon_Arc}(b). The graph for the mildly underdoped regime ($-2\Delta\mu_{eff}<W$) shows that a smaller Fermi pocket with high concentrations of holons and low concentration of bound spinon pairs undergoes a superconducting d-wave transition where the Fermi volume shrinks to a point. In the optimally doped regime ($-2\Delta\mu_{eff}\sim W$) on the brink of transiting into the superconducting dome, we observe a larger gapless Fermi pocket with a peak in the concentration of Cooper pairs (in the gapped parts of the Fermi surface) and an intermediate concentration of holons. On the brink of entering the dome at overdoping ($-2\Delta\mu_{eff}> W$), we find a reduction in the Cooper pair density at the neighborhood of the antinodes coinciding with a lowering hole concentration on a fully connected Fermi surface. This shows that optimality arises from an interplay of a peak in the density of Cooper pairs in the gapped parts of the Fermi surface (as well as the superfluid density $\rho_{s}$)and the critical fluctuations associated with a quantum critical Fermi surface (arising from the QCP). The dome-like structure of $T_{C}$ is observed to be inherited from the similar structure of $T_{ons}$ (as discussed earlier; also see Appendix \ref{symmbreakAppendix}), and offers an explanation for the quadratic dependence of $T_{C}$ on the hole-doping concentration commonly observed in the cuprates~\cite{presland1991general}. 
\par\noindent
We now establish the origin of the nodal structure of d-wave superconducting order as the QCP at $\mu^{*}_{eff}=-4$. We have shown above that the nodal points support a gapless NFL metal, as the irrelevance of all gapping mechanisms at the nodes arise from the topological signature of the doublon propagator. Thus, the onset of superconductivity at critical doping takes place in the backdrop of this protection for the nodal gapless states. We have also seen that the spin-gap at optimal doping has d-wave structure, but without a sign change across the nodes. The $U(1)$ phase rotation symmetry breaking RG calculation presented in Appendix \ref{symmbreakAppendix} further reveals that the nodal points act as domain walls for the growth of the superconducting order upon scaling down to low energies: the RG-integrated superconducting order parameter $\langle c^{\dagger}_{\Lambda -T\hat{s}\downarrow}c^{\dagger}_{\Lambda\hat{s}\uparrow}\rangle = - \langle c^{\dagger}_{\Lambda -\hat{s}\downarrow}c^{\dagger}_{\Lambda T\hat{s}\uparrow}\rangle \propto e^{i\gamma_{\hat{s}, \hat{s}_{node}}}$~, where the relative phase $\gamma_{\hat{s}, \hat{s}_{node}}=-i\ln(sgn\left[ 2(\epsilon_{\Lambda\hat{s}}-\epsilon_{\Lambda\hat{s}_{node}})\right])$ and $2(\epsilon_{\Lambda\hat{s}}-\epsilon_{\Lambda\hat{s}_{node}})$ is the energy scale for a Cooper pair (with respect to the nodes). The alternation of sign in this energy scale is, thus, the alternation in sign of the pairing order parameter. 
\par\noindent
The suppression of d-wave superconductivity is clearly observed in the presence of the CDW order (Fig.\ref{phasediagwithsymmbreak}(a)), while the charge-gap dominated Mott liquid (dark blue region in Fig.\ref{phasediagwithsymmbreak}(a)) leads to further suppression of superconductivity. These results obtained are in consonance with the finding of two distinct antinodal energy-gap scales in recent ARPES and STM experiments carried out on the PG phase of the cuprate La-Bi2201 \cite{madhavan-PhysRevLett.101.207002}, one of which appears to be linked with the onset of superconductivity and the other with charge ordering. The black curve in the RG phase diagram corresponds to the onset of spin-nematic fluctuations, and appears to be confined within a range of chemical potential $-1.4\geq \Delta\mu_{eff}\geq -4$ associated with moderate to light underdoping. The highest scale (in $4-\omega$) for the onset of spin-nematic fluctuations lies well within the pseudogapped part of the phase diagram, in broad agreement with recent experimental findings in the cuprates~\cite{fauque2006,hinkov2008,
ando2002,daou2010,hinkov2007}.
For $-4\geq \Delta\mu_{eff}\geq -6$, the competition between the nodal NFL metal, the spin-pseudogapped parts of the FS and the tangential scattering leads to the reconstruction of the FS. This is apparent in the suppression of superconducting fluctuations (dashed line in Fig.\ref{phasediagwithsymmbreak}) beyond critical doping and shows how competition with the CFL suppresses d-wave superconductivity. Variations of the resistivity ($\rho$) and inverse superfluid stiffness ($\sigma_{SC}^{-1}$) upon tuning $\omega$ towards the QCP is shown in Figs.~\ref{SCdomeAndCDW}(a,b) (see also Videos S5, S6). Maps of the single-particle spectral function $A(E)$ within and in the neighbourhood of the d-wave SC phase are also presented in Figs.~\ref{MapsInMLandAboveSC} and \ref{MapsInCFLandSC}. These results are in excellent agreement with several transport and spectroscopy measurements made in various parts of the cuprates phase diagram \cite{keimer2015quantum, chatterjee2011electronic, tallon2001doping,
he2014, fujita2014,balakirev2003, kanigel2006evolution, ong-PhysRevB.73.024510, madhavan-PhysRevLett.101.207002,fauque2006,hinkov2008,
ando2002,daou2010,hinkov2007}. 
\par\noindent
Finally, we close this section by computing numerical values for some of the temperature scales, e.g., $T_{ons}$ (the onset of the formation of pre-formed Cooper pairs at the gapped antinodal regions) and $T_{ML}$ (the scale at which the entire FS is gapped during the formation of the Mott liquid), for the materials HgBa$_{2}$CuO$_{4}$ and La$_{2}$CuO$_{4}$ by using the values for the parameters of the effective one-band Hubbard model found in Ref.\cite{hirayama2018ab}. First, in the limit of large system size $N\to \infty$, using the formula for $\tilde{\omega}^{sc}_{onset}$ (eqs.\eqref{frequency_onset_sc} and \eqref{fixed_point_three_particle}), we obtain 
\begin{eqnarray}
\tilde{\omega} ^{sc}_{onset}\leq \omega_{PG}&=&t\frac{\Lambda^{**}N^{2}}{8\pi\sqrt{2}}\frac{16\pi^{2}}{N^{2}}=\sqrt{2}t\pi\Lambda^{**}\nonumber\\
&=& \sqrt{2}t\pi\arcsin\left(\frac{\bar{\omega}}{2t}\right)=0.067t~,
\end{eqnarray}
where $\bar{\omega} \approx \omega_{PG} = 0.034t$ as given earlier. Then, using the values of the hopping parameter $t$ for HgBa$_{2}$CuO$_{4}$ and La$_{2}$CuO$_{4}$ with $t=-0.461,-0.482eV$~\cite{hirayama2018ab}, we find the onset temperature scale of pair formation $T_{ons}$ is bounded (on the upper side) by the pseudogap temperature $T_{PG}$ are given by $T_{PG}=0.067t\times 11605K~=~358K$ and $374K$ for HBCO and LCO respectively. This is in fair agreement with experimental estimates of $T_{PG}\sim 200K-300K$ for most of the cuprates~\cite{timusk1999pseudogap}. That the onset temperature scale for pairing $T_{ons}$ is less than $T_{PG}$ is consistent with the findings from Nernst measurements of Ref.\cite{ong-PhysRevB.73.024510}. Further, by using the formula for $T_{ML}$ (eq.\eqref{frequency_ML}) for the Mott liquid, we find $T_{ML}\approx\frac{\frac{W}{2}-\omega_{ML}}{\frac{W}{2}-\omega_{PG}}T_{ons} = 119K$ and $124K$ for HBCO and LCO respectively. Recalling the relation between $T_{ML}$ and the superconducting transition temperature $T_{C}$ obtained in eq.\eqref{TCTMLrelation}, we see that the upper bound for $T_{C}$ is provided by $T_{ML}$. An upper bound of around $120K$ for the $T_{C}$ is reasonable for the experimentally known $T_{C}$ of $40K$ for LCO and $90K$ for HBCO. Importantly, as $T_{C}$ is also found to be bounded (on the upper side) by $T_{ons}$ (see eq.\eqref{critOns}), it appears plausible to search for mechanisms that can raise $T_{C}$ further towards $T_{ons}$. We will further discuss this point briefly in the concluding section.
\section{Conclusions and perspectives}\label{ConclusionsSection}
\noindent
In conclusion, an RG analysis of the 2D Hubbard model on the square lattice reveals the nature of the FS topology-changing Mott metal-insulator transition at half-filling. It pinpoints the marginal Fermi liquid as the parent metal of the insulating Mott liquid state, and highlights the existence of a pseudogap phase as the pathway from the metallic to the insulating phase. The marginal Fermi liquid is found to arise from singular forward scattering in directions normal to the Fermi surface, and causes the destruction of Landau quasiparticles  leading to a linear variation of resistivity with temperature. The pseudogap itself arises from the electronic differentiation encoded within the nested Fermi surface of the half-filled tight-binding model, and involves the gradual gapping of the Fermi surface (from antinodes to nodes) via charge and spin excitations that are mutually entangled. The resultant Mott liquid is
observed to possess topological order with a two-fold degeneracy of the ground state on the torus, and fractionally-charged topological excitations that interpolate between them. Upon including the possibility of symmetry breaking within the RG, the Mott liquid is found to turn into the familiar Neel spin-ordered charge insulating Mott insulator. Interestingly, the Mott liquid state at half-filling is also observed to possess subdominant Cooper pairing, a feature that becomes dominant with hole doping. 
\par
The collapse of the pseudogap for charge excitations (Mottness) upon hole doping is seen to lead to a QCP lying between Mott liquid and correlated Fermi liquid phases. The QCP involves a drastic change in the nature of the ground state as well as the many-body spectrum: the underdoped side of the QCP is a gapped and hole-doped Mott liquid, while the overdoped side involves the appearance of electronic quasiparticles that gradually lead to the topological reconstruction of a fully connected Fermi surface. Precisely at the QCP, we find the existence of nodal marginal Fermi liquid with gapless 2e-1h composite excitations. The pseudogapped state at underdoping is formed from a dominant charge gapping of the antinodes, as well as possesses strong spin-nematic fluctuations and sub-dominant superconducting phase fluctuations. The underdoped Mott liquid leads, upon including the effects of symmetry breaking, to Neel SDW and chequerboard CDW orders. 
\par
In the v-shaped quantum critical region within phase diagram Fig.~\ref{Phase_diagram_with_doping-1}, we find spin-gapped antinodal regions of the Fermi surface containing pre-formed Cooper pairs co-existing with gapless stretches of marginal Fermi liquid on arcs centered about the nodes. This appears to confirm dominant spin excitations emerging from the collapse of Mottness as the mechanism for the formation of hole pairs. Upon including the possibility of phase stiffness within the RG, we find the existence of a dome of d-wave superconductivity that surrounds the QCP. Remarkably,  
even as the superconducting phase shields the QCP, it possesses properties of that criticality (e.g., gapless nodes, gap with d-wave symmetry). In this way, we find that the pseudogap is both friend and foe to the emergent superconductivity: the underdoped Mottness-related pseudogap contains fluctuations that can nucleate various orders (SDW, CDW and spin-nematicity) inimical to superconducting order, its collapse unveils the spin-pseudogapped state of matter that finally leads to a state with pre-formed Cooper pairs (and eventually phase stiffness and the ODLRO pertaining to superconductivity). 
\par
A striking feature of our results is the qualitative agreement of RG phase diagram, Fig.\ref{phasediagwithsymmbreak}, with the experimentally obtained temperature versus doping phase diagram for the cuprates~\cite{keimer2015quantum}. We believe that this arises from the fact that the RG unveils an entire heirarchy of energy scales on the quantum fluctuation axis $\omega$ related to the metallic, pseudogap, Mott liquid and symmetry broken phases. Further, at various points in this work, we have shown analytic relations between these $T=0$ energy scales and equivalent temperature scales at which these phases can be observed. Importantly, we have also established that the heirarchy of temperature scales for the pseudogap ($T_{PG}$), onset temperature for pairing ($T_{ons}$), formation of the Mott liquid ($T_{ML}$) and superconductivity ($T_{C}$) obtained from our analysis is quantitatively consistent with that observed experimentally for some members of the cuprates.
\par
The effective Hamiltonians and low-energy wavefunctions obtained for the fixed points of the RG formalism has afforded considerable insight into the nature of the Mott liquid at half-filling, as well as with hole doping. This is evidenced by the remarkable consistency between the numbers obtained for the ground state energy per site and double occupancy fraction with those obtained from various numerical methods in Ref.\cite{leblanc2015solutions,ehlers2017hybridDMRG,dagotto1992}. This benchmarking gives us confidence in the nature of the Mott liquid state as well as in the quantum phase transition that it undergoes upon doping. The effective Hamiltonians have also enabled an understanding of the essence of various universal features of the large body of experimental results obtained for the cuprates, e.g., Homes law, Planckian dissipation and the T-linear resistivity of the normal state, the mixed nature of the optical conductivity at overdoping, optimality with doping and the dome-like structure of the superconducting phase etc. In seeking further comparisons with the extensive body of experimental data available for the cuprates, it appears plausible to carry out a numerical simulation of these effective Hamiltonians at finite temperature. We leave this for a future work.
\par
Even as these results offer considerable evidence that the strong correlation physics of the one-band Hubbard model at, and away from, 1/2-filling is pertinent to the physics of high-temperature superconductivity~\cite{zhangrice-PhysRevB.37.3759}, they also open several new directions for further investigation. Foremost among these lies the search for answers to questions on what makes certain members of the cuprate family special in the search for higher superconducting $T_{C}$, as well as what could enable a raise in $T_{C}$. In reaching some conclusions on the former, the results obtained from recent DFT$+$downfolding study of Hirayama et al.~\cite{hirayama2018ab} on La$_{2}$CuO$_{4}$ and HgBa$_{2}$CuO$_{4}$ offer some insight. This study appears to conclude that the latter member of the cuprate family has a higher $T_{C}$ than the former as it is better described by an effective one-band Hubbard model in two dimensions, whereas the former is likely to have a larger hybridisation with a second dispersive band (arising from the Cu $3d_{3z^{2}-r^{2}}$ orbital) near the putative Fermi surface. It appears quite plausible that such hybridisation will be harmful to the physics of the Mott liquid and resulting emergent superconductivity we find from our studies, and can be studied in future. This suggests that materials that afford the isolation of Cu-O planes described by an effective 2D one-band Hubbard model are more likely to offer a higher superconducting $T_{C}$.
\par
In seeking answers to the question on how to further raise the superconducting $T_{C}$, it is important to make an increasingly realistic model pertinent to the cuprates. Thus, one should first investigate the role played by next-nearest neighbour hopping within the Cu-O plane. Our results predict, for instance, that shortening the extent of the pseudogap will likely enhance the optimal quantum fluctuation energy scale for the onset of d-wave superconducting order. This can be achieved, for instance, by tuning the curvature of the Fermi surface via next-nearest neighbour hopping~\cite{pavarini-PhysRevLett.87.047003}. 
Similarly, a study of the effects of an inter-plane electron hopping element appears relevant, as the presence of other Cu-O planes is observed in some members of the cuprates (e.g., Bi$_{2}$Sr$_{2}$Ca$_{2}$Cu$_{3}$O$_{10}$ (Bi2223)~\cite{chen2010enhancement}, HgBa$_{2}$Ca$_{2}$Cu$_{3}$O$_{9}$ (Hg1223)~\cite{gao1994superconductivity} and YBa$_{2}$Cu$_{3}$O$_{7}$ (Y123)~\cite{muramatsu2011possible} as being important in leading to higher $T_{C}$ upon the application of pressure.
Our findings are also likely to be pertinent to the ubiquitous presence of superconductivity in several other forms of strongly correlated quantum matter, e.g., the heavy-fermion systems, where there exist proposals for how the collapse of Mottness can lead to superconductivity~\cite{pepin-PhysRevB.77.245129}. Finally, given that we now have wavefunctions available for the ground and low-lying excited states of the doped Mott liquid, a direction worth pursuing is to understand the nature of many-body entanglement in this system. Given the effort presently invested in answering such questions, any progress in this direction will likely help usher new ways in which to think about the many-particle physics of strongly correlated electronic systems.
\vspace*{0.5cm}
\par \noindent {\bf Acknowledgments}\\
The authors thank Sourav Pal, Apoorva Patel, H.R.Krishnamurthy, T.V. Ramakrishnan, S. Mukherjee, T. Das, A. Garg, R. K. Singh, A. Dasgupta, A. Taraphder, S. Sinha, M. S. Laad, G. Baskaran, A. M. Srivastava, N. S. Vidhyadhiraja, S. Kumar, B. Bansal, S. Raj, P. Majumdar, S. Pal, S. Patra and M. Patra for several discussions and feedback. A. M. thanks the CSIR, Govt. of India for funding through a junior and senior research fellowship. S. L. thanks the DST, Govt. of India for funding through a Ramanujan Fellowship during which a part of this work was carried out. 
\appendix
\section{The Hamiltonian renormalization group flow}\label{derv_RG}
Starting from the Hamiltonian RG flow eq.\eqref{isogeometric_curve_RG_flow},
and using the expressions for the unitary operator eqs.\eqref{Unitary_operator_prod_decoupling_curve} and\eqref{unitary_op_def}, we obtain
\begin{eqnarray}
\hspace*{-0.6cm}\Delta H_{(j)} &=& \sum_{l}Tr_{j,l}(c^{\dagger}_{j,l}H_{(j)})c_{j,l}G_{(j),l}c^{\dagger}_{j,l}Tr_{j,l}(H_{(j)}c_{j,l})~,
\label{renormalization of Hamiltonian}
\end{eqnarray}
where $G_{(j),l}= (\omega - Tr_{j}(H^{D}_{(j)}\hat{n}_{j,l}))^{-1}$ and the diagonal Hamiltonian $H^{D}_{(j)}$ is given by
\begin{eqnarray}
Tr_{j, l}(H^{D}_{(j)}\hat{n}_{j, l}) &=& 
\sum_{l}\epsilon_{j, l}\hat{n}_{j, l}\nonumber\\
&+&\sum_{l}V^{(j)}_{l}(\delta)\bigg(\hat{n}_{j, l}\hat{n}_{j, l'}-\hat{n}_{j, l}(1-\hat{n}_{j, l'})\nonumber\\
&+&(1-\hat{n}_{j, l})(1-\hat{n}_{j, l'})-\hat{n}_{j, l'}(1-\hat{n}_{j, l})\bigg)\nonumber\\
&+&\sum_{l, l''}L^{(j)}(\delta)\hat{n}_{j, l}\hat{n}_{j', l''}\nonumber\\
&+&\sum_{l, l''}R^{(j)}_{\delta\delta}\hat{n}_{j, l}\hat{n}_{j, l'}(1-\hat{n}_{j, l''})+\ldots ~.
\end{eqnarray}
The various terms in $H^{D}_{(j)}$ are: the first term is the electronic dispersion ($\epsilon_{j, l}$), the term with coupling $V_{l}^{(j)}(\delta)$ is the longitudinal density-density interaction in the (ee-hh) and (eh-he) channels, the term with coupling $L^{(j)}(\delta)$ is the tangential density-density interaction and, finally, the term with coupling $R^{(j)}_{\delta\delta}$ is the 2e-1h interaction. 
The renormalization $\Delta H^{F}_{(j)}$ for the longitudinal forward-scattering terms in ee, hh, eh and he channels are obtained from eq.\eqref{renormalization of Hamiltonian} as
\begin{widetext}
\begin{eqnarray}
\Delta H^{F}_{(j)} &=& \sum_{k, k', l'}
\bigg[\underbrace{c^{\dagger}_{k', l}c^{\dagger}_{k', l'}c_{j, l'}c_{j, l}\frac{(V^{(j)}_{l}(\delta))^{2}}{[G_{j, l}]^{-1}-V^{(j)}_{l}(\delta)\tau_{j, l}\tau_{j, l'}}c^{\dagger}_{j, l}c^{\dagger}_{j, l'}c_{k, l'}c_{k, l}}_{\text{ee scattering channel}}+\underbrace{c^{\dagger}_{j, l}c^{\dagger}_{j, l'}c_{k', l'}c_{k', l}\frac{(V^{(j)}_{l}(\delta))^{2}}{[G_{j, l}]^{-1}-V^{(j)}_{l}(\delta)\tau_{j, l}\tau_{j, l'}}c^{\dagger}_{k, l}c^{\dagger}_{k, l'}c_{j, l'}c_{j, l}}_{\text{hh scattering channel}}\nonumber\\
&+&\underbrace{c^{\dagger}_{k', l}c_{\bar{k}', l'}c^{\dagger}_{j, l'}c_{j, l}\frac{(V^{(j)}_{l}(\delta))^{2}}{[G_{j, l}]^{-1}-V^{(j)}_{l}(\delta)\tau_{j, l}\tau_{j, l'}}c^{\dagger}_{k', l}c_{\bar{k}', l'}c^{\dagger}_{j, l'}c_{j, l}}_{\text{eh scattering channel}}+\underbrace{c^{\dagger}_{k', l}c_{\bar{k}', l'}c^{\dagger}_{j, l}c_{j, l'}\frac{(V^{(j)}_{l}(\delta))^{2}}{[G_{j, l}]^{-1}-V^{(j)}_{l}(\delta)\tau_{j, l}\tau_{j, l'}}c^{\dagger}_{j, l'}c_{j, l}c^{\dagger}_{k', l}c_{\bar{k}', l'}}_{\text{eh scattering channel}}
\bigg]\nonumber\\
&=& \sum_{k, k', l'}c^{\dagger}_{k', l}c^{\dagger}_{k', l'}\frac{4(V^{(j)}_{l}(\delta))^{2}\tau_{j, l}\tau_{j, l'}}{[G_{j, l}]^{-1}-V^{(j)}_{l}(\delta)\tau_{j, l}\tau_{j, l'}}c_{k, l'}c_{k, l}~.
\end{eqnarray}
\end{widetext}
Here, $(G_{j, l})^{-1}=\hat{\omega} - \epsilon_{j, l}\tau_{j, l}-\epsilon_{j, l'}\tau_{j, l'}$ is the inverse Greens function operator.
Similarly, the renormalization $\Delta H^{B}_{(j)}$ for backscattering terms is given by
\begin{equation}
\Delta H^{B}_{(j)} = \sum_{k, k', l'}c^{\dagger}_{k', p}c^{\dagger}_{k', p'}\frac{4V^{(j)}_{l}(\delta)K^{(j)}_{l}(\delta)\tau_{j, l}\tau_{j, l'}}{[G_{j, l}]^{-1}-V^{(j)}_{l}(\delta)\tau_{j, l}\tau_{j, l'}}c_{k, l'}c_{k, l}~,
\end{equation}
where the same Greens function operator has been used.
The renormalization of the tangential scattering terms in the Hamiltonian can be similarly obtained by decoupling a collective configuration of states on the jth isogeometric curve
\begin{eqnarray}
\Delta H^{T}_{(j)} = \sum_{kk'}c^{\dagger}_{k, m}c^{\dagger}_{k, m'}\frac{(L^{(j)})^{2}L_{j}^{+}L_{j}^{-}}{\hat{\omega}-\tilde{\epsilon}^{c}_{j, avg}L^{z}_{j}-L^{(j)}L^{z2}_{j}}c_{k', n'}c_{k', n}.\nonumber\\
\end{eqnarray}
Here, $L^{+}_{j} = \sum_{m}c^{\dagger}_{j, m}c^{\dagger}_{j, m'}$, $L^{z}_{j} = 2^{-1}\sum_{m}(\hat{n}_{j, m}+\hat{n}_{j, m'}-1)$ and $L^{-}_{j}$ is the hermitian conjugate to $L^{+}_{j}$. The intermediate configurations of the states involved in the tangential scattering processes are labeled by \[(j, m)=\mathbf{k}_{\Lambda_{j}, \hat{s}}, \sigma), (j, m')=\mathbf{k}_{-\Lambda_{j}, T\hat{s}}, -\sigma), \] \[(j, n)=\mathbf{k}_{\Lambda_{j}, \hat{s}'}, \sigma), (j, n')=\mathbf{k}_{-\Lambda_{j}, T\hat{s}'}, -\sigma).\] Using the angular momentum algebra $L_{j}^{+}L_{j}^{-} = L^{2}_{j}-L^{z2}_{j}-L^{z}$, we obtain
\begin{eqnarray}
\Delta H^{T}_{(j)} = \sum_{kk', m, n}c^{\dagger}_{k, m}c^{\dagger}_{k, m'}\frac{(L^{(j)})^{2}(L^{2}_{j}-L^{z2}_{j}-L^{z}_{j})}{\hat{\omega}-\tilde{\epsilon}^{c}_{j, avg}L^{z}_{j}-L^{(j)}L^{z2}_{j}}c_{k', n'}c_{k', n}~.~~~~~~\nonumber\\
\end{eqnarray}
\begin{widetext}
The Hamiltonian RG for the three particle scattering vertices terms can also be obtained as 
\begin{eqnarray}
\Delta H^{3}_{(j)} &=& \sum_{k'', k', l', l''}c^{\dagger}_{k', l}c^{\dagger}_{k', l'}c_{j, l'}\frac{V^{(j)}_{l}(\delta)V^{(j)}_{l}(\delta')\tau_{j, l}}{\omega - \tilde{\epsilon}_{j, l}\tau_{j, l}}c^{\dagger}_{j, l''}c_{k'', l}c_{k'', l'}\nonumber\\
&+&\sum_{k'', p', l', l''}c^{\dagger}_{p', l}c^{\dagger}_{p', l'}c_{j, l'}\frac{K^{(j)}_{l}(\delta)K^{(j)}_{l}(\delta')\tau_{j, l}}{\omega - \tilde{\epsilon}_{j, l}\tau_{j, l}}c^{\dagger}_{j, l''}c_{k'', l}c_{k'', l'}\nonumber\\
 &+&\sum_{\Lambda'<\Lambda_{j}, p', k''}c^{\dagger}_{p', l}c^{\dagger}_{p', l'}c_{j', l'}\frac{8R^{(j)}_{l, \delta\delta''}R^{(j)}_{l, \delta'\delta''}\prod_{i=1}^{3}\tau_{i}}{[G_{j, l, 3}]^{-1}-R^{(j)}_{l, \delta'', \delta''}\prod_{i=1}^{3}\tau_{i}}c^{\dagger}_{j', l''}c_{k'', l}c_{k'', l'}~,\label{Three-particle-vertex-derv-RG}
\end{eqnarray}
\end{widetext}
where the states are labeled by $i=1:(\mathbf{k}_{\Lambda_{j}, \hat{s}}, \sigma)~,~ 2:(\mathbf{k}_{-\Lambda_{j}+\delta'', T\hat{s}}, -\sigma)~,~ 3:(\mathbf{k}_{\Lambda'\hat{s}}, \sigma)$~.
\section{The renormalized Hamiltonian at RG step $j$}
\label{renm_H}
The Hamiltonian at RG step $j$ using flow equations eq.\eqref{Long RG equations with doping}, eq.\eqref{tang_scatt_RG_flows} and eq.\eqref{six_point_RG_flows} is given by
\begin{widetext}
\begin{eqnarray}
H_{(j)}(\omega) &=& \sum_{\Lambda <\Lambda_{j}}(\epsilon_{\Lambda\hat{s}}-\Delta\mu_{eff})\left(\hat{n}_{\Lambda, \hat{s}, \sigma}-\frac{1}{2}\right)+\sum_{\delta, \hat{s}\neq \pm \hat{s}'}\underbrace{L^{(j)(\omega)}c^{\dagger}_{\Lambda, \hat{s}, \sigma}c^{\dagger}_{-\Lambda, T\hat{s},-\sigma}c_{-\Lambda', T\hat{s}', -\sigma}c_{\Lambda', \hat{s}', \sigma}}_{\text{2e-charged opposite spin-pair tangential scattering}}\nonumber\\
&+&\sum_{\substack{\delta, \hat{s}, \\ \Lambda <\Lambda_{j}}}(\underbrace{V^{(j)}_{c, l}(\omega)c^{\dagger}_{\Lambda,\hat{s}, \sigma}c^{\dagger}_{-\Lambda+\delta , T\hat{s}, -\sigma}c_{-\Lambda'+\delta , T\hat{s}, -\sigma}c_{\Lambda', \hat{s}, \sigma}}_{\text{2e-charged opposite spin-pair forward scattering}}+\underbrace{K^{(j)}_{c, l}(\omega)c^{\dagger}_{\Lambda, \hat{s}, \sigma}c^{\dagger}_{-\Lambda+\delta, T\hat{s}, -\sigma}c_{-\Lambda'-\delta, -T\hat{s}, -\sigma}c_{\Lambda', -\hat{s},\sigma}}_{\text{2e-charged opposite spin-pair Umklapp scattering}})
\nonumber\\
&-&\sum_{\substack{\delta, \hat{s},\\ \Lambda<\Lambda_{j}}}(\underbrace{V^{(j)}_{s, l}(\omega)c^{\dagger}_{\Lambda,\hat{s}, \sigma}c_{-\Lambda +\delta , T\hat{s}, -\sigma}c^{\dagger}_{\Lambda' , T\hat{s}, -\sigma}c_{2\Lambda+\Lambda'-\delta, \hat{s},\sigma}}_{\text{charge-neutral opposite spin-pair forward scattering}}+\underbrace{K^{(j)}_{s, l}(\omega)c^{\dagger}_{\Lambda, \hat{s}, \sigma}c_{-\Lambda +\delta , T\hat{s}, -\sigma}c^{\dagger}_{\Lambda', -\hat{s}, -\sigma}c_{\Lambda'+\delta-2\Lambda , -T\hat{s}, \sigma}}_{\text{charge-neutral opposite spin-pair backscattering}})\nonumber\\
&+&\sum_{\substack{\delta,\hat{s},\\ \Lambda,\Lambda',\Lambda''<\Lambda_{j}}}\underbrace{R^{(j)}_{l, \delta\delta'}(\omega)c^{\dagger}_{\Lambda , \hat{s},\sigma}c^{\dagger}_{-\Lambda +\delta, T\hat{s}, -\sigma}c_{-\Lambda'+\delta, \hat{s}, -\sigma}c^{\dagger}_{-\Lambda'+\delta', \hat{s}, \sigma}c_{-\Lambda''+\delta', T\hat{s}, -\sigma}c_{\Lambda'', \hat{s}, \sigma}}_{\text{three-particle forward scattering}}~.
\end{eqnarray}
\end{widetext}
\section{Algorithm for numerical simulations of the RG equations}\label{AlgoSimRG}
On a given momentum space XY lattice of size $N\times N$, we define the dispersion $E[kx,ky]=-2t(\cos kx+\cos ky)$ at every point ($kx,ky$). The momentum space lattice is a pair of 2D arrays. 
Then, the Fermi surface contour (i.e., the collection of $k_{Fx}$, $k_{Fy}$ values) is collected from an array $F:E[kx,ky]=0$. 
From the collection of $kFx$ and $kFy$ coordinates, we can compute the velocity vector at every point, $\mathbf{v}=(\frac{\partial E[kx,ky]}{\partial kx}|_{F},\frac{\partial E[kx,ky]}{\partial ky}|_{F})$. This then allows us to obtain the isogeometric coordinate normals $\hat{s}$'s at every point on the FS, i.e., the set $\lbrace\hat{s}_{i}=s_{x}\hat{x}+s_{y}\hat{y}, s_{x}=v_{Fx}/|\mathbf{v}|, s_{y}=v_{Fy}/|\mathbf{v}|,\hat{s}_{i}\in F\rbrace$. We take equally spaced normal vectors $N$s from the AN ($kF=(\pi,0)$) to the N ($kF=(\pi/2,\pi/2)$).
\par\noindent
The starting value of $\Lambda_{0}$, i.e., the outward distance from the Fermi surface is then determined from $N_{s}$ as follows. The bare $\Lambda_{0}$ is such that the state along $\delta\hat{s}$, i.e., the second normal $\hat{s}_{2}||(\pi,0)$, falls within the first Brillouin zone
\begin{eqnarray}
\mathbf{k}_{F\hat{s}_{AN}} &=& (k_{Fx}=\pi, k_{Fy}=0)\nonumber\\
\mathbf{k}_{F\hat{s}_{2}}&=&(k_{Fx}=\pi-\frac{\pi}{Ns}, k_{Fy}=\frac{\pi}{Ns})\nonumber\\
k_{Fx}+\frac{1}{\sqrt{2}}\Lambda\leq \pi &&\implies \Lambda_{0}=\frac{\sqrt{2}\pi}{Ns}~.
\end{eqnarray}
A numpy meshgrid $(\Lambda,\hat{s})$ is constructed from the given $\Lambda_{0}$ to $\Lambda=0$ for all the normal directions, and the RG equations are then solved numerically on this meshgrid. 
\begin{widetext}
\section{RG equation of the symmetry breaking orders}\label{symmbreakAppendix}
In the doped Mott liquid Hamiltonian (eq.\eqref{dopedMlQCP}, we include $(\pi,\pi)$ charge density wave, $(\pi,\pi)$ spin density wave symmetry-breaking fields together with the Hamiltonian for the 1-e and 2e-1h composites (eq.\eqref{decoupled_states_Hamiltonian_2nd_level})
\begin{eqnarray}
\hat{H}&=& \sum_{\hat{s}, \Lambda}\Delta A^{x}_{\Lambda,\hat{s}}+\sum_{\hat{s}, \Lambda}\Delta S^{x}_{\Lambda,\hat{s}}+\sum_{\hat{s},\Lambda}\epsilon_{\Lambda\hat{s}}-\Delta\mu_{eff})\left(\hat{n}_{\Lambda, \hat{s}, \sigma}-\frac{1}{2}\right)+\sum_{\hat{s},\Lambda,\delta}R^{(j)}_{l\delta}\hat{n}_{k, l}\hat{n}_{-k, l'}(1-\hat{n}_{k', l'})\nonumber\\
&+&\sum_{\hat{s}}\bar{U}_{0}[\mathbf{A}_{*,\hat{s}}\cdot\mathbf{A}_{*,-\hat{s}}-\mathbf{S}_{*,\hat{s}}\cdot\mathbf{S}_{*,-\hat{s}}]+\Delta\mu_{eff}^{*}\sum_{\hat{s}}B^{z}_{*,\hat{s}}+(\Delta\mu_{eff}-\Delta\mu_{eff}^{*})\sum_{\hat{s}}A^{z}_{*,\hat{s}}~,\label{dopedMLwithSymBreak}
\end{eqnarray}
where $A^{x}_{\Lambda,\hat{s}}$ and $S^{x}_{\Lambda,\hat{s}}$ are the $x$-components of the charge and spin pseudospins respectively given by eq\eqref{pseudospins}. We now perform a second level of the RG calculation for the symmetry-breaking instabilities by block- diagonalizing in the eigenbasis of the single-particle piece of the Hamiltonian eq.\eqref{dopedMLwithSymBreak}
\begin{eqnarray}
\hat{H}_{1} = \sum_{\hat{s},\Lambda}\epsilon_{\Lambda,\hat{s}}-\Delta\mu_{eff}+\Delta\mu_{eff}^{*}+\Sigma(\hat{s}))\left(\hat{n}_{\Lambda, \hat{s}, \sigma}-\frac{1}{2}\right)+\Delta_{c,\Lambda}A^{x}_{*}+\Delta_{s}S^{x}_{*}\Rightarrow \hat{H}_{1} = \sum_{\hat{s},\Lambda}\tilde{\epsilon}_{c,\Lambda\hat{s}}\tilde{A}^{z}_{\Lambda,\hat{s}}+\tilde{\epsilon}_{s,\Lambda\hat{s}}\tilde{S}^{z}_{\Lambda,\hat{s}}~.\label{rotated basis}
\end{eqnarray}
Here, the self energy $\Sigma(\hat{s})$ incorporates the effects of the 2e-1h dispersion (eq.\eqref{1-p self energy}). With this, the renormalized charge/spin pseudospin disperion is given by $\tilde{\epsilon}_{c/s,\Lambda,\hat{s}}=\sqrt{(\epsilon_{\Lambda,\hat{s}}\pm\epsilon_{-\Lambda,T\hat{s}}+\Delta\mu_{eFF}-\Delta\mu_{eff}^{*}+\Sigma(\hat{s}))^{2}+\Delta^{2}}$. 
\par\noindent
The RG equations for the CDW and SDW instabilities are given by
\begin{eqnarray}
\frac{\Delta K_{c/s,\hat{s}}}{\Delta\log\Lambda} &=& \frac{K_{c/s,\hat{s}}^{2}}{\omega-\tilde{\epsilon}_{c/s,\Lambda,\hat{s}}-\frac{1}{4}K_{c/s,\hat{s}}}~.
\end{eqnarray}
The fixed point values of the SDW and CDW gaps can be obtained from the fixed points reached by solving the RG equations
\begin{eqnarray}
\omega-\tilde{\epsilon}_{c/s,\Lambda^{*},\hat{s}}-\frac{1}{4}K^{*}_{c/s,\hat{s}}=0\to \Delta_{c}^{*} = \frac{1}{\sum_{\hat{s}}1}\sum_{\hat{s}}K_{c,\hat{s}}^{*}\langle A^{x}_{*,\hat{s}}\rangle, \Delta_{s}^{*} = \frac{1}{\sum_{\hat{s}}1}\sum_{\hat{s}}K^{*}_{s,\hat{s}}\langle S^{x}_{*,\hat{s}}\rangle~.  
\end{eqnarray}
Finally, in order to achieve the phase stiffness and ODLRO associated with superconductivity, we add a $U(1)$ symmetry-breaking field to the Hamiltonian eq.\eqref{effHamforQCPregion} 
\begin{eqnarray}
H_{SB} &=& 
\sum_{\Lambda,\hat{s}}\Delta_{sc}B_{\Lambda,\hat{s}}^{x}~,
\end{eqnarray}
where $B_{\Lambda,\hat{s}}^{+}=c^{\dagger}_{\mathbf{k}_{\Lambda\hat{s}},\sigma}c^{\dagger}_{-\mathbf{k}_{\Lambda\hat{s}},-\sigma}$ is defined in eq\eqref{anderson_pseudo}. 
By performing the block-diagonalization in the rotated single-particle basis (i.e., similar to eq.\eqref{rotated basis}) leads to a modified RG equation (eq.\eqref{spin_cooper_RG}) for the superconducting fluctuation terms in eq.\eqref{effHamforQCPregion}
\begin{eqnarray}
\Delta K^{(j)}_{s, 0, \hat{s}}(\delta) &=& \frac{-(1-p)(K^{(j)}_{s, 0, \hat{s}}(\delta))^{2}}{\omega - \frac{p}{2}\sqrt{\Delta_{sc}^{2}+(\epsilon_{\Lambda,\hat{s}}+\epsilon_{-\Lambda, T\hat{s}}+|\Delta\mu_{eff}-\Delta\mu_{eff}^{*}|+2\Sigma(\hat{s}))^{2}}-\frac{(1-p)}{2}(\epsilon_{\Lambda\hat{s}}-\epsilon_{-\Lambda T\hat{s}})-\frac{K_{p, 0, \hat{s}}^{(j)}(\delta)}{4}}\nonumber\\
&+&\frac{(K^{(j)}_{s, 0, \hat{s}}(\delta))^{2}}{\omega -\frac{1}{2}\sqrt{\Delta_{sc}^{2}+(\epsilon_{\mathbf{k}_{\Lambda\hat{s}}}+\epsilon_{-\mathbf{k}_{\Lambda \hat{s}}}+|\Delta\mu_{eff}-\Delta\mu^{*}_{eff}|+2\Sigma(\hat{s}))^{2}}-\frac{K^{(j)}_{p,0,\hat{s}}(\delta)}{4}}\nonumber\\
&-& \frac{N_{j}^{2}(L^{(j)}_{\delta})^{2}}{\omega +sgn(\Delta\mu_{eff})W+(\epsilon^{s}_{j, avg}-\Delta\mu_{eff})-L^{(j)}(\delta)}~.
\end{eqnarray}
The RG equation at critical doping $\Delta\mu_{eff}=\Delta\mu_{eff}^{*}$ is given by 
\begin{eqnarray}
\Delta K^{(j)}_{s, 0, \hat{s}}(\delta) &=& \frac{(K^{(j)}_{s, 0, \hat{s}}(\delta))^{2}}{\omega -\frac{1}{2}\sqrt{\Delta_{sc}^{2}+(\epsilon_{\mathbf{k}_{\Lambda\hat{s}}}+\epsilon_{-\mathbf{k}_{\Lambda\hat{s}}}+2\Sigma(\hat{s}))^{2}}-\frac{(1-p)K^{(j)}_{p,0,\hat{s}}(\delta)}{4}}-\frac{(K^{(j)}_{s, 0, \hat{s}}(\delta))^{2}}{\omega -\frac{1-p}{2}(\epsilon_{\Lambda\hat{s}}-\epsilon_{-\Lambda T\hat{s}})-\frac{p}{2}(\epsilon_{\Lambda\hat{s}}+\epsilon_{-\Lambda T\hat{s}})-\frac{K_{p, 0, \hat{s}}^{(j)}(\delta)}{4}}~.
\end{eqnarray}
\par\noindent
In what follows, we show how this RG equation unveils the d-wave symmetry of the SC order. For $\omega<\omega^{*} = 2^{-1}\max\limits_{\hat{s}}(\epsilon_{\Lambda,\hat{s}}+\epsilon_{\Lambda,-T\hat{s}})=2^{-1}(\epsilon_{\Lambda,\hat{s}_{N}}+\epsilon_{\Lambda,-T\hat{s}_{N}})=4\sin(\Lambda_{0}/\sqrt{2})$ (the highest kinetic energy of pairwise states present along the nodes at a distance $\Lambda_{0}$) in an momentum-space arc centered about the nodes, we find that the RG equation for superconducting fluctuations is RG irrelevant:~ 
$\Delta K^{(j)}_{s,0,\hat{s}_{N}}<0$~, as $\omega-\frac{1}{2}\sqrt{\Delta_{sc}^{2}+(\epsilon_{\mathbf{k}_{\Lambda\hat{s}}}+\epsilon_{-\mathbf{k}_{\Lambda\hat{s}}}+\Sigma(\hat{s}))^{2}}<0$. On the other hand, the RG equation for superconducting fluctuations for all the other normal directions to the FS are RG relevant. 
\par\noindent
Beyond $\omega\geq\omega^{*}$, a gapless nodal stretch extends on both sides of the FS from $\hat{s}_{N}$ to 
$\hat{s}^{*}\to\min\limits_{\hat{s}^{*}}(\omega^{*}-\frac{1}{2}\sqrt{\Delta^{2}_{sc}+(\epsilon_{\mathbf{k}_{\Lambda\hat{s}^{*}}}+\epsilon_{-\mathbf{k}_{\Lambda\hat{s}^{*}}}+\Sigma(\hat{s}))^{2}}>0$. The fixed point Hamiltonian for $\Delta\mu_{eff}=\Delta\mu_{eff}^{*}$ and for $2^{-1}W>\omega\geq \omega^{*}$ (with the initial $\Delta_{sc} \to 0$) is given by
\begin{eqnarray}
H_{sc}^{*} &=& \sum_{\Lambda,\hat{s}=\hat{s}_{N]}}^{\hat{s}^{*}}\epsilon_{\Lambda,\hat{s}}B^{z} +\sum_{\hat{s}_{N},\Lambda,\delta}^{\hat{s}^{*}}R^{(j)}_{l\delta}\hat{n}_{k, l}\hat{n}_{-k, l'}(1-\hat{n}_{k', l'})+\sum^{\Lambda_{0},\hat{s}^{*}}_{\Lambda>\Lambda^{*}(\hat{s}),\hat{s}=\hat{s}_{AN}}\sqrt{(\epsilon_{\Lambda,\hat{s}}+\Sigma(\hat{s}))^{2}+(\Delta^{*}_{\hat{s}}(\omega))^{2}}\tilde{B}^{z}_{\Lambda,\hat{s}}\nonumber\\
&+&\sum^{\Lambda^{*}(\hat{s}),\hat{s}^{*}}_{\Lambda = 0,\hat{s}=\hat{s}_{AN}}\epsilon_{\Lambda,\hat{s}}\tilde{B}^{z}_{\Lambda,\hat{s}}+\frac{1}{2}\sum_{\Lambda , \hat{s}=\hat{s}_{AN}}^{\hat{s}^{*}}(\sum_{\delta }[K^{*}_{s,0,\hat{s}}(\delta)\langle B^{-}_{-2\Lambda^{*}+\Lambda+\delta, T\hat{s}}\rangle)B^{+}_{\Lambda,\hat{s}}+h.c.]\nonumber\\
&+&\sum_{\Lambda , \hat{s}=\hat{s}_{AN}, \delta}^{\hat{s}^{*}}K^{*}_{s,0,\hat{s}}(\delta)(B^{-}_{-2\Lambda^{*}+\Lambda+\delta, T\hat{s}} - \langle B^{-}_{-2\Lambda^{*}+\Lambda+\delta, T\hat{s}}\rangle)(B^{+}_{\Lambda , T\hat{s}} - \langle B^{+}_{\Lambda, T\hat{s}}\rangle)~,
\label{U(1) symmetry broken theory}
\end{eqnarray}  
where $\Delta^{*}_{\hat{s}}(\omega)=(\sum_{\delta }K^{*}_{s,0,\hat{s}}(\delta)\langle B^{-}_{-2\Lambda^{*}+\Lambda+\delta, T\hat{s}}\rangle)$ and $K^{*}_{s,0,\hat{s}}(\delta)=4(\omega -\frac{1}{2}(\epsilon_{\mathbf{k}_{\Lambda\hat{s}}}+\epsilon_{-\mathbf{k}_{\Lambda\hat{s}}}+\Sigma_{\Lambda,\hat{s}}))$. 
Here ,the first and second terms are the 1e and 2e-1h dispersions of the gapless nodal stretches respectively. The third and fourth terms denote the dispersion of the antinodal stretch residing outside and inside the emergent window respectively. The single-particle states residing outside the emergent window, i.e., $\Lambda>\Lambda^{*}_{\hat{s}}$, corresponds to gapped marginal Fermi liquid quasiparticles with self energy given by eq.\eqref{1-p self energy}. Finally, the fifth and sixth terms describe the effective $U(1)$ symmetry-breaking mean-field and the (emergent Nambu-Goldstone) superconducting fluctuations respectively. 
\par\noindent
In this way, we obtain the self consistency equations for the superconducting order parameter (with d-wave symmetry) from the eigenstates of the symmetry-broken Hamiltonian lying within the emergent window 
\begin{eqnarray}
\langle c^{\dagger}_{\Lambda -\hat{s}\downarrow}c^{\dagger}_{\Lambda\hat{s}\uparrow}\rangle &=& \frac{\Delta^{*}_{\hat{s}}(\omega)}{\sqrt{\Delta^{*2}_{\hat{s}}(\omega)+(\epsilon_{\Lambda,\hat{s}}+\Sigma_{\Lambda,\hat{s}})^{2}}}~,~\nonumber\\
\langle c^{\dagger}_{\Lambda ,-T\hat{s}\downarrow}c^{\dagger}_{\Lambda ,T\hat{s}\uparrow}\rangle &=& \frac{\Delta^{*}_{T\hat{s}}(\omega)}{\sqrt{\Delta^{*2}_{\hat{s}}(\omega)+(\epsilon_{\Lambda,\hat{s}}+\Sigma_{\Lambda,\hat{s}})^{2}}}~,\nonumber\\
\langle c^{\dagger}_{\Lambda -\hat{s}_{N}\downarrow}c^{\dagger}_{\Lambda T\hat{s}_{N}\uparrow}\rangle &=& 0~.~\label{self-consistency}
\end{eqnarray}
For minimizing the energy, we require $\langle c^{\dagger}_{\Lambda -\hat{s}\downarrow}c^{\dagger}_{\Lambda\hat{s}\uparrow}\rangle = -\langle c^{\dagger}_{\Lambda -T\hat{s}\downarrow}c^{\dagger}_{\Lambda T\hat{s}\uparrow}\rangle$, such that $K^{*}_{s,0,\hat{s}}(\delta)(\langle c^{\dagger}_{\Lambda -\hat{s}\downarrow}c^{\dagger}_{\Lambda\hat{s}\uparrow}\rangle\langle c^{\dagger}_{\Lambda -T\hat{s}\downarrow}c^{\dagger}_{\Lambda T\hat{s}\uparrow}\rangle)<0$. The vanishing of the nodal order parameter $\langle c^{\dagger}_{\Lambda -\hat{s}_{N}\downarrow}c^{\dagger}_{\Lambda T\hat{s}_{N}\uparrow}\rangle$, along with the above displayed change of sign in the order parameter upon crossing the nodes, manifests in the d-wave nature of the superconductivity.
\par\noindent
From the above analysis, we can now extract a temperature scale for the symmetry-broken d-wave superconductor using the relation between the quantum fluctuation scale $\omega$ and temperature (see Sec.\ref{meaning_of_omega}) and the effective 1e self-energy of the states residing outside the emergent window. For this, we first expand the gapped dispersion energy about the one-particle gap $\Delta^{*}_{\hat{s}}$ 
\begin{eqnarray}
\sqrt{(\epsilon_{\Lambda^{*},\hat{s}}+\Sigma(\hat{s}))^{2}+\Delta^{*2}_{\hat{s}}} =\Delta^{*}_{\hat{s}}\sqrt{1+\frac{(\epsilon_{\Lambda^{*},\hat{s}}+\Sigma_{\Lambda^{*},\hat{s}})^{2}}{\Delta^{*2}_{\hat{s}}}} \approx 
\Delta^{*}_{\hat{s}}+\frac{(\epsilon_{\Lambda^{*},\hat{s}}+\Sigma(\hat{s}))^{2}}{2\Delta^{*}_{\hat{s}}}\approx\Delta^{*}_{\hat{s}}+\frac{\epsilon_{\Lambda^{*},\hat{s}}\Sigma(\hat{s})}{\Delta^{*}_{\hat{s}}} ~.
\end{eqnarray}
We thus obtain the renormalized self-energy along the nodal direction by choosing $\omega = \omega^{*}$ and using eq.\eqref{forward_scattering_magnitude}, eq.\eqref{lowest_energy_end_of_spectrum}, eq.\eqref{fixed_point_three_particle} and eq.\eqref{1-p self energy} (along with $\epsilon_{\mathbf{k}_{\Lambda,\hat{s}}}=\epsilon_{-\mathbf{k}_{\Lambda,\hat{s}}}$) 
\begin{eqnarray}
\Sigma^{ren}(\hat{s}) = \Delta^{*}_{\hat{s}} &=& 4(\omega^{*}-\epsilon_{\Lambda^{*},\hat{s}}-\Sigma(\hat{s}))f(\hat{s},\omega^{*}) =
2\tilde{\omega}^{*}_{\hat{s}}\frac{f(\hat{s},\omega^{*})}{N^{*}(\hat{s},\omega^{*})}\ln\left\vert\frac{N^{*}(\hat{s},\omega^{*}))\bar{\omega}}{\tilde{\omega}^{*}_{\hat{s}}}\right\vert~,
\end{eqnarray}
where $f(\hat{s},\omega) =\sum_{\delta}\langle B^{-}_{-\Lambda^{*}+\delta,T\hat{s}}\rangle$ is the net spectral weight of the Cooper pairs along the normal $\hat{s}$ obtained from the self-consistency equations eq.\eqref{self-consistency}. In the above expression, we can determine $\tilde{\omega}^{*}_{\hat{s}}$ by following the steps leading to the derivation of eq.\eqref{frequency_onset_sc}
\begin{eqnarray}
\tilde{\omega}^{*}_{\hat{s}} = \frac{N^{*}(\hat{s}_{1},\omega^{*})}{2}(\epsilon_{\Lambda_{0},\hat{s}_{N}}-\epsilon_{\Lambda^{**},\hat{s}})~. 
\end{eqnarray}
Here, $\Lambda^{**}$ is determined from eq.\eqref{fixed_point_three_particle}.
Using the relation between the quantum fluctuation scale $\omega$ and an equivalent temperature scale (eq.\eqref{equivalent_Thermal scale}), we obtain the $T_{C}$ for the d-wave superconductivity as follows
\begin{eqnarray}
T_{c} = \frac{2\hbar}{k_{B}}\max_{\hat{s}}\tilde{\omega}_{\hat{s}}^{*}f(\hat{s},\omega^{*}) =\frac{2\hbar}{k_{B}}(\epsilon_{\Lambda_{0},\hat{s}_{N}}-\epsilon_{\Lambda^{**},\hat{s}_{1}})f(\hat{s}_{1},\omega^{*}) ~,
\end{eqnarray}
where $\hat{s}_{1}=\hat{s}_{AN}\left(1-\frac{\Lambda_{0}}{\sqrt{2}\pi}\right)$. The ratio of 
$T_{C}$ with onset scale of superconductivity $T_{ons}$ is then obtained as
\begin{eqnarray}
\frac{T_{c}}{T_{ons}} =2\frac{\epsilon_{\Lambda_{0},\hat{s}_{N}}-\epsilon_{\Lambda^{**}(\omega^{*}),\hat{s}_{1}}}{\epsilon_{\Lambda_{0},\hat{s}_{N}}-\epsilon_{\Lambda^{**}(0),\hat{s}_{1}}}\frac{f(\hat{s}_{1},\omega^{*})}{N^{*}(\hat{s}_{1},0)}~.\label{critOns}
\end{eqnarray} 
For the Mott liquid, the temparature scale $T_{ML}$ is given by eq.\eqref{frequency_ML}. As $\epsilon_{\Lambda_{0},\hat{s}}+\epsilon_{\Lambda_{0},-\hat{s}} \simeq \epsilon_{\Lambda_{0},\hat{s}}-\epsilon_{-\Lambda_{0},T\hat{s}}$, we find $\omega_{ins}\simeq \omega^{*}$, and we can write the ratio of $T_{c}$ and $T_{ML}$ as
\begin{eqnarray}
\frac{T_{c}}{T_{ML}} = 2\frac{f(\hat{s}_{1},\omega^{*})}{N^{*}(\hat{s}_{1},0)}~.
\label{TCTMLrelation}
\end{eqnarray}
\par\noindent
Finally, the Hamiltonian that describes the inclusion of spin-nematic ordering for the spins is given by
\begin{eqnarray}
 H_{nm} = Q(S^{x}_{tot})^{2}+Q(S^{y}_{tot})^{2}-2Q(S^{z}_{tot})^{2}~,
\end{eqnarray}
where $S^{i}_{tot}=\sum_{\mathbf{r}}S^{i}_{\mathbf{r}}$, $i=(x, y, z)$ are the net spin angular-momentum operators along the $x$, $y$ and $z$ directions. The RG equation for spin-nematic ordering has the form
\begin{eqnarray}
\Delta Q_{(j)} &=& \frac{Q_{(j)}^{2}}{\omega -\frac{Q_{(j)}}{4}}-\frac{p_{s}(K^{(j)}_{s, 0})^{2}}{\omega - \epsilon_{p, \Lambda_{j}}-K^{(j)}_{p, 0}-\frac{Q_{(j)}}{4}},
\end{eqnarray} 
where the first term denotes the contribution from spin-nematic fluctuations, while the second term denotes the contributions from superconducting fluctuations that suppress nematicity.
\end{widetext}
\section{Benchmarking results for $2\leq U/t\leq 12$}\label{further benchmarking}
\begin{figure}
\includegraphics[width=0.45\textwidth, height=0.35\textwidth]{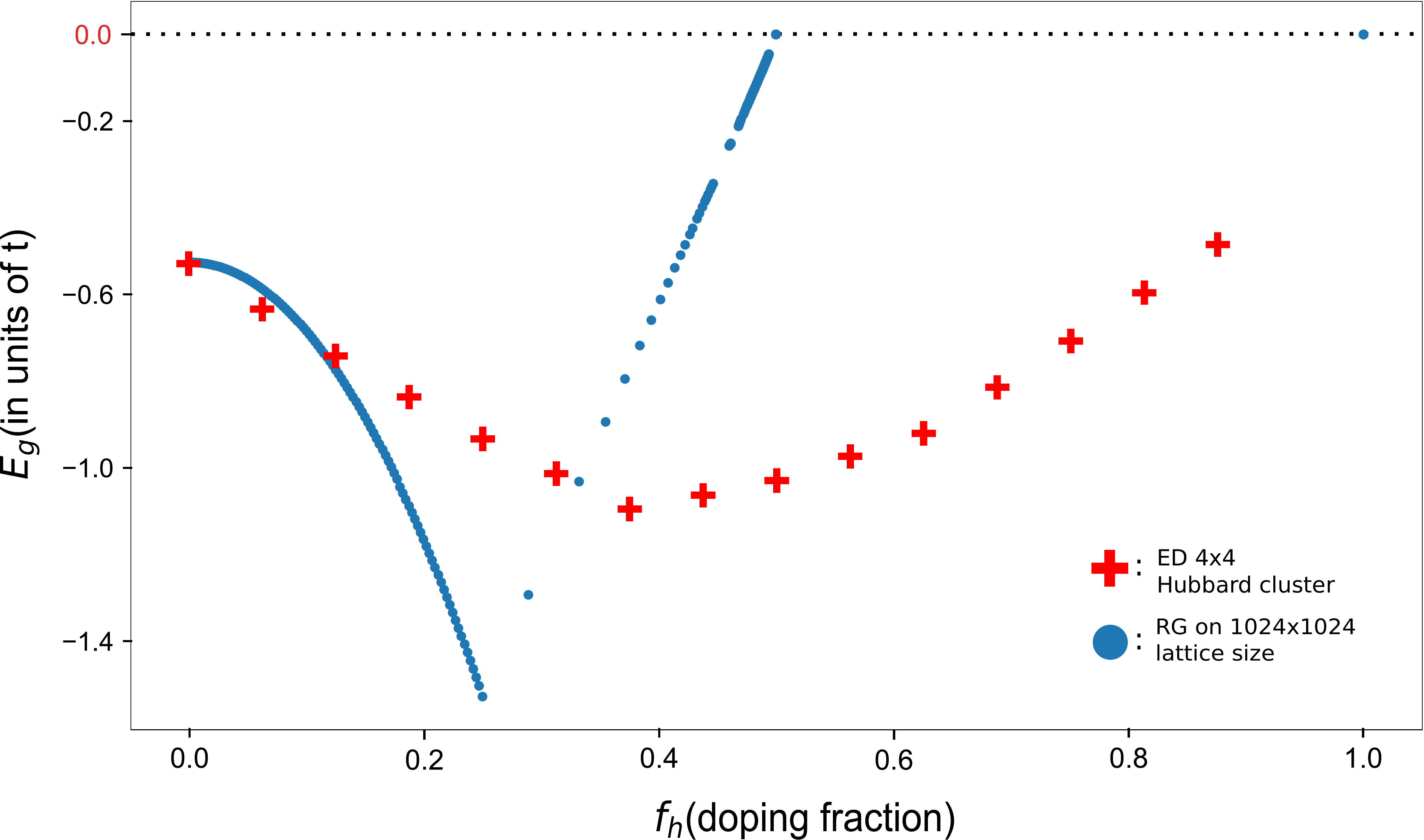}
\label{egwithdopingplused}
\caption{(Colour Online) Plot of the ground state energy per site ($E_{g}$) obtained for $U/t=8$ from the RG method for a $k$-space grid of size $1024\times 1024$ (blue circles) and exact diagonalisation (ED) on a $4\times 4$ cluster~(from data in Ref.\cite{dagotto1992}). The two approaches are in close agreement for small hole-doping ($f_{h}\leq 12.5\%$). However, while the RG enveils a QCP upon increasing doping, the ED calculations show a crossover behaviour expected for a small system.} \label{egwithdoping}
\end{figure}

Here we present results for the ground state energy per site $E_{g}$ obtained from the RG fixed point theories for various values of $U/t$ at $f_{h}=0$ and $f_{h}=0.125$ hole doping (see main text for details of the method). The values in the third and fifth columns for $U/t = 2,4,6,8,12$ are obtained from several different numerical methods, as presented in LeBlanc et. al.~\cite{leblanc2015solutions} and Ehlers et al.~\cite{ehlers2017hybridDMRG}. The ground state energy values for $U/t=10$ are obtained from exact diagonalization studies ~\cite{dagotto1992} of a $4\times 4$ Hubbard-cluster. The value in the fifth column for $U/t=12$ and $f_{h}=0.125$ is absent, as no values are available in the literature to compare against. Further, in Fig.\ref{egwithdopingplused}, we present a plot of the ground state energy per site ($E_{g}$) obtained for $U/t=8$ from the RG method for a $k$-space grid of size $1024\times 1024$ (blue circles) and exact diagonalisation (ED) on a $4\times 4$ cluster~(from data in Ref.\cite{dagotto1992}). The two approaches are in close agreement for small hole-doping ($f_{h}\leq 12.5\%$). However, while the RG unveils a QCP upon increasing doping, the ED calculations show a rounded-out minima. We have checked that the precisely the same QCP is attained at $U/t=8$ from the RG for $k$-space grid sizes down to $256\times 256$. This reveals the fact that while ED calculations show the crossover behaviour expected for a small system, the RG captures well the physics arising from divergent fluctuations of the doped Mott liquid near critical hole-doping $f_{h}^{*}$.
\begin{table*} 
 \caption{Ground state energy per site values obtained from the RG fixed points for $U/t=2,4,6,8,10,12$. The error bar for all data obtained from the RG is O($10^{-4}t$). Third and fifth columns for U/t = 2, 4, 6, 8 and 12 represent the range of values obtained for the ground state per site from several different numerical methods (presented in Refs.~\cite{leblanc2015solutions}, \cite{ehlers2017hybridDMRG} and 
\cite{dagotto1992}) for the half-filled ($f_{h} = 0$) and the doped ($f_{h} = 1/8$) Hubbard model respectively.}
   \label{tab:table1}
    \begin{tabular}{c|c|c|c|c} % <-- Alignments: 1st column left, 2nd middle and 3rd right, with vertical lines in between
    \\
      $U_{0}/t$ & $E_{g}$ from RG  & $E_{g}$ from various  & $E_{g}$ from RG & $E_{g}$ from other\\
      & $f_{h}=0$ &  numerical methods($f_{h}=0$) & at $f_{h}=\frac{1}{8}$ & numerical methods($f_{h}=\frac{1}{8}$)\\
      \hline
      2 & -1.199 &(-1.176)--(-1.16)  &-1.28 & (-1.285)--(-1.267)\\
      4 & -0.854 & (-0.864)--(-0.85) & -0.996 & (-1.026)--(-1.0)\\
      6 & -0.652 & (-0.658)--(-0.651)&-0.857 & (-0.863)--(-0.829)\\
      8 &-0.526&(-0.53)--(-0.51)&-0.777&(-0.766)--(-0.709)\\
      10 &-0.439&-0.439&-0.753&-0.675\\
      12 &-0.367&(-0.369)--(-0.362)&-0.744&--\\
      \hline
    \end{tabular}
\end{table*}
\bibliography{Bibliography}

\begin{thebibliography}{156}%
\makeatletter
\providecommand \@ifxundefined [1]{%
 \@ifx{#1\undefined}
}%
\providecommand \@ifnum [1]{%
 \ifnum #1\expandafter \@firstoftwo
 \else \expandafter \@secondoftwo
 \fi
}%
\providecommand \@ifx [1]{%
 \ifx #1\expandafter \@firstoftwo
 \else \expandafter \@secondoftwo
 \fi
}%
\providecommand \natexlab [1]{#1}%
\providecommand \enquote  [1]{``#1''}%
\providecommand \bibnamefont  [1]{#1}%
\providecommand \bibfnamefont [1]{#1}%
\providecommand \citenamefont [1]{#1}%
\providecommand \href@noop [0]{\@secondoftwo}%
\providecommand \href [0]{\begingroup \@sanitize@url \@href}%
\providecommand \@href[1]{\@@startlink{#1}\@@href}%
\providecommand \@@href[1]{\endgroup#1\@@endlink}%
\providecommand \@sanitize@url [0]{\catcode `\\12\catcode `\$12\catcode
  `\&12\catcode `\#12\catcode `\^12\catcode `\_12\catcode `\%12\relax}%
\providecommand \@@startlink[1]{}%
\providecommand \@@endlink[0]{}%
\providecommand \url  [0]{\begingroup\@sanitize@url \@url }%
\providecommand \@url [1]{\endgroup\@href {#1}{\urlprefix }}%
\providecommand \urlprefix  [0]{URL }%
\providecommand \Eprint [0]{\href }%
\providecommand \doibase [0]{http://dx.doi.org/}%
\providecommand \selectlanguage [0]{\@gobble}%
\providecommand \bibinfo  [0]{\@secondoftwo}%
\providecommand \bibfield  [0]{\@secondoftwo}%
\providecommand \translation [1]{[#1]}%
\providecommand \BibitemOpen [0]{}%
\providecommand \bibitemStop [0]{}%
\providecommand \bibitemNoStop [0]{.\EOS\space}%
\providecommand \EOS [0]{\spacefactor3000\relax}%
\providecommand \BibitemShut  [1]{\csname bibitem#1\endcsname}%
\let\auto@bib@innerbib\@empty
%</preamble>
\bibitem [{\citenamefont {Lieb}\ and\ \citenamefont
  {Wu}(1968)}]{lieb1968absence}%
  \BibitemOpen
  \bibfield  {author} {\bibinfo {author} {\bibfnamefont {E.~H.}\ \bibnamefont
  {Lieb}}\ and\ \bibinfo {author} {\bibfnamefont {F.~Y.}\ \bibnamefont {Wu}},\
  }\href@noop {} {\bibfield  {journal} {\bibinfo  {journal} {Phys. Rev. Lett.}\
  }\textbf {\bibinfo {volume} {20}},\ \bibinfo {pages} {1445} (\bibinfo {year}
  {1968})}\BibitemShut {NoStop}%
\bibitem [{\citenamefont {Imada}\ \emph {et~al.}(1998)\citenamefont {Imada},
  \citenamefont {Fujimori},\ and\ \citenamefont {Tokura}}]{imada1998metal}%
  \BibitemOpen
  \bibfield  {author} {\bibinfo {author} {\bibfnamefont {M.}~\bibnamefont
  {Imada}}, \bibinfo {author} {\bibfnamefont {A.}~\bibnamefont {Fujimori}}, \
  and\ \bibinfo {author} {\bibfnamefont {Y.}~\bibnamefont {Tokura}},\
  }\href@noop {} {\bibfield  {journal} {\bibinfo  {journal} {Reviews of modern
  physics}\ }\textbf {\bibinfo {volume} {70}},\ \bibinfo {pages} {1039}
  (\bibinfo {year} {1998})}\BibitemShut {NoStop}%
\bibitem [{\citenamefont {Anderson}(1987{\natexlab{a}})}]{anderson1987RVB}%
  \BibitemOpen
  \bibfield  {author} {\bibinfo {author} {\bibfnamefont {P.~W.}\ \bibnamefont
  {Anderson}},\ }\href@noop {} {\bibfield  {journal} {\bibinfo  {journal}
  {science}\ }\textbf {\bibinfo {volume} {235}},\ \bibinfo {pages} {1196}
  (\bibinfo {year} {1987}{\natexlab{a}})}\BibitemShut {NoStop}%
\bibitem [{\citenamefont {Edegger}\ \emph {et~al.}(2007)\citenamefont
  {Edegger}, \citenamefont {Muthukumar},\ and\ \citenamefont
  {Gros}}]{grosEdgerMuthukumar2007RVB}%
  \BibitemOpen
  \bibfield  {author} {\bibinfo {author} {\bibfnamefont {B.}~\bibnamefont
  {Edegger}}, \bibinfo {author} {\bibfnamefont {V.~N.}\ \bibnamefont
  {Muthukumar}}, \ and\ \bibinfo {author} {\bibfnamefont {C.}~\bibnamefont
  {Gros}},\ }\href@noop {} {\bibfield  {journal} {\bibinfo  {journal} {Advances
  in Physics}\ }\textbf {\bibinfo {volume} {56}},\ \bibinfo {pages} {927}
  (\bibinfo {year} {2007})}\BibitemShut {NoStop}%
\bibitem [{\citenamefont {Paramekanti}\ \emph {et~al.}(2004)\citenamefont
  {Paramekanti}, \citenamefont {Randeria},\ and\ \citenamefont
  {Trivedi}}]{paramekanti2004RVB}%
  \BibitemOpen
  \bibfield  {author} {\bibinfo {author} {\bibfnamefont {A.}~\bibnamefont
  {Paramekanti}}, \bibinfo {author} {\bibfnamefont {M.}~\bibnamefont
  {Randeria}}, \ and\ \bibinfo {author} {\bibfnamefont {N.}~\bibnamefont
  {Trivedi}},\ }\href@noop {} {\bibfield  {journal} {\bibinfo  {journal}
  {Physical Review B}\ }\textbf {\bibinfo {volume} {70}},\ \bibinfo {pages}
  {054504} (\bibinfo {year} {2004})}\BibitemShut {NoStop}%
\bibitem [{\citenamefont {Anderson}\ \emph {et~al.}(2004)\citenamefont
  {Anderson}, \citenamefont {Lee}, \citenamefont {Randeria}, \citenamefont
  {Rice}, \citenamefont {Trivedi},\ and\ \citenamefont {Zhang}}]{PWA2004}%
  \BibitemOpen
  \bibfield  {author} {\bibinfo {author} {\bibfnamefont {P.~W.}\ \bibnamefont
  {Anderson}}, \bibinfo {author} {\bibfnamefont {P.}~\bibnamefont {Lee}},
  \bibinfo {author} {\bibfnamefont {M.}~\bibnamefont {Randeria}}, \bibinfo
  {author} {\bibfnamefont {T.}~\bibnamefont {Rice}}, \bibinfo {author}
  {\bibfnamefont {N.}~\bibnamefont {Trivedi}}, \ and\ \bibinfo {author}
  {\bibfnamefont {F.}~\bibnamefont {Zhang}},\ }\href@noop {} {\bibfield
  {journal} {\bibinfo  {journal} {Journal of Physics: Condensed Matter}\
  }\textbf {\bibinfo {volume} {16}},\ \bibinfo {pages} {R755} (\bibinfo {year}
  {2004})}\BibitemShut {NoStop}%
\bibitem [{\citenamefont {Kurosaki}\ \emph {et~al.}(2005)\citenamefont
  {Kurosaki}, \citenamefont {Shimizu}, \citenamefont {Miyagawa}, \citenamefont
  {Kanoda},\ and\ \citenamefont {Saito}}]{kurosaki2005mott}%
  \BibitemOpen
  \bibfield  {author} {\bibinfo {author} {\bibfnamefont {Y.}~\bibnamefont
  {Kurosaki}}, \bibinfo {author} {\bibfnamefont {Y.}~\bibnamefont {Shimizu}},
  \bibinfo {author} {\bibfnamefont {K.}~\bibnamefont {Miyagawa}}, \bibinfo
  {author} {\bibfnamefont {K.}~\bibnamefont {Kanoda}}, \ and\ \bibinfo {author}
  {\bibfnamefont {G.}~\bibnamefont {Saito}},\ }\href@noop {} {\bibfield
  {journal} {\bibinfo  {journal} {Physical review letters}\ }\textbf {\bibinfo
  {volume} {95}},\ \bibinfo {pages} {177001} (\bibinfo {year}
  {2005})}\BibitemShut {NoStop}%
\bibitem [{\citenamefont {Helton}\ \emph {et~al.}(2010)\citenamefont {Helton},
  \citenamefont {Matan}, \citenamefont {Shores}, \citenamefont {Nytko},
  \citenamefont {Bartlett}, \citenamefont {Qiu}, \citenamefont {Nocera},\ and\
  \citenamefont {Lee}}]{helton2010dynamic}%
  \BibitemOpen
  \bibfield  {author} {\bibinfo {author} {\bibfnamefont {J.}~\bibnamefont
  {Helton}}, \bibinfo {author} {\bibfnamefont {K.}~\bibnamefont {Matan}},
  \bibinfo {author} {\bibfnamefont {M.}~\bibnamefont {Shores}}, \bibinfo
  {author} {\bibfnamefont {E.}~\bibnamefont {Nytko}}, \bibinfo {author}
  {\bibfnamefont {B.}~\bibnamefont {Bartlett}}, \bibinfo {author}
  {\bibfnamefont {Y.}~\bibnamefont {Qiu}}, \bibinfo {author} {\bibfnamefont
  {D.}~\bibnamefont {Nocera}}, \ and\ \bibinfo {author} {\bibfnamefont
  {Y.}~\bibnamefont {Lee}},\ }\href@noop {} {\bibfield  {journal} {\bibinfo
  {journal} {Physical review letters}\ }\textbf {\bibinfo {volume} {104}},\
  \bibinfo {pages} {147201} (\bibinfo {year} {2010})}\BibitemShut {NoStop}%
\bibitem [{\citenamefont {Dalla~Piazza}\ \emph {et~al.}(2015)\citenamefont
  {Dalla~Piazza}, \citenamefont {Mourigal}, \citenamefont {Christensen},
  \citenamefont {Nilsen}, \citenamefont {Tregenna-Piggott}, \citenamefont
  {Perring}, \citenamefont {Enderle}, \citenamefont {McMorrow}, \citenamefont
  {Ivanov},\ and\ \citenamefont {R{\o}nnow}}]{piazza2015TopologicalOrder}%
  \BibitemOpen
  \bibfield  {author} {\bibinfo {author} {\bibfnamefont {B.}~\bibnamefont
  {Dalla~Piazza}}, \bibinfo {author} {\bibfnamefont {M.}~\bibnamefont
  {Mourigal}}, \bibinfo {author} {\bibfnamefont {N.~B.}\ \bibnamefont
  {Christensen}}, \bibinfo {author} {\bibfnamefont {G.}~\bibnamefont {Nilsen}},
  \bibinfo {author} {\bibfnamefont {P.}~\bibnamefont {Tregenna-Piggott}},
  \bibinfo {author} {\bibfnamefont {T.}~\bibnamefont {Perring}}, \bibinfo
  {author} {\bibfnamefont {M.}~\bibnamefont {Enderle}}, \bibinfo {author}
  {\bibfnamefont {D.~F.}\ \bibnamefont {McMorrow}}, \bibinfo {author}
  {\bibfnamefont {D.}~\bibnamefont {Ivanov}}, \ and\ \bibinfo {author}
  {\bibfnamefont {H.~M.}\ \bibnamefont {R{\o}nnow}},\ }\href@noop {} {\bibfield
   {journal} {\bibinfo  {journal} {Nature physics}\ }\textbf {\bibinfo {volume}
  {11}},\ \bibinfo {pages} {62} (\bibinfo {year} {2015})}\BibitemShut {NoStop}%
\bibitem [{\citenamefont {Iazzi}\ \emph {et~al.}(2016)\citenamefont {Iazzi},
  \citenamefont {Soluyanov},\ and\ \citenamefont
  {Troyer}}]{iazzi2016topological}%
  \BibitemOpen
  \bibfield  {author} {\bibinfo {author} {\bibfnamefont {M.}~\bibnamefont
  {Iazzi}}, \bibinfo {author} {\bibfnamefont {A.~A.}\ \bibnamefont
  {Soluyanov}}, \ and\ \bibinfo {author} {\bibfnamefont {M.}~\bibnamefont
  {Troyer}},\ }\href@noop {} {\bibfield  {journal} {\bibinfo  {journal}
  {Physical Review B}\ }\textbf {\bibinfo {volume} {93}},\ \bibinfo {pages}
  {115102} (\bibinfo {year} {2016})}\BibitemShut {NoStop}%
\bibitem [{\citenamefont {LeBlanc}\ \emph {et~al.}(2015)\citenamefont
  {LeBlanc}, \citenamefont {Antipov}, \citenamefont {Becca}, \citenamefont
  {Bulik}, \citenamefont {Chan}, \citenamefont {Chung}, \citenamefont {Deng},
  \citenamefont {Ferrero}, \citenamefont {Henderson}, \citenamefont
  {Jim{\'e}nez-Hoyos} \emph {et~al.}}]{leblanc2015solutions}%
  \BibitemOpen
  \bibfield  {author} {\bibinfo {author} {\bibfnamefont {J.}~\bibnamefont
  {LeBlanc}}, \bibinfo {author} {\bibfnamefont {A.~E.}\ \bibnamefont
  {Antipov}}, \bibinfo {author} {\bibfnamefont {F.}~\bibnamefont {Becca}},
  \bibinfo {author} {\bibfnamefont {I.~W.}\ \bibnamefont {Bulik}}, \bibinfo
  {author} {\bibfnamefont {G.~K.-L.}\ \bibnamefont {Chan}}, \bibinfo {author}
  {\bibfnamefont {C.-M.}\ \bibnamefont {Chung}}, \bibinfo {author}
  {\bibfnamefont {Y.}~\bibnamefont {Deng}}, \bibinfo {author} {\bibfnamefont
  {M.}~\bibnamefont {Ferrero}}, \bibinfo {author} {\bibfnamefont {T.~M.}\
  \bibnamefont {Henderson}}, \bibinfo {author} {\bibfnamefont {C.~A.}\
  \bibnamefont {Jim{\'e}nez-Hoyos}},  \emph {et~al.},\ }\href@noop {}
  {\bibfield  {journal} {\bibinfo  {journal} {Physical Review X}\ }\textbf
  {\bibinfo {volume} {5}},\ \bibinfo {pages} {041041} (\bibinfo {year}
  {2015})}\BibitemShut {NoStop}%
\bibitem [{\citenamefont {Georges}\ \emph
  {et~al.}(1996{\natexlab{a}})\citenamefont {Georges}, \citenamefont {Kotliar},
  \citenamefont {Krauth},\ and\ \citenamefont {Rozenberg}}]{kotliar1992}%
  \BibitemOpen
  \bibfield  {author} {\bibinfo {author} {\bibfnamefont {A.}~\bibnamefont
  {Georges}}, \bibinfo {author} {\bibfnamefont {G.}~\bibnamefont {Kotliar}},
  \bibinfo {author} {\bibfnamefont {W.}~\bibnamefont {Krauth}}, \ and\ \bibinfo
  {author} {\bibfnamefont {M.~J.}\ \bibnamefont {Rozenberg}},\ }\href@noop {}
  {\bibfield  {journal} {\bibinfo  {journal} {Reviews of Modern Physics}\
  }\textbf {\bibinfo {volume} {68}},\ \bibinfo {pages} {13} (\bibinfo {year}
  {1996}{\natexlab{a}})}\BibitemShut {NoStop}%
\bibitem [{\citenamefont {Georges}\ and\ \citenamefont
  {Kotliar}(1992)}]{kotliar1996}%
  \BibitemOpen
  \bibfield  {author} {\bibinfo {author} {\bibfnamefont {A.}~\bibnamefont
  {Georges}}\ and\ \bibinfo {author} {\bibfnamefont {G.}~\bibnamefont
  {Kotliar}},\ }\href@noop {} {\bibfield  {journal} {\bibinfo  {journal}
  {Physical Review B}\ }\textbf {\bibinfo {volume} {45}},\ \bibinfo {pages}
  {6479} (\bibinfo {year} {1992})}\BibitemShut {NoStop}%
\bibitem [{\citenamefont {Werner}\ and\ \citenamefont
  {Millis}(2007)}]{werner2007}%
  \BibitemOpen
  \bibfield  {author} {\bibinfo {author} {\bibfnamefont {P.}~\bibnamefont
  {Werner}}\ and\ \bibinfo {author} {\bibfnamefont {A.~J.}\ \bibnamefont
  {Millis}},\ }\href@noop {} {\bibfield  {journal} {\bibinfo  {journal} {Phys.
  Rev. B}\ }\textbf {\bibinfo {volume} {75}},\ \bibinfo {pages} {085108}
  (\bibinfo {year} {2007})}\BibitemShut {NoStop}%
\bibitem [{\citenamefont {Joo}\ and\ \citenamefont
  {Oudovenko}(2001)}]{joo2001}%
  \BibitemOpen
  \bibfield  {author} {\bibinfo {author} {\bibfnamefont {J.}~\bibnamefont
  {Joo}}\ and\ \bibinfo {author} {\bibfnamefont {V.}~\bibnamefont
  {Oudovenko}},\ }\href@noop {} {\bibfield  {journal} {\bibinfo  {journal}
  {Phys. Rev. B}\ }\textbf {\bibinfo {volume} {64}},\ \bibinfo {pages} {193102}
  (\bibinfo {year} {2001})}\BibitemShut {NoStop}%
\bibitem [{\citenamefont {Moukouri}\ and\ \citenamefont
  {Jarrell}(2001{\natexlab{a}})}]{jarell2001}%
  \BibitemOpen
  \bibfield  {author} {\bibinfo {author} {\bibfnamefont {S.}~\bibnamefont
  {Moukouri}}\ and\ \bibinfo {author} {\bibfnamefont {M.}~\bibnamefont
  {Jarrell}},\ }\href@noop {} {\bibfield  {journal} {\bibinfo  {journal}
  {Physical review letters}\ }\textbf {\bibinfo {volume} {87}},\ \bibinfo
  {pages} {167010} (\bibinfo {year} {2001}{\natexlab{a}})}\BibitemShut
  {NoStop}%
\bibitem [{\citenamefont {Gull}\ \emph {et~al.}(2013)\citenamefont {Gull},
  \citenamefont {Parcollet},\ and\ \citenamefont {Millis}}]{gull2013}%
  \BibitemOpen
  \bibfield  {author} {\bibinfo {author} {\bibfnamefont {E.}~\bibnamefont
  {Gull}}, \bibinfo {author} {\bibfnamefont {O.}~\bibnamefont {Parcollet}}, \
  and\ \bibinfo {author} {\bibfnamefont {A.~J.}\ \bibnamefont {Millis}},\
  }\href@noop {} {\bibfield  {journal} {\bibinfo  {journal} {Physical review
  letters}\ }\textbf {\bibinfo {volume} {110}},\ \bibinfo {pages} {216405}
  (\bibinfo {year} {2013})}\BibitemShut {NoStop}%
\bibitem [{\citenamefont {Merino}\ and\ \citenamefont
  {Gunnarsson}(2014)}]{merino2014}%
  \BibitemOpen
  \bibfield  {author} {\bibinfo {author} {\bibfnamefont {J.}~\bibnamefont
  {Merino}}\ and\ \bibinfo {author} {\bibfnamefont {O.}~\bibnamefont
  {Gunnarsson}},\ }\href@noop {} {\bibfield  {journal} {\bibinfo  {journal}
  {Physical Review B}\ }\textbf {\bibinfo {volume} {89}},\ \bibinfo {pages}
  {245130} (\bibinfo {year} {2014})}\BibitemShut {NoStop}%
\bibitem [{\citenamefont {Park}\ \emph {et~al.}(2008)\citenamefont {Park},
  \citenamefont {Haule},\ and\ \citenamefont {Kotliar}}]{park2008}%
  \BibitemOpen
  \bibfield  {author} {\bibinfo {author} {\bibfnamefont {H.}~\bibnamefont
  {Park}}, \bibinfo {author} {\bibfnamefont {K.}~\bibnamefont {Haule}}, \ and\
  \bibinfo {author} {\bibfnamefont {G.}~\bibnamefont {Kotliar}},\ }\href@noop
  {} {\bibfield  {journal} {\bibinfo  {journal} {Physical review letters}\
  }\textbf {\bibinfo {volume} {101}},\ \bibinfo {pages} {186403} (\bibinfo
  {year} {2008})}\BibitemShut {NoStop}%
\bibitem [{\citenamefont {Toschi}\ \emph {et~al.}(2007)\citenamefont {Toschi},
  \citenamefont {Katanin},\ and\ \citenamefont {Held}}]{toschi2007}%
  \BibitemOpen
  \bibfield  {author} {\bibinfo {author} {\bibfnamefont {A.}~\bibnamefont
  {Toschi}}, \bibinfo {author} {\bibfnamefont {A.}~\bibnamefont {Katanin}}, \
  and\ \bibinfo {author} {\bibfnamefont {K.}~\bibnamefont {Held}},\ }\href@noop
  {} {\bibfield  {journal} {\bibinfo  {journal} {Physical Review B}\ }\textbf
  {\bibinfo {volume} {75}},\ \bibinfo {pages} {045118} (\bibinfo {year}
  {2007})}\BibitemShut {NoStop}%
\bibitem [{\citenamefont {Held}\ \emph {et~al.}(2008)\citenamefont {Held},
  \citenamefont {Katanin},\ and\ \citenamefont {Toschi}}]{held2008}%
  \BibitemOpen
  \bibfield  {author} {\bibinfo {author} {\bibfnamefont {K.}~\bibnamefont
  {Held}}, \bibinfo {author} {\bibfnamefont {A.~A.}\ \bibnamefont {Katanin}}, \
  and\ \bibinfo {author} {\bibfnamefont {A.}~\bibnamefont {Toschi}},\
  }\href@noop {} {\bibfield  {journal} {\bibinfo  {journal} {Progress of
  Theoretical Physics Supplement}\ }\textbf {\bibinfo {volume} {176}},\
  \bibinfo {pages} {117} (\bibinfo {year} {2008})}\BibitemShut {NoStop}%
\bibitem [{\citenamefont {Sch{\"a}fer}\ \emph {et~al.}(2015)\citenamefont
  {Sch{\"a}fer}, \citenamefont {Geles}, \citenamefont {Rost}, \citenamefont
  {Rohringer}, \citenamefont {Arrigoni}, \citenamefont {Held}, \citenamefont
  {Bl{\"u}mer}, \citenamefont {Aichhorn},\ and\ \citenamefont
  {Toschi}}]{schafer2015}%
  \BibitemOpen
  \bibfield  {author} {\bibinfo {author} {\bibfnamefont {T.}~\bibnamefont
  {Sch{\"a}fer}}, \bibinfo {author} {\bibfnamefont {F.}~\bibnamefont {Geles}},
  \bibinfo {author} {\bibfnamefont {D.}~\bibnamefont {Rost}}, \bibinfo {author}
  {\bibfnamefont {G.}~\bibnamefont {Rohringer}}, \bibinfo {author}
  {\bibfnamefont {E.}~\bibnamefont {Arrigoni}}, \bibinfo {author}
  {\bibfnamefont {K.}~\bibnamefont {Held}}, \bibinfo {author} {\bibfnamefont
  {N.}~\bibnamefont {Bl{\"u}mer}}, \bibinfo {author} {\bibfnamefont
  {M.}~\bibnamefont {Aichhorn}}, \ and\ \bibinfo {author} {\bibfnamefont
  {A.}~\bibnamefont {Toschi}},\ }\href@noop {} {\bibfield  {journal} {\bibinfo
  {journal} {Physical Review B}\ }\textbf {\bibinfo {volume} {91}},\ \bibinfo
  {pages} {125109} (\bibinfo {year} {2015})}\BibitemShut {NoStop}%
\bibitem [{\citenamefont {Blankenbecler}(1981)}]{bss1981}%
  \BibitemOpen
  \bibfield  {author} {\bibinfo {author} {\bibfnamefont {D.}~\bibnamefont
  {Blankenbecler}},\ }\href@noop {} {\bibfield  {journal} {\bibinfo  {journal}
  {Phys. Rev. D}\ }\textbf {\bibinfo {volume} {24}},\ \bibinfo {pages} {2278}
  (\bibinfo {year} {1981})}\BibitemShut {NoStop}%
\bibitem [{\citenamefont {Zheng}\ and\ \citenamefont {Chan}(2016)}]{zheng2016}%
  \BibitemOpen
  \bibfield  {author} {\bibinfo {author} {\bibfnamefont {B.-X.}\ \bibnamefont
  {Zheng}}\ and\ \bibinfo {author} {\bibfnamefont {G.~K.-L.}\ \bibnamefont
  {Chan}},\ }\href@noop {} {\bibfield  {journal} {\bibinfo  {journal} {Physical
  Review B}\ }\textbf {\bibinfo {volume} {93}},\ \bibinfo {pages} {035126}
  (\bibinfo {year} {2016})}\BibitemShut {NoStop}%
\bibitem [{\citenamefont {van Loon}\ \emph {et~al.}(2018)\citenamefont {van
  Loon}, \citenamefont {Hafermann},\ and\ \citenamefont
  {Katsnelson}}]{loon2018}%
  \BibitemOpen
  \bibfield  {author} {\bibinfo {author} {\bibfnamefont {E.~G.}\ \bibnamefont
  {van Loon}}, \bibinfo {author} {\bibfnamefont {H.}~\bibnamefont {Hafermann}},
  \ and\ \bibinfo {author} {\bibfnamefont {M.~I.}\ \bibnamefont {Katsnelson}},\
  }\href@noop {} {\bibfield  {journal} {\bibinfo  {journal} {Physical Review
  B}\ }\textbf {\bibinfo {volume} {97}},\ \bibinfo {pages} {085125} (\bibinfo
  {year} {2018})}\BibitemShut {NoStop}%
\bibitem [{\citenamefont {Tocchio}\ \emph {et~al.}(2008)\citenamefont
  {Tocchio}, \citenamefont {Becca}, \citenamefont {Parola},\ and\ \citenamefont
  {Sorella}}]{tocchio2008}%
  \BibitemOpen
  \bibfield  {author} {\bibinfo {author} {\bibfnamefont {L.~F.}\ \bibnamefont
  {Tocchio}}, \bibinfo {author} {\bibfnamefont {F.}~\bibnamefont {Becca}},
  \bibinfo {author} {\bibfnamefont {A.}~\bibnamefont {Parola}}, \ and\ \bibinfo
  {author} {\bibfnamefont {S.}~\bibnamefont {Sorella}},\ }\href@noop {}
  {\bibfield  {journal} {\bibinfo  {journal} {Physical Review B}\ }\textbf
  {\bibinfo {volume} {78}},\ \bibinfo {pages} {041101} (\bibinfo {year}
  {2008})}\BibitemShut {NoStop}%
\bibitem [{\citenamefont {Cosentini}\ \emph {et~al.}(1998)\citenamefont
  {Cosentini}, \citenamefont {Capone}, \citenamefont {Guidoni},\ and\
  \citenamefont {Bachelet}}]{cosentini1998}%
  \BibitemOpen
  \bibfield  {author} {\bibinfo {author} {\bibfnamefont {A.}~\bibnamefont
  {Cosentini}}, \bibinfo {author} {\bibfnamefont {M.}~\bibnamefont {Capone}},
  \bibinfo {author} {\bibfnamefont {L.}~\bibnamefont {Guidoni}}, \ and\
  \bibinfo {author} {\bibfnamefont {G.}~\bibnamefont {Bachelet}},\ }\href@noop
  {} {\bibfield  {journal} {\bibinfo  {journal} {Physical Review B}\ }\textbf
  {\bibinfo {volume} {58}},\ \bibinfo {pages} {R14685} (\bibinfo {year}
  {1998})}\BibitemShut {NoStop}%
\bibitem [{\citenamefont {Becca}\ \emph {et~al.}(2000)\citenamefont {Becca},
  \citenamefont {Capone},\ and\ \citenamefont {Sorella}}]{becca2000}%
  \BibitemOpen
  \bibfield  {author} {\bibinfo {author} {\bibfnamefont {F.}~\bibnamefont
  {Becca}}, \bibinfo {author} {\bibfnamefont {M.}~\bibnamefont {Capone}}, \
  and\ \bibinfo {author} {\bibfnamefont {S.}~\bibnamefont {Sorella}},\
  }\href@noop {} {\bibfield  {journal} {\bibinfo  {journal} {Physical Review
  B}\ }\textbf {\bibinfo {volume} {62}},\ \bibinfo {pages} {12700} (\bibinfo
  {year} {2000})}\BibitemShut {NoStop}%
\bibitem [{\citenamefont {Van~Bemmel}\ \emph {et~al.}(1994)\citenamefont
  {Van~Bemmel}, \citenamefont {Ten~Haaf}, \citenamefont {Van~Saarloos},
  \citenamefont {Van~Leeuwen},\ and\ \citenamefont {An}}]{bemmel1994}%
  \BibitemOpen
  \bibfield  {author} {\bibinfo {author} {\bibfnamefont {H.}~\bibnamefont
  {Van~Bemmel}}, \bibinfo {author} {\bibfnamefont {D.}~\bibnamefont
  {Ten~Haaf}}, \bibinfo {author} {\bibfnamefont {W.}~\bibnamefont
  {Van~Saarloos}}, \bibinfo {author} {\bibfnamefont {J.}~\bibnamefont
  {Van~Leeuwen}}, \ and\ \bibinfo {author} {\bibfnamefont {G.}~\bibnamefont
  {An}},\ }\href@noop {} {\bibfield  {journal} {\bibinfo  {journal} {Physical
  review letters}\ }\textbf {\bibinfo {volume} {72}},\ \bibinfo {pages} {2442}
  (\bibinfo {year} {1994})}\BibitemShut {NoStop}%
\bibitem [{\citenamefont {Zhang}\ \emph {et~al.}(1997)\citenamefont {Zhang},
  \citenamefont {Carlson},\ and\ \citenamefont {Gubernatis}}]{zhang1997}%
  \BibitemOpen
  \bibfield  {author} {\bibinfo {author} {\bibfnamefont {S.}~\bibnamefont
  {Zhang}}, \bibinfo {author} {\bibfnamefont {J.}~\bibnamefont {Carlson}}, \
  and\ \bibinfo {author} {\bibfnamefont {J.}~\bibnamefont {Gubernatis}},\
  }\href@noop {} {\bibfield  {journal} {\bibinfo  {journal} {Physical Review
  B}\ }\textbf {\bibinfo {volume} {55}},\ \bibinfo {pages} {7464} (\bibinfo
  {year} {1997})}\BibitemShut {NoStop}%
\bibitem [{\citenamefont {Chang}\ and\ \citenamefont
  {Zhang}(2008)}]{chang2008}%
  \BibitemOpen
  \bibfield  {author} {\bibinfo {author} {\bibfnamefont {C.-C.}\ \bibnamefont
  {Chang}}\ and\ \bibinfo {author} {\bibfnamefont {S.}~\bibnamefont {Zhang}},\
  }\href@noop {} {\bibfield  {journal} {\bibinfo  {journal} {Physical Review
  B}\ }\textbf {\bibinfo {volume} {78}},\ \bibinfo {pages} {165101} (\bibinfo
  {year} {2008})}\BibitemShut {NoStop}%
\bibitem [{\citenamefont {Chang}\ and\ \citenamefont
  {Zhang}(2010)}]{chang2010}%
  \BibitemOpen
  \bibfield  {author} {\bibinfo {author} {\bibfnamefont {C.-C.}\ \bibnamefont
  {Chang}}\ and\ \bibinfo {author} {\bibfnamefont {S.}~\bibnamefont {Zhang}},\
  }\href@noop {} {\bibfield  {journal} {\bibinfo  {journal} {Physical review
  letters}\ }\textbf {\bibinfo {volume} {104}},\ \bibinfo {pages} {116402}
  (\bibinfo {year} {2010})}\BibitemShut {NoStop}%
\bibitem [{\citenamefont {Yokoyama}\ and\ \citenamefont
  {Shiba}(1987)}]{yokoyama1987}%
  \BibitemOpen
  \bibfield  {author} {\bibinfo {author} {\bibfnamefont {H.}~\bibnamefont
  {Yokoyama}}\ and\ \bibinfo {author} {\bibfnamefont {H.}~\bibnamefont
  {Shiba}},\ }\href@noop {} {\bibfield  {journal} {\bibinfo  {journal} {Journal
  of the Physical Society of Japan}\ }\textbf {\bibinfo {volume} {56}},\
  \bibinfo {pages} {1490} (\bibinfo {year} {1987})}\BibitemShut {NoStop}%
\bibitem [{\citenamefont {Eichenberger}\ and\ \citenamefont
  {Baeriswyl}(2007)}]{eichenberger2007}%
  \BibitemOpen
  \bibfield  {author} {\bibinfo {author} {\bibfnamefont {D.}~\bibnamefont
  {Eichenberger}}\ and\ \bibinfo {author} {\bibfnamefont {D.}~\bibnamefont
  {Baeriswyl}},\ }\href@noop {} {\bibfield  {journal} {\bibinfo  {journal}
  {Physical Review B}\ }\textbf {\bibinfo {volume} {76}},\ \bibinfo {pages}
  {180504} (\bibinfo {year} {2007})}\BibitemShut {NoStop}%
\bibitem [{\citenamefont {Yamaji}\ \emph {et~al.}(1998)\citenamefont {Yamaji},
  \citenamefont {Yanagisawa}, \citenamefont {Nakanishi},\ and\ \citenamefont
  {Koike}}]{yamaji1998}%
  \BibitemOpen
  \bibfield  {author} {\bibinfo {author} {\bibfnamefont {K.}~\bibnamefont
  {Yamaji}}, \bibinfo {author} {\bibfnamefont {T.}~\bibnamefont {Yanagisawa}},
  \bibinfo {author} {\bibfnamefont {T.}~\bibnamefont {Nakanishi}}, \ and\
  \bibinfo {author} {\bibfnamefont {S.}~\bibnamefont {Koike}},\ }\href@noop {}
  {\bibfield  {journal} {\bibinfo  {journal} {Physica C: Superconductivity}\
  }\textbf {\bibinfo {volume} {304}},\ \bibinfo {pages} {225} (\bibinfo {year}
  {1998})}\BibitemShut {NoStop}%
\bibitem [{\citenamefont {Giamarchi}\ and\ \citenamefont
  {Lhuillier}(1991)}]{giamarchi1991}%
  \BibitemOpen
  \bibfield  {author} {\bibinfo {author} {\bibfnamefont {T.}~\bibnamefont
  {Giamarchi}}\ and\ \bibinfo {author} {\bibfnamefont {C.}~\bibnamefont
  {Lhuillier}},\ }\href@noop {} {\bibfield  {journal} {\bibinfo  {journal}
  {Physical Review B}\ }\textbf {\bibinfo {volume} {43}},\ \bibinfo {pages}
  {12943} (\bibinfo {year} {1991})}\BibitemShut {NoStop}%
\bibitem [{\citenamefont {White}\ and\ \citenamefont
  {Scalapino}(2000)}]{white2000}%
  \BibitemOpen
  \bibfield  {author} {\bibinfo {author} {\bibfnamefont {S.~R.}\ \bibnamefont
  {White}}\ and\ \bibinfo {author} {\bibfnamefont {D.}~\bibnamefont
  {Scalapino}},\ }\href@noop {} {\bibfield  {journal} {\bibinfo  {journal}
  {Physical review B}\ }\textbf {\bibinfo {volume} {61}},\ \bibinfo {pages}
  {6320} (\bibinfo {year} {2000})}\BibitemShut {NoStop}%
\bibitem [{\citenamefont {Scalapino}\ and\ \citenamefont
  {White}(2001)}]{scalapino2001}%
  \BibitemOpen
  \bibfield  {author} {\bibinfo {author} {\bibfnamefont {D.~J.}\ \bibnamefont
  {Scalapino}}\ and\ \bibinfo {author} {\bibfnamefont {S.~R.}\ \bibnamefont
  {White}},\ }\href@noop {} {\bibfield  {journal} {\bibinfo  {journal}
  {Foundations of Physics}\ }\textbf {\bibinfo {volume} {31}},\ \bibinfo
  {pages} {27} (\bibinfo {year} {2001})}\BibitemShut {NoStop}%
\bibitem [{\citenamefont {White}\ and\ \citenamefont
  {Scalapino}(2003)}]{white2003}%
  \BibitemOpen
  \bibfield  {author} {\bibinfo {author} {\bibfnamefont {S.~R.}\ \bibnamefont
  {White}}\ and\ \bibinfo {author} {\bibfnamefont {D.}~\bibnamefont
  {Scalapino}},\ }\href@noop {} {\bibfield  {journal} {\bibinfo  {journal}
  {Physical review letters}\ }\textbf {\bibinfo {volume} {91}},\ \bibinfo
  {pages} {136403} (\bibinfo {year} {2003})}\BibitemShut {NoStop}%
\bibitem [{\citenamefont {Hettler}\ \emph {et~al.}(1998)\citenamefont
  {Hettler}, \citenamefont {Tahvildar-Zadeh}, \citenamefont {Jarrell},
  \citenamefont {Pruschke},\ and\ \citenamefont {Krishnamurthy}}]{hettler1998}%
  \BibitemOpen
  \bibfield  {author} {\bibinfo {author} {\bibfnamefont {M.}~\bibnamefont
  {Hettler}}, \bibinfo {author} {\bibfnamefont {A.}~\bibnamefont
  {Tahvildar-Zadeh}}, \bibinfo {author} {\bibfnamefont {M.}~\bibnamefont
  {Jarrell}}, \bibinfo {author} {\bibfnamefont {T.}~\bibnamefont {Pruschke}}, \
  and\ \bibinfo {author} {\bibfnamefont {H.}~\bibnamefont {Krishnamurthy}},\
  }\href@noop {} {\bibfield  {journal} {\bibinfo  {journal} {Physical Review
  B}\ }\textbf {\bibinfo {volume} {58}},\ \bibinfo {pages} {R7475} (\bibinfo
  {year} {1998})}\BibitemShut {NoStop}%
\bibitem [{\citenamefont {Hettler}\ \emph {et~al.}(2000)\citenamefont
  {Hettler}, \citenamefont {Mukherjee}, \citenamefont {Jarrell},\ and\
  \citenamefont {Krishnamurthy}}]{hettler2000}%
  \BibitemOpen
  \bibfield  {author} {\bibinfo {author} {\bibfnamefont {M.}~\bibnamefont
  {Hettler}}, \bibinfo {author} {\bibfnamefont {M.}~\bibnamefont {Mukherjee}},
  \bibinfo {author} {\bibfnamefont {M.}~\bibnamefont {Jarrell}}, \ and\
  \bibinfo {author} {\bibfnamefont {H.}~\bibnamefont {Krishnamurthy}},\
  }\href@noop {} {\bibfield  {journal} {\bibinfo  {journal} {Physical Review
  B}\ }\textbf {\bibinfo {volume} {61}},\ \bibinfo {pages} {12739} (\bibinfo
  {year} {2000})}\BibitemShut {NoStop}%
\bibitem [{\citenamefont {Khatami}\ \emph {et~al.}(2010)\citenamefont
  {Khatami}, \citenamefont {Mikelsons}, \citenamefont {Galanakis},
  \citenamefont {Macridin}, \citenamefont {Moreno}, \citenamefont {Scalettar},\
  and\ \citenamefont {Jarrell}}]{khatami2010quantum}%
  \BibitemOpen
  \bibfield  {author} {\bibinfo {author} {\bibfnamefont {E.}~\bibnamefont
  {Khatami}}, \bibinfo {author} {\bibfnamefont {K.}~\bibnamefont {Mikelsons}},
  \bibinfo {author} {\bibfnamefont {D.}~\bibnamefont {Galanakis}}, \bibinfo
  {author} {\bibfnamefont {A.}~\bibnamefont {Macridin}}, \bibinfo {author}
  {\bibfnamefont {J.}~\bibnamefont {Moreno}}, \bibinfo {author} {\bibfnamefont
  {R.}~\bibnamefont {Scalettar}}, \ and\ \bibinfo {author} {\bibfnamefont
  {M.}~\bibnamefont {Jarrell}},\ }\href@noop {} {\bibfield  {journal} {\bibinfo
   {journal} {Physical Review B}\ }\textbf {\bibinfo {volume} {81}},\ \bibinfo
  {pages} {201101} (\bibinfo {year} {2010})}\BibitemShut {NoStop}%
\bibitem [{\citenamefont {Vidhyadhiraja}\ \emph {et~al.}(2009)\citenamefont
  {Vidhyadhiraja}, \citenamefont {Macridin}, \citenamefont {{\c{S}}en},
  \citenamefont {Jarrell},\ and\ \citenamefont
  {Ma}}]{vidhyadhiraja2009quantum}%
  \BibitemOpen
  \bibfield  {author} {\bibinfo {author} {\bibfnamefont {N.}~\bibnamefont
  {Vidhyadhiraja}}, \bibinfo {author} {\bibfnamefont {A.}~\bibnamefont
  {Macridin}}, \bibinfo {author} {\bibfnamefont {C.}~\bibnamefont {{\c{S}}en}},
  \bibinfo {author} {\bibfnamefont {M.}~\bibnamefont {Jarrell}}, \ and\
  \bibinfo {author} {\bibfnamefont {M.}~\bibnamefont {Ma}},\ }\href@noop {}
  {\bibfield  {journal} {\bibinfo  {journal} {Physical review letters}\
  }\textbf {\bibinfo {volume} {102}},\ \bibinfo {pages} {206407} (\bibinfo
  {year} {2009})}\BibitemShut {NoStop}%
\bibitem [{\citenamefont {Mikelsons}\ \emph {et~al.}(2009)\citenamefont
  {Mikelsons}, \citenamefont {Khatami}, \citenamefont {Galanakis},
  \citenamefont {Macridin}, \citenamefont {Moreno},\ and\ \citenamefont
  {Jarrell}}]{mikelsons2009thermodynamics}%
  \BibitemOpen
  \bibfield  {author} {\bibinfo {author} {\bibfnamefont {K.}~\bibnamefont
  {Mikelsons}}, \bibinfo {author} {\bibfnamefont {E.}~\bibnamefont {Khatami}},
  \bibinfo {author} {\bibfnamefont {D.}~\bibnamefont {Galanakis}}, \bibinfo
  {author} {\bibfnamefont {A.}~\bibnamefont {Macridin}}, \bibinfo {author}
  {\bibfnamefont {J.}~\bibnamefont {Moreno}}, \ and\ \bibinfo {author}
  {\bibfnamefont {M.}~\bibnamefont {Jarrell}},\ }\href@noop {} {\bibfield
  {journal} {\bibinfo  {journal} {Physical Review B}\ }\textbf {\bibinfo
  {volume} {80}},\ \bibinfo {pages} {140505} (\bibinfo {year}
  {2009})}\BibitemShut {NoStop}%
\bibitem [{\citenamefont {Lichtenstein}\ and\ \citenamefont
  {Katsnelson}(2000{\natexlab{a}})}]{lichtenstein2000}%
  \BibitemOpen
  \bibfield  {author} {\bibinfo {author} {\bibfnamefont {A.}~\bibnamefont
  {Lichtenstein}}\ and\ \bibinfo {author} {\bibfnamefont {M.}~\bibnamefont
  {Katsnelson}},\ }\href@noop {} {\bibfield  {journal} {\bibinfo  {journal}
  {Physical Review B}\ }\textbf {\bibinfo {volume} {62}},\ \bibinfo {pages}
  {R9283} (\bibinfo {year} {2000}{\natexlab{a}})}\BibitemShut {NoStop}%
\bibitem [{\citenamefont {Kotliar}\ \emph {et~al.}(2001)\citenamefont
  {Kotliar}, \citenamefont {Savrasov}, \citenamefont {P{\'a}lsson},\ and\
  \citenamefont {Biroli}}]{kotliar2001}%
  \BibitemOpen
  \bibfield  {author} {\bibinfo {author} {\bibfnamefont {G.}~\bibnamefont
  {Kotliar}}, \bibinfo {author} {\bibfnamefont {S.~Y.}\ \bibnamefont
  {Savrasov}}, \bibinfo {author} {\bibfnamefont {G.}~\bibnamefont
  {P{\'a}lsson}}, \ and\ \bibinfo {author} {\bibfnamefont {G.}~\bibnamefont
  {Biroli}},\ }\href@noop {} {\bibfield  {journal} {\bibinfo  {journal}
  {Physical review letters}\ }\textbf {\bibinfo {volume} {87}},\ \bibinfo
  {pages} {186401} (\bibinfo {year} {2001})}\BibitemShut {NoStop}%
\bibitem [{\citenamefont {Civelli}\ \emph {et~al.}(2008)\citenamefont
  {Civelli}, \citenamefont {Capone}, \citenamefont {Georges}, \citenamefont
  {Haule}, \citenamefont {Parcollet}, \citenamefont {Stanescu},\ and\
  \citenamefont {Kotliar}}]{civelli2008nodal}%
  \BibitemOpen
  \bibfield  {author} {\bibinfo {author} {\bibfnamefont {M.}~\bibnamefont
  {Civelli}}, \bibinfo {author} {\bibfnamefont {M.}~\bibnamefont {Capone}},
  \bibinfo {author} {\bibfnamefont {A.}~\bibnamefont {Georges}}, \bibinfo
  {author} {\bibfnamefont {K.}~\bibnamefont {Haule}}, \bibinfo {author}
  {\bibfnamefont {O.}~\bibnamefont {Parcollet}}, \bibinfo {author}
  {\bibfnamefont {T.}~\bibnamefont {Stanescu}}, \ and\ \bibinfo {author}
  {\bibfnamefont {G.}~\bibnamefont {Kotliar}},\ }\href@noop {} {\bibfield
  {journal} {\bibinfo  {journal} {Physical review letters}\ }\textbf {\bibinfo
  {volume} {100}},\ \bibinfo {pages} {046402} (\bibinfo {year}
  {2008})}\BibitemShut {NoStop}%
\bibitem [{\citenamefont {Ferrero}\ \emph {et~al.}(2009)\citenamefont
  {Ferrero}, \citenamefont {Cornaglia}, \citenamefont {De~Leo}, \citenamefont
  {Parcollet}, \citenamefont {Kotliar},\ and\ \citenamefont
  {Georges}}]{ferrero2009pseudogap}%
  \BibitemOpen
  \bibfield  {author} {\bibinfo {author} {\bibfnamefont {M.}~\bibnamefont
  {Ferrero}}, \bibinfo {author} {\bibfnamefont {P.~S.}\ \bibnamefont
  {Cornaglia}}, \bibinfo {author} {\bibfnamefont {L.}~\bibnamefont {De~Leo}},
  \bibinfo {author} {\bibfnamefont {O.}~\bibnamefont {Parcollet}}, \bibinfo
  {author} {\bibfnamefont {G.}~\bibnamefont {Kotliar}}, \ and\ \bibinfo
  {author} {\bibfnamefont {A.}~\bibnamefont {Georges}},\ }\href@noop {}
  {\bibfield  {journal} {\bibinfo  {journal} {Physical Review B}\ }\textbf
  {\bibinfo {volume} {80}},\ \bibinfo {pages} {064501} (\bibinfo {year}
  {2009})}\BibitemShut {NoStop}%
\bibitem [{\citenamefont {Sakai}\ \emph {et~al.}(2010)\citenamefont {Sakai},
  \citenamefont {Motome},\ and\ \citenamefont {Imada}}]{sakai2010doped}%
  \BibitemOpen
  \bibfield  {author} {\bibinfo {author} {\bibfnamefont {S.}~\bibnamefont
  {Sakai}}, \bibinfo {author} {\bibfnamefont {Y.}~\bibnamefont {Motome}}, \
  and\ \bibinfo {author} {\bibfnamefont {M.}~\bibnamefont {Imada}},\
  }\href@noop {} {\bibfield  {journal} {\bibinfo  {journal} {Physical Review
  B}\ }\textbf {\bibinfo {volume} {82}},\ \bibinfo {pages} {134505} (\bibinfo
  {year} {2010})}\BibitemShut {NoStop}%
\bibitem [{\citenamefont {Sakai}\ \emph {et~al.}(2009)\citenamefont {Sakai},
  \citenamefont {Motome},\ and\ \citenamefont
  {Imada}}]{imada2010unconventional}%
  \BibitemOpen
  \bibfield  {author} {\bibinfo {author} {\bibfnamefont {S.}~\bibnamefont
  {Sakai}}, \bibinfo {author} {\bibfnamefont {Y.}~\bibnamefont {Motome}}, \
  and\ \bibinfo {author} {\bibfnamefont {M.}~\bibnamefont {Imada}},\
  }\href@noop {} {\bibfield  {journal} {\bibinfo  {journal} {Phys. Rev. Lett.}\
  }\textbf {\bibinfo {volume} {102}},\ \bibinfo {pages} {056404} (\bibinfo
  {year} {2009})}\BibitemShut {NoStop}%
\bibitem [{\citenamefont {Gull}\ \emph {et~al.}(2010)\citenamefont {Gull},
  \citenamefont {Ferrero}, \citenamefont {Parcollet}, \citenamefont {Georges},\
  and\ \citenamefont {Millis}}]{gull2010momentum}%
  \BibitemOpen
  \bibfield  {author} {\bibinfo {author} {\bibfnamefont {E.}~\bibnamefont
  {Gull}}, \bibinfo {author} {\bibfnamefont {M.}~\bibnamefont {Ferrero}},
  \bibinfo {author} {\bibfnamefont {O.}~\bibnamefont {Parcollet}}, \bibinfo
  {author} {\bibfnamefont {A.}~\bibnamefont {Georges}}, \ and\ \bibinfo
  {author} {\bibfnamefont {A.~J.}\ \bibnamefont {Millis}},\ }\href@noop {}
  {\bibfield  {journal} {\bibinfo  {journal} {Physical Review B}\ }\textbf
  {\bibinfo {volume} {82}},\ \bibinfo {pages} {155101} (\bibinfo {year}
  {2010})}\BibitemShut {NoStop}%
\bibitem [{\citenamefont {Potthoff}\ \emph {et~al.}(2003)\citenamefont
  {Potthoff}, \citenamefont {Aichhorn},\ and\ \citenamefont
  {Dahnken}}]{potthoff2003}%
  \BibitemOpen
  \bibfield  {author} {\bibinfo {author} {\bibfnamefont {M.}~\bibnamefont
  {Potthoff}}, \bibinfo {author} {\bibfnamefont {M.}~\bibnamefont {Aichhorn}},
  \ and\ \bibinfo {author} {\bibfnamefont {C.}~\bibnamefont {Dahnken}},\
  }\href@noop {} {\bibfield  {journal} {\bibinfo  {journal} {Physical review
  letters}\ }\textbf {\bibinfo {volume} {91}},\ \bibinfo {pages} {206402}
  (\bibinfo {year} {2003})}\BibitemShut {NoStop}%
\bibitem [{\citenamefont {Dahnken}\ \emph {et~al.}(2004)\citenamefont
  {Dahnken}, \citenamefont {Aichhorn}, \citenamefont {Hanke}, \citenamefont
  {Arrigoni},\ and\ \citenamefont {Potthoff}}]{dahnken2003}%
  \BibitemOpen
  \bibfield  {author} {\bibinfo {author} {\bibfnamefont {C.}~\bibnamefont
  {Dahnken}}, \bibinfo {author} {\bibfnamefont {M.}~\bibnamefont {Aichhorn}},
  \bibinfo {author} {\bibfnamefont {W.}~\bibnamefont {Hanke}}, \bibinfo
  {author} {\bibfnamefont {E.}~\bibnamefont {Arrigoni}}, \ and\ \bibinfo
  {author} {\bibfnamefont {M.}~\bibnamefont {Potthoff}},\ }\href@noop {}
  {\bibfield  {journal} {\bibinfo  {journal} {Physical Review B}\ }\textbf
  {\bibinfo {volume} {70}},\ \bibinfo {pages} {245110} (\bibinfo {year}
  {2004})}\BibitemShut {NoStop}%
\bibitem [{\citenamefont {Schmitt}(2010)}]{schmitt2010}%
  \BibitemOpen
  \bibfield  {author} {\bibinfo {author} {\bibfnamefont {S.}~\bibnamefont
  {Schmitt}},\ }\href@noop {} {\bibfield  {journal} {\bibinfo  {journal}
  {Physical Review B}\ }\textbf {\bibinfo {volume} {82}},\ \bibinfo {pages}
  {155126} (\bibinfo {year} {2010})}\BibitemShut {NoStop}%
\bibitem [{\citenamefont {Huang}\ \emph {et~al.}(2018)\citenamefont {Huang},
  \citenamefont {Sheppard}, \citenamefont {Moritz},\ and\ \citenamefont
  {Devereaux}}]{huang2018}%
  \BibitemOpen
  \bibfield  {author} {\bibinfo {author} {\bibfnamefont {E.~W.}\ \bibnamefont
  {Huang}}, \bibinfo {author} {\bibfnamefont {R.}~\bibnamefont {Sheppard}},
  \bibinfo {author} {\bibfnamefont {B.}~\bibnamefont {Moritz}}, \ and\ \bibinfo
  {author} {\bibfnamefont {T.~P.}\ \bibnamefont {Devereaux}},\ }\href@noop {}
  {\bibfield  {journal} {\bibinfo  {journal} {arXiv preprint arXiv:1806.08346}\
  } (\bibinfo {year} {2018})}\BibitemShut {NoStop}%
\bibitem [{\citenamefont {Kaczmarczyk}\ \emph {et~al.}(2016)\citenamefont
  {Kaczmarczyk}, \citenamefont {Schickling},\ and\ \citenamefont
  {B{\"u}nemann}}]{kaczmarczyk2016}%
  \BibitemOpen
  \bibfield  {author} {\bibinfo {author} {\bibfnamefont {J.}~\bibnamefont
  {Kaczmarczyk}}, \bibinfo {author} {\bibfnamefont {T.}~\bibnamefont
  {Schickling}}, \ and\ \bibinfo {author} {\bibfnamefont {J.}~\bibnamefont
  {B{\"u}nemann}},\ }\href@noop {} {\bibfield  {journal} {\bibinfo  {journal}
  {Physical Review B}\ }\textbf {\bibinfo {volume} {94}},\ \bibinfo {pages}
  {085152} (\bibinfo {year} {2016})}\BibitemShut {NoStop}%
\bibitem [{\citenamefont {S{\'e}n{\'e}chal}\ \emph {et~al.}(2005)\citenamefont
  {S{\'e}n{\'e}chal}, \citenamefont {Lavertu}, \citenamefont {Marois},\ and\
  \citenamefont {Tremblay}}]{senechal2005}%
  \BibitemOpen
  \bibfield  {author} {\bibinfo {author} {\bibfnamefont {D.}~\bibnamefont
  {S{\'e}n{\'e}chal}}, \bibinfo {author} {\bibfnamefont {P.-L.}\ \bibnamefont
  {Lavertu}}, \bibinfo {author} {\bibfnamefont {M.-A.}\ \bibnamefont {Marois}},
  \ and\ \bibinfo {author} {\bibfnamefont {A.-M.}\ \bibnamefont {Tremblay}},\
  }\href@noop {} {\bibfield  {journal} {\bibinfo  {journal} {Physical review
  letters}\ }\textbf {\bibinfo {volume} {94}},\ \bibinfo {pages} {156404}
  (\bibinfo {year} {2005})}\BibitemShut {NoStop}%
\bibitem [{\citenamefont {Aichhorn}\ \emph {et~al.}(2006)\citenamefont
  {Aichhorn}, \citenamefont {Arrigoni}, \citenamefont {Potthoff},\ and\
  \citenamefont {Hanke}}]{aichhorn2006}%
  \BibitemOpen
  \bibfield  {author} {\bibinfo {author} {\bibfnamefont {M.}~\bibnamefont
  {Aichhorn}}, \bibinfo {author} {\bibfnamefont {E.}~\bibnamefont {Arrigoni}},
  \bibinfo {author} {\bibfnamefont {M.}~\bibnamefont {Potthoff}}, \ and\
  \bibinfo {author} {\bibfnamefont {W.}~\bibnamefont {Hanke}},\ }\href@noop {}
  {\bibfield  {journal} {\bibinfo  {journal} {Physical Review B}\ }\textbf
  {\bibinfo {volume} {74}},\ \bibinfo {pages} {024508} (\bibinfo {year}
  {2006})}\BibitemShut {NoStop}%
\bibitem [{\citenamefont {Halboth}\ and\ \citenamefont
  {Metzner}(2000)}]{halboth2000}%
  \BibitemOpen
  \bibfield  {author} {\bibinfo {author} {\bibfnamefont {C.~J.}\ \bibnamefont
  {Halboth}}\ and\ \bibinfo {author} {\bibfnamefont {W.}~\bibnamefont
  {Metzner}},\ }\href@noop {} {\bibfield  {journal} {\bibinfo  {journal}
  {Physical review letters}\ }\textbf {\bibinfo {volume} {85}},\ \bibinfo
  {pages} {5162} (\bibinfo {year} {2000})}\BibitemShut {NoStop}%
\bibitem [{\citenamefont {Schulz}(1990)}]{schulz1990}%
  \BibitemOpen
  \bibfield  {author} {\bibinfo {author} {\bibfnamefont {H.}~\bibnamefont
  {Schulz}},\ }\href@noop {} {\bibfield  {journal} {\bibinfo  {journal}
  {Physical review letters}\ }\textbf {\bibinfo {volume} {64}},\ \bibinfo
  {pages} {1445} (\bibinfo {year} {1990})}\BibitemShut {NoStop}%
\bibitem [{\citenamefont {White}\ \emph {et~al.}(1989)\citenamefont {White},
  \citenamefont {Scalapino}, \citenamefont {Sugar}, \citenamefont {Loh},
  \citenamefont {Gubernatis},\ and\ \citenamefont {Scalettar}}]{white1989}%
  \BibitemOpen
  \bibfield  {author} {\bibinfo {author} {\bibfnamefont {S.~R.}\ \bibnamefont
  {White}}, \bibinfo {author} {\bibfnamefont {D.~J.}\ \bibnamefont
  {Scalapino}}, \bibinfo {author} {\bibfnamefont {R.~L.}\ \bibnamefont
  {Sugar}}, \bibinfo {author} {\bibfnamefont {E.}~\bibnamefont {Loh}}, \bibinfo
  {author} {\bibfnamefont {J.~E.}\ \bibnamefont {Gubernatis}}, \ and\ \bibinfo
  {author} {\bibfnamefont {R.~T.}\ \bibnamefont {Scalettar}},\ }\href@noop {}
  {\bibfield  {journal} {\bibinfo  {journal} {Physical Review B}\ }\textbf
  {\bibinfo {volume} {40}},\ \bibinfo {pages} {506} (\bibinfo {year}
  {1989})}\BibitemShut {NoStop}%
\bibitem [{\citenamefont {Chubukov}\ and\ \citenamefont
  {Musaelian}(1995)}]{chubukov1995}%
  \BibitemOpen
  \bibfield  {author} {\bibinfo {author} {\bibfnamefont {A.~V.}\ \bibnamefont
  {Chubukov}}\ and\ \bibinfo {author} {\bibfnamefont {K.~A.}\ \bibnamefont
  {Musaelian}},\ }\href@noop {} {\bibfield  {journal} {\bibinfo  {journal}
  {Physical Review B}\ }\textbf {\bibinfo {volume} {51}},\ \bibinfo {pages}
  {12605} (\bibinfo {year} {1995})}\BibitemShut {NoStop}%
\bibitem [{\citenamefont {Capone}\ and\ \citenamefont
  {Kotliar}(2006)}]{capone2006}%
  \BibitemOpen
  \bibfield  {author} {\bibinfo {author} {\bibfnamefont {M.}~\bibnamefont
  {Capone}}\ and\ \bibinfo {author} {\bibfnamefont {G.}~\bibnamefont
  {Kotliar}},\ }\href@noop {} {\bibfield  {journal} {\bibinfo  {journal}
  {Physical Review B}\ }\textbf {\bibinfo {volume} {74}},\ \bibinfo {pages}
  {054513} (\bibinfo {year} {2006})}\BibitemShut {NoStop}%
\bibitem [{\citenamefont {Imada}\ \emph {et~al.}(2013)\citenamefont {Imada},
  \citenamefont {Sakai}, \citenamefont {Yamaji},\ and\ \citenamefont
  {Motome}}]{imada2013}%
  \BibitemOpen
  \bibfield  {author} {\bibinfo {author} {\bibfnamefont {M.}~\bibnamefont
  {Imada}}, \bibinfo {author} {\bibfnamefont {S.}~\bibnamefont {Sakai}},
  \bibinfo {author} {\bibfnamefont {Y.}~\bibnamefont {Yamaji}}, \ and\ \bibinfo
  {author} {\bibfnamefont {Y.}~\bibnamefont {Motome}},\ }in\ \href@noop {}
  {\emph {\bibinfo {booktitle} {Journal of Physics: Conference Series}}},\
  Vol.\ \bibinfo {volume} {449}\ (\bibinfo {organization} {IOP Publishing},\
  \bibinfo {year} {2013})\ p.\ \bibinfo {pages} {012005}\BibitemShut {NoStop}%
\bibitem [{\citenamefont {Lin}\ \emph {et~al.}(2010)\citenamefont {Lin},
  \citenamefont {Gull},\ and\ \citenamefont {Millis}}]{lin2010}%
  \BibitemOpen
  \bibfield  {author} {\bibinfo {author} {\bibfnamefont {N.}~\bibnamefont
  {Lin}}, \bibinfo {author} {\bibfnamefont {E.}~\bibnamefont {Gull}}, \ and\
  \bibinfo {author} {\bibfnamefont {A.~J.}\ \bibnamefont {Millis}},\
  }\href@noop {} {\bibfield  {journal} {\bibinfo  {journal} {Phys. Rev. B}\
  }\textbf {\bibinfo {volume} {82}},\ \bibinfo {pages} {045104} (\bibinfo
  {year} {2010})}\BibitemShut {NoStop}%
\bibitem [{\citenamefont {Wang}\ \emph {et~al.}(2009)\citenamefont {Wang},
  \citenamefont {Zhai},\ and\ \citenamefont {Lee}}]{wang2009}%
  \BibitemOpen
  \bibfield  {author} {\bibinfo {author} {\bibfnamefont {F.}~\bibnamefont
  {Wang}}, \bibinfo {author} {\bibfnamefont {H.}~\bibnamefont {Zhai}}, \ and\
  \bibinfo {author} {\bibfnamefont {D.-H.}\ \bibnamefont {Lee}},\ }\href@noop
  {} {\bibfield  {journal} {\bibinfo  {journal} {EPL (Europhysics Letters)}\
  }\textbf {\bibinfo {volume} {85}},\ \bibinfo {pages} {37005} (\bibinfo {year}
  {2009})}\BibitemShut {NoStop}%
\bibitem [{\citenamefont {Gull}\ and\ \citenamefont {Millis}(2012)}]{gull2012}%
  \BibitemOpen
  \bibfield  {author} {\bibinfo {author} {\bibfnamefont {E.}~\bibnamefont
  {Gull}}\ and\ \bibinfo {author} {\bibfnamefont {A.~J.}\ \bibnamefont
  {Millis}},\ }\href@noop {} {\bibfield  {journal} {\bibinfo  {journal} {Phys.
  Rev. B}\ }\textbf {\bibinfo {volume} {86}},\ \bibinfo {pages} {241106}
  (\bibinfo {year} {2012})}\BibitemShut {NoStop}%
\bibitem [{\citenamefont {Maier}\ \emph {et~al.}(2005)\citenamefont {Maier},
  \citenamefont {Jarrell}, \citenamefont {Schulthess}, \citenamefont {Kent},\
  and\ \citenamefont {White}}]{maier2005}%
  \BibitemOpen
  \bibfield  {author} {\bibinfo {author} {\bibfnamefont {T.~A.}\ \bibnamefont
  {Maier}}, \bibinfo {author} {\bibfnamefont {M.}~\bibnamefont {Jarrell}},
  \bibinfo {author} {\bibfnamefont {T.~C.}\ \bibnamefont {Schulthess}},
  \bibinfo {author} {\bibfnamefont {P.~R.~C.}\ \bibnamefont {Kent}}, \ and\
  \bibinfo {author} {\bibfnamefont {J.~B.}\ \bibnamefont {White}},\ }\href@noop
  {} {\bibfield  {journal} {\bibinfo  {journal} {Phys. Rev. Lett.}\ }\textbf
  {\bibinfo {volume} {95}},\ \bibinfo {pages} {237001} (\bibinfo {year}
  {2005})}\BibitemShut {NoStop}%
\bibitem [{\citenamefont {Keimer}\ \emph {et~al.}(2015)\citenamefont {Keimer},
  \citenamefont {Kivelson}, \citenamefont {Norman}, \citenamefont {Uchida},\
  and\ \citenamefont {Zaanen}}]{keimer2015quantum}%
  \BibitemOpen
  \bibfield  {author} {\bibinfo {author} {\bibfnamefont {B.}~\bibnamefont
  {Keimer}}, \bibinfo {author} {\bibfnamefont {S.~A.}\ \bibnamefont
  {Kivelson}}, \bibinfo {author} {\bibfnamefont {M.~R.}\ \bibnamefont
  {Norman}}, \bibinfo {author} {\bibfnamefont {S.}~\bibnamefont {Uchida}}, \
  and\ \bibinfo {author} {\bibfnamefont {J.}~\bibnamefont {Zaanen}},\
  }\href@noop {} {\bibfield  {journal} {\bibinfo  {journal} {Nature}\ }\textbf
  {\bibinfo {volume} {518}},\ \bibinfo {pages} {179} (\bibinfo {year}
  {2015})}\BibitemShut {NoStop}%
\bibitem [{\citenamefont {Metzner}\ \emph {et~al.}(2012)\citenamefont
  {Metzner}, \citenamefont {Salmhofer}, \citenamefont {Honerkamp},
  \citenamefont {Meden},\ and\ \citenamefont {Sch{\"o}nhammer}}]{metzner2012}%
  \BibitemOpen
  \bibfield  {author} {\bibinfo {author} {\bibfnamefont {W.}~\bibnamefont
  {Metzner}}, \bibinfo {author} {\bibfnamefont {M.}~\bibnamefont {Salmhofer}},
  \bibinfo {author} {\bibfnamefont {C.}~\bibnamefont {Honerkamp}}, \bibinfo
  {author} {\bibfnamefont {V.}~\bibnamefont {Meden}}, \ and\ \bibinfo {author}
  {\bibfnamefont {K.}~\bibnamefont {Sch{\"o}nhammer}},\ }\href@noop {}
  {\bibfield  {journal} {\bibinfo  {journal} {Reviews of Modern Physics}\
  }\textbf {\bibinfo {volume} {84}},\ \bibinfo {pages} {299} (\bibinfo {year}
  {2012})}\BibitemShut {NoStop}%
\bibitem [{\citenamefont {Tagliavini}\ \emph {et~al.}(2019)\citenamefont
  {Tagliavini}, \citenamefont {Hille}, \citenamefont {Kugler}, \citenamefont
  {Andergassen}, \citenamefont {Toschi},\ and\ \citenamefont
  {Honerkamp}}]{tagliavini2019}%
  \BibitemOpen
  \bibfield  {author} {\bibinfo {author} {\bibfnamefont {A.}~\bibnamefont
  {Tagliavini}}, \bibinfo {author} {\bibfnamefont {C.}~\bibnamefont {Hille}},
  \bibinfo {author} {\bibfnamefont {F.}~\bibnamefont {Kugler}}, \bibinfo
  {author} {\bibfnamefont {S.}~\bibnamefont {Andergassen}}, \bibinfo {author}
  {\bibfnamefont {A.}~\bibnamefont {Toschi}}, \ and\ \bibinfo {author}
  {\bibfnamefont {C.}~\bibnamefont {Honerkamp}},\ }\href@noop {} {\bibfield
  {journal} {\bibinfo  {journal} {SciPost Physics}\ }\textbf {\bibinfo {volume}
  {6}},\ \bibinfo {pages} {009} (\bibinfo {year} {2019})}\BibitemShut {NoStop}%
\bibitem [{\citenamefont {Fu}\ and\ \citenamefont {Lee}(2006)}]{fu2006}%
  \BibitemOpen
  \bibfield  {author} {\bibinfo {author} {\bibfnamefont {H.}~\bibnamefont
  {Fu}}\ and\ \bibinfo {author} {\bibfnamefont {D.-H.}\ \bibnamefont {Lee}},\
  }\href@noop {} {\bibfield  {journal} {\bibinfo  {journal} {Physical Review
  B}\ }\textbf {\bibinfo {volume} {74}},\ \bibinfo {pages} {174513} (\bibinfo
  {year} {2006})}\BibitemShut {NoStop}%
\bibitem [{\citenamefont {Vilardi}\ \emph {et~al.}(2019)\citenamefont
  {Vilardi}, \citenamefont {Taranto},\ and\ \citenamefont
  {Metzner}}]{vilardi2019}%
  \BibitemOpen
  \bibfield  {author} {\bibinfo {author} {\bibfnamefont {D.}~\bibnamefont
  {Vilardi}}, \bibinfo {author} {\bibfnamefont {C.}~\bibnamefont {Taranto}}, \
  and\ \bibinfo {author} {\bibfnamefont {W.}~\bibnamefont {Metzner}},\
  }\href@noop {} {\bibfield  {journal} {\bibinfo  {journal} {Physical Review
  B}\ }\textbf {\bibinfo {volume} {99}},\ \bibinfo {pages} {104501} (\bibinfo
  {year} {2019})}\BibitemShut {NoStop}%
\bibitem [{\citenamefont {Lichtenstein}\ and\ \citenamefont
  {Katsnelson}(2000{\natexlab{b}})}]{lichenstein2000}%
  \BibitemOpen
  \bibfield  {author} {\bibinfo {author} {\bibfnamefont {A.}~\bibnamefont
  {Lichtenstein}}\ and\ \bibinfo {author} {\bibfnamefont {M.}~\bibnamefont
  {Katsnelson}},\ }\href@noop {} {\bibfield  {journal} {\bibinfo  {journal}
  {Physical Review B}\ }\textbf {\bibinfo {volume} {62}},\ \bibinfo {pages}
  {R9283} (\bibinfo {year} {2000}{\natexlab{b}})}\BibitemShut {NoStop}%
\bibitem [{\citenamefont {Yamase}\ \emph {et~al.}(2016)\citenamefont {Yamase},
  \citenamefont {Eberlein},\ and\ \citenamefont {Metzner}}]{yamase2016}%
  \BibitemOpen
  \bibfield  {author} {\bibinfo {author} {\bibfnamefont {H.}~\bibnamefont
  {Yamase}}, \bibinfo {author} {\bibfnamefont {A.}~\bibnamefont {Eberlein}}, \
  and\ \bibinfo {author} {\bibfnamefont {W.}~\bibnamefont {Metzner}},\
  }\href@noop {} {\bibfield  {journal} {\bibinfo  {journal} {Physical review
  letters}\ }\textbf {\bibinfo {volume} {116}},\ \bibinfo {pages} {096402}
  (\bibinfo {year} {2016})}\BibitemShut {NoStop}%
\bibitem [{\citenamefont {Husemann}\ and\ \citenamefont
  {Metzner}(2012)}]{husemann2012}%
  \BibitemOpen
  \bibfield  {author} {\bibinfo {author} {\bibfnamefont {C.}~\bibnamefont
  {Husemann}}\ and\ \bibinfo {author} {\bibfnamefont {W.}~\bibnamefont
  {Metzner}},\ }\href@noop {} {\bibfield  {journal} {\bibinfo  {journal}
  {Physical Review B}\ }\textbf {\bibinfo {volume} {86}},\ \bibinfo {pages}
  {085113} (\bibinfo {year} {2012})}\BibitemShut {NoStop}%
\bibitem [{\citenamefont {Zeyher}\ and\ \citenamefont
  {Greco}(2018)}]{zeyher2018}%
  \BibitemOpen
  \bibfield  {author} {\bibinfo {author} {\bibfnamefont {R.}~\bibnamefont
  {Zeyher}}\ and\ \bibinfo {author} {\bibfnamefont {A.}~\bibnamefont {Greco}},\
  }\href@noop {} {\bibfield  {journal} {\bibinfo  {journal} {Physical Review
  B}\ }\textbf {\bibinfo {volume} {98}},\ \bibinfo {pages} {224504} (\bibinfo
  {year} {2018})}\BibitemShut {NoStop}%
\bibitem [{\citenamefont {Vanhala}\ and\ \citenamefont
  {T{\"o}rm{\"a}}(2018)}]{tuomas2018}%
  \BibitemOpen
  \bibfield  {author} {\bibinfo {author} {\bibfnamefont {T.~I.}\ \bibnamefont
  {Vanhala}}\ and\ \bibinfo {author} {\bibfnamefont {P.}~\bibnamefont
  {T{\"o}rm{\"a}}},\ }\href@noop {} {\bibfield  {journal} {\bibinfo  {journal}
  {Physical Review B}\ }\textbf {\bibinfo {volume} {97}},\ \bibinfo {pages}
  {075112} (\bibinfo {year} {2018})}\BibitemShut {NoStop}%
\bibitem [{\citenamefont {Lange}\ \emph {et~al.}(2017)\citenamefont {Lange},
  \citenamefont {Tsyplyatyev},\ and\ \citenamefont {Kopietz}}]{lange2017}%
  \BibitemOpen
  \bibfield  {author} {\bibinfo {author} {\bibfnamefont {P.}~\bibnamefont
  {Lange}}, \bibinfo {author} {\bibfnamefont {O.}~\bibnamefont {Tsyplyatyev}},
  \ and\ \bibinfo {author} {\bibfnamefont {P.}~\bibnamefont {Kopietz}},\
  }\href@noop {} {\bibfield  {journal} {\bibinfo  {journal} {Physical Review
  B}\ }\textbf {\bibinfo {volume} {96}},\ \bibinfo {pages} {064506} (\bibinfo
  {year} {2017})}\BibitemShut {NoStop}%
\bibitem [{\citenamefont {Liu}\ \emph {et~al.}(2017)\citenamefont {Liu},
  \citenamefont {Wang}, \citenamefont {Wang}, \citenamefont {Zhang},\ and\
  \citenamefont {Rice}}]{liu2017}%
  \BibitemOpen
  \bibfield  {author} {\bibinfo {author} {\bibfnamefont {Y.-H.}\ \bibnamefont
  {Liu}}, \bibinfo {author} {\bibfnamefont {W.-S.}\ \bibnamefont {Wang}},
  \bibinfo {author} {\bibfnamefont {Q.-H.}\ \bibnamefont {Wang}}, \bibinfo
  {author} {\bibfnamefont {F.-C.}\ \bibnamefont {Zhang}}, \ and\ \bibinfo
  {author} {\bibfnamefont {T.~M.}\ \bibnamefont {Rice}},\ }\href@noop {}
  {\bibfield  {journal} {\bibinfo  {journal} {Physical Review B}\ }\textbf
  {\bibinfo {volume} {96}},\ \bibinfo {pages} {014522} (\bibinfo {year}
  {2017})}\BibitemShut {NoStop}%
\bibitem [{\citenamefont {Giering}\ and\ \citenamefont
  {Salmhofer}(2012)}]{giering2012}%
  \BibitemOpen
  \bibfield  {author} {\bibinfo {author} {\bibfnamefont {K.-U.}\ \bibnamefont
  {Giering}}\ and\ \bibinfo {author} {\bibfnamefont {M.}~\bibnamefont
  {Salmhofer}},\ }\href@noop {} {\bibfield  {journal} {\bibinfo  {journal}
  {Physical Review B}\ }\textbf {\bibinfo {volume} {86}},\ \bibinfo {pages}
  {245122} (\bibinfo {year} {2012})}\BibitemShut {NoStop}%
\bibitem [{\citenamefont {G{\l}azek}\ and\ \citenamefont
  {Wilson}(1993)}]{glazek1993}%
  \BibitemOpen
  \bibfield  {author} {\bibinfo {author} {\bibfnamefont {S.~D.}\ \bibnamefont
  {G{\l}azek}}\ and\ \bibinfo {author} {\bibfnamefont {K.~G.}\ \bibnamefont
  {Wilson}},\ }\href@noop {} {\bibfield  {journal} {\bibinfo  {journal}
  {Physical Review D}\ }\textbf {\bibinfo {volume} {48}},\ \bibinfo {pages}
  {5863} (\bibinfo {year} {1993})}\BibitemShut {NoStop}%
\bibitem [{\citenamefont {Glazek}\ and\ \citenamefont
  {Wilson}(1994)}]{glazek1994}%
  \BibitemOpen
  \bibfield  {author} {\bibinfo {author} {\bibfnamefont {S.~D.}\ \bibnamefont
  {Glazek}}\ and\ \bibinfo {author} {\bibfnamefont {K.~G.}\ \bibnamefont
  {Wilson}},\ }\href@noop {} {\bibfield  {journal} {\bibinfo  {journal}
  {Physical Review D}\ }\textbf {\bibinfo {volume} {49}},\ \bibinfo {pages}
  {4214} (\bibinfo {year} {1994})}\BibitemShut {NoStop}%
\bibitem [{\citenamefont {Wegner}(1994)}]{wegner1994}%
  \BibitemOpen
  \bibfield  {author} {\bibinfo {author} {\bibfnamefont {F.}~\bibnamefont
  {Wegner}},\ }\href@noop {} {\bibfield  {journal} {\bibinfo  {journal}
  {Annalen der physik}\ }\textbf {\bibinfo {volume} {506}},\ \bibinfo {pages}
  {77} (\bibinfo {year} {1994})}\BibitemShut {NoStop}%
\bibitem [{\citenamefont {Grote}\ \emph {et~al.}(2002)\citenamefont {Grote},
  \citenamefont {K{\"o}rding},\ and\ \citenamefont {Wegner}}]{grote2002}%
  \BibitemOpen
  \bibfield  {author} {\bibinfo {author} {\bibfnamefont {I.}~\bibnamefont
  {Grote}}, \bibinfo {author} {\bibfnamefont {E.}~\bibnamefont {K{\"o}rding}},
  \ and\ \bibinfo {author} {\bibfnamefont {F.}~\bibnamefont {Wegner}},\
  }\href@noop {} {\bibfield  {journal} {\bibinfo  {journal} {Journal of low
  temperature physics}\ }\textbf {\bibinfo {volume} {126}},\ \bibinfo {pages}
  {1385} (\bibinfo {year} {2002})}\BibitemShut {NoStop}%
\bibitem [{\citenamefont {Ma}(1979)}]{ma1979sk}%
  \BibitemOpen
  \bibfield  {author} {\bibinfo {author} {\bibfnamefont {S.}~\bibnamefont
  {Ma}},\ }\href@noop {} {\bibfield  {journal} {\bibinfo  {journal} {Phys. Rev.
  Lett.}\ }\textbf {\bibinfo {volume} {43}},\ \bibinfo {pages} {1434} (\bibinfo
  {year} {1979})}\BibitemShut {NoStop}%
\bibitem [{\citenamefont {Fisher}(1992)}]{fisher1992random}%
  \BibitemOpen
  \bibfield  {author} {\bibinfo {author} {\bibfnamefont {D.~S.}\ \bibnamefont
  {Fisher}},\ }\href@noop {} {\bibfield  {journal} {\bibinfo  {journal}
  {Physical review letters}\ }\textbf {\bibinfo {volume} {69}},\ \bibinfo
  {pages} {534} (\bibinfo {year} {1992})}\BibitemShut {NoStop}%
\bibitem [{\citenamefont {Rademaker}\ and\ \citenamefont
  {Ortuno}(2016)}]{rademaker2016explicit}%
  \BibitemOpen
  \bibfield  {author} {\bibinfo {author} {\bibfnamefont {L.}~\bibnamefont
  {Rademaker}}\ and\ \bibinfo {author} {\bibfnamefont {M.}~\bibnamefont
  {Ortuno}},\ }\href@noop {} {\bibfield  {journal} {\bibinfo  {journal}
  {Physical review letters}\ }\textbf {\bibinfo {volume} {116}},\ \bibinfo
  {pages} {010404} (\bibinfo {year} {2016})}\BibitemShut {NoStop}%
\bibitem [{\citenamefont {You}\ \emph {et~al.}(2016)\citenamefont {You},
  \citenamefont {Qi},\ and\ \citenamefont {Xu}}]{you2016entanglement}%
  \BibitemOpen
  \bibfield  {author} {\bibinfo {author} {\bibfnamefont {Y.-Z.}\ \bibnamefont
  {You}}, \bibinfo {author} {\bibfnamefont {X.-L.}\ \bibnamefont {Qi}}, \ and\
  \bibinfo {author} {\bibfnamefont {C.}~\bibnamefont {Xu}},\ }\href@noop {}
  {\bibfield  {journal} {\bibinfo  {journal} {Physical Review B}\ }\textbf
  {\bibinfo {volume} {93}},\ \bibinfo {pages} {104205} (\bibinfo {year}
  {2016})}\BibitemShut {NoStop}%
\bibitem [{\citenamefont {Pal}\ \emph {et~al.}(2019)\citenamefont {Pal},
  \citenamefont {Mukherjee},\ and\ \citenamefont {Lal}}]{pal2019}%
  \BibitemOpen
  \bibfield  {author} {\bibinfo {author} {\bibfnamefont {S.}~\bibnamefont
  {Pal}}, \bibinfo {author} {\bibfnamefont {A.}~\bibnamefont {Mukherjee}}, \
  and\ \bibinfo {author} {\bibfnamefont {S.}~\bibnamefont {Lal}},\ }\href@noop
  {} {\bibfield  {journal} {\bibinfo  {journal} {New Journal of Physics}\
  }\textbf {\bibinfo {volume} {21}},\ \bibinfo {pages} {023019} (\bibinfo
  {year} {2019})}\BibitemShut {NoStop}%
\bibitem [{\citenamefont {Ehlers}\ \emph {et~al.}(2017)\citenamefont {Ehlers},
  \citenamefont {White},\ and\ \citenamefont {Noack}}]{ehlers2017hybridDMRG}%
  \BibitemOpen
  \bibfield  {author} {\bibinfo {author} {\bibfnamefont {G.}~\bibnamefont
  {Ehlers}}, \bibinfo {author} {\bibfnamefont {S.~R.}\ \bibnamefont {White}}, \
  and\ \bibinfo {author} {\bibfnamefont {R.~M.}\ \bibnamefont {Noack}},\
  }\href@noop {} {\bibfield  {journal} {\bibinfo  {journal} {Phys. Rev. B}\
  }\textbf {\bibinfo {volume} {95}},\ \bibinfo {pages} {125125} (\bibinfo
  {year} {2017})}\BibitemShut {NoStop}%
\bibitem [{\citenamefont {Dagotto}\ \emph {et~al.}(1992)\citenamefont
  {Dagotto}, \citenamefont {Moreo}, \citenamefont {Ortolani}, \citenamefont
  {Poilblanc},\ and\ \citenamefont {Riera}}]{dagotto1992}%
  \BibitemOpen
  \bibfield  {author} {\bibinfo {author} {\bibfnamefont {E.}~\bibnamefont
  {Dagotto}}, \bibinfo {author} {\bibfnamefont {A.}~\bibnamefont {Moreo}},
  \bibinfo {author} {\bibfnamefont {F.}~\bibnamefont {Ortolani}}, \bibinfo
  {author} {\bibnamefont {Poilblanc}}, \ and\ \bibinfo {author} {\bibfnamefont
  {J.}~\bibnamefont {Riera}},\ }\href@noop {} {\bibfield  {journal} {\bibinfo
  {journal} {Physical Review B}\ }\textbf {\bibinfo {volume} {45}},\ \bibinfo
  {pages} {10741} (\bibinfo {year} {1992})}\BibitemShut {NoStop}%
\bibitem [{\citenamefont {Homes}\ \emph {et~al.}(2004)\citenamefont {Homes},
  \citenamefont {Dordevic}, \citenamefont {Strongin}, \citenamefont {Bonn},
  \citenamefont {Liang}, \citenamefont {Hardy}, \citenamefont {Komiya},
  \citenamefont {Ando}, \citenamefont {Yu}, \citenamefont {Kaneko} \emph
  {et~al.}}]{homes2004}%
  \BibitemOpen
  \bibfield  {author} {\bibinfo {author} {\bibfnamefont {C.}~\bibnamefont
  {Homes}}, \bibinfo {author} {\bibfnamefont {S.}~\bibnamefont {Dordevic}},
  \bibinfo {author} {\bibfnamefont {M.}~\bibnamefont {Strongin}}, \bibinfo
  {author} {\bibfnamefont {D.}~\bibnamefont {Bonn}}, \bibinfo {author}
  {\bibfnamefont {R.}~\bibnamefont {Liang}}, \bibinfo {author} {\bibfnamefont
  {W.}~\bibnamefont {Hardy}}, \bibinfo {author} {\bibfnamefont
  {S.}~\bibnamefont {Komiya}}, \bibinfo {author} {\bibfnamefont
  {Y.}~\bibnamefont {Ando}}, \bibinfo {author} {\bibfnamefont {G.}~\bibnamefont
  {Yu}}, \bibinfo {author} {\bibfnamefont {N.}~\bibnamefont {Kaneko}},  \emph
  {et~al.},\ }\href@noop {} {\bibfield  {journal} {\bibinfo  {journal}
  {Nature}\ }\textbf {\bibinfo {volume} {430}},\ \bibinfo {pages} {539}
  (\bibinfo {year} {2004})}\BibitemShut {NoStop}%
\bibitem [{\citenamefont {Legros}\ \emph {et~al.}(2019)\citenamefont {Legros},
  \citenamefont {Benhabib}, \citenamefont {Tabis}, \citenamefont
  {Lalibert{\'e}}, \citenamefont {Dion}, \citenamefont {Lizaire}, \citenamefont
  {Vignolle}, \citenamefont {Vignolles}, \citenamefont {Raffy}, \citenamefont
  {Li} \emph {et~al.}}]{legros2019}%
  \BibitemOpen
  \bibfield  {author} {\bibinfo {author} {\bibfnamefont {A.}~\bibnamefont
  {Legros}}, \bibinfo {author} {\bibfnamefont {S.}~\bibnamefont {Benhabib}},
  \bibinfo {author} {\bibfnamefont {W.}~\bibnamefont {Tabis}}, \bibinfo
  {author} {\bibfnamefont {F.}~\bibnamefont {Lalibert{\'e}}}, \bibinfo {author}
  {\bibfnamefont {M.}~\bibnamefont {Dion}}, \bibinfo {author} {\bibfnamefont
  {M.}~\bibnamefont {Lizaire}}, \bibinfo {author} {\bibfnamefont
  {B.}~\bibnamefont {Vignolle}}, \bibinfo {author} {\bibfnamefont
  {D.}~\bibnamefont {Vignolles}}, \bibinfo {author} {\bibfnamefont
  {H.}~\bibnamefont {Raffy}}, \bibinfo {author} {\bibfnamefont
  {Z.}~\bibnamefont {Li}},  \emph {et~al.},\ }\href@noop {} {\bibfield
  {journal} {\bibinfo  {journal} {Nature Physics}\ }\textbf {\bibinfo {volume}
  {15}},\ \bibinfo {pages} {142} (\bibinfo {year} {2019})}\BibitemShut
  {NoStop}%
\bibitem [{\citenamefont {Varma}\ \emph {et~al.}(1989)\citenamefont {Varma},
  \citenamefont {Littlewood}, \citenamefont {Schmitt-Rink}, \citenamefont
  {Abrahams},\ and\ \citenamefont {Ruckenstein}}]{varma-PhysRevLett.63.1996}%
  \BibitemOpen
  \bibfield  {author} {\bibinfo {author} {\bibfnamefont {C.}~\bibnamefont
  {Varma}}, \bibinfo {author} {\bibfnamefont {P.~B.}\ \bibnamefont
  {Littlewood}}, \bibinfo {author} {\bibfnamefont {S.}~\bibnamefont
  {Schmitt-Rink}}, \bibinfo {author} {\bibfnamefont {E.}~\bibnamefont
  {Abrahams}}, \ and\ \bibinfo {author} {\bibfnamefont {A.}~\bibnamefont
  {Ruckenstein}},\ }\href@noop {} {\bibfield  {journal} {\bibinfo  {journal}
  {Physical Review Letters}\ }\textbf {\bibinfo {volume} {63}},\ \bibinfo
  {pages} {1996} (\bibinfo {year} {1989})}\BibitemShut {NoStop}%
\bibitem [{\citenamefont
  {Anderson}(1987{\natexlab{b}})}]{anderson-science-1987}%
  \BibitemOpen
  \bibfield  {author} {\bibinfo {author} {\bibfnamefont {P.~W.}\ \bibnamefont
  {Anderson}},\ }\href@noop {} {\bibfield  {journal} {\bibinfo  {journal}
  {science}\ }\textbf {\bibinfo {volume} {235}},\ \bibinfo {pages} {1196}
  (\bibinfo {year} {1987}{\natexlab{b}})}\BibitemShut {NoStop}%
\bibitem [{\citenamefont {Mukherjee}\ and\ \citenamefont
  {Lal}(2019)}]{anirbanhubbard2019code}%
  \BibitemOpen
  \bibfield  {author} {\bibinfo {author} {\bibfnamefont {A.}~\bibnamefont
  {Mukherjee}}\ and\ \bibinfo {author} {\bibfnamefont {S.}~\bibnamefont
  {Lal}},\ }\href@noop {} {\  (\bibinfo {year} {2019})},\ \bibinfo {note}
  {codes available at
  "https://github.com/xp7orer/scaling-Mott-Hubbard"}\BibitemShut {NoStop}%
\bibitem [{\citenamefont {Phillips}(2011)}]{phillips2011mottness}%
  \BibitemOpen
  \bibfield  {author} {\bibinfo {author} {\bibfnamefont {P.}~\bibnamefont
  {Phillips}},\ }\href@noop {} {\bibfield  {journal} {\bibinfo  {journal}
  {Philosophical Transactions of the Royal Society A: Mathematical, Physical
  and Engineering Sciences}\ }\textbf {\bibinfo {volume} {369}},\ \bibinfo
  {pages} {1574} (\bibinfo {year} {2011})}\BibitemShut {NoStop}%
\bibitem [{\citenamefont {Zaanen}\ and\ \citenamefont
  {Overbosch}(2011)}]{zaanen2011mottness}%
  \BibitemOpen
  \bibfield  {author} {\bibinfo {author} {\bibfnamefont {J.}~\bibnamefont
  {Zaanen}}\ and\ \bibinfo {author} {\bibfnamefont {B.}~\bibnamefont
  {Overbosch}},\ }\href@noop {} {\bibfield  {journal} {\bibinfo  {journal}
  {Philosophical Transactions of the Royal Society A: Mathematical, Physical
  and Engineering Sciences}\ }\textbf {\bibinfo {volume} {369}},\ \bibinfo
  {pages} {1599} (\bibinfo {year} {2011})}\BibitemShut {NoStop}%
\bibitem [{\citenamefont {Emery}\ and\ \citenamefont
  {Kivelson}(1995)}]{emery1995}%
  \BibitemOpen
  \bibfield  {author} {\bibinfo {author} {\bibfnamefont {V.}~\bibnamefont
  {Emery}}\ and\ \bibinfo {author} {\bibfnamefont {S.}~\bibnamefont
  {Kivelson}},\ }\href@noop {} {\bibfield  {journal} {\bibinfo  {journal}
  {Nature}\ }\textbf {\bibinfo {volume} {374}},\ \bibinfo {pages} {434}
  (\bibinfo {year} {1995})}\BibitemShut {NoStop}%
\bibitem [{\citenamefont {Anderson}(1997)}]{anderson-advphys-1997}%
  \BibitemOpen
  \bibfield  {author} {\bibinfo {author} {\bibfnamefont {P.}~\bibnamefont
  {Anderson}},\ }\href@noop {} {\ \textbf {\bibinfo {volume} {46}},\ \bibinfo
  {pages} {3} (\bibinfo {year} {1997})}\BibitemShut {NoStop}%
\bibitem [{\citenamefont {Van Der~Marel}\ \emph {et~al.}(2003)\citenamefont
  {Van Der~Marel}, \citenamefont {Molegraaf}, \citenamefont {Zaanen},
  \citenamefont {Nussinov}, \citenamefont {Carbone}, \citenamefont
  {Damascelli}, \citenamefont {Eisaki}, \citenamefont {Greven}, \citenamefont
  {Kes},\ and\ \citenamefont {Li}}]{van2003}%
  \BibitemOpen
  \bibfield  {author} {\bibinfo {author} {\bibfnamefont {D.}~\bibnamefont {Van
  Der~Marel}}, \bibinfo {author} {\bibfnamefont {H.}~\bibnamefont {Molegraaf}},
  \bibinfo {author} {\bibfnamefont {J.}~\bibnamefont {Zaanen}}, \bibinfo
  {author} {\bibfnamefont {Z.}~\bibnamefont {Nussinov}}, \bibinfo {author}
  {\bibfnamefont {F.}~\bibnamefont {Carbone}}, \bibinfo {author} {\bibfnamefont
  {A.}~\bibnamefont {Damascelli}}, \bibinfo {author} {\bibfnamefont
  {H.}~\bibnamefont {Eisaki}}, \bibinfo {author} {\bibfnamefont
  {M.}~\bibnamefont {Greven}}, \bibinfo {author} {\bibfnamefont
  {P.}~\bibnamefont {Kes}}, \ and\ \bibinfo {author} {\bibfnamefont
  {M.}~\bibnamefont {Li}},\ }\href@noop {} {\bibfield  {journal} {\bibinfo
  {journal} {Nature}\ }\textbf {\bibinfo {volume} {425}},\ \bibinfo {pages}
  {271} (\bibinfo {year} {2003})}\BibitemShut {NoStop}%
\bibitem [{\citenamefont {Krishnamurthy}\ \emph {et~al.}(1990)\citenamefont
  {Krishnamurthy}, \citenamefont {Jayaprakash}, \citenamefont {Sarker},\ and\
  \citenamefont {Wenzel}}]{krishnamurthy-PhysRevLett.64.950}%
  \BibitemOpen
  \bibfield  {author} {\bibinfo {author} {\bibfnamefont {H.}~\bibnamefont
  {Krishnamurthy}}, \bibinfo {author} {\bibfnamefont {C.}~\bibnamefont
  {Jayaprakash}}, \bibinfo {author} {\bibfnamefont {S.}~\bibnamefont {Sarker}},
  \ and\ \bibinfo {author} {\bibfnamefont {W.}~\bibnamefont {Wenzel}},\
  }\href@noop {} {\bibfield  {journal} {\bibinfo  {journal} {Physical review
  letters}\ }\textbf {\bibinfo {volume} {64}},\ \bibinfo {pages} {950}
  (\bibinfo {year} {1990})}\BibitemShut {NoStop}%
\bibitem [{\citenamefont {Kemeny}\ and\ \citenamefont
  {Caron}(1967)}]{kemeny1967}%
  \BibitemOpen
  \bibfield  {author} {\bibinfo {author} {\bibfnamefont {G.}~\bibnamefont
  {Kemeny}}\ and\ \bibinfo {author} {\bibfnamefont {L.~G.}\ \bibnamefont
  {Caron}},\ }\href@noop {} {\bibfield  {journal} {\bibinfo  {journal}
  {Physical Review}\ }\textbf {\bibinfo {volume} {159}},\ \bibinfo {pages}
  {768} (\bibinfo {year} {1967})}\BibitemShut {NoStop}%
\bibitem [{\citenamefont {Shavitt}\ and\ \citenamefont
  {Redmon}(1980)}]{shavitt1980quasidegenerate}%
  \BibitemOpen
  \bibfield  {author} {\bibinfo {author} {\bibfnamefont {I.}~\bibnamefont
  {Shavitt}}\ and\ \bibinfo {author} {\bibfnamefont {L.~T.}\ \bibnamefont
  {Redmon}},\ }\href@noop {} {\bibfield  {journal} {\bibinfo  {journal} {The
  Journal of Chemical Physics}\ }\textbf {\bibinfo {volume} {73}},\ \bibinfo
  {pages} {5711} (\bibinfo {year} {1980})}\BibitemShut {NoStop}%
\bibitem [{\citenamefont {Suzuki}(1982)}]{suzuki1982construction}%
  \BibitemOpen
  \bibfield  {author} {\bibinfo {author} {\bibfnamefont {K.}~\bibnamefont
  {Suzuki}},\ }\href@noop {} {\bibfield  {journal} {\bibinfo  {journal}
  {Progress of Theoretical Physics}\ }\textbf {\bibinfo {volume} {68}},\
  \bibinfo {pages} {246} (\bibinfo {year} {1982})}\BibitemShut {NoStop}%
\bibitem [{\citenamefont {Savitz}\ and\ \citenamefont
  {Refael}(2017)}]{savitz2017stable}%
  \BibitemOpen
  \bibfield  {author} {\bibinfo {author} {\bibfnamefont {S.}~\bibnamefont
  {Savitz}}\ and\ \bibinfo {author} {\bibfnamefont {G.}~\bibnamefont
  {Refael}},\ }\href@noop {} {\bibfield  {journal} {\bibinfo  {journal}
  {Physical Review B}\ }\textbf {\bibinfo {volume} {96}},\ \bibinfo {pages}
  {115129} (\bibinfo {year} {2017})}\BibitemShut {NoStop}%
\bibitem [{\citenamefont {Monthus}(2016)}]{monthus2016flow}%
  \BibitemOpen
  \bibfield  {author} {\bibinfo {author} {\bibfnamefont {C.}~\bibnamefont
  {Monthus}},\ }\href@noop {} {\bibfield  {journal} {\bibinfo  {journal}
  {Journal of Physics A: Mathematical and Theoretical}\ }\textbf {\bibinfo
  {volume} {49}},\ \bibinfo {pages} {305002} (\bibinfo {year}
  {2016})}\BibitemShut {NoStop}%
\bibitem [{\citenamefont {Sahin}\ \emph {et~al.}(2017)\citenamefont {Sahin},
  \citenamefont {Schmidt},\ and\ \citenamefont
  {Or{\'u}s}}]{sahin2017entanglement}%
  \BibitemOpen
  \bibfield  {author} {\bibinfo {author} {\bibfnamefont {S.}~\bibnamefont
  {Sahin}}, \bibinfo {author} {\bibfnamefont {K.~P.}\ \bibnamefont {Schmidt}},
  \ and\ \bibinfo {author} {\bibfnamefont {R.}~\bibnamefont {Or{\'u}s}},\
  }\href@noop {} {\bibfield  {journal} {\bibinfo  {journal} {EPL (Europhysics
  Letters)}\ }\textbf {\bibinfo {volume} {117}},\ \bibinfo {pages} {20002}
  (\bibinfo {year} {2017})}\BibitemShut {NoStop}%
\bibitem [{\citenamefont {G{\l}azek}\ and\ \citenamefont
  {Wilson}(2004)}]{Glazek2004}%
  \BibitemOpen
  \bibfield  {author} {\bibinfo {author} {\bibfnamefont {S.~D.}\ \bibnamefont
  {G{\l}azek}}\ and\ \bibinfo {author} {\bibfnamefont {K.~G.}\ \bibnamefont
  {Wilson}},\ }\href@noop {} {\bibfield  {journal} {\bibinfo  {journal}
  {Physical Review B}\ }\textbf {\bibinfo {volume} {69}},\ \bibinfo {pages}
  {094304} (\bibinfo {year} {2004})}\BibitemShut {NoStop}%
\bibitem [{\citenamefont {Haag}\ \emph {et~al.}(1967)\citenamefont {Haag},
  \citenamefont {Hugenholtz},\ and\ \citenamefont {Winnink}}]{haag1967}%
  \BibitemOpen
  \bibfield  {author} {\bibinfo {author} {\bibfnamefont {R.}~\bibnamefont
  {Haag}}, \bibinfo {author} {\bibfnamefont {N.~M.}\ \bibnamefont
  {Hugenholtz}}, \ and\ \bibinfo {author} {\bibfnamefont {M.}~\bibnamefont
  {Winnink}},\ }\href@noop {} {\ \textbf {\bibinfo {volume} {5}},\ \bibinfo
  {pages} {215} (\bibinfo {year} {1967})}\BibitemShut {NoStop}%
\bibitem [{\citenamefont {Anderson}(1958)}]{anderson1958random}%
  \BibitemOpen
  \bibfield  {author} {\bibinfo {author} {\bibfnamefont {P.~W.}\ \bibnamefont
  {Anderson}},\ }\href@noop {} {\bibfield  {journal} {\bibinfo  {journal}
  {Physical Review}\ }\textbf {\bibinfo {volume} {112}},\ \bibinfo {pages}
  {1900} (\bibinfo {year} {1958})}\BibitemShut {NoStop}%
\bibitem [{\citenamefont {Anderson}(1970)}]{anderson1970poor}%
  \BibitemOpen
  \bibfield  {author} {\bibinfo {author} {\bibfnamefont {P.}~\bibnamefont
  {Anderson}},\ }\href@noop {} {\bibfield  {journal} {\bibinfo  {journal}
  {Journal of Physics C: Solid State Physics}\ }\textbf {\bibinfo {volume}
  {3}},\ \bibinfo {pages} {2436} (\bibinfo {year} {1970})}\BibitemShut
  {NoStop}%
\bibitem [{\citenamefont {Tallon}\ and\ \citenamefont
  {Loram}(2001)}]{tallon2001doping}%
  \BibitemOpen
  \bibfield  {author} {\bibinfo {author} {\bibfnamefont {J.}~\bibnamefont
  {Tallon}}\ and\ \bibinfo {author} {\bibfnamefont {J.}~\bibnamefont {Loram}},\
  }\href@noop {} {\bibfield  {journal} {\bibinfo  {journal} {Physica C:
  Superconductivity}\ }\textbf {\bibinfo {volume} {349}},\ \bibinfo {pages}
  {53} (\bibinfo {year} {2001})}\BibitemShut {NoStop}%
\bibitem [{\citenamefont {Volovik}(2003)}]{volovik2009universe}%
  \BibitemOpen
  \bibfield  {author} {\bibinfo {author} {\bibfnamefont {G.~E.}\ \bibnamefont
  {Volovik}},\ }\href@noop {} {\emph {\bibinfo {title} {The universe in a
  helium droplet}}},\ Vol.\ \bibinfo {volume} {117}\ (\bibinfo  {publisher}
  {Oxford University Press on Demand},\ \bibinfo {year} {2003})\BibitemShut
  {NoStop}%
\bibitem [{\citenamefont {Imada}\ \emph {et~al.}(2010)\citenamefont {Imada},
  \citenamefont {Misawa},\ and\ \citenamefont
  {Yamaji}}]{imada2010unconventional0}%
  \BibitemOpen
  \bibfield  {author} {\bibinfo {author} {\bibfnamefont {M.}~\bibnamefont
  {Imada}}, \bibinfo {author} {\bibfnamefont {T.}~\bibnamefont {Misawa}}, \
  and\ \bibinfo {author} {\bibfnamefont {Y.}~\bibnamefont {Yamaji}},\
  }\href@noop {} {\bibfield  {journal} {\bibinfo  {journal} {Journal of
  Physics: Condensed Matter}\ }\textbf {\bibinfo {volume} {22}},\ \bibinfo
  {pages} {164206} (\bibinfo {year} {2010})}\BibitemShut {NoStop}%
\bibitem [{\citenamefont {Moukouri}\ and\ \citenamefont
  {Jarrell}(2001{\natexlab{b}})}]{moukouri2001absence}%
  \BibitemOpen
  \bibfield  {author} {\bibinfo {author} {\bibfnamefont {S.}~\bibnamefont
  {Moukouri}}\ and\ \bibinfo {author} {\bibfnamefont {M.}~\bibnamefont
  {Jarrell}},\ }\href@noop {} {\bibfield  {journal} {\bibinfo  {journal}
  {Physical review letters}\ }\textbf {\bibinfo {volume} {87}},\ \bibinfo
  {pages} {167010} (\bibinfo {year} {2001}{\natexlab{b}})}\BibitemShut
  {NoStop}%
\bibitem [{\citenamefont {Rozenberg}\ \emph {et~al.}(1994)\citenamefont
  {Rozenberg}, \citenamefont {Kotliar},\ and\ \citenamefont
  {Zhang}}]{rozenberg1994}%
  \BibitemOpen
  \bibfield  {author} {\bibinfo {author} {\bibfnamefont {M.}~\bibnamefont
  {Rozenberg}}, \bibinfo {author} {\bibfnamefont {G.}~\bibnamefont {Kotliar}},
  \ and\ \bibinfo {author} {\bibfnamefont {X.}~\bibnamefont {Zhang}},\
  }\href@noop {} {\bibfield  {journal} {\bibinfo  {journal} {Physical Review
  B}\ }\textbf {\bibinfo {volume} {49}},\ \bibinfo {pages} {10181} (\bibinfo
  {year} {1994})}\BibitemShut {NoStop}%
\bibitem [{\citenamefont {Georges}\ \emph
  {et~al.}(1996{\natexlab{b}})\citenamefont {Georges}, \citenamefont {Kotliar},
  \citenamefont {Krauth},\ and\ \citenamefont {Rozenberg}}]{georges1996}%
  \BibitemOpen
  \bibfield  {author} {\bibinfo {author} {\bibfnamefont {A.}~\bibnamefont
  {Georges}}, \bibinfo {author} {\bibfnamefont {G.}~\bibnamefont {Kotliar}},
  \bibinfo {author} {\bibfnamefont {W.}~\bibnamefont {Krauth}}, \ and\ \bibinfo
  {author} {\bibfnamefont {M.~J.}\ \bibnamefont {Rozenberg}},\ }\href@noop {}
  {\bibfield  {journal} {\bibinfo  {journal} {Reviews of Modern Physics}\
  }\textbf {\bibinfo {volume} {68}},\ \bibinfo {pages} {13} (\bibinfo {year}
  {1996}{\natexlab{b}})}\BibitemShut {NoStop}%
\bibitem [{\citenamefont {Martin}(1982)}]{martin-PhysRevLett.48.362}%
  \BibitemOpen
  \bibfield  {author} {\bibinfo {author} {\bibfnamefont {R.~M.}\ \bibnamefont
  {Martin}},\ }\href@noop {} {\bibfield  {journal} {\bibinfo  {journal}
  {Physical Review Letters}\ }\textbf {\bibinfo {volume} {48}},\ \bibinfo
  {pages} {362} (\bibinfo {year} {1982})}\BibitemShut {NoStop}%
\bibitem [{\citenamefont {Cooper}(1956)}]{cooper1956}%
  \BibitemOpen
  \bibfield  {author} {\bibinfo {author} {\bibfnamefont {L.~N.}\ \bibnamefont
  {Cooper}},\ }\href@noop {} {\bibfield  {journal} {\bibinfo  {journal}
  {Physical Review}\ }\textbf {\bibinfo {volume} {104}},\ \bibinfo {pages}
  {1189} (\bibinfo {year} {1956})}\BibitemShut {NoStop}%
\bibitem [{\citenamefont {EDMONDS}(1985)}]{edmond1985}%
  \BibitemOpen
  \bibfield  {author} {\bibinfo {author} {\bibfnamefont {A.~R.}\ \bibnamefont
  {EDMONDS}},\ }\enquote {\bibinfo {title} {The coupling of angular momentum
  vectors},}\ in\ \href@noop {} {\emph {\bibinfo {booktitle} {Angular Momentum
  in Quantum Mechanics}}}\ (\bibinfo  {publisher} {Princeton University
  Press},\ \bibinfo {year} {1985})\ pp.\ \bibinfo {pages} {31--52}\BibitemShut
  {NoStop}%
\bibitem [{\citenamefont {Wen}(1990)}]{wen1990}%
  \BibitemOpen
  \bibfield  {author} {\bibinfo {author} {\bibfnamefont {X.-G.}\ \bibnamefont
  {Wen}},\ }\href@noop {} {\bibfield  {journal} {\bibinfo  {journal}
  {International Journal of Modern Physics B}\ }\textbf {\bibinfo {volume}
  {4}},\ \bibinfo {pages} {239} (\bibinfo {year} {1990})}\BibitemShut {NoStop}%
\bibitem [{\citenamefont {Oshikawa}\ and\ \citenamefont
  {Senthil}(2006)}]{oshikawa2006}%
  \BibitemOpen
  \bibfield  {author} {\bibinfo {author} {\bibfnamefont {M.}~\bibnamefont
  {Oshikawa}}\ and\ \bibinfo {author} {\bibfnamefont {T.}~\bibnamefont
  {Senthil}},\ }\href@noop {} {\bibfield  {journal} {\bibinfo  {journal}
  {Physical review letters}\ }\textbf {\bibinfo {volume} {96}},\ \bibinfo
  {pages} {060601} (\bibinfo {year} {2006})}\BibitemShut {NoStop}%
\bibitem [{\citenamefont {Chen}\ \emph
  {et~al.}(2010{\natexlab{a}})\citenamefont {Chen}, \citenamefont {Gu},\ and\
  \citenamefont {Wen}}]{chen2010}%
  \BibitemOpen
  \bibfield  {author} {\bibinfo {author} {\bibfnamefont {X.}~\bibnamefont
  {Chen}}, \bibinfo {author} {\bibfnamefont {Z.-C.}\ \bibnamefont {Gu}}, \ and\
  \bibinfo {author} {\bibfnamefont {X.-G.}\ \bibnamefont {Wen}},\ }\href@noop
  {} {\bibfield  {journal} {\bibinfo  {journal} {Physical review b}\ }\textbf
  {\bibinfo {volume} {82}},\ \bibinfo {pages} {155138} (\bibinfo {year}
  {2010}{\natexlab{a}})}\BibitemShut {NoStop}%
\bibitem [{\citenamefont {Salmhofer}\ \emph {et~al.}(2004)\citenamefont
  {Salmhofer}, \citenamefont {Honerkamp}, \citenamefont {Metzner},\ and\
  \citenamefont {Lauscher}}]{salmhofer2004}%
  \BibitemOpen
  \bibfield  {author} {\bibinfo {author} {\bibfnamefont {M.}~\bibnamefont
  {Salmhofer}}, \bibinfo {author} {\bibfnamefont {C.}~\bibnamefont
  {Honerkamp}}, \bibinfo {author} {\bibfnamefont {W.}~\bibnamefont {Metzner}},
  \ and\ \bibinfo {author} {\bibfnamefont {O.}~\bibnamefont {Lauscher}},\
  }\href@noop {} {\bibfield  {journal} {\bibinfo  {journal} {Progress of
  theoretical physics}\ }\textbf {\bibinfo {volume} {112}},\ \bibinfo {pages}
  {943} (\bibinfo {year} {2004})}\BibitemShut {NoStop}%
\bibitem [{\citenamefont {Fradkin}(2013)}]{fradkin2013field}%
  \BibitemOpen
  \bibfield  {author} {\bibinfo {author} {\bibfnamefont {E.}~\bibnamefont
  {Fradkin}},\ }\href@noop {} {\emph {\bibinfo {title} {Field theories of
  condensed matter physics}}}\ (\bibinfo  {publisher} {Cambridge University
  Press},\ \bibinfo {year} {2013})\BibitemShut {NoStop}%
\bibitem [{\citenamefont {Tocchio}\ \emph {et~al.}(2016)\citenamefont
  {Tocchio}, \citenamefont {Becca},\ and\ \citenamefont
  {Sorella}}]{tocchio2016}%
  \BibitemOpen
  \bibfield  {author} {\bibinfo {author} {\bibfnamefont {L.~F.}\ \bibnamefont
  {Tocchio}}, \bibinfo {author} {\bibfnamefont {F.}~\bibnamefont {Becca}}, \
  and\ \bibinfo {author} {\bibfnamefont {S.}~\bibnamefont {Sorella}},\
  }\href@noop {} {\bibfield  {journal} {\bibinfo  {journal} {Physical Review
  B}\ }\textbf {\bibinfo {volume} {94}},\ \bibinfo {pages} {195126} (\bibinfo
  {year} {2016})}\BibitemShut {NoStop}%
\bibitem [{\citenamefont {Chakraborty}\ \emph {et~al.}(2010)\citenamefont
  {Chakraborty}, \citenamefont {Galanakis},\ and\ \citenamefont
  {Phillips}}]{phillips-PhysRevB.82.214503}%
  \BibitemOpen
  \bibfield  {author} {\bibinfo {author} {\bibfnamefont {S.}~\bibnamefont
  {Chakraborty}}, \bibinfo {author} {\bibfnamefont {D.}~\bibnamefont
  {Galanakis}}, \ and\ \bibinfo {author} {\bibfnamefont {P.}~\bibnamefont
  {Phillips}},\ }\href@noop {} {\bibfield  {journal} {\bibinfo  {journal}
  {Physical Review B}\ }\textbf {\bibinfo {volume} {82}},\ \bibinfo {pages}
  {214503} (\bibinfo {year} {2010})}\BibitemShut {NoStop}%
\bibitem [{\citenamefont {Yang}\ \emph {et~al.}(2011)\citenamefont {Yang},
  \citenamefont {Fotso}, \citenamefont {Su}, \citenamefont {Galanakis},
  \citenamefont {Khatami}, \citenamefont {She}, \citenamefont {Moreno},
  \citenamefont {Zaanen},\ and\ \citenamefont
  {Jarrell}}]{zaanenjarrell-PRL-2011}%
  \BibitemOpen
  \bibfield  {author} {\bibinfo {author} {\bibfnamefont {S.-X.}\ \bibnamefont
  {Yang}}, \bibinfo {author} {\bibfnamefont {H.}~\bibnamefont {Fotso}},
  \bibinfo {author} {\bibfnamefont {S.-Q.}\ \bibnamefont {Su}}, \bibinfo
  {author} {\bibfnamefont {D.}~\bibnamefont {Galanakis}}, \bibinfo {author}
  {\bibfnamefont {E.}~\bibnamefont {Khatami}}, \bibinfo {author} {\bibfnamefont
  {J.-H.}\ \bibnamefont {She}}, \bibinfo {author} {\bibfnamefont
  {J.}~\bibnamefont {Moreno}}, \bibinfo {author} {\bibfnamefont
  {J.}~\bibnamefont {Zaanen}}, \ and\ \bibinfo {author} {\bibfnamefont
  {M.}~\bibnamefont {Jarrell}},\ }\href@noop {} {\bibfield  {journal} {\bibinfo
   {journal} {Physical review letters}\ }\textbf {\bibinfo {volume} {106}},\
  \bibinfo {pages} {047004} (\bibinfo {year} {2011})}\BibitemShut {NoStop}%
\bibitem [{\citenamefont {He}\ \emph {et~al.}(2014)\citenamefont {He},
  \citenamefont {Yin}, \citenamefont {Zech}, \citenamefont {Soumyanarayanan},
  \citenamefont {Yee}, \citenamefont {Williams}, \citenamefont {Boyer},
  \citenamefont {Chatterjee}, \citenamefont {Wise}, \citenamefont {Zeljkovic}
  \emph {et~al.}}]{he2014}%
  \BibitemOpen
  \bibfield  {author} {\bibinfo {author} {\bibfnamefont {Y.}~\bibnamefont
  {He}}, \bibinfo {author} {\bibfnamefont {Y.}~\bibnamefont {Yin}}, \bibinfo
  {author} {\bibfnamefont {M.}~\bibnamefont {Zech}}, \bibinfo {author}
  {\bibfnamefont {A.}~\bibnamefont {Soumyanarayanan}}, \bibinfo {author}
  {\bibfnamefont {M.~M.}\ \bibnamefont {Yee}}, \bibinfo {author} {\bibfnamefont
  {T.}~\bibnamefont {Williams}}, \bibinfo {author} {\bibfnamefont
  {M.}~\bibnamefont {Boyer}}, \bibinfo {author} {\bibfnamefont
  {K.}~\bibnamefont {Chatterjee}}, \bibinfo {author} {\bibfnamefont
  {W.}~\bibnamefont {Wise}}, \bibinfo {author} {\bibfnamefont {I.}~\bibnamefont
  {Zeljkovic}},  \emph {et~al.},\ }\href@noop {} {\bibfield  {journal}
  {\bibinfo  {journal} {Science}\ }\textbf {\bibinfo {volume} {344}},\ \bibinfo
  {pages} {608} (\bibinfo {year} {2014})}\BibitemShut {NoStop}%
\bibitem [{\citenamefont {Fujita}\ \emph {et~al.}(2014)\citenamefont {Fujita},
  \citenamefont {Kim}, \citenamefont {Lee}, \citenamefont {Lee}, \citenamefont
  {Hamidian}, \citenamefont {Firmo}, \citenamefont {Mukhopadhyay},
  \citenamefont {Eisaki}, \citenamefont {Uchida}, \citenamefont {Lawler} \emph
  {et~al.}}]{fujita2014}%
  \BibitemOpen
  \bibfield  {author} {\bibinfo {author} {\bibfnamefont {K.}~\bibnamefont
  {Fujita}}, \bibinfo {author} {\bibfnamefont {C.~K.}\ \bibnamefont {Kim}},
  \bibinfo {author} {\bibfnamefont {I.}~\bibnamefont {Lee}}, \bibinfo {author}
  {\bibfnamefont {J.}~\bibnamefont {Lee}}, \bibinfo {author} {\bibfnamefont
  {M.}~\bibnamefont {Hamidian}}, \bibinfo {author} {\bibfnamefont
  {I.}~\bibnamefont {Firmo}}, \bibinfo {author} {\bibfnamefont
  {S.}~\bibnamefont {Mukhopadhyay}}, \bibinfo {author} {\bibfnamefont
  {H.}~\bibnamefont {Eisaki}}, \bibinfo {author} {\bibfnamefont
  {S.}~\bibnamefont {Uchida}}, \bibinfo {author} {\bibfnamefont
  {M.}~\bibnamefont {Lawler}},  \emph {et~al.},\ }\href@noop {} {\bibfield
  {journal} {\bibinfo  {journal} {Science}\ }\textbf {\bibinfo {volume}
  {344}},\ \bibinfo {pages} {612} (\bibinfo {year} {2014})}\BibitemShut
  {NoStop}%
\bibitem [{\citenamefont {Balakirev}\ \emph {et~al.}(2003)\citenamefont
  {Balakirev}, \citenamefont {Betts}, \citenamefont {Migliori}, \citenamefont
  {Ono}, \citenamefont {Ando},\ and\ \citenamefont
  {Boebinger}}]{balakirev2003}%
  \BibitemOpen
  \bibfield  {author} {\bibinfo {author} {\bibfnamefont {F.~F.}\ \bibnamefont
  {Balakirev}}, \bibinfo {author} {\bibfnamefont {J.~B.}\ \bibnamefont
  {Betts}}, \bibinfo {author} {\bibfnamefont {A.}~\bibnamefont {Migliori}},
  \bibinfo {author} {\bibfnamefont {S.}~\bibnamefont {Ono}}, \bibinfo {author}
  {\bibfnamefont {Y.}~\bibnamefont {Ando}}, \ and\ \bibinfo {author}
  {\bibfnamefont {G.~S.}\ \bibnamefont {Boebinger}},\ }\href@noop {} {\bibfield
   {journal} {\bibinfo  {journal} {Nature}\ }\textbf {\bibinfo {volume}
  {424}},\ \bibinfo {pages} {912} (\bibinfo {year} {2003})}\BibitemShut
  {NoStop}%
\bibitem [{\citenamefont {Wang}\ \emph {et~al.}(2015)\citenamefont {Wang},
  \citenamefont {Liu}, \citenamefont {Imri{\v{s}}ka}, \citenamefont {Ma},\ and\
  \citenamefont {Troyer}}]{wang2015fidelity}%
  \BibitemOpen
  \bibfield  {author} {\bibinfo {author} {\bibfnamefont {L.}~\bibnamefont
  {Wang}}, \bibinfo {author} {\bibfnamefont {Y.-H.}\ \bibnamefont {Liu}},
  \bibinfo {author} {\bibfnamefont {J.}~\bibnamefont {Imri{\v{s}}ka}}, \bibinfo
  {author} {\bibfnamefont {P.~N.}\ \bibnamefont {Ma}}, \ and\ \bibinfo {author}
  {\bibfnamefont {M.}~\bibnamefont {Troyer}},\ }\href@noop {} {\bibfield
  {journal} {\bibinfo  {journal} {Physical Review X}\ }\textbf {\bibinfo
  {volume} {5}},\ \bibinfo {pages} {031007} (\bibinfo {year}
  {2015})}\BibitemShut {NoStop}%
\bibitem [{\citenamefont {You}\ \emph {et~al.}(2007)\citenamefont {You},
  \citenamefont {Li},\ and\ \citenamefont {Gu}}]{you2007fidelity}%
  \BibitemOpen
  \bibfield  {author} {\bibinfo {author} {\bibfnamefont {W.-L.}\ \bibnamefont
  {You}}, \bibinfo {author} {\bibfnamefont {Y.-W.}\ \bibnamefont {Li}}, \ and\
  \bibinfo {author} {\bibfnamefont {S.-J.}\ \bibnamefont {Gu}},\ }\href@noop {}
  {\bibfield  {journal} {\bibinfo  {journal} {Physical Review E}\ }\textbf
  {\bibinfo {volume} {76}},\ \bibinfo {pages} {022101} (\bibinfo {year}
  {2007})}\BibitemShut {NoStop}%
\bibitem [{\citenamefont {Anderson}(1987{\natexlab{c}})}]{anderson1987}%
  \BibitemOpen
  \bibfield  {author} {\bibinfo {author} {\bibfnamefont {P.~W.}\ \bibnamefont
  {Anderson}},\ }\href@noop {} {\bibfield  {journal} {\bibinfo  {journal}
  {science}\ }\textbf {\bibinfo {volume} {235}},\ \bibinfo {pages} {1196}
  (\bibinfo {year} {1987}{\natexlab{c}})}\BibitemShut {NoStop}%
\bibitem [{\citenamefont {Kanigel}\ \emph {et~al.}(2006)\citenamefont
  {Kanigel}, \citenamefont {Norman}, \citenamefont {Randeria}, \citenamefont
  {Chatterjee}, \citenamefont {Souma}, \citenamefont {Kaminski}, \citenamefont
  {Fretwell}, \citenamefont {Rosenkranz}, \citenamefont {Shi}, \citenamefont
  {Sato} \emph {et~al.}}]{kanigel2006evolution}%
  \BibitemOpen
  \bibfield  {author} {\bibinfo {author} {\bibfnamefont {A.}~\bibnamefont
  {Kanigel}}, \bibinfo {author} {\bibfnamefont {M.}~\bibnamefont {Norman}},
  \bibinfo {author} {\bibfnamefont {M.}~\bibnamefont {Randeria}}, \bibinfo
  {author} {\bibfnamefont {U.}~\bibnamefont {Chatterjee}}, \bibinfo {author}
  {\bibfnamefont {S.}~\bibnamefont {Souma}}, \bibinfo {author} {\bibfnamefont
  {A.}~\bibnamefont {Kaminski}}, \bibinfo {author} {\bibfnamefont
  {H.}~\bibnamefont {Fretwell}}, \bibinfo {author} {\bibfnamefont
  {S.}~\bibnamefont {Rosenkranz}}, \bibinfo {author} {\bibfnamefont
  {M.}~\bibnamefont {Shi}}, \bibinfo {author} {\bibfnamefont {T.}~\bibnamefont
  {Sato}},  \emph {et~al.},\ }\href@noop {} {\bibfield  {journal} {\bibinfo
  {journal} {Nature Physics}\ }\textbf {\bibinfo {volume} {2}},\ \bibinfo
  {pages} {447} (\bibinfo {year} {2006})}\BibitemShut {NoStop}%
\bibitem [{\citenamefont {Wang}\ \emph {et~al.}(2006)\citenamefont {Wang},
  \citenamefont {Li},\ and\ \citenamefont {Ong}}]{ong-PhysRevB.73.024510}%
  \BibitemOpen
  \bibfield  {author} {\bibinfo {author} {\bibfnamefont {Y.}~\bibnamefont
  {Wang}}, \bibinfo {author} {\bibfnamefont {L.}~\bibnamefont {Li}}, \ and\
  \bibinfo {author} {\bibfnamefont {N.}~\bibnamefont {Ong}},\ }\href@noop {}
  {\bibfield  {journal} {\bibinfo  {journal} {Physical Review B}\ }\textbf
  {\bibinfo {volume} {73}},\ \bibinfo {pages} {024510} (\bibinfo {year}
  {2006})}\BibitemShut {NoStop}%
\bibitem [{\citenamefont {Pushp}\ \emph {et~al.}(2009)\citenamefont {Pushp},
  \citenamefont {Parker}, \citenamefont {Pasupathy}, \citenamefont {Gomes},
  \citenamefont {Ono}, \citenamefont {Wen}, \citenamefont {Xu}, \citenamefont
  {Gu},\ and\ \citenamefont {Yazdani}}]{pushp2009}%
  \BibitemOpen
  \bibfield  {author} {\bibinfo {author} {\bibfnamefont {A.}~\bibnamefont
  {Pushp}}, \bibinfo {author} {\bibfnamefont {C.~V.}\ \bibnamefont {Parker}},
  \bibinfo {author} {\bibfnamefont {A.~N.}\ \bibnamefont {Pasupathy}}, \bibinfo
  {author} {\bibfnamefont {K.~K.}\ \bibnamefont {Gomes}}, \bibinfo {author}
  {\bibfnamefont {S.}~\bibnamefont {Ono}}, \bibinfo {author} {\bibfnamefont
  {J.}~\bibnamefont {Wen}}, \bibinfo {author} {\bibfnamefont {Z.}~\bibnamefont
  {Xu}}, \bibinfo {author} {\bibfnamefont {G.}~\bibnamefont {Gu}}, \ and\
  \bibinfo {author} {\bibfnamefont {A.}~\bibnamefont {Yazdani}},\ }\href@noop
  {} {\bibfield  {journal} {\bibinfo  {journal} {Science}\ }\textbf {\bibinfo
  {volume} {324}},\ \bibinfo {pages} {1689} (\bibinfo {year}
  {2009})}\BibitemShut {NoStop}%
\bibitem [{\citenamefont {Sebastian}\ \emph {et~al.}(2010)\citenamefont
  {Sebastian}, \citenamefont {Harrison}, \citenamefont {Altarawneh},
  \citenamefont {Mielke}, \citenamefont {Liang}, \citenamefont {Bonn},\ and\
  \citenamefont {Lonzarich}}]{sebastian2010}%
  \BibitemOpen
  \bibfield  {author} {\bibinfo {author} {\bibfnamefont {S.~E.}\ \bibnamefont
  {Sebastian}}, \bibinfo {author} {\bibfnamefont {N.}~\bibnamefont {Harrison}},
  \bibinfo {author} {\bibfnamefont {M.}~\bibnamefont {Altarawneh}}, \bibinfo
  {author} {\bibfnamefont {C.}~\bibnamefont {Mielke}}, \bibinfo {author}
  {\bibfnamefont {R.}~\bibnamefont {Liang}}, \bibinfo {author} {\bibfnamefont
  {D.}~\bibnamefont {Bonn}}, \ and\ \bibinfo {author} {\bibfnamefont
  {G.}~\bibnamefont {Lonzarich}},\ }\href@noop {} {\bibfield  {journal}
  {\bibinfo  {journal} {Proceedings of the National Academy of Sciences}\
  }\textbf {\bibinfo {volume} {107}},\ \bibinfo {pages} {6175} (\bibinfo {year}
  {2010})}\BibitemShut {NoStop}%
\bibitem [{\citenamefont {Presland}\ \emph {et~al.}(1991)\citenamefont
  {Presland}, \citenamefont {Tallon}, \citenamefont {Buckley}, \citenamefont
  {Liu},\ and\ \citenamefont {Flower}}]{presland1991general}%
  \BibitemOpen
  \bibfield  {author} {\bibinfo {author} {\bibfnamefont {M.}~\bibnamefont
  {Presland}}, \bibinfo {author} {\bibfnamefont {J.}~\bibnamefont {Tallon}},
  \bibinfo {author} {\bibfnamefont {R.}~\bibnamefont {Buckley}}, \bibinfo
  {author} {\bibfnamefont {R.}~\bibnamefont {Liu}}, \ and\ \bibinfo {author}
  {\bibfnamefont {N.}~\bibnamefont {Flower}},\ }\href@noop {} {\bibfield
  {journal} {\bibinfo  {journal} {Physica C: Superconductivity}\ }\textbf
  {\bibinfo {volume} {176}},\ \bibinfo {pages} {95} (\bibinfo {year}
  {1991})}\BibitemShut {NoStop}%
\bibitem [{\citenamefont {Ma}\ \emph {et~al.}(2008)\citenamefont {Ma},
  \citenamefont {Pan}, \citenamefont {Niestemski}, \citenamefont {Neupane},
  \citenamefont {Xu}, \citenamefont {Richard}, \citenamefont {Nakayama},
  \citenamefont {Sato}, \citenamefont {Takahashi}, \citenamefont {Luo} \emph
  {et~al.}}]{madhavan-PhysRevLett.101.207002}%
  \BibitemOpen
  \bibfield  {author} {\bibinfo {author} {\bibfnamefont {J.-H.}\ \bibnamefont
  {Ma}}, \bibinfo {author} {\bibfnamefont {Z.-H.}\ \bibnamefont {Pan}},
  \bibinfo {author} {\bibfnamefont {F.}~\bibnamefont {Niestemski}}, \bibinfo
  {author} {\bibfnamefont {M.}~\bibnamefont {Neupane}}, \bibinfo {author}
  {\bibfnamefont {Y.-M.}\ \bibnamefont {Xu}}, \bibinfo {author} {\bibfnamefont
  {P.}~\bibnamefont {Richard}}, \bibinfo {author} {\bibfnamefont
  {K.}~\bibnamefont {Nakayama}}, \bibinfo {author} {\bibfnamefont
  {T.}~\bibnamefont {Sato}}, \bibinfo {author} {\bibfnamefont {T.}~\bibnamefont
  {Takahashi}}, \bibinfo {author} {\bibfnamefont {H.-Q.}\ \bibnamefont {Luo}},
  \emph {et~al.},\ }\href@noop {} {\bibfield  {journal} {\bibinfo  {journal}
  {Physical Review Letters}\ }\textbf {\bibinfo {volume} {101}},\ \bibinfo
  {pages} {207002} (\bibinfo {year} {2008})}\BibitemShut {NoStop}%
\bibitem [{\citenamefont {Fauqu{\'e}}\ \emph {et~al.}(2006)\citenamefont
  {Fauqu{\'e}}, \citenamefont {Sidis}, \citenamefont {Hinkov}, \citenamefont
  {Pailhes}, \citenamefont {Lin}, \citenamefont {Chaud},\ and\ \citenamefont
  {Bourges}}]{fauque2006}%
  \BibitemOpen
  \bibfield  {author} {\bibinfo {author} {\bibfnamefont {B.}~\bibnamefont
  {Fauqu{\'e}}}, \bibinfo {author} {\bibfnamefont {Y.}~\bibnamefont {Sidis}},
  \bibinfo {author} {\bibfnamefont {V.}~\bibnamefont {Hinkov}}, \bibinfo
  {author} {\bibfnamefont {S.}~\bibnamefont {Pailhes}}, \bibinfo {author}
  {\bibfnamefont {C.}~\bibnamefont {Lin}}, \bibinfo {author} {\bibfnamefont
  {X.}~\bibnamefont {Chaud}}, \ and\ \bibinfo {author} {\bibfnamefont
  {P.}~\bibnamefont {Bourges}},\ }\href@noop {} {\bibfield  {journal} {\bibinfo
   {journal} {Physical Review Letters}\ }\textbf {\bibinfo {volume} {96}},\
  \bibinfo {pages} {197001} (\bibinfo {year} {2006})}\BibitemShut {NoStop}%
\bibitem [{\citenamefont {Hinkov}\ \emph {et~al.}(2008)\citenamefont {Hinkov},
  \citenamefont {Haug}, \citenamefont {Fauqu{\'e}}, \citenamefont {Bourges},
  \citenamefont {Sidis}, \citenamefont {Ivanov}, \citenamefont {Bernhard},
  \citenamefont {Lin},\ and\ \citenamefont {Keimer}}]{hinkov2008}%
  \BibitemOpen
  \bibfield  {author} {\bibinfo {author} {\bibfnamefont {V.}~\bibnamefont
  {Hinkov}}, \bibinfo {author} {\bibfnamefont {D.}~\bibnamefont {Haug}},
  \bibinfo {author} {\bibfnamefont {B.}~\bibnamefont {Fauqu{\'e}}}, \bibinfo
  {author} {\bibfnamefont {P.}~\bibnamefont {Bourges}}, \bibinfo {author}
  {\bibfnamefont {Y.}~\bibnamefont {Sidis}}, \bibinfo {author} {\bibfnamefont
  {A.}~\bibnamefont {Ivanov}}, \bibinfo {author} {\bibfnamefont
  {C.}~\bibnamefont {Bernhard}}, \bibinfo {author} {\bibfnamefont
  {C.}~\bibnamefont {Lin}}, \ and\ \bibinfo {author} {\bibfnamefont
  {B.}~\bibnamefont {Keimer}},\ }\href@noop {} {\bibfield  {journal} {\bibinfo
  {journal} {Science}\ }\textbf {\bibinfo {volume} {319}},\ \bibinfo {pages}
  {597} (\bibinfo {year} {2008})}\BibitemShut {NoStop}%
\bibitem [{\citenamefont {Ando}\ \emph {et~al.}(2002)\citenamefont {Ando},
  \citenamefont {Segawa}, \citenamefont {Komiya},\ and\ \citenamefont
  {Lavrov}}]{ando2002}%
  \BibitemOpen
  \bibfield  {author} {\bibinfo {author} {\bibfnamefont {Y.}~\bibnamefont
  {Ando}}, \bibinfo {author} {\bibfnamefont {K.}~\bibnamefont {Segawa}},
  \bibinfo {author} {\bibfnamefont {S.}~\bibnamefont {Komiya}}, \ and\ \bibinfo
  {author} {\bibfnamefont {A.}~\bibnamefont {Lavrov}},\ }\href@noop {}
  {\bibfield  {journal} {\bibinfo  {journal} {Physical review letters}\
  }\textbf {\bibinfo {volume} {88}},\ \bibinfo {pages} {137005} (\bibinfo
  {year} {2002})}\BibitemShut {NoStop}%
\bibitem [{\citenamefont {Daou}\ \emph {et~al.}(2010)\citenamefont {Daou},
  \citenamefont {Chang}, \citenamefont {LeBoeuf}, \citenamefont
  {Cyr-Choiniere}, \citenamefont {Lalibert{\'e}}, \citenamefont
  {Doiron-Leyraud}, \citenamefont {Ramshaw}, \citenamefont {Liang},
  \citenamefont {Bonn}, \citenamefont {Hardy} \emph {et~al.}}]{daou2010}%
  \BibitemOpen
  \bibfield  {author} {\bibinfo {author} {\bibfnamefont {R.}~\bibnamefont
  {Daou}}, \bibinfo {author} {\bibfnamefont {J.}~\bibnamefont {Chang}},
  \bibinfo {author} {\bibfnamefont {D.}~\bibnamefont {LeBoeuf}}, \bibinfo
  {author} {\bibfnamefont {O.}~\bibnamefont {Cyr-Choiniere}}, \bibinfo {author}
  {\bibfnamefont {F.}~\bibnamefont {Lalibert{\'e}}}, \bibinfo {author}
  {\bibfnamefont {N.}~\bibnamefont {Doiron-Leyraud}}, \bibinfo {author}
  {\bibfnamefont {B.}~\bibnamefont {Ramshaw}}, \bibinfo {author} {\bibfnamefont
  {R.}~\bibnamefont {Liang}}, \bibinfo {author} {\bibfnamefont
  {D.}~\bibnamefont {Bonn}}, \bibinfo {author} {\bibfnamefont {W.}~\bibnamefont
  {Hardy}},  \emph {et~al.},\ }\href@noop {} {\bibfield  {journal} {\bibinfo
  {journal} {Nature}\ }\textbf {\bibinfo {volume} {463}},\ \bibinfo {pages}
  {519} (\bibinfo {year} {2010})}\BibitemShut {NoStop}%
\bibitem [{\citenamefont {Hinkov}\ \emph {et~al.}(2007)\citenamefont {Hinkov},
  \citenamefont {Bourges}, \citenamefont {Pailhes}, \citenamefont {Sidis},
  \citenamefont {Ivanov}, \citenamefont {Frost}, \citenamefont {Perring},
  \citenamefont {Lin}, \citenamefont {Chen},\ and\ \citenamefont
  {Keimer}}]{hinkov2007}%
  \BibitemOpen
  \bibfield  {author} {\bibinfo {author} {\bibfnamefont {V.}~\bibnamefont
  {Hinkov}}, \bibinfo {author} {\bibfnamefont {P.}~\bibnamefont {Bourges}},
  \bibinfo {author} {\bibfnamefont {S.}~\bibnamefont {Pailhes}}, \bibinfo
  {author} {\bibfnamefont {Y.}~\bibnamefont {Sidis}}, \bibinfo {author}
  {\bibfnamefont {A.}~\bibnamefont {Ivanov}}, \bibinfo {author} {\bibfnamefont
  {C.}~\bibnamefont {Frost}}, \bibinfo {author} {\bibfnamefont
  {T.}~\bibnamefont {Perring}}, \bibinfo {author} {\bibfnamefont
  {C.}~\bibnamefont {Lin}}, \bibinfo {author} {\bibfnamefont {D.}~\bibnamefont
  {Chen}}, \ and\ \bibinfo {author} {\bibfnamefont {B.}~\bibnamefont
  {Keimer}},\ }\href@noop {} {\bibfield  {journal} {\bibinfo  {journal} {Nature
  Physics}\ }\textbf {\bibinfo {volume} {3}},\ \bibinfo {pages} {780} (\bibinfo
  {year} {2007})}\BibitemShut {NoStop}%
\bibitem [{\citenamefont {Chatterjee}\ \emph {et~al.}(2011)\citenamefont
  {Chatterjee}, \citenamefont {Ai}, \citenamefont {Zhao}, \citenamefont
  {Rosenkranz}, \citenamefont {Kaminski}, \citenamefont {Raffy}, \citenamefont
  {Li}, \citenamefont {Kadowaki}, \citenamefont {Randeria}, \citenamefont
  {Norman} \emph {et~al.}}]{chatterjee2011electronic}%
  \BibitemOpen
  \bibfield  {author} {\bibinfo {author} {\bibfnamefont {U.}~\bibnamefont
  {Chatterjee}}, \bibinfo {author} {\bibfnamefont {D.}~\bibnamefont {Ai}},
  \bibinfo {author} {\bibfnamefont {J.}~\bibnamefont {Zhao}}, \bibinfo {author}
  {\bibfnamefont {S.}~\bibnamefont {Rosenkranz}}, \bibinfo {author}
  {\bibfnamefont {A.}~\bibnamefont {Kaminski}}, \bibinfo {author}
  {\bibfnamefont {H.}~\bibnamefont {Raffy}}, \bibinfo {author} {\bibfnamefont
  {Z.}~\bibnamefont {Li}}, \bibinfo {author} {\bibfnamefont {K.}~\bibnamefont
  {Kadowaki}}, \bibinfo {author} {\bibfnamefont {M.}~\bibnamefont {Randeria}},
  \bibinfo {author} {\bibfnamefont {M.~R.}\ \bibnamefont {Norman}},  \emph
  {et~al.},\ }\href@noop {} {\bibfield  {journal} {\bibinfo  {journal}
  {Proceedings of the National Academy of Sciences}\ }\textbf {\bibinfo
  {volume} {108}},\ \bibinfo {pages} {9346} (\bibinfo {year}
  {2011})}\BibitemShut {NoStop}%
\bibitem [{\citenamefont {Hirayama}\ \emph {et~al.}(2018)\citenamefont
  {Hirayama}, \citenamefont {Yamaji}, \citenamefont {Misawa},\ and\
  \citenamefont {Imada}}]{hirayama2018ab}%
  \BibitemOpen
  \bibfield  {author} {\bibinfo {author} {\bibfnamefont {M.}~\bibnamefont
  {Hirayama}}, \bibinfo {author} {\bibfnamefont {Y.}~\bibnamefont {Yamaji}},
  \bibinfo {author} {\bibfnamefont {T.}~\bibnamefont {Misawa}}, \ and\ \bibinfo
  {author} {\bibfnamefont {M.}~\bibnamefont {Imada}},\ }\href@noop {}
  {\bibfield  {journal} {\bibinfo  {journal} {Physical Review B}\ }\textbf
  {\bibinfo {volume} {98}},\ \bibinfo {pages} {134501} (\bibinfo {year}
  {2018})}\BibitemShut {NoStop}%
\bibitem [{\citenamefont {Timusk}\ and\ \citenamefont
  {Statt}(1999)}]{timusk1999pseudogap}%
  \BibitemOpen
  \bibfield  {author} {\bibinfo {author} {\bibfnamefont {T.}~\bibnamefont
  {Timusk}}\ and\ \bibinfo {author} {\bibfnamefont {B.}~\bibnamefont {Statt}},\
  }\href@noop {} {\bibfield  {journal} {\bibinfo  {journal} {Reports on
  Progress in Physics}\ }\textbf {\bibinfo {volume} {62}},\ \bibinfo {pages}
  {61} (\bibinfo {year} {1999})}\BibitemShut {NoStop}%
\bibitem [{\citenamefont {Zhang}\ and\ \citenamefont
  {Rice}(1988)}]{zhangrice-PhysRevB.37.3759}%
  \BibitemOpen
  \bibfield  {author} {\bibinfo {author} {\bibfnamefont {F.}~\bibnamefont
  {Zhang}}\ and\ \bibinfo {author} {\bibfnamefont {T.}~\bibnamefont {Rice}},\
  }\href@noop {} {\bibfield  {journal} {\bibinfo  {journal} {Physical Review
  B}\ }\textbf {\bibinfo {volume} {37}},\ \bibinfo {pages} {3759} (\bibinfo
  {year} {1988})}\BibitemShut {NoStop}%
\bibitem [{\citenamefont {Pavarini}\ \emph {et~al.}(2001)\citenamefont
  {Pavarini}, \citenamefont {Dasgupta}, \citenamefont {Saha-Dasgupta},
  \citenamefont {Jepsen},\ and\ \citenamefont
  {Andersen}}]{pavarini-PhysRevLett.87.047003}%
  \BibitemOpen
  \bibfield  {author} {\bibinfo {author} {\bibfnamefont {E.}~\bibnamefont
  {Pavarini}}, \bibinfo {author} {\bibfnamefont {I.}~\bibnamefont {Dasgupta}},
  \bibinfo {author} {\bibfnamefont {T.}~\bibnamefont {Saha-Dasgupta}}, \bibinfo
  {author} {\bibfnamefont {O.}~\bibnamefont {Jepsen}}, \ and\ \bibinfo {author}
  {\bibfnamefont {O.}~\bibnamefont {Andersen}},\ }\href@noop {} {\bibfield
  {journal} {\bibinfo  {journal} {Physical review letters}\ }\textbf {\bibinfo
  {volume} {87}},\ \bibinfo {pages} {047003} (\bibinfo {year}
  {2001})}\BibitemShut {NoStop}%
\bibitem [{\citenamefont {Chen}\ \emph
  {et~al.}(2010{\natexlab{b}})\citenamefont {Chen}, \citenamefont {Struzhkin},
  \citenamefont {Yu}, \citenamefont {Goncharov}, \citenamefont {Lin},
  \citenamefont {Mao},\ and\ \citenamefont {Hemley}}]{chen2010enhancement}%
  \BibitemOpen
  \bibfield  {author} {\bibinfo {author} {\bibfnamefont {X.-J.}\ \bibnamefont
  {Chen}}, \bibinfo {author} {\bibfnamefont {V.~V.}\ \bibnamefont {Struzhkin}},
  \bibinfo {author} {\bibfnamefont {Y.}~\bibnamefont {Yu}}, \bibinfo {author}
  {\bibfnamefont {A.~F.}\ \bibnamefont {Goncharov}}, \bibinfo {author}
  {\bibfnamefont {C.-T.}\ \bibnamefont {Lin}}, \bibinfo {author} {\bibfnamefont
  {H.-k.}\ \bibnamefont {Mao}}, \ and\ \bibinfo {author} {\bibfnamefont
  {R.~J.}\ \bibnamefont {Hemley}},\ }\href@noop {} {\bibfield  {journal}
  {\bibinfo  {journal} {Nature}\ }\textbf {\bibinfo {volume} {466}},\ \bibinfo
  {pages} {950} (\bibinfo {year} {2010}{\natexlab{b}})}\BibitemShut {NoStop}%
\bibitem [{\citenamefont {Gao}\ \emph {et~al.}(1994)\citenamefont {Gao},
  \citenamefont {Xue}, \citenamefont {Chen}, \citenamefont {Xiong},
  \citenamefont {Meng}, \citenamefont {Ramirez}, \citenamefont {Chu},
  \citenamefont {Eggert},\ and\ \citenamefont
  {Mao}}]{gao1994superconductivity}%
  \BibitemOpen
  \bibfield  {author} {\bibinfo {author} {\bibfnamefont {L.}~\bibnamefont
  {Gao}}, \bibinfo {author} {\bibfnamefont {Y.}~\bibnamefont {Xue}}, \bibinfo
  {author} {\bibfnamefont {F.}~\bibnamefont {Chen}}, \bibinfo {author}
  {\bibfnamefont {Q.}~\bibnamefont {Xiong}}, \bibinfo {author} {\bibfnamefont
  {R.}~\bibnamefont {Meng}}, \bibinfo {author} {\bibfnamefont {D.}~\bibnamefont
  {Ramirez}}, \bibinfo {author} {\bibfnamefont {C.}~\bibnamefont {Chu}},
  \bibinfo {author} {\bibfnamefont {J.}~\bibnamefont {Eggert}}, \ and\ \bibinfo
  {author} {\bibfnamefont {H.}~\bibnamefont {Mao}},\ }\href@noop {} {\bibfield
  {journal} {\bibinfo  {journal} {Physical Review B}\ }\textbf {\bibinfo
  {volume} {50}},\ \bibinfo {pages} {4260} (\bibinfo {year}
  {1994})}\BibitemShut {NoStop}%
\bibitem [{\citenamefont {Muramatsu}\ \emph {et~al.}(2011)\citenamefont
  {Muramatsu}, \citenamefont {Pham},\ and\ \citenamefont
  {Chu}}]{muramatsu2011possible}%
  \BibitemOpen
  \bibfield  {author} {\bibinfo {author} {\bibfnamefont {T.}~\bibnamefont
  {Muramatsu}}, \bibinfo {author} {\bibfnamefont {D.}~\bibnamefont {Pham}}, \
  and\ \bibinfo {author} {\bibfnamefont {C.}~\bibnamefont {Chu}},\ }\href@noop
  {} {\bibfield  {journal} {\bibinfo  {journal} {Applied Physics Letters}\
  }\textbf {\bibinfo {volume} {99}},\ \bibinfo {pages} {052508} (\bibinfo
  {year} {2011})}\BibitemShut {NoStop}%
\bibitem [{\citenamefont {P{\'e}pin}(2008)}]{pepin-PhysRevB.77.245129}%
  \BibitemOpen
  \bibfield  {author} {\bibinfo {author} {\bibfnamefont {C.}~\bibnamefont
  {P{\'e}pin}},\ }\href@noop {} {\bibfield  {journal} {\bibinfo  {journal}
  {Physical Review B}\ }\textbf {\bibinfo {volume} {77}},\ \bibinfo {pages}
  {245129} (\bibinfo {year} {2008})}\BibitemShut {NoStop}%
\end{thebibliography}
\end{document}